\documentclass[a4paper,pdftex,12pt]{scrbook}
\usepackage{a4wide}
\usepackage{listings}
\usepackage{graphicx}
\usepackage{hyperref}
\usepackage{subfigure}
\usepackage{placeins}
\usepackage{palatino}
\usepackage{fancyhdr}
\usepackage{appendix}
\usepackage[Conny]{fncychap}

\ChNameUpperCase
\ChTitleUpperCase
\ChNameVar{\Huge\rm\bfseries}
\ChNumVar{\Huge}
\ChTitleVar{\Huge\rm}

\begin{document}

\makeatletter
\gdef\lst@PrintOtherKeyword#1\@empty{%
    \lst@XPrintToken
    \begingroup
      \lst@modetrue \lsthk@TextStyle
      \let\lst@ProcessDigit\lst@ProcessLetter
      \let\lst@ProcessOther\lst@ProcessLetter
      \lst@lettertrue
      #1%
      \lst@SaveToken
    \endgroup
        \lst@RestoreToken
        \global\let\lst@savedcurrstyle\lst@currstyle
        \let\lst@currstyle\lst@gkeywords@sty
    \lst@Output
        \let\lst@currstyle\lst@savedcurrstyle}
\makeatother

\lstdefinelanguage{XOCL}{
        morekeywords={context,self,throw,try,catch,let,then,
                      if,else,elseif,and,or,implies,
                      parserImport,in,end,import,not,true,false,when,do},
        otherkeywords={@,|>,<|,<,>,|,::=,{,},[|,|],[,],;,:,
                       .,*,-,+,->,=,<>,>=,<=},
        sensitive=false,
        comment=[l]//,
        morecomment=[s]{/*}{*/},
        basicstyle=\ttfamily\small,
        stringstyle=\ttfamily\small,
        frame=lines,
        aboveskip=20pt,
        belowskip=20pt,
        breaklines=true,
        showspaces=false,
        showtabs=false
}

\lstset{language=XOCL}

\frontmatter


\begin{titlepage}
 
\begin{center}

 
\textsc{}\\[2cm] 
 
\textsc{\LARGE \textsf{Super-Languages}}\\[0.5cm]
 
\textsc{\Large \textsf{Developing Languages and Applications with XMF}}\\[1.0cm]
\textsc{\Large \textsf{Second Edition}}\\[2cm]

\includegraphics[width=0.3\textwidth]{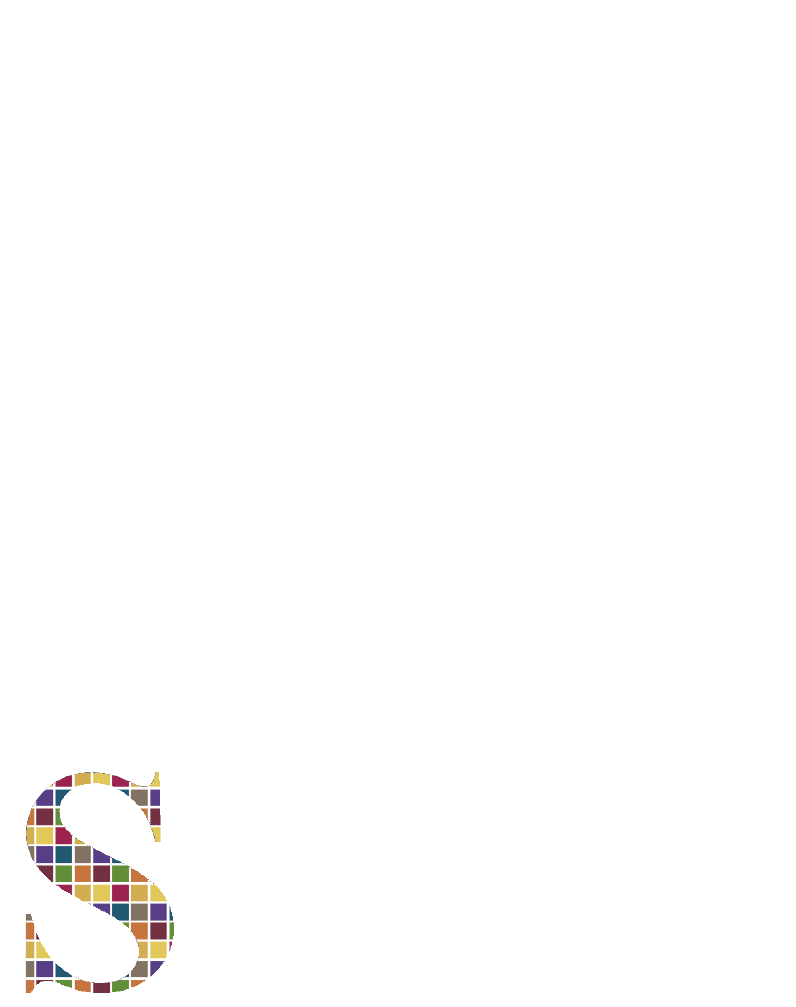}\\[2cm]
 
\large
\textsf{Tony Clark, Paul Sammut, James Willans}\\[0.7cm]
 
\vfill
 
 
\end{center}
 
\end{titlepage}

\tableofcontents

\mainmatter

\chapter{Introduction}

The days of monolithic languages are over. The idea that one size
fits all is dead. Long live diversity! 

Applications should not, and indeed cannot, be written in a single
language these days. We have lots of technologies that are specific
to different aspects of complete application. We have SQL for databases,
we have JSP for web presentation, we have XML everywhere, we have
Java and J2EE, we have frameworks and libraries. So everything is
great - just map you application to an appropriate technology and
away you go! \normalsize

\begin{quotation}
{\em But wait a minute, perhaps this diversity is not such a good
thing. Representing our application in these technologies can be quite
hard. I understand how the application breaks down into aspects, but
those aspects don't seem to map down onto different aspects of the
technology very well. OK, so, I'll put a bit of each application feature
into the appropriate technologies. It takes a bit of translation,
but our guys are experts in the technologies so there should be no
problem. Great - the system works, everyone is happy.

Uh, oh - the requirements have changed. No matter - we'll just
talk it through with the customer and modify the implementation accordingly.
It will take a bit of time to trace the requirements through to the
implementation on all those different bits of technology, er how did
we implement that customer transaction again? Never mind, we can put
that other project on hold while the original developers are brought
back to reverse engineer the code.

Yikes! the original developers have left. Just great, now I
have to pick through all those technologies and try to do it myself.
Crikey - I don't really understand this stuff, it talks about accrual
periods and debentures. I know, I'll show it to the customer, surely
they can extract some meaning from this mess. Aargh - the customer
won't touch it and the new development team just wants to rewrite
it!}
\end{quotation}

\normalsize

\section{The Representation Chasm}

We are drowning in implementation technology. Software language engineers
have been very busy over the last 10 years inventing languages that
control the new frontiers of technology. In particular there are a
huge number of languages that provide control over distributed applications,
databases and web-content. Basic technology issues are very well provided
for. This can be seen in the rise of open-source technologies and
the standardisation of architectures for information processing applications. 

And yet there are still problems. Little attention has been paid to
closing the gap between the languages and the applications that they
implement: the \textbf{representation chasm}. The chasm is bridged
by highly skilled developers who take designs and produce the application.
But what if your developers are not as highly skilled as you would
like? This leads to errors and delays arising from misunderstandings.
Iterative development is compromised by the representation chasm since
various project stakeholders become disenfranchised depending on which
side of the chasm they live. Maintenance become difficult: the customers,
architects and designers shout new requirements from one side of the
chasm only for the development team to mis-hear them on the other.

How do we bridge the representation chasm? The answer is in moving
towards a new breed of language technology: \textbf{super-languages}.
A super-language provides a number of technical features that are
specifically designed to close the gap between the concepts in the
application domain and the technologies that are used to implement
the application. Super-languages can be used to empower development
teams by increasing the sophistication of the tools they use to implement
the application. 

Super-languages can be used to de-skill various aspects of the implementation
by offering restricted languages suitable for the application. Super-languages
can be made accessible to project stakeholders that would usually
be unable to engage with implementation technologies. Super-languages
can radically increase the efficiency of the development team and
can radically increase the quality of the delivered products, derisk
the development process, and increase the ability of the supplier
to maintain the products.

\section{Super-Languages}

What is a super-language and how does it differ from the current breed
of implementation technologies? An idealised super-language provides
control over all aspects of representation and execution. A super-language
can be extended with new features that make it easy to represent concepts
from the application that a customer would understand. These new concepts
can be weaved into the existing features of a super-language or may
be kept separate (as a different aspect for example). Existing execution
mechanisms in the super-language can be changed to reflect the needs
of each new application. Each new feature that is added to a super-language
has a description of how it should execute and how it integrates with
other features (where that is appropriate). 

A super-language must provide super-features that address: \emph{usability;
expressiveness; extensibility}. These are described in more detail
in the following sections.

\subsection{Usability}

A super-language must be fit for purpose. It should support the developer
in capturing features of the application and place no unnecessary
barriers in the way of development. The following features contribute
to the usability of a super-language:

\begin{description}
\item [{interactive}] A super-language cannot be a batch execution language.
Some form of top-level command interpreter is essential. Developers
want to interact with their programs and to deny them this pleasure
is cruel. Interactivity aids development and debugging, increases
developer productivity, and facilitates a deep understanding of execution
mechanisms. If a super-language is used for language engineering then
the developer will be implementing execution mechanisms which will
require a good understanding of how languages tick.
\item [{dynamic}] A super-language must be dynamic. It must be possible
to load new definitions into the language at any stage during execution.
Why wait to recompile and restart an application when it is not necessary?
A super-language may be used as the glue that binds an application
made up of many different modules written in a variety of languages.
In order to be glue, the super-language must provide mechanisms for
dynamically loading and processing the modules. This is closely related
to the idea of aspects.
\item [{reflection}] A super-language must be able to reflect on its own
structure and behaviour. It should be possible for the language to
query its own state and to represent and manipulate its own syntax.
Many of the domain specific languages that are represented by super-languages
will ultimately be represented in a basic core language (an alternative
is to be translated to some external language). Syntax translations
rely on having a complete representation of the super-language syntax
available to the language engineer.
\item [{interfaces}] Super-languages must not be closed worlds. It is unrealistic
to expect mainstream programming languages to become super-languages
overnight. In addition there are practical reasons why certain aspects
of a system may need to be implemented using conventional technology.
Therefore, until standard languages become super-languages, a super-language
should support interfaces to conventional technology. Super-languages
should support provision for running as embedded applications.
\end{description}

\subsection{Expressiveness}

Where possible, a super-language should abstract away from implementation
details in programs. It should be possible to quickly and efficiently
use a super-language for prototyping an application in addition to
product development. Key features are:

\begin{description}
\item [{high-level}] There is no excuse for a language which does not provide
features that make representation and computation easy to achieve.
These include things like heterogeneous lists; closures; pattern matching;
symbolic data; rich type systems etc. High-level features make a language
a pleasure to use and contribute to the readability of the code. Closures
are particularly important since they can be used to engineer applications
that would be complex and difficult to understand without them (particularly
search related applications).
\item [{dynamic(duck)typing}] Languages that are limited to static type
systems are too restrictive to be super-languages. It is difficult
to achieve interaction and meta-features with exclusively statically
typed languages. Trust the developer - they really do know what they
are doing! The type-system of a super-language should be extensible
just like all other aspects of the language. A super-language may
optionally impose static typing (static typing is a good thing), but
should not mandate it.
\item [{garbage-collection}] The days of direct pointer manipulation and
user defined memory management are surely behind us. All but the very
specialized real-time or hardware related applications cannot justify
the exposure of raw machine-level details in this way. 
\end{description}

\subsection{Extensibility}

A super-language must be extensible. In principle, all aspects of
the language from the syntax to the execution engine should be available
to the developer. Extensibility features must be fullsome and integrated
with the rest ofthe language. It \textbf{does not} count if you provide
a couple of system calls that changes the mode of the execution engine.
It \textbf{does} count if you can write a complete new execution algorithm
and plug it in. It \textbf{does not} count if you write a pre-processor
for your language. It \textbf{does} count if you can write a source
code translator and integrate it into the language compiler using
all of the context that the compiler manipulates. Key features are:

\begin{description}
\item [{aspects}] A super-language must support the idea of aspects whereby
new definitions can be added to existing language components. Ideally
the super-language will be written in itself and must be extensible
so that it can be tailored through extension to suit different application
domains. Groups of definitions added to existing components are grouped
into logically related packages or \emph{aspects}. It should be possible
to manipulate aspects as a unit, minimally in terms of addition, deletion
and debugging.
\item [{reflexive}] A super-language should be reflexive. This is related
to reflection, but differs in that the language \emph{can affect}
its own behaviour. Debuggers for languages rely on the ability to
be reflexive since they have to change the trace state of the operations
that they are tracing. An ideal super-language will be meta-circular
since this is the best way of ensuring (given the characteristic features
listed here) that the language will be sufficiently expressive.
\item [{meta-protocols}] Execution engines for languages are like software
frameworks. To plug into the framework you supply program code written
in a language that the framework understands. The framework executes
and processes the program code. The key steps in execution forms an
execution procotol for the framework. The execution protocol for a
super-language must be clearly defined. A \emph{meta-protocol }is
a collection of features that can be used to extend or redefine the
execution protocol. A super-language should offer a meta-protocol
that allows new languages to hook into the existing execution framework,
but which allows them to change it. A typical excample is the protocol
for an object-oriented language that includes: send; setSlot; getSlot
etc. A \emph{meta-object-protocol} (or MOP) allows you to replace
the definition of these features for specific types of object. 
\item [{lightweight-extensible-syntax}] The syntax of a super-language
must be extensible and ultimately replaceable. In order to be a power-tool
for developers, the super-language must allow new language features
to be added and weaved into the existing language. New language features
must be able to take full advantage of the execution framework for
the super-language including accesing context information such as
local variables etc. Super-languages should support the definition
of new language features that support standard programming idioms
and also support the publication of such definitions for others to
reuse.
\item [{heavyweight-extensible-syntax}] A super-language should provide
features that support the definition of complete external languages.
Such language should support a development method that allows different
aspects of a system to be represented by DSLs that are accessible
to external stakeholders and which may also deskill the development
process. These languages are not necessarily weaved into the main
super-language, instead they are used externally to the main language
and loaded on demand.
\end{description}

\section{The Pros and Cons of Super-Languages}

The benefits of super-languages have been outlined above. They empower
the developer to be more productive by raising the abstraction level
of the programming language and inventing standard representations
for common idioms. This technique has been used in many programming
languages from the C-preprocessor through to Lisp. New language constructs
can be standardized and shared.

Super-languages also provide an unprecedented degree of control over
the language engine via meta-features. Access to meta-features provides
a way of tailoring a language in a modular way without polluting programs.
This separation of concerns is very powerful and can be used to support
a mixed paradigm application on the same language platform without
compromising the readability or maintainability of the application.

If used to implement multiple heavyweight or external languages within
the same application, super-languages provide the basis for addressing
some of the significant concerns raised at the beginning of the chapter
relating to current implementation technology and the representation
chasm. This can lead to significant commercial benefits arising from
the ability to engage more stakeholders will all aspects of an application
and the ability to control and deskill aspects of the application
development.

Super-languages are not without their problems. Since they offer advanced
features they can be used inappropriately and can be challenging for
the novice programmer. The techniques used to implement super-languages
are not as mature as those for more conventional languages and therefore
efficiency and useability issues can be an issue. It is important
to put these issues in perspective. Super-languages represent a technology
goal for current generation programming languages to aim for. The
software industry must embrace super-languages and help to develop
the technologies and methods since they represent a very significant
step forward in software application development. Existing mainstream
languages should strive to adopt super-features and thereby move forward.

\section{Related Concepts}

The concept of super-languages is related to several other terms used
by the software industry. These are:

\begin{description}
\item [{LOP}] Language Oriented Programming is a term that is used to mean
the process of defining new programming concepts in order to address
a given application. A super-language supports LOP although not all
systems that support LOP need be super-languages. It is not necessary
to be a programming language in the first place to support LOP - you
just need to be a system that allows the definition of programming
languages. It is certainly not necessary to be meta-circular to provide
LOP.
\item [{DSL}] Domain Specific Languages is a term that is used to mean
a language that has been specifically designed to represent concepts
from an application domain. A super-language supports DSLs, however
the term DSL can mean languages that do not involve programming in
any way. The term includes graphical languages, such as profiles in
UML, that are used to express designs in terms of concepts and relationships.
\item [{LDD}] Language Driven Development is a method that can be used
to produce software applications. LDD involves the use of DSLs at
various points in the development life-cycle. If LDD is used as part
of the implementation phase of the project then the use of super-languages
is almost essential.
\item [{MDA}] Model Driven Architecture is a term used to mean the generation
of program code from UML models, possibly involving the use of DSLs
or UML profiles. Super-languages are not specifically required for
this process (although they may help in the transformation process
- see QVT).
\item [{Software-Factories}] Software factories is a term used to mean
the application of standard industry practices to software development.
More specifically it has been used to refer to an approach supported
by Visual Studio to develop languages that support the construction
of software components. This is similar in many ways to MDA, but is
not limited to using UML.
\item [{Language-Workbenches}] The term language workbench was proposed
by Martin Fowler in http://martinfowler.com/articles/languageWorkbench.html
to mean the IDE support for language oriented programming. Examples
of language workbenches include Intentional Software, the Meta programming
System from JetBrains and the Visual Studio provision for Software
Factories. A super-language is \emph{not} an IDE. Programming and
development processes that are based on super-languages do not require
IDE support. However, given that super-languages are by nature more
sophisticated than conventional languages, IDE support and language
workbenches are an important tool.
\end{description}

\section{Examples of Super-Languages}

The perfect super-language does not exist. Such a language would be
arbitrarily extensible, as efficient and usable as current languages,
have IDE support for all aspects of the language and would integrate
with anything. The perfect super-language is an ideal that should
motivate current language developers to strive for better features
that can address the representation chasm.

Languages with super-features do exist and have done so for quite
a while. Smalltalk and Lisp offer a significant number of super-features
although both frighten the horses in terms of scary syntax. 

C\# and Java offer a degree of reflection, but offer no support for
language extension and high-level programming features. It would be
very interesting to chart out a course of language evolution for these
languages which incrementally adds super-features.

Languages such as Ruby and Python offer a range of super-features,
including dynamic typing and meta-features, but are generally fixed
in their syntax. It is interesting to note that this style of language
(often referred to as \emph{scripting languages}) are generally popular
with technology lead programmers. This is a good indicator that languages
with super-features are likely to find their way into the mainstream.

\section{The Super-language XMF}

This book is about a language system called XMF. XMF has been designed
as a super-language and offers a full range of super-features. XMF
is not the perfect super-language, but it claims to be the language
that exhibits more super-features than any other language currently
available. XMF has been in development since around 2000 and has gone
through a number of development cycles. Its roots are in the Object
Management Group initiative to define the UML 2.0 standard when an
early version of XMF was developed in order to process fragments of
UML language definition. In case anyone is wondering, this explains
the terms OCL and XOCL that occur in XMF and also explains some of
the syntax used in XMF. The relationship between XMF and UML/OCL was
dropped many moons ago, leaving some of the features that contribute
to being a super-language and dropping all others.

Since 2003, XMF has been actively developed as a commercial product
and used by real paying customers as part of a larger product called
XMF-Mosaic. The Mosaic component was a large library of XMF code (over
100K lines of XOCL) that implemented diagram tools and property editors
generated from XMF meta-descriptions. Mosaic allowed XMF to run inside
Eclipse. At the time, Mosaic was ahead of the game, but since we started
development on Mosaic, systems like EMF/GMF and Software Factories
have caught up and overtaken Mosaic in terms of development effort.
In any case, Mosaic was a sophisticated working example of a fairly
large XMF application and shook most of the bugs out of the system.

The key super-strengths of XMF:

\begin{itemize}
\item it is written in itself and therefore it is highly self consistent.
Since extensibility is part of the language, this makes XMF completely
extensible.
\item it provides extensive support for working with syntax, both concrete
and abstract. XMF is a true LOP engine.
\item it provides very high-level programming features including pattern
matching, undoability, daemons, heterogeneous lists.
\item it implements a MOP and is truly meta-circular. The meta-language
is called XCore which defines all the data types and behaviour for
the rest of the system. There is no fudging of meta-classes as occurs
in some language systems.
\item it is interactive, dynamic and multi-threaded.
\item it can be embedded in Java both calling out to Java and being called
from Java.
\end{itemize}
The weaknesses of XMF as a super-language are:

\begin{itemize}
\item it is not as fast as a mainstream language. For many super-applications
this will not be a problem.
\item error handling is adequate, but no special support is provided for
new languages engineered with XMF.
\item there is no static typing (although this is planned for the next major
release).
\end{itemize}

\section{Guide to Book}

The aim of this book is to introduce you to the super-language XMF.
This will be done by defining the language, providing some examples of
applications that can be written directly in the XOCL language that
comes with XMF, and then by showing how you can use XMF for language
engineering.

The main focus of this book is on language engineering by example.
Although we introduce the XOCL language in general terms, we quickly focus
on the features that make XMF good for writing new languages and then
give lots of examples.

There are many super-features of XMF that we do not cover. These features include:
Java interface; daemons; undo. We don't spend very much time on the
MOP and the meta-interfaces of XMF. 

XMF is open-source technology. You can get a copy of XMF from {\tt www.ceteva.com}.
Since XMF is written in itself, it is one of the best examples of 
how to write XMF programs. In addition to the material in this book
we would encourage you to check out the XMF source code for examples
of how to write languages and how to program in XMF. 

This book is divided into the following parts:

\begin{description}
\item [{XOCL}] The XMF programming system consists of a
language called XOCL that runs on the XMF VM. In order to program
in the XMF system you will need to learn XOCL. This part includes
a tutorial on XOCL and the main data that runs on the XMF VM.
\item [{Programming}] This part of the book describes a number of
applications written in XOCL. None of the applications involve language
engineering, just using XOCL to produce a program that runs on XMF.
\item [{Language-Engineering}] Super-languages allow new language constructs
and even complete languages to be defined. This part of the book provides
a number of examples of how you can use XMF to engineer languages.
\end{description}

\section{References:}

\begin{itemize}
\item http://www.cse.dmu.ac.uk/\textasciitilde{}mward/martin/papers/middle-out-t.pdf
\item http://www.onboard.jetbrains.com/articles/04/10/lop/
\item http://martinfowler.com/articles/languageWorkbench.html
\item http://www.infoq.com/presentations/domain-specific-languages
\item http://www.martinfowler.com/
\end{itemize}

\part{XMF}

\chapter{Languages}

In order for a language to be at all useful it must \textit{mean something.}
Meaning is a difficult thing to grasp since in order to talk about
it at all it is necessary to use language which itself involves a
question of meaning. However, it is possible to set down some broad
ground-rules that can be used to establish a basic framework. 

First of all, a language has a syntax which comes in two flavours:
concrete and abstract. Concrete syntax is how you read or write a
language, it tends to be human-centric. The following is an example
of concrete syntax:

\begin{lstlisting}
x + 1
\end{lstlisting}
Abstract syntax deals with how a language is represented as data;
it tends to be computer-centric. Here is the equivalent example represented
as abstract syntax (think of the Java data structure that is created
rather than the characters making up the program):

\begin{lstlisting}
new BinExp(new Var("x"),PLUS,new Int(1))
\end{lstlisting}
In order to build systems we need a language with a syntax. Rather
than use any old language, this book is about constructing languages
that are a good fit for the required system. In doing so we need control
over the concrete syntax (to make it easy to work with for a human)
and control over the abstract syntax (to make it easy for a computer
to work with it).

The process by which concrete is turned into abstract syntax is called
\textit{parsing}. This book deals with models of languages and provides
a great deal of examples of grammars that specify how to turn concrete
into abstract syntax language representations.

Next, the syntax of a language does not define its meaning any more
than you can judge a book by its cover. In order to have meaning,
a language must have a \textit{semantic domain}. A semantic domain
is a model of the elements that the language denotes. A semantic domain
for natural language (such as English) contains all the real-world
concepts you can think of (such as Elephants and houses) in addition
to all the conceptual things you can think of; in fact, it contains
anything you can think of. 

Languages used for system development are much more controlled in
terms of syntax and semantics than natural languages. The semantic
domains of such languages are defined by data types for elements appropriate
to the systems domain. The particular data types depend on the system
domain, for example a language for expressing telecomms will have a
semantic domain containing networks, routers and devices. Languages
defined for general use have general purpose semantic domains containing
such things as records, integers, strings, events etc.

Finally, the meaning of a language is not defined by its syntax or
its semantics alone; the meaning is the \textit{mapping} that links
the syntax to the semantics. In terms of modelling languages the mapping
is often static, for example linking class definitions to sets of
objects that are the instances of the classes. This of UML-style class
models. What is the meaning of such a model? Since the class diagram
does not specify any behaviour, neither does the semantics. a suitable
semantics might be all the objects that can be constructed whose structure
matches that specified in the class model. For programming languages
the mapping is dynamic since it links programs to execution traces
and all the execution machinery that lives in the traces. 

To understand a language you must be fluent in its syntax, its semantic
domain and the mapping between them.

This book defines the super-language XMF and describes how you use
it to bridge the representation chasm. This book is \emph{not} about
how you analyse the systems themselves in order to model them; there
are plenty of books about that topic. This book aims to provide a
collection of techniques that allow you to raise the abstraction-bar
when addressing system development issues. 

When modelling systems it is desirable to focus on \emph{what} is
being expressed rather than \emph{how} the information is being represented.
This is the essence of a language driven approach: we can use high-level
abstractions to capture the information without worrying about unnecessary
implementation detail. The language that will be used to represent
the information in this book is designed to provide a wide range of
high-level constructs for modelling information. 

\begin{figure} \begin{center}
\includegraphics[width=12cm]{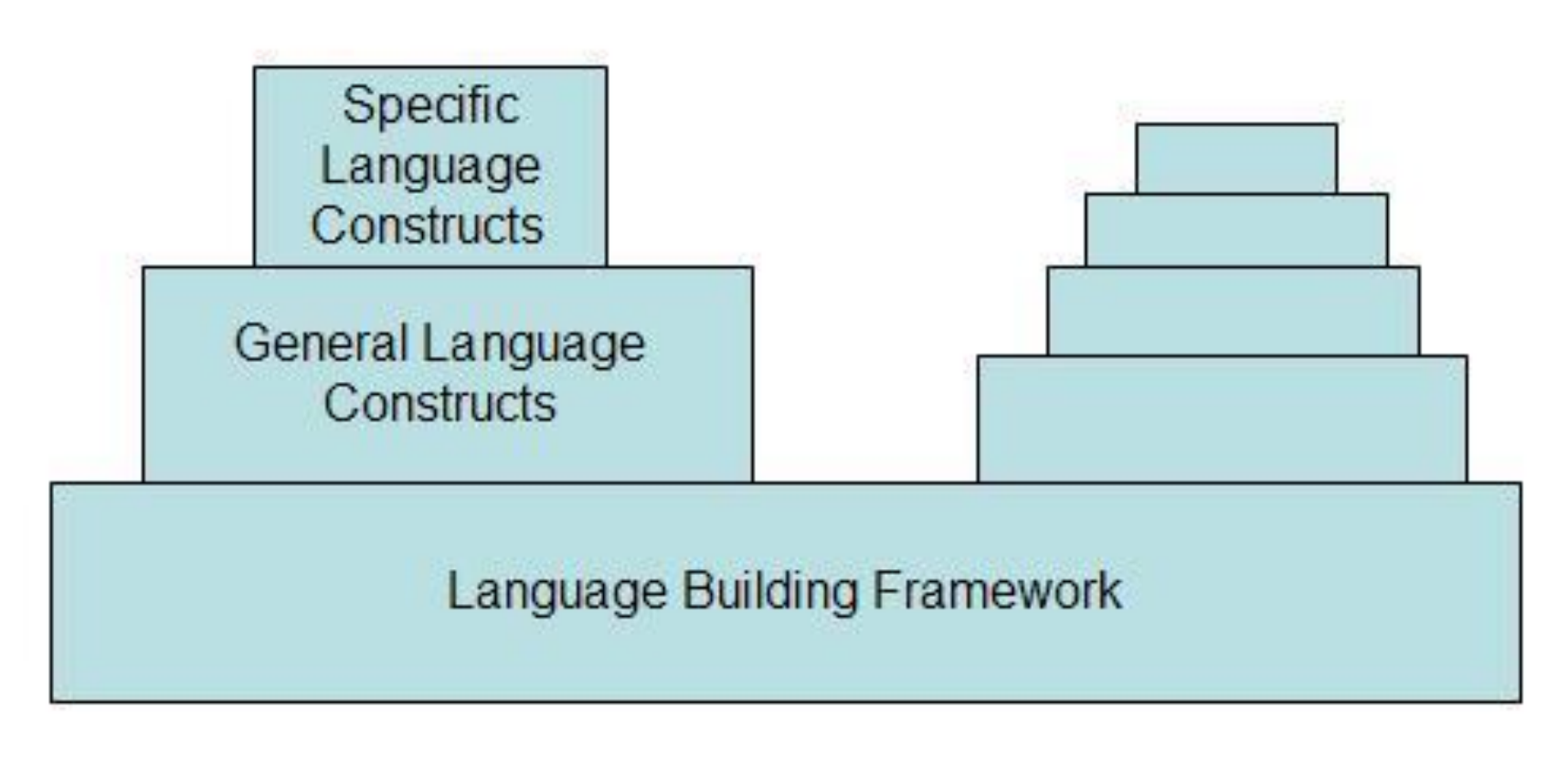}
\caption{Towers of Languages\label{fig:Towers-of-Languages}}
\end{center} \end{figure}

Futhermore, the language has been specifically designed to be extensible:
if no language construct is provided to express the required information
then the user can extend the language with a new construct that does
the job. This is the \emph{ball-of-mud} model first pioneered by the
Lisp family of programming languages. Rather than being a pejorative
term, a language is like a ball of mud when new features can be added
to the language in a way that the new features become part of the
language and are indistinguishable from the existing features. No
special actions need to be taken to use the new extensions: they have
been seamlessly integrated into the language.

Rather than being a ball of mud, it is possible to view the different
languages as towers as shown in figure \ref{fig:Towers-of-Languages}.
The base level is a very general language that supports the construction
of other language (ideally it will support the definition of itself).
A language is grown from the basic framework by adding successive
layers of language constructs; each layer is built from all those
layers below it and the layers become progressively more tailored
to specific application domains. The ball of mud principle is maintained
when the addition of a layer does not preclude the use of constructs
from lower layers and when languages build in different towers can
be used in the same application.

This chapter defines the semantic domain for the language that is
used throughout the rest of the book. The next chapter describes the
syntax of the language XOCL. The chapter after than describes language
building features. On reading the rest of this chapter you should
expect to appreciate the key data values that can be expressed and
the kind of operations that are applied to the values. Each major
category of value has a section to itself.

The element types are divided into groups described in the following
sections. In each case the types that form a coherent aspect of the
semantic domain are grouped together. An overview of the aspect is
given along with a collection of operations that make up an interface
for the group. Note that the interface is very general; it is intended
that a given application will extend this interface by dynamically
adding more operations to the domain types as required. In addition,
the operations describes in each section are an overview. The gold
standard of operations for each type is the source code of XMF itself.

\section{Everything is an Element}

\begin{figure} \begin{center}
\hfill{}\includegraphics[width=8cm]{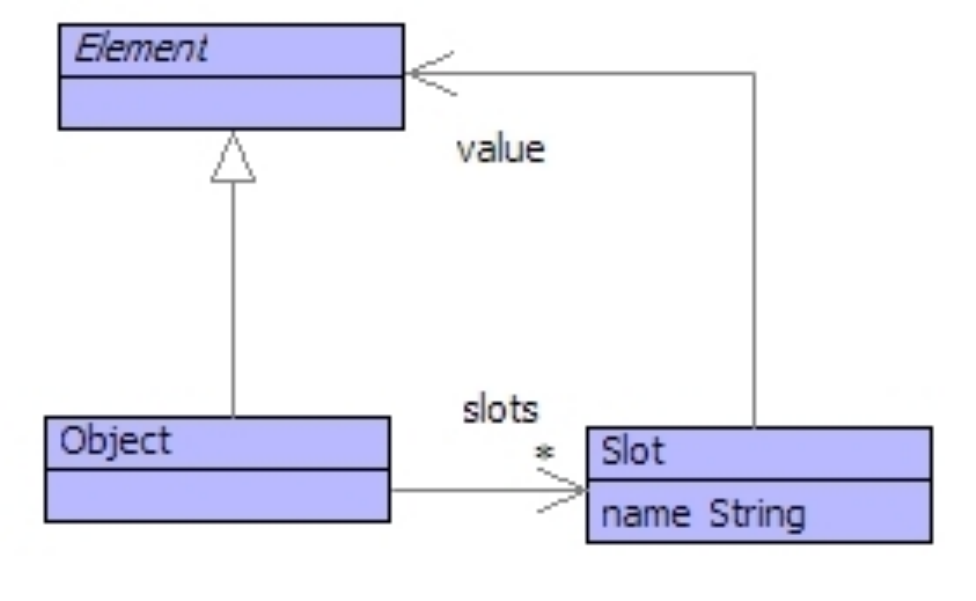}\hfill{}

\caption{Element and Object\label{fig:Element-and-Object}}

\end{center} \end{figure}

Figure \ref{fig:Element-and-Object} shows the root class of the value
domain: everything is an element - some things are objects. When designing
the value domain we had a choice: should everything be an object?
Objects have state represented as slots; a slot has a name and a value.
The slots of an object can be updated. Objects have an identity: when
you create an object it will be different to every other object that
you have ever created (even if the slots are the same).

But not all values have changeable state and an identity. For example:
the integer 3. Does it make sense to talk of the state of 3? Does
it make sense to change the state of 3? Given two occurrences of 3,
are they the same or not?

The distinction between Element and Object allows us to make the distinction
between values whose state cannot change and which have no identity.
Therefore, 3 is an element, but not an object. A set is an element
and not an object, as is a string and a boolean value.

Our language is object-oriented and computation proceeds by message
passing. When an element is sent a message, if its type defines an
operation with the same name as the message then the operation defines
the actions that are performed. If T defines an operation o expecting
arguments (x,y,z) then T::o(x,y,z) is the operation. In the rest of
this chapter each section defines some types; the key operations for
the types are described; where the type of arguments or the return
type of an operation is Element then the type is omitted.

The main operations defined by Element are:

\begin{lstlisting}
Element::copy()
  // Returns a copy of the receiver.
Element::equals(other):Boolean
  // Returns true when the arg is equal
  // to the receiver.
Element::init()
  // Initializes the receiver.
Element::isKindOf(type:Classifier):Boolean
  // Returns true when the receiver is an
  // instance of the supplied type.
Element::of():Classifier
  // Returns the type of the receiver.
Element::toString():String
  // Returns a string representation 
  // of the receiver.
\end{lstlisting}
The main operations defined by Object are:

\begin{lstlisting}
Object::get(name:String)
  // Return the value of the slot or throw
  // an error if the slot does not exist.
Object::hasSlot(name:String):Boolean
  // Returns true when the receiver has a
  // slot with the given name.
Object::set(name:String,value)
  // Updates the given slot or throws an
  // error if the slot does not exist.
\end{lstlisting}
\section{Indexing Things}

\begin{figure} \begin{center}
\hfill{}\includegraphics[width=12cm]{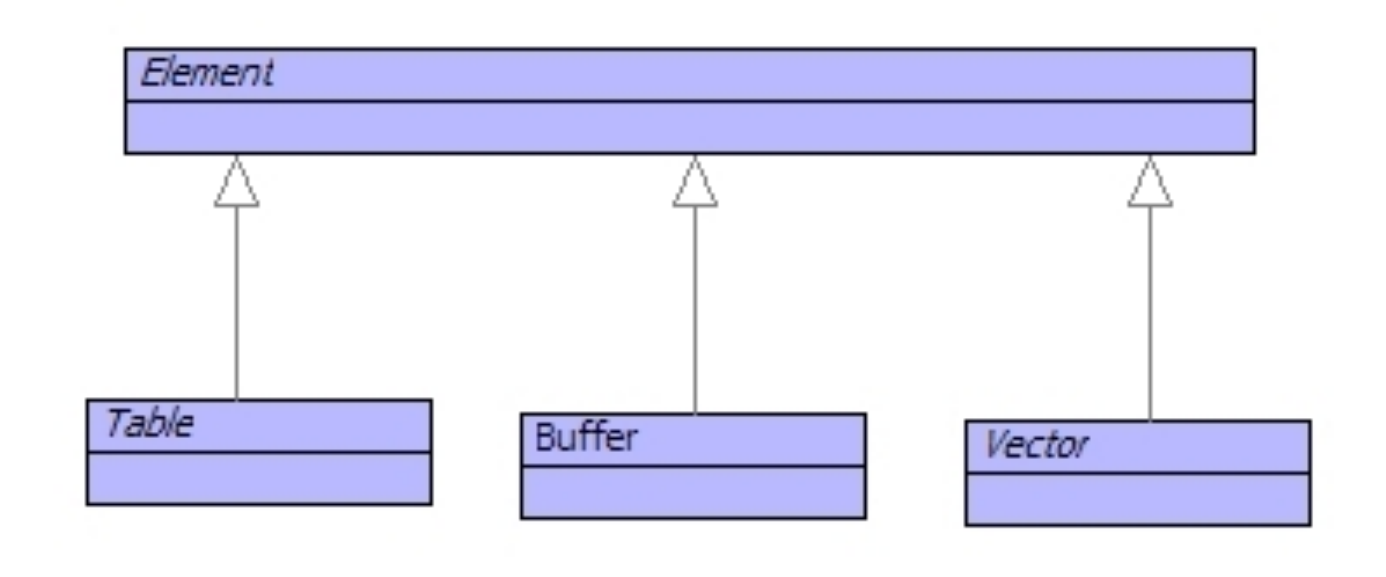}\hfill{}

\caption{Indexed Collections\label{fig:Indexed-Collections}}

\end{center} \end{figure}

When dealing with values it is desirable to build collections where
the elements in the collection are indexed by a key. The key is specified
when the element is placed into the collection and the key is used
to look the element up in the collection. The value associated with
a key in a collection may be changed. 

Figure \ref{fig:Indexed-Collections} shows three classes that are
used to represent basic indexed collections. Notice these classes
extend Element, not Object, therefore these indexed collections have
no slots. A table uses any element as a key, buffers and vectors use
integers as keys. The difference between a buffer and a vector is
that the former grows to accommodate elements as they are added whereas
the latter is a fixed size.

What would you use these for? A table is useful when building arbitrary
associations, for example associating names of personnel with their
employment records. Vectors are useful when you know the size of the
collection in advance, for example the locations on a chess board.
Buffers are useful when you need the collection to grow as required
and want to index elements by their position, for example building
an output string.

The main table operations are as follows:

\begin{lstlisting}
Table(n:Integer):Table
  // Returns a table. The argument indicates the
  // likely max keys.
Table::clear()
  // Empties the receiver.
Table::get(key)
  // Returns the value of the key or raises 
  // an error if the key is not present.
Table::hasKey(key):Boolean
  // Returns true when the key is in the table.
Table::keys():Set(Element)
  // Returns the keys in the table.
Table::put(key,value)
  // Updates the table.
Table::remove(key)
  // Removes the key from the table.
Table::values():Set(Element)
  // Returns the set of values in the table.
\end{lstlisting}
The main buffer operations are as follows:

\begin{lstlisting}
Buffer(n:Integer,isString:Boolean):Buffer
  // Returns a buffer. The first arg is the size to 
  // grow by each time the buffer is extended. The 
  // second indicates whether the buffer is a string 
  // buffer or a general buffer.
Buffer::add(element)
  // Adds the element to the end of the buffer.
Buffer::append(s:String)
  // Appends the chars to the end of a string buffer.
Buffer::asSeq():Seq(Element)
  // returns the buffer as a sequence.
Buffer::at(i:Integer)
  // Returns the element at index i or throws an error
  // if the buffer has no element at position i.
Buffer::size():Integer
  // Returns the number of elements currently in 
  // the buffer.
\end{lstlisting}
The main vector operations are:

\begin{lstlisting}
Vector(n:Integer):Vector
  // Returns a new vector of the specified size.
Vector::asSeq():Seq(Element)
  // Returns the vector as a sequence.
Vector::put(element,index:Integer)
  // Updates the vector at the specified index or
  // throws an error if the index is out of range.
Vector::ref(i:Integer)
  // Returns the element at the specified index.
Vector::size():Integer
  // Returns the size of the vector.
\end{lstlisting}
\section{Naming Things and Looking Them Up}

\begin{figure} \begin{center}
\hfill{}\includegraphics[width=12cm]{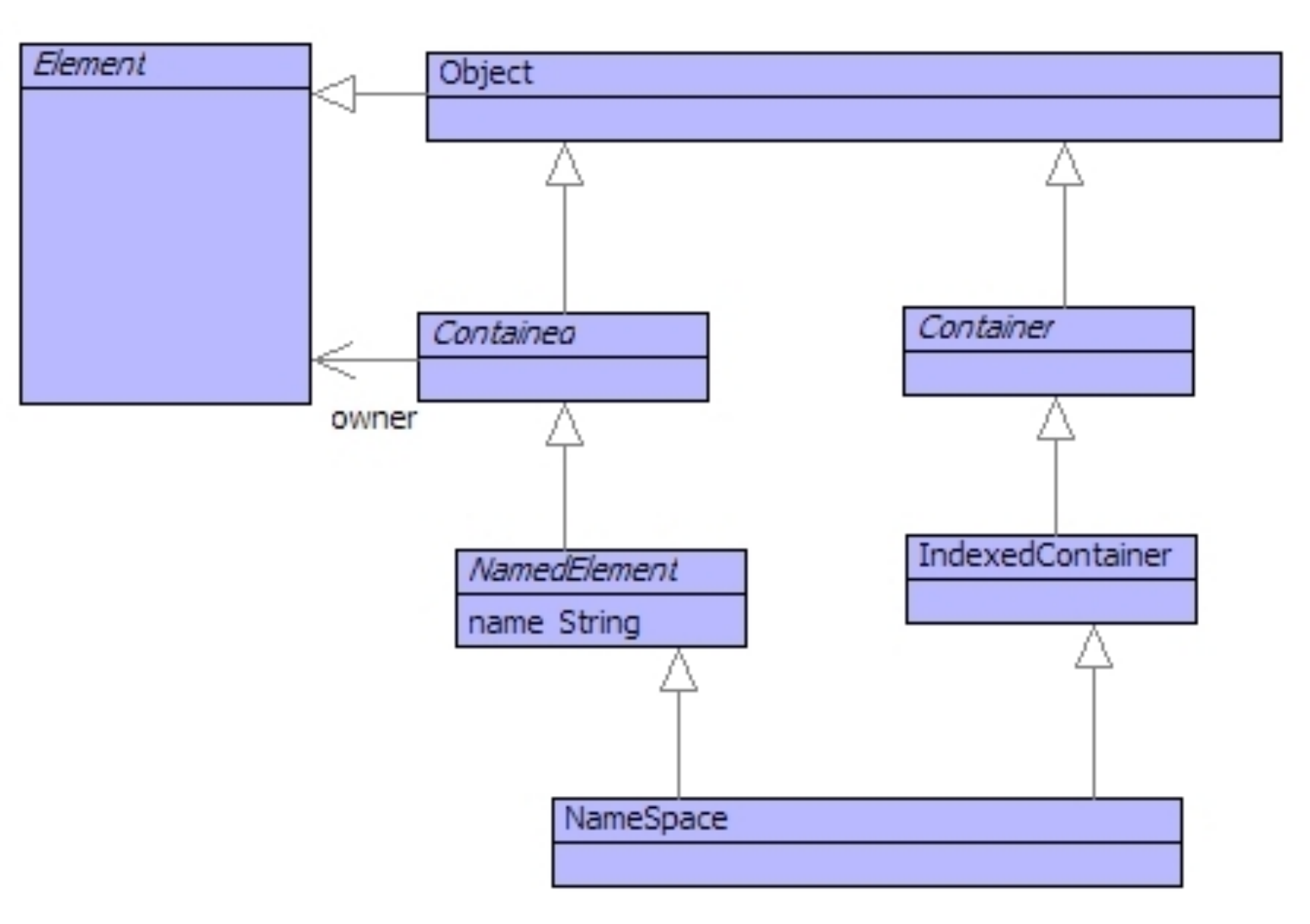}\hfill{}

\caption{Named Elements live in Name Spaces\label{fig:Named-Elements-live}}

\end{center} \end{figure}

Elements are often identified by their name and the value domain provides
special support for named elements and their containers. Figure \ref{fig:Named-Elements-live}
shows the classes that define these elements. A contained element
has a link to its owner and a named element is a contained element
with a name. A container is a class that wraps up internal storage
implemented by tables, vectors and the like. An indexed container
is a wrapper for a table (allowing users to extend the class). A name
space is a named element that contains other named elements.

Notice that the structure of a name space allows name spaces to contain
other name spaces. This means that references to named elements may
cross a number of nested name spaces, leading to the idea of a \emph{path}
to an element. There is always a global name space called Root in
the value domain: everything is contained in Root. A path is a sequence
of names separated by '::'. So Root::X references the named element
with the name X in the name space Root. If X is a name space containing
a named element Y then the path is Root::X::Y. Since Root is special,
it can be omitted: X::Y.

The main operations defined by Contained are:

\begin{lstlisting}
Contained::owner()
  // Returns the owner of the receiver.
Contained::setOwner(owner)
  // Updates the owner.
\end{lstlisting}
The main operations defined by Container are:

\begin{lstlisting}
Container::add(element)
  // Adds the element to the receiver.
Container::allContents():Set(Element)
  // Returns all the elements directly
  // or indirectly contained by the receiver.
Container::contents():Set(Element)
  // Returns the direct contents of the
  // receiver.
Container::includes(element):Boolean
  // Returns true when the receiver
  // contains the element.
Container::remove(element)
  // Removes the element from the
  // receiver.
\end{lstlisting}
The main elements defined by NamedElement are:

\begin{lstlisting}
NamedElement::name():String
  // Returns the name.
NamedElement::path():String
  // Returns the path through name-spaces to
  // the receiver.
NamedElement::pathSeq():Seq(String)
  // Returns the path through name-spaces to
  // the receiver as a sequence of strings.
NamedElement::setName(name:String)
  // Updates the name of the receiver.
\end{lstlisting}
The main operations defined by IndexedContainer are:

\begin{lstlisting}
IndexedContainer():IndexedContainer
  // Creates and returns an indexed container.
IndexedContainer::add(key,value)
  // Updates the indexed container.
IndexedContainer::index(key)
  // Returns the indexed value or throws
  // an error.
\end{lstlisting}
A package is a special type of name-space that contains operations,
classes and sub-packages. Most of XMF is constructed around packages.
When you write your own applications in XMF you will probably define
one or more packages to contain the definitions.

\section{Classifying Things}

\begin{figure} \begin{center}
\hfill{}\includegraphics[width=12cm]{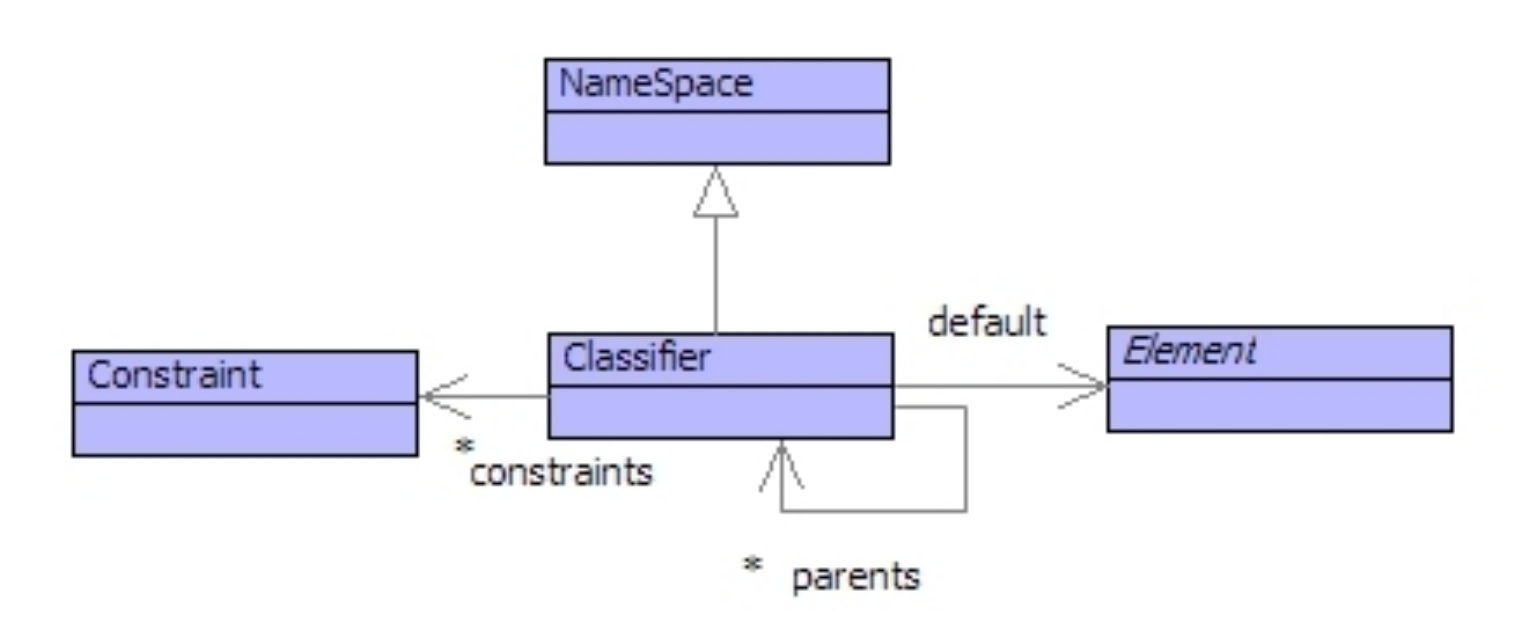}\hfill{}

\caption{Classifiers\label{fig:Classifiers}}

\end{center} \end{figure}

The values in a value domain fall into different groups: the integers,
the objects, the personnel records, the update events etc. Each of
these groups is a type and there is an element that represents the
type: a \emph{classifier}. A classifier is a single value that defines
characterising features of a group of values: its instances. Each
instance in the group refers to the classifier as its type. For example
the integer 3 has Integer as its type. Furthermore, sub-groups can
often be identified such that elements of a sub-group have more characterising
features - this is specialization.

Figure \ref{fig:Classifiers} shows the classes that define classification.
A classifier defines a collection of constraints: rules that govern
the characterising features of its instances. A classifier has parents,
the child parents are spcializations of the parent classifiers. All
the constraint rules of the parents are used to classify the instances
of the child classifier which may add rules of its own. Finally, a
classifier has a default element, for example the classifier Integer
has a default value of 0.

Classifier defines the following operations:

\begin{lstlisting}
Classifier::new()
  // Return a new instance of the receiver.
Classifier::new(args:Seq(Element))
  // Return a new instance of the receiver
  // after it is initialized using the args.
Classifier::invoke(target,args:Seq(Element))
  // Same as using new.
\end{lstlisting}
Each element in figure \ref{fig:Classifiers} is a classifier. For
example NameSpace classifies all name spaces in the value domain such
that a name space must have a collection of named elements and must
itself be a named element. It is always possible to ask an element
in the value domin what its classifier is. The operation Element::of()
returns a classifier; since this is provided by Element which classifies
everything, then all values in the domain support this operation.

\begin{figure} \begin{center}
\hfill{}\includegraphics[width=12cm]{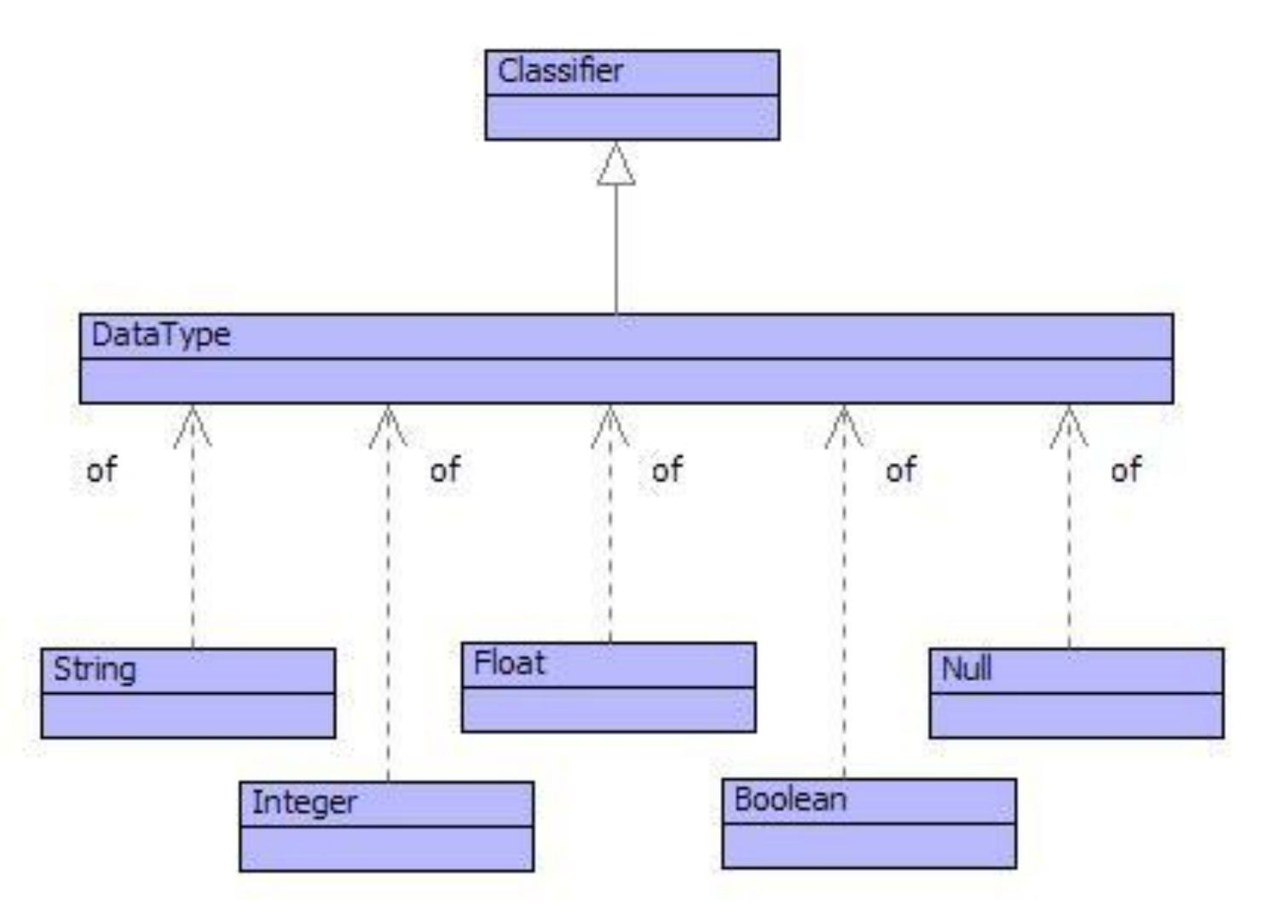}\hfill{}

\caption{DataTypes\label{fig:DataTypes}}

\end{center} \end{figure}

The basic data values such as 10, true and {}``a string'' are all
classified by distinguished classifiers Integer, Boolean and string.
These classifiers, in turn are classified by the classifier DataType
as shown in figure \ref{fig:DataTypes}. The special value null is
classified by Null; it is used as the default value for all values
that have internal state.

The types String, Integer, Float, Boolean and Null each define their
own collection of operations that are listed in the appendix.

\section{Collections of Things}

\begin{figure} \begin{center}
\hfill{}\includegraphics[width=12cm]{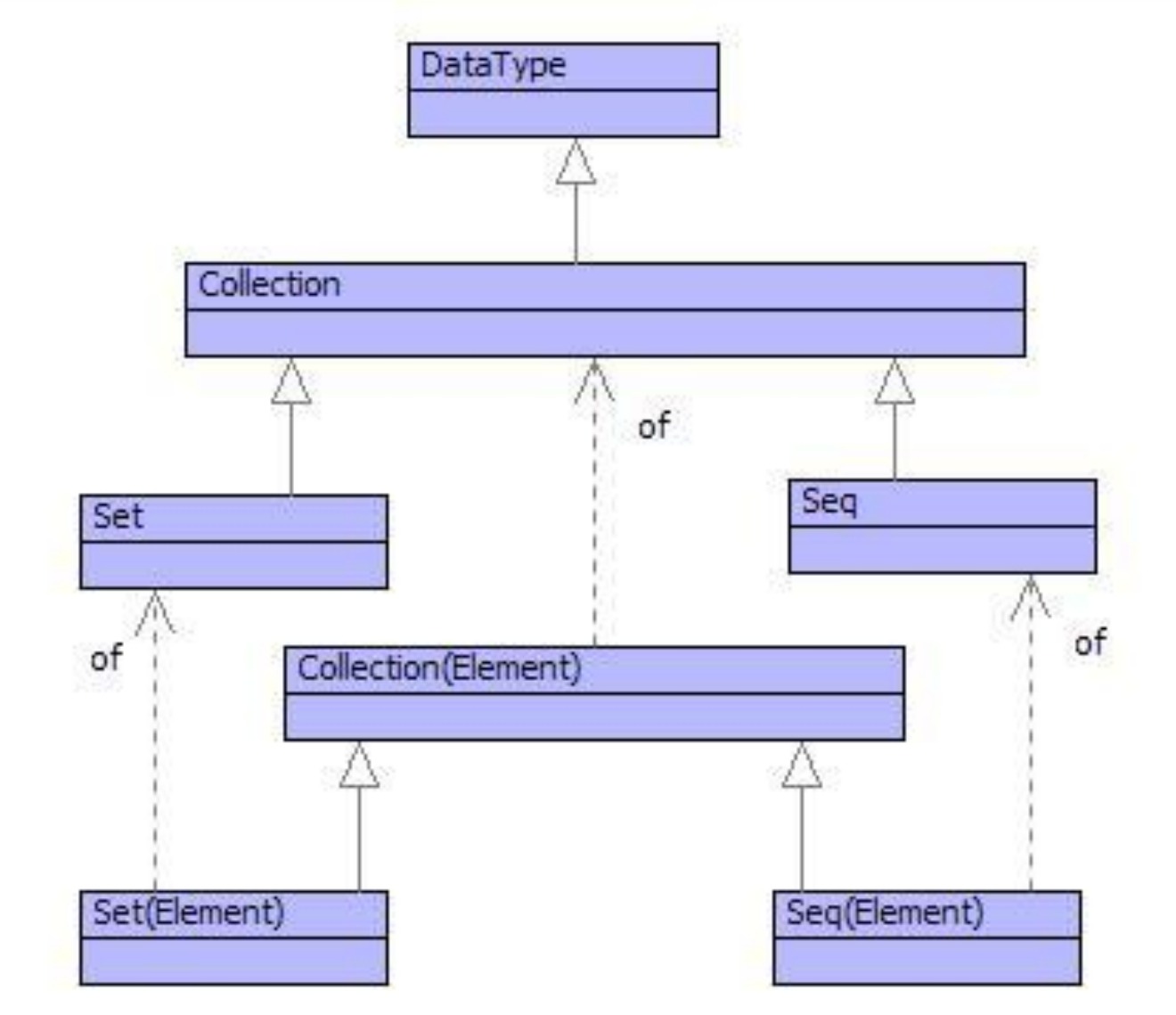}\hfill{}

\caption{Collection Classifiers\label{fig:Collection-Classifiers}}

\end{center} \end{figure}

Two types of non-indexed collections are very useful: sets and sequences.
A set is an unordered collection, adding the same element to a set
has no effect and it is not possible to rely on the order of elements
in a set. The following is a set of integers: Set\{1,2,3,4\}, it is
the same set as Set\{4,3,2,1\} and the set Set\{1,1,3,2,3,4\}. Figure
\ref{fig:Collection-Classifiers} shows the classifiers that apply
to sets and sequences. In general the classifier of a set of elements
of type T is Set(T); since everything is classified by Element then
the most general set-based classifier is Set(Element).

\begin{figure} \begin{center}
\hfill{}\includegraphics[width=5cm]{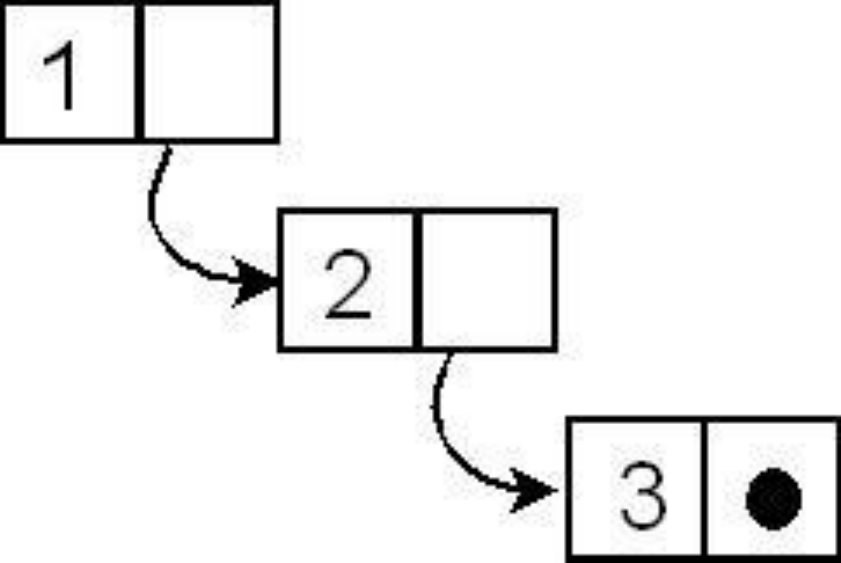}\hfill{}

\caption{A Sequence\label{fig:A-Sequence}}

\end{center} \end{figure}

A (non fixed size) sequence is either empty Seq\{\} or is a value
v followed by a sequence s: Seq\{v | s\}. A sequence has a head, v,
and a tail, s. Sequences lend themselves to recursive processing by
case analysis on the empty sequence and the non-empty sequence. Sequences
of elements of type T are classified by type Seq(T) and the most general
sequence classifier is Seq(Element).

Figure \ref{fig:A-Sequence} shows the sequence Seq\{1,2,3\} which
is equivalent to the sequence Seq\{1 | Seq\{2 | Seq\{3 | Seq\{\}\}\}\}.
The head of the sequence is the integer 1 and the tail of the sequence
is Seq\{2,3\}. The final dot in the figure represents the empty sequence
Seq\{\}.

Sequences of element are used heavily in this book to represent collections.
They are built-in to the implementation of the language and cannot
be extended. The appendix lists many of the operations that are provided
for manipulating sequences.

\section{Classifying Objects}

\begin{figure} \begin{center}
\includegraphics[width=12cm]{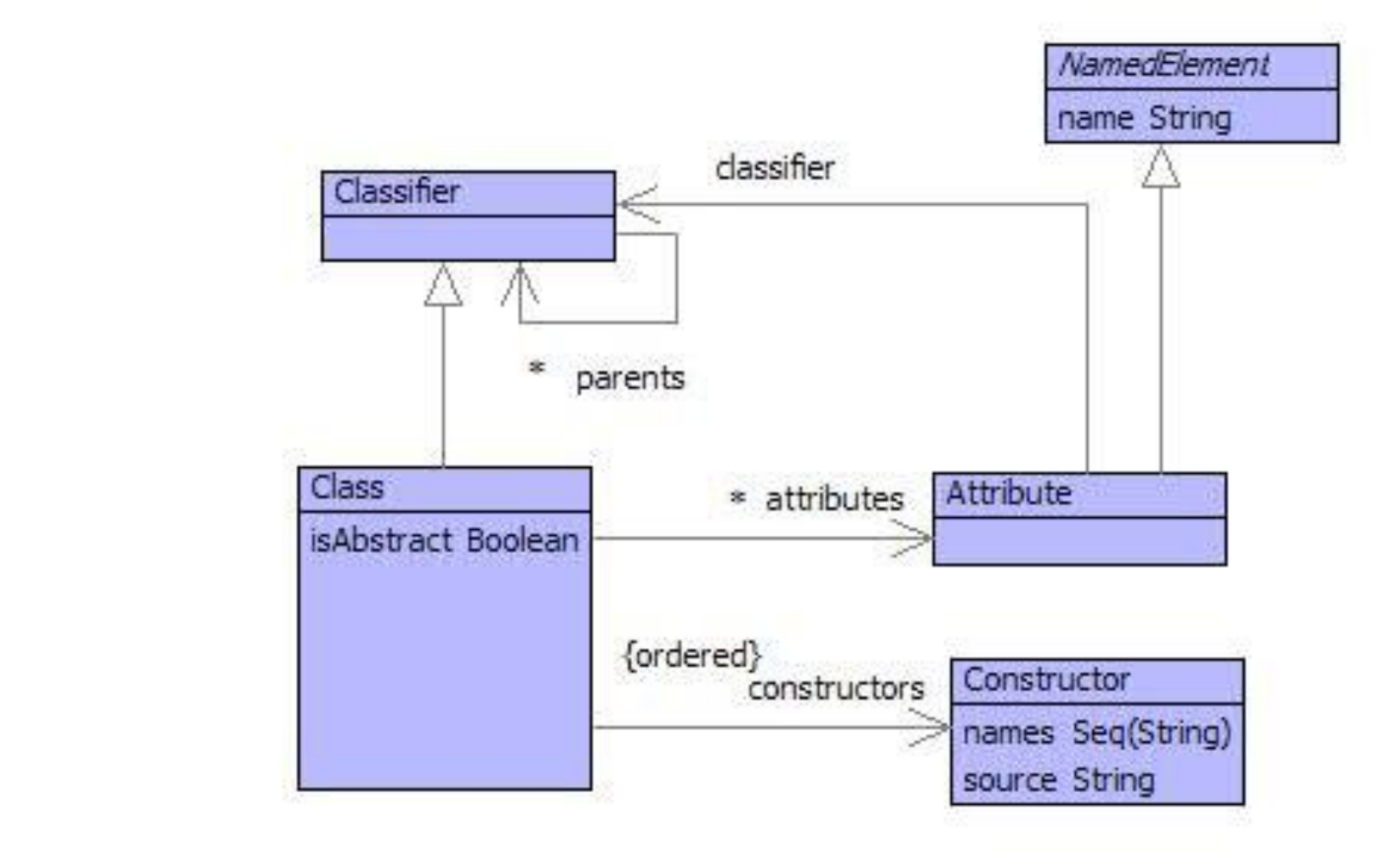}

\caption{Classifying Objects\label{fig:Classifying-Objects}}

\end{center} \end{figure}

Objects are values with identity and state. The identity is set when
the object is created and the state is represented as a collection
of slots. Each slot has a name and a value. Objects are used to represent
elements with properties, for example personnel records have personnel
properties including the name of the employee, salary and department.
Objects can be used to represent real-world things such as companies
or products and can be used to represent abstract things such as colours
(with red, green and blue properties) and system models.

Figure \ref{fig:Classifying-Objects} shows the classes that are used
to classify objects. The classifier of an object is a class with attributes
that classify the slots of the object. Each attribute has a name (defining
the name of the slot that the attribute classifies) and a type that
classifies the value of the slot. A class may be abstract in which
case it does not directly classify any objects, but may have sub-classes
(linked via the parents attribute from Classifier) whose direct instances
it classifies.

Each class has a number of constructors. Each constructor is a rule
that defines how a new instance of the class is created. A constructor
has a sequence of names that must correspond to the names of the attributes
defined (or inherited) by the class. When a constructor is used to
create an instance of a class it is supplied with values for each
of its names. A new instance of the class is created and the slots
with the corresponding names are initialized from the supplied values.
Each new instance will have a new identity. Constructors are inherited
from the parents of a class and there is always a constructor (inherited
from Object) that has no names.

For example, you can create a new object using a basic constructor
with no arguments as:

\begin{lstlisting}
Object()
\end{lstlisting}
The constructor for class (inherited from named element) allows a
new class to be constructed by supplying a name:

\begin{lstlisting}
Class("Animal")
\end{lstlisting}
The constructor for Table allows you to supply the approximate number
of elements to be stored in the table:

\begin{lstlisting}
Table(100)
\end{lstlisting}
\section{Performing Things}

The domain described so far contains elements that represent data,
but nothing that actually \emph{does} anything. It is possible to
inspect the value of an object's slot and update it, but there is
no domain element that represents what happens when an object receives
a message. the value domain contains operations that, upon being supplied
with argument values, perform actions. The actions that can be performed
are described in the next section; this section describes the basic
structure of an operation.

\begin{figure} \begin{center}
\includegraphics[width=12cm]{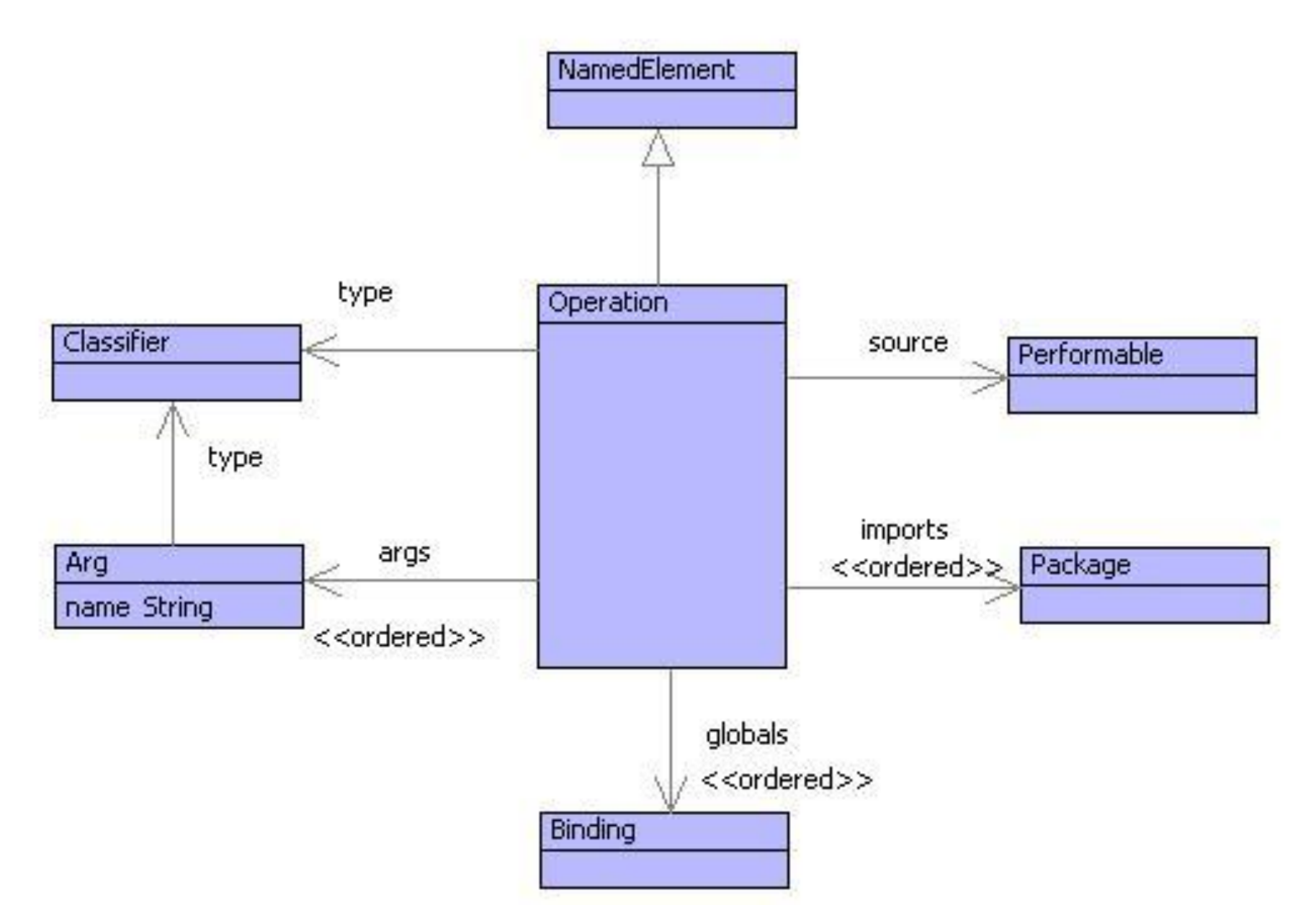}

\caption{Operations \label{fig:Operations}}

\end{center} \end{figure}

Figure \ref{fig:Operations} shows the classes that define the tructure
of an operation. Each operation is a named element that can be placed
into a name space. An operation has a return type and a sequence of
typed arguments. The source of an operation is a performable element;
this is the source code of the operation that is performed then the
operation is invoked. 

Within the source there may be variable references. A variable reference
may correspond to an argument, a name from an imported name space
(called a dynamic variable), or a variable that was in scope when
the operation was created (called a global variable). Each of these
categories of variable are described with respect to examples in the
next section.

Operations give us the ability to run programs. Furthermore, there
is no restriction on where an operation can live in the value domain:
operations may be placed as the values of slots, may be added to collections
or may be placed in tables. All data values have to come into existence
somehow: they are produced by performing the actions in an operation.
Even operations come into existence this way.

Operation defines the following operations:

\begin{lstlisting}
Operation::arity():Integer
  // The number of arguments.
Operation::fork()
  // Start a new thread by calling the operations.
Operation::invoke(target,args:Seq(Element))
  // Call the operation with the target and args.
Operation::sig():Seq(Element)
  // Return the type signature of the operation.
Operation::source():String
  // Return the source code of the operation.
Operation::trace()
  // Trace calls of the operation.
Operation::untrace()
  // Opposite of trace.
\end{lstlisting}

\chapter{The Executable Object Command Language (XOCL)}

The super-language XMF runs programs. XMF can run many different programming
languages, some interleaved, some completely separate. The basic language
provided with XMF is called XOCL - the executable object command language.
XOCL is very high-level and provides many features that make developing
applications easy. In addition to application development, XOCL is
intended for language engineering. You are likely to be using XMF
for its language engineering facilities and as such are going to use
XOCL to write languages. In order for your languages to do something
you are likely to link them to XOCL in some way, by translating them
to XOCL or by writing an language interpreter for them in XOCL.

This chapter describes the basic features of the XOCL language. XOCL
executes in terms of the concepts that were introduced in the previous
chapter. In order to ensure that we practice what we preach, XOCL
is completely implemented in XOCL: it has a grammar written in XOCL,
a compiler written in XOCL and an interpreter written in XOCL. That
property should not concern us when introducing the language, but
should give you confidence that XOCL is a good language for doing
language engineering!

\section{Working With Basic Elements}

The simplest components of XOCL are constants (notice that comments
follow // on a single line or within /{*} and {*}/ over multiple lines
as in Java):

\begin{lstlisting}
10             // An integer.
3.145          // A float.
true           // A boolean.
"Hello World"  // A string.
\end{lstlisting}
Basic operators are: not, (unary)-, + (binary)-, {*}, /, <, <=, >,
>=, =, and, or, andthen, orelse. Here is an operation definition in
XOCL: 

\begin{lstlisting}
context Root
  @Operation fact(n)
    if n = 0
    then 1
    else n * fact(n - 1)
    end
  end
\end{lstlisting}
The operator is named fact, takes a single argument n and is defined
in the global context Root which means that the name fact is available
everywhere. The argument n is local to the body of fact. All constructs
in XOCL return a value; the if-expression returns 1 or calls fact
again to return a value. If you want to follow along with a live XMF
system then you can type the definitions into a file and load them
and run the examples. To keep things short, we have omitted information
that would be in a source file. The file would actually look as follows:

\begin{lstlisting}
parserImport XOCL;

context Root
  @Operation fact(n)
    if n = 0
    then 1
    else n * fact(n - 1)
    end
  end
\end{lstlisting}
To load this into your system you would do the following:

\begin{center}
\includegraphics[width=15cm]{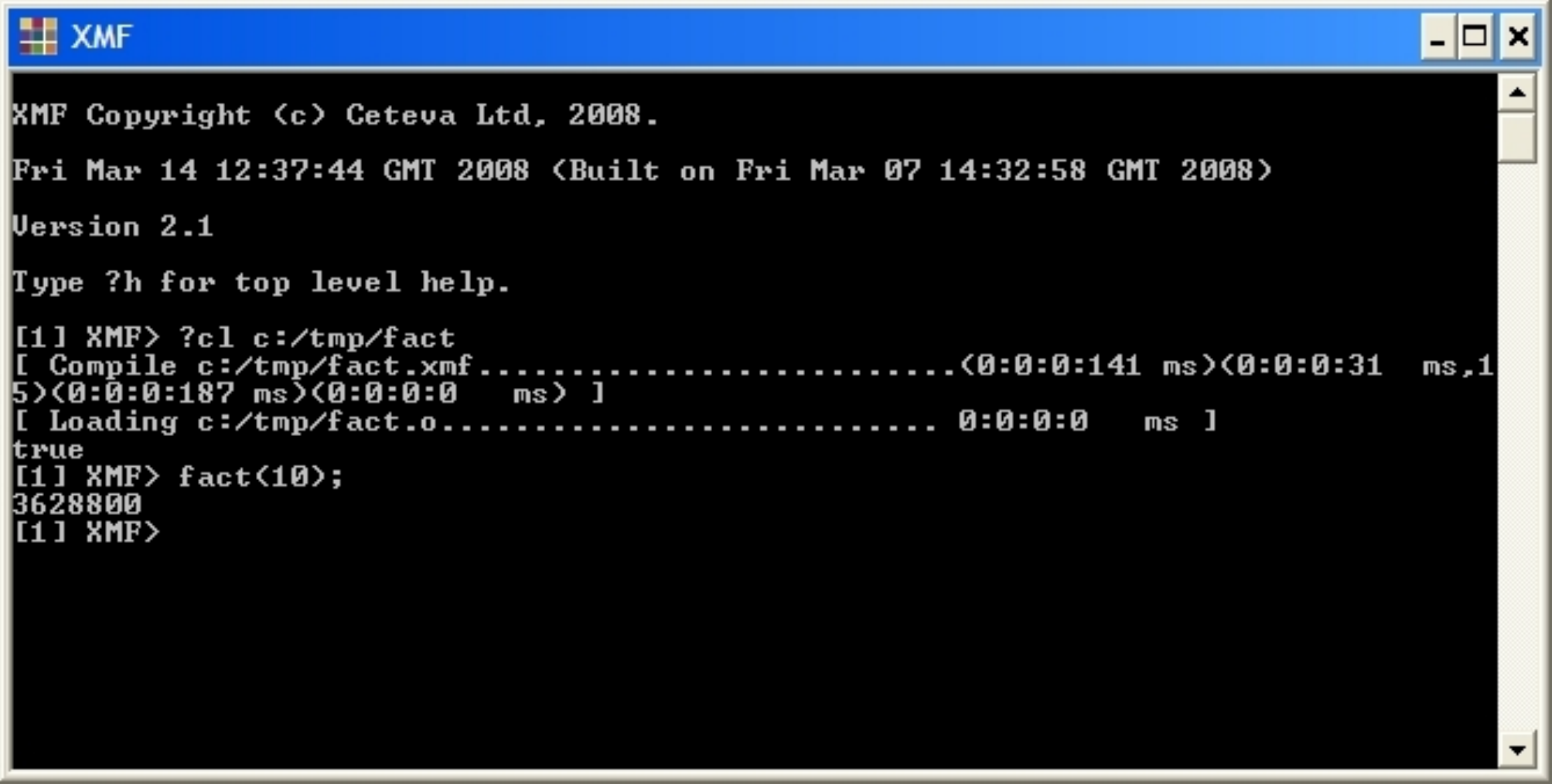}
\end{center}

Another example of a global operation definition is gcd below that
computes the greatest common divisor for a pair of positive integers.
The example shows that operations can optionally have argument and
return types:

\begin{lstlisting}
context Root
  @Operation gcd(m:Integer,n:Integer):Integer
    if m = n 
    then n
    elseif m > n 
    then gcd(m-n,n)
    else gcd(m,n-m)
    end
  end
\end{lstlisting}
Integers are arbitrary precision in XMF. 

The operators 'and' and 'or' are overloaded to perform bit comparison
when supplied with integers. The operators lsh and rsh are defined
on integers. They take the number of bits to shift left and right
respectively. The following operation adds up the number of bits in
an integer:

\begin{lstlisting}
context Root
  @Operation addBits(n:Integer):Integer
    if n = 0 or n < 0
    then 0
    else (n and 1) + (addBits(n.rsh(1)))
    end
  end
\end{lstlisting}
In addition, integers support the following operators: mod, abs, bit,
max, min and byte. XOCL supports floating point numbers. All numeric
operators described above are defined for floating point numbers.
In general integers and floats can be mixed in expressions and the
integer is translated to the equivalent float. You can construct a
float using the constructor Float so that 3.14 is created by Float(''3'',''14'').
Floats are translated to integers by the operations round (upwards)
and floor (downwards).

Integer division is performed by the operation div and floating point
division is performed by the infix operator / (translating integers
to floats as necessary). The operators sqrt, sin and cos are defined
for floats. 

All values in XMF support the operation 'of' that returns the most
specific class of the receiver. A value can be asked whether it is
of a given class using the operator isKindOf. Classes can be compared
using inheritsFrom where a class is considered to inherit from itself.
We could define isKindOf as:

\begin{lstlisting}
context Element
  @Operation isKindOf(type:Classifier):Boolean
    self.of().inheritsFrom(type)
  end
\end{lstlisting}
The distinguished value null, of type Null, is special in that it
returns true when asked whether it isKindOf any class. It is used
as the default value for all non-basic classes in XMF.

XMF strings are of type String. The following operation wraps the
string \char`\"{}The\char`\"{} and \char`\"{}.\char`\"{} around a
supplied string:

\begin{lstlisting}
context Root
  @Operation makeSentence(noun:String):String
    "The " + noun + "."
  end
\end{lstlisting}
Strings are sequences of characters indexed starting from 0. Equality
of strings is defined by = on a character by character comparison.
Characters are represented as integer ASCII codes. The following operation
checks whether a string starts with an upper case char:

\begin{lstlisting}
context Root
  @Operation startsUpperCase(s:String):Boolean
    if s->size > 0
    then
      let c = s->at(0)
      in "A"->at(0) <= c and c <= "Z"->at(0)
      end
    else false
    end
  end
\end{lstlisting}
Strings can be compared using <, <=, > and >= in which case the usual
lexicographic ordering applies.

Since strings are compared on a character by character basis this
makes string comparison relatively inefficient when performing many
comparisons. Strings are often used as the keys in lookup tables (for
example as the names of named elements). In order to speed up comparison,
XMF provides a specialization of String called Symbol. A symbol is
the same as a string except that two symbols with the same sequence
of characters have the same identity. Comparison of symbols by identity
rather than character by character is much more efficient. A string
s can be converted into a symbol by Symbol(s).

Any value can be converted into a string using the operation toString.
To get the string representation of a number for example: 100.toString()).

\section{Tables}

A table is used to associate keys with values. A table has operations
to add associations between keys and values and lookup a given key.
A table is created using the Table constructor supplying a single
argument that indicates the approximate number of elements to be stored
in the table. Suppose that a library maintains records on borrowers:

\begin{lstlisting}
context Root
  @Class Library
    @Attribute borrowers : Table = Table(100) end
  end
\end{lstlisting}
A new borrower is added by supplying the id, name and address. When
the new borrower is added we check that no borrower has been allocated
the same id. If the id is not already in use then we register the
new borrower by associating the id with a borrower record in the table:

\begin{lstlisting}
context Library
  @Operation newBorrower(id:String,name:String,address:String)
    if not borrowers.hasKey(id)
    then
      let borrower = Borrower(id,name,address)
      in borrowers.put(id,borrower)
      end
    else self.error("Borrower with id = " + id + " already exists.")
    end
  end
\end{lstlisting}
The library also provides an operation that gets a borrower record:

\begin{lstlisting}
context Library
  @Operation getBorrower(id:String):Borrower
    if borrowers.hasKey(id)
    then borrowers.get(id)
    else self.error("No borrower with id = " + id)
    end
  end
\end{lstlisting}
Tables provide operations to get all the keys and all the values:

\begin{lstlisting}
context Library
  @Operation idsInUse():Set(String)
    borrowers.keys()
  end
context Library
  @Operation allBorrowers():Set(Borrower)
    borrowers.values()
  end
\end{lstlisting}
\begin{itemize}
\item Constants, BinExp, Negate
\item Operations, Imports, Apply
\end{itemize}

\section{Sets}

A set is an unordered collection of elements. The elements of a set
need not all be of the same type. When T is the type of each element
of the set then the set is of type Set(T). Operations are provided
to add and remove elements from sets and to combine sets. Sets can
be used in iterate expressions.

Sets are created by evaluating a set expression of the form: Set\{x,y,z,...\}
where x, y, z etc are element expressions. For example:

\begin{lstlisting}
Set{1,true,Set{"a","b","c"},Borrower("1","Fred","3 The Cuttings")}
\end{lstlisting}
The expression Set\{\} produces the empty set. A set is unordered.
An element can be selected at random from a non-empty set by performing
S->sel. A set is empty when S->isEmpty produces true (or when it is
= to Set\{\}). An element e is added to a set by S->including(e) and
removed from a set by S->excluding(e). The union of two sets is produced
by S1 + S2 and the difference is constructed by S1 - S2. An element
is contained in a set when S->includes(e).

Suppose that the set operation includes was not provided as part of
XOCL. It could be defined by:

\begin{lstlisting}
context Set(Element)
  @Operation includes(e:Element):Boolean
    if self->isEmpty
    then false
    else 
      let x = self->sel
      in if x = e
         then true
         else self->excluding(x)->includes(e)
         end
      end
    end
  end
\end{lstlisting}
\section{Sequences}

A sequence is an ordered collection of elements. The elements in the
sequence need not all be of the same type. When T is the type of each
element in the sequence then the sequence is of type Seq(T). Sequences
can be used in iterate expressions as described in section iterate.

Sequences are created by evaluating a sequence expression or by translating
an existing element into a sequence. Sets, strings, integers and vectors
can be translated to sequences of elements, characters, bits and elements
respectively by performing e.asSeq(). 

The following operations are defined on sequences: + appends sequences;
asSet transforms a sequence into a set; asString transforms a sequence
of character codes into a string; asVector transforms a sequence into
a vector; at takes an index and returns the element at that position
in the sequence, it could be defined as:

\begin{lstlisting}
context Seq(Element)
  @Operation at(n:Integer):Element
    if self->size = 0
    then self.error("Seq(Element).at: empty sequence.")
    else if n <= 0 
         then self->head
         else self->tail.at(n - 1)
         end
    end
  end  
\end{lstlisting}
The operation butLast returns all elements in a sequence but the last
element. It could have been defined as follows, note the use of Seq\{head
| tail\} to construct a sequence with the given head and tail:

\begin{lstlisting}
context Seq(Element)
  @Operation butLast():Seq(Element)
    if self->size = 0
    then self.error("Seq(Element)::butLast: empty sequence.")
    elseif self->size = 1
    then Seq{}
    else Seq{self->head | self->tail->butLast}
    end
  end
\end{lstlisting}
The operation contains returns true when a sequence contains a supplied
element; drop takes an integer and returns a sequence that is the
result of dropping the supplied number of elements; flatten maps a
sequence of sequences to a sequence:

\begin{lstlisting}
context Seq(Element)
  @Operation flatten():Seq(Element)
    if self->isEmpty
    then self
    else self->head + self->tail->flatten
    end
  end
\end{lstlisting}
The operation hasPrefix takes a sequence as an argument and returns
true when the receiver has a prefix that is equal to the argument;
including takes an element and returns a new sequence, this is the
receiver if it contains to the argument or the argument prepended
to the receiver; indexOf takes an element and returns the index of
the argument in the receiver:

\begin{lstlisting}
context SeqOfElement
  @Operation indexOf(element:Element):Integer
    if self = Seq{}
    then -1
    elseif self->head = element
    then 0
    else self->tail->indexOf(element) + 1
    end
  end
\end{lstlisting}
The operation insertAt takes an element and an index and inserts the
element at the given index; last returns the last element in a sequence;
hasSuffix takes a sequence as an argument and returns true when the
receiver ends with the argument:

\begin{lstlisting}
context Seq(Element)
  @Operation hasSuffix(suffix):Boolean
    self->reverse->hasPrefix(suffix->reverse)
  end
\end{lstlisting}
The operation head returns the head of a non-empty sequence; includes
returns true when the argument is included in the receiver; isEmpty
returns true when the receiver is empty; isProperSequence returns
true when the final element in the sequence is a pair whose tail is
Seq\{\}; map applies an operation to each element of a sequence, the
definition of map shows the varargs feature of XOCL operations where
the last argument may be preceded by a '.' indicating that any further
supplied arguments are bundled up into a sequence and supplied as
a single value:

\begin{lstlisting}
context Seq(Element)
  @Operation map(message:String . args:Seq(Element)):Element
    self->collect(x | x.send(message,args))
  end
\end{lstlisting}
The operation max finds the maximum of a sequence of integers; prepend
adds an element to the head of a sequence; qsort takes a binary predicate
operation and sorts the receiver into an order that satisfies the
predicate using the quicksort algorithm:

\begin{lstlisting}
context Seq(Element)
  @Operation qsort(pred):Seq(Element)
    if self->isEmpty
    then self
    else 
      let e = self->head
      in let pre = self->select(x | pred(x,e));
             post = self->select(x | x <> e and not pred(x,e))
         in pre->sort(pred) + Seq{e} + post->sort(pred)
         end
      end
    end
  end
\end{lstlisting}
The operation ref can be used to lookup a name-space path represented
as a sequence of strings to the element found at the path. The operation
takes a sequence of name-spaces as an argument; the name-space arguments
are used as the basis for the lookup, for example:

\begin{lstlisting}
Seq{"Root","EMOF","Class","attributes"}->ref(Seq{Root})
\end{lstlisting}
returns the attribute named attributes. Note that name-space Root contains
itself.

The operation reverse reverses the receiver:

\begin{lstlisting}
context Seq(Element)
  @Operation reverse():Seq(Element)
    if self->isEmpty
    then Seq{}
    else self->tail->reverse + Seq{self->head}
    end
  end
\end{lstlisting}
The operation separateWith takes a string as an argument and returns
the string that is formed by placing the argument between each element
of the receiver after the element is transformed into a string using
toString.

The operation subst takes three arguments: new, old and all; it returns
the result of replacing element old with new in the receiver. If all
is true then all elements are replaced otherwise just the first element
is replaced.

The operation subSequence takes a staring and terminating indices
and returns the appropriate subsequence; take takes an integer argument
and returns the prefix of the receiver with that number of elements;
tail returns the tail of a non-empty sequence.

Sequences have identity in XOCL; the head and tail of a sequence can
be updated using:

\begin{lstlisting}
S->head := e
\end{lstlisting}
and

\begin{lstlisting}
S->tail := e
\end{lstlisting}
This makes sequences very flexible and can be used for efficient storage
and update of large collections (otherwise each time a sequence was
updated it would be necessary to copy the sequence).

\section{A-Lists}

An a-list is a sequence of pairs; each pair has a head that is a key
and a tail that is the value associated with the key in the a-list.
A-lists are used as simple lookup tables. They are much more lightweight
than instances of the class Table and have the advantage that the
large number of builtin sequence operations apply to a-lists. The
following class shows how an a-list can be used to store the age of
a collection of people:

\begin{lstlisting}
context Root
  @Class People
    @Attribute table : Seq(Element) end
    @Operation newPerson(name:String,age:Integer)
      self.table := table->bind(name,age)
    end
    @Operation getAge(name:String):Integer
      table->lookup(name)
    end
    @Operation hasPerson(name:String):Boolean
      table->binds(name)
    end
    @Operation birthday(name:String)
      // Assumes name is in table:
      table->set(name,table->lookup(name) + 1)
    end
  end
\end{lstlisting}
\section{Iteration Expressions}

Iteration expressions in XOCL allow collections (sets and sequences)
to be manipulated in a convenient way. Iteration expressions are a
shorthand for higher-order operations that take an operation as an
argument and apply the argument operation to each element of the collection
in turn. As such, iteration expressions can be viewed as sugar for
the invocation of the equivaent higher-order operations.

A collection can be filtered using a select expression:

\begin{lstlisting}
S->select(x | e)
\end{lstlisting}
where x is a variable and e is a predicate expression. The result
is the sub-collection of the receiver where each element y in the
sub-collection satisfies the predicate when x is bound to y. This
can be defined as follows for sequences:

\begin{lstlisting}
context Seq(Element)
  @Operation select(pred:Operation):Seq(Element)
    if self->isEmpty
    then self
    elseif pred(self->head)
    then Seq{self->head | self->tail.select(pred)}
    else self->tail.select(pred)
    end
  end
\end{lstlisting}
The reject expression is like select except that it produces all elements
that fail the predicate; here is the definition of reject for sets:

\begin{lstlisting}
context Set(Element)
  @Operation reject(pred:Operation):Set(Element)
    if self->isEmpty
    then self
    else 
      let x = self->sel
      in if pred(x)
         then self->excluding(x)->reject(pred)
         else self->excluding(x)->reject(pred)->including(x)
         end
      end
    end
  end
\end{lstlisting}
The collect expression maps a unary operation over a collection:

\begin{lstlisting}
S->collect(x | e)
\end{lstlisting}
It is defined for sequences as follows:

\begin{lstlisting}
context Seq(Element)
  @Operation collect(map:Operation):Seq(Element)
    if not self->isEmpty
    then 
      let x = self->sel
      in self->excluding(x)->select(map)->including(map(x))
      end
    else self
    end
  end
\end{lstlisting}
The expression iterate has the form:

\begin{lstlisting}
S->iterate(x y = v | e)
\end{lstlisting}
This expression steps through the elements of S and repeatedly sets
the value of y to be the value of e where e may refer to x\} and y.
The initial value of y is the value v. The following shows an example
that adds up the value of a sequence of integers:

\begin{lstlisting}
Seq{1,2,3,4,5}->iterate(i sum = 0 | sum + i)
\end{lstlisting}
The definition of iterate as a higher order operation on sequences
is as follows:

\begin{lstlisting}
context Seq(Element)
  @Operation iterate(y:Element,map:Operation):Element
    if self->isEmpty
    then y
    else self->tail.iterate(map(self->head,y),map)
    end
  end
\end{lstlisting}
\section{Variables and Scope}

Variables in XMF are of three types: slots, dynamic and lexical. When
a message is sent to an object, all of the slots defined by the object
are bound to the slot values, For example:

\begin{lstlisting}
context Root
  @Class Point
    @Attribute x : Integer end
    @Attribute y : Integer end
    @Constructor(x,y) ! end
    @Operation getX():Integer
      x
    end
    @Operation getY():Integer
      y
    end
  end
\end{lstlisting}
The references to x and y in the accessor operations defined in Point
are variables of type slot. The class definition given for Point is
equivalent to the shorthand definition:

\begin{lstlisting}
context Root
  @Class Point
    @Attribute x : Integer (?) end
    @Attribute y : Integer (?) end
    @Constructor(x,y) ! end
  end
\end{lstlisting}
Slot variables can always be replaced with a qualified reference using
self as in self.x and self.y. Using qualified references is a matter
of taste: sometimes it is more readable to use a qualified reference.
Note that it is not possible to update slots using variable syntax:
x := 0; it is always necessary to update slots using a qualified update
as follows self.x := 0.

Dynamic variables are typically values associated with names in name-spaces.
Dynamic variables are established when the association between the
variable name and the value is created and typically persist for the
rest of the lifetime of the XMF session. Lexical variables are typically
created when values are supplied as arguments to an operation or when
local definitions are executed. The association between the lexical
variable name and the value persist for the duration of the operation
definition or the execution of the body of the local block. In both
cases, as the name suggests, variable values can change by side-effect.

A dynamic variable added to the Root name-space has global scope because
Root is imported everywhere. A new dynamic variable in a name-space
N is created as follows:

\begin{lstlisting}
N::v := e;
\end{lstlisting}
where v is the name of the variable and e is an expression that produces
the initial value. In order to make the names in a name-space visible
you need to import the name-space. Imports are declared at the top
of a source file which has the general form:

\begin{lstlisting}
// Parser imports...
// Name-space imports...
// Definitions...
\end{lstlisting}
The following is a typical example:

\begin{lstlisting}
parserImport XOCL;

import MyPackage;

// We can reference MyClass without qualification
// since it is defined in MyPackage which is imported...
context MyClass
  @Operation myOperation()
     // Imported names from MyPackage are available here.
     // Note that MyClass is not imported - the context
     // does not mean import...
end
\end{lstlisting}
Lexical variables are established when arguments are passed to an
operation or using a let expression. In both cases the variable can
be referenced in the body of the expression, but not outside the body.
In both cases the variables can be updated using v := e. Suppose we
require an operation that takes two integers and returns a pair where
the head is the smallest integer and the tail is the other integer:

\begin{lstlisting}
context Root
  @Operation orderedPair(x,y)
    let min = 0;
        max = 0
    in if x < y then min := x else min := y end;
       if x > y then max := x else max := y end;
       Seq{min | max}
    end
  end
\end{lstlisting}
The definition of orderedPair shows how a let expression can introduce
a number of variables (in this case min and max). If the let-bindings
are separated using';' then the bindings are established in-parallel
meaning that the variables cannot affect each other (i.e. the value
for max cannot refer to min and vice versa). If the bindings are separated
using then they are established in-series meaning that values in subsequent
bindings can refer to variables in earlier bindings, for example:

\begin{lstlisting}
context Root
  @Operation orderedPair(x,y)
    let min = if x < y then x else y end then
        max = if min = x then y else x end
    in Seq{min | max}
    end
  end
\end{lstlisting}
\section{Loops}

XMF provides While and For for looping through collections and provides
Find for selecting an element in a collection that satisfies a condition.
A While loop performs an action until a condition is satisfied (not
a named element may use a symbol for its name so we ensure the name
is a string using the toString operation):

\begin{lstlisting}
context Root
  @Operation findElement(N:Set(NamedElement),name:String)
    let found = null
    in @While not N->isEmpty do
         let n = N->sel
         in if n.name().toString() = name
            then found := n
            else N := N->excluding(n)
            end
         end
       end;
       found
    end
  end
\end{lstlisting}
It is often the case that While loops are used to iterate through
a collection. This pattern is captured by a For loop:

\begin{lstlisting}
context Root
  @Operation findElement(N:Set(NamedElement),name:String)
    let found = null
    in @For n in N do
         if n.name().toString() = name
         then found := n
         end
       end;
       found
    end
  end
\end{lstlisting} 
In general a For loop @For x in S do e end is equivalent to the following
While loop:

\begin{lstlisting}
let forColl = S;
    isFirst = true
in @While not forColl->isEmpty do
     let x = forColl->sel
     in forColl := forColl->excluding(x);
        let isLast = forColl->isEmpty
        in e;
           isFirst := false
        end
     end
   end
end
\end{lstlisting}
where the variables forColl, isFirst and isLast are scoped over the
body of the loop e. These can be useful if we want the body action
to depend on

whether this is the first or last iteration, for example turning a
sequence into a string:

\begin{lstlisting}
context Seq(Operation)
  @Operation toString()
    let s = "Seq{"
    in @For e in self do
         s := s + e.toString();
         if not isLast then s := s + "," end
       end;
       s + "}"
    end
  end
\end{lstlisting}
A For loop may return a result. The keyword do states that the body
of the For loop is an action and that the result of performing the
entire loop will be

ignored when the loop exits. Alternatively, the keyword produce states
that the loop will return a sequence of values. The values are the
results returned by the loop body each time it is performed. For example,
suppose we want to calculate the sequence of names from a sequence
of people:

\begin{lstlisting}
context Root
  @Operation getNames(people:Seq(Person)):Seq(String)
    @For person in people produce 
      person.name 
    end
  end
\end{lstlisting}
The keyword in is a For-loop \emph{directive}. After in the loop expects
one or more collections. The in directive supports multiple variables.
This feature is useful when stepping through multiple collections
in sync, as in:

\begin{lstlisting}
context Root
  @Operation createTable(names:Seq(String),addresses:Seq(String),telNos:Seq(String))
    @For name,address,telNo in names,addresses,telNos produce
      Seq{name,address,telNo}
    end
  end
\end{lstlisting}
A For-loop can be used to iterate through a table. Often we want to
iterate either through the table values or the table keys. If we use
the in

directive to iterate through either of these then we will create an
intermediate collection:

\begin{lstlisting}
context Root
  @Operation addToAll(n:Integer,t:Table)
    @For k in table.keys() do
      k.put(k,t.get(k) + n)
    end
  end
\end{lstlisting}
The expression table.keys() creates an intermediate collection of
all the keys in the table. The collection is not returned and cannot
be referenced independently within the body of the loop. Therefore
the collection is transient. The allocation of transient table collections
can be avoided using the For-loop directives inTableValues and inTableKeys:

\begin{lstlisting}
context Root
  @Operation addToAll(n:Integer,t:Table)
    @For k inTableKeys table do
      k.put(k,t.get(k) + n)
    end
  end
\end{lstlisting}
\section{Operations}

XMF operations are used to implement both procedures and functions.
An operation has an optional name, some parameters, a return type
and a body. Operations are objects with internal state; part of the
internal state is the name, parameter information, type and body.
Operations also have property lists that can be used to attach information
to the operation for use by XMF programs.

Operations can be created and stored in XMF data items. In particular,
operations can be added to name spaces and then referenced via the
name space (either where the name space is imported or directly by
giving the path to the operation). We have seen many examples of adding
operations to the name space called Root. The syntax:

\begin{lstlisting}
context Root
  @Operation add(x,y) x + y end
\end{lstlisting}
can occur at the top-level of an XMF source file, compiled and loaded.
It is equivalent to the following expression:

\begin{lstlisting}
Root.add(@Operation add(x,y) x + y end);
\end{lstlisting}
Unlike the context expression, the call to add may occur anywhere
in XMF code. Operations are performed by sending them a message invoke
with two arguments: the value of self (or target) to be used in the
body of the operation and a sequence of argument values. The target
of the invocation is important because it provides the value of self
in the body of the operation and supplies the values of the slot-bound
variables. The add operation can be invoked by:

\begin{lstlisting}
add.invoke(null,Seq{1,2})
\end{lstlisting}
produces the value 3. Note that in this case there is no reference
to self or slot-bound variables in the body and therefore the target
of the invocation is null. A shorthand for invocation is provided:

\begin{lstlisting}
add(1,2)
\end{lstlisting}
however, note that no target can be supplied with the shorthand. In
this case the target will default to the value of self that was in
scope when the operation was created.

Lexically bound variables that are scoped over an operation are available
within the body of the operation event though the operation is returned
from the lexical context. This is often referred to as \emph{closing}
the lexical variable into the operation (or \emph{closure}). This
feature is very useful when generating behaviour that differs only
in terms of context. Suppose that transition machine states have an
action that is implemented as an operation and that the action is
to be performed when the state is entered:

\begin{lstlisting}
context StateMachines
  @Class State
    @Attribute name : String end
    @Attribute action : Operation end
    @Constructor(name,action) end
    @Operation enter()
      action()
    end
  end
\end{lstlisting}
\section{Exception Handling}

When an error occurs in XOCL, the source of the error \emph{throws}
an exception. The exception is a value that, in general, contains
a description of the problem and any data that might help explain
the reason why the problem occurred. 

An exception is thrown to the most recently established handler; intermediate
code is discarded. If no handler exists then the XOCL VM will terminate.
In most cases, the exception is \emph{caught} by a user defined handler
or, for example in the case of the XMF console, a handler established
by the top level command interpreter.

When an exception is caught, the handler can inspect the contents
of the exception and decide what to do. For example it may be necessary
to re-throw the exception to the next-most recently established handler,
since it cannot be dealt with. On the other hand, it is usual to catch
the exception, print a message, patch up the problem, or just give
up on the requested action.

For example, suppose that you have an application that reads data
from a file. If the file does not exist then an exception is raised.
This can be done as follows:

\begin{lstlisting}
@Operation readData(file:String) 
  if file.fileExists() 
  then // Read the data... 
  else self.error("The file does not exist.") 
  end 
end
\end{lstlisting}
In the example given above, the operation 'error' is used to raise
the exception. The operation Exception::error(message:String) is defined
for all data elements and just creates a general type of exception
and throws it. The class XCore::Exception contains a message. The
above example is equivalent to:

\begin{lstlisting}
@Operation readData(file:String) 
  if file.fileExists() 
  then // Read the data... 
  else throw Exception("The file does not exist.") 
  end 
end
\end{lstlisting}
The throw expression takes a single data element and throws it. The
thrown data element should be an instance of Exception or a sub-class
of Exception. The thrown exception can be caught in a try-expression:

\begin{lstlisting}
try 
  readData(someFile) 
catch(x:Exception) 
  // Do something with x... 
end
\end{lstlisting}
The class Exception has various sub-classes that can be used to capture
specific information about the exception when it occurs. For example,
the class Exceptions::FileNotFound can be used to create an exception
when a file is not present:

\begin{lstlisting}
@Operation readData(file:String) 
  if file.fileExists() 
  then // Read the data... 
  else throw FileNotFound(File) 
  end 
end
\end{lstlisting}
Then specific types of exception can be caught and dealt with:

\begin{lstlisting}
try readData(someFile) 
  catch(x:Exception) 
    @TypeCase(x) 
      FileNotFound do 
        // OK use a default data file... 
        readData(defaultFile) 
      end 
      else 
        // We cannot handle the exception so 
        // re-throw it to a less-specific handler... 
        throw x 
    end 
end
\end{lstlisting}
\section{Patterns}

A pattern is matched against a value. The pattern match may succeed
or fail in a given matching context. A matching context keeps track
of any variable bindings generated by the match and maintains choice
points for backtracking if the current match fails.

Pattern matching can be viewed as being performed by a pattern matching
engine that maintains the current pattern matching context as its
state. The engine state consists of a stack of patterns to be matched
against a stack of values, a collection of variable bindings and a
stack of choice points. A choice point is a machine state. At any
given time there is a pattern at the head of the pattern stack and
a value at the head of the value stack. The machine executes by performing
state transitions driven by the head of the pattern stack: if the
outer structure of the pattern matches that of the value at the head
of the value stack then:

\begin{itemize}
\item 0 or more values are bound.
\item 0 or more choice points are added to the choice point stack.
\item 0 or more component patterns are pushed onto the pattern stack.
\item 0 or more component values are pushed onto the value stack.
\end{itemize}
If the machine fails to match the pattern and value at the head of
the respective stacks then the most recently created choice point
is popped and becomes the new machine state. Execution continues until
either the pattern stack is exhausted or the machine fails when the
choice stack is empty.

A \emph{variable pattern} consists of a name, optionally another pattern
and optionally a type. The simplest form of variable pattern is just
a name, for example, the formal parameter x is a variable pattern:

\begin{lstlisting}
let add1 = @Operation(x) x + 1 end in ...
\end{lstlisting}
Matching a simple variable pattern such as that shown above always
succeeds and causes the name to be bound to the corresponding value.
A variable may be qualified with a type declaration:

\begin{lstlisting}
let add1 = @Operation(x:Integer) x + 1 end in ...
\end{lstlisting}
which has no effect on pattern matching. A variable may be qualified
with a pattern as in x = <Pattern> where the pattern must occur before
any type declaration. Such a qualified variable matches a value when
the pattern also matches the value. Any variables in the pattern \emph{and}
x are bound in the process.

A \emph{constant pattern} is either a string, an integer, a boolean
or an expression (in the case of an expression the pattern consists
of {[} followed by an expression followed by ]). A constant pattern
matches a value when the values is equal to the constant (in the case
of an expression the matching process evaluates the expression each
time the match occurs). For example:

\begin{lstlisting}
let fourArgs = @Operation(1,true,"three",x = [2 + 2]) x end in ...
\end{lstlisting}
is an operation that succeeds in the case:

\begin{lstlisting}
fourArgs(1,true,"three",4)
\end{lstlisting}
and returns 4.

A \emph{sequence pattern} consists of either a pair of patterns or
a sequence of patterns. In the case of a pair:

\begin{lstlisting}
let head = @Operation(Seq{head | tail}) head end in ...
\end{lstlisting}
the pattern matches a non-empty sequence whose head must match the
head pattern and whose tail must match the tail pattern. In the case
of a sequence of patterns:

\begin{lstlisting}
let add3 = @Operation(Seq{x,y,z}) x + y + z end in ...
\end{lstlisting}
the pattern matches a sequence of exactly the same size where each
element matches the corresponding pattern.

A \emph{constructor pattern} matches an object. A constructor pattern
may be either a by-order-of-arguments constructor pattern (or BOA-constructor
pattern) or a keyword constructor pattern. A \emph{BOA-constructor
pattern} is linked with the constructors of a class. It has the form:

\begin{lstlisting}
let flatten = @Operation(C(x,y,z)) Seq{x,y,z} end in ...
\end{lstlisting}
where the class C must define a 3-argument constructor. A BOA-constructor
pattern matches an object when the object is an instance of the class
(here C but in general defined using a path) and when the object's
slot values identified by the constructor of the class with the appropriate
arity match the corresponding sub-patterns (here x, y and z).

A \emph{keyword constructor pattern} has the form:

\begin{lstlisting}
let flatten = 
      @Operation(C[name=y,age=x,address=y]) 
        Seq{x,y,z} 
      end 
in ...
\end{lstlisting}
where the names of the slots are explicitly defined in any order (and
may be repeated). Such a pattern matches an object when it is an instance
of the given class and when the values of the named slots match the
appropriate sub-patterns.

A conditional pattern consists of a pattern and a predicate expression.
It matches a value when the value matches the sub-pattern and when
the expression evaluates to true in the resulting variable context.
For example:

\begin{lstlisting}
let repeat = @Operation(Seq{x,y} when x = y) Seq{x} end in ...
\end{lstlisting}
Note that the above example will fail (and probably throw an error
depending on the context) if it is supplied with a pair whose values
are different.

\emph{Set patterns} consist of an element pattern and a residual pattern.
A set matches a pattern when an element can be chosen that matches
the element pattern and where the rest of the set matches the residual
pattern. For example:

\begin{lstlisting}
let choose = @Operation(S->including(x)) x end in ...
\end{lstlisting}
which matches any non-empty set and selects a value from it at random.
Set patterns introduce choice into the current context because often
there is more

than one way to choose a value from the set that matches the element
pattern. For example:

\begin{lstlisting}
let chooseBigger = 
      @Operation(S->including(x),y where x > y) 
        x 
      end 
in ...
\end{lstlisting}
Pattern matching in chooseBigger, for example:

\begin{lstlisting}
chooseBigger(Set{1,2,3},2)
\end{lstlisting}
starts by selecting an element and binding it to x and binding S to
the rest. In this case suppose that x = 1 and S = Set\{2,3\}. The
pattern y matches and binds 2 and then the condition is applied. At
this point, in general, there may be choices left in the context due
to there being more than one element in the set supplied as the first
parameter. If the condition x > y fails then the matching process
jumps to the most recent choice point (which in this cases causes
the next element in the set to be chosen and bound to x). Suppose
that 3 is chosen this time; the condition is satisfied and the call
returns 3.

The following is an example that sorts a set of integers into descending
order:

\begin{lstlisting}
context Root
  @Operation sort(S)
    @Case S of
      Set{} do 
        Seq{} 
      end
      S->including(x) when S->forAll(y | y <= x) do 
        Seq{x | Q} where Q = sort(S) 
      end
    end
  end
\end{lstlisting}
\emph{Sequence patterns} use the infix + operator to combine two patterns
that match against two sub-sequences. For example the following operation

removes a sequence of 0's occurring in a sequence:

\begin{lstlisting}
context Root
  @Operation remove0s(x) 
    @Case x of 
      (S1 when S1->forAll(x | x <> 0)) + 
        (S2 when S2->forAll(x | x = 0)) + 
          (S3 when S3->forAll(x | x <> 0)) do 
        S1 + S3 
      end 
    end
  end
\end{lstlisting}
Patterns may be used in the following contexts:

\begin{itemize}
\item Operation Parameters. Each parameter in an operation definition is
a pattern. Parameter patterns are useful when defining an operation
that must deconstruct one or more values passed as arguments. Note
that if the pattern match fails then the operation invocation will
raise an error. Operations defined in the same class and with the
same name are merged into a single operation in which each operation
is tried in turn when the operation is called via an instance of the
class. Therefore in the following example: \begin{lstlisting}
@Class P
  @Operation f(Seq{}) 0 end
  @Operation f(Seq{x | t}) x + self.f(t) end
end
\end{lstlisting}an instance of P has a single operation f that adds up all the elements
of a sequence.
\item Case Arms. A case expression consists of a number of arms each of
which has a sequence of patterns and an expression. A case expression
dispatches on a sequence of values and attempts to match them against
the corresponding patterns in each arm in turn. For example, suppose
we want to calculate the set of duplicated elements in a pair of sets:
\begin{lstlisting}
context Root
  @Operation dups(s1,s2)
    @Case s1,s2 of
      s1->including(x),s2->including(y) when x = y do 
        Set{x} + dups(s1,s2) 
      end
      s1->including(x),s2 do 
        dups(s1,s2) 
      end
      s1,s2->including(y) do 
        dups(s1,s2) 
      end
      Set{},Set{} do 
        Set{} 
      end
    end
  end
\end{lstlisting}
\end{itemize}

\section{Working with Syntax}

\begin{figure} 
\begin{center}
\includegraphics[width=12cm]{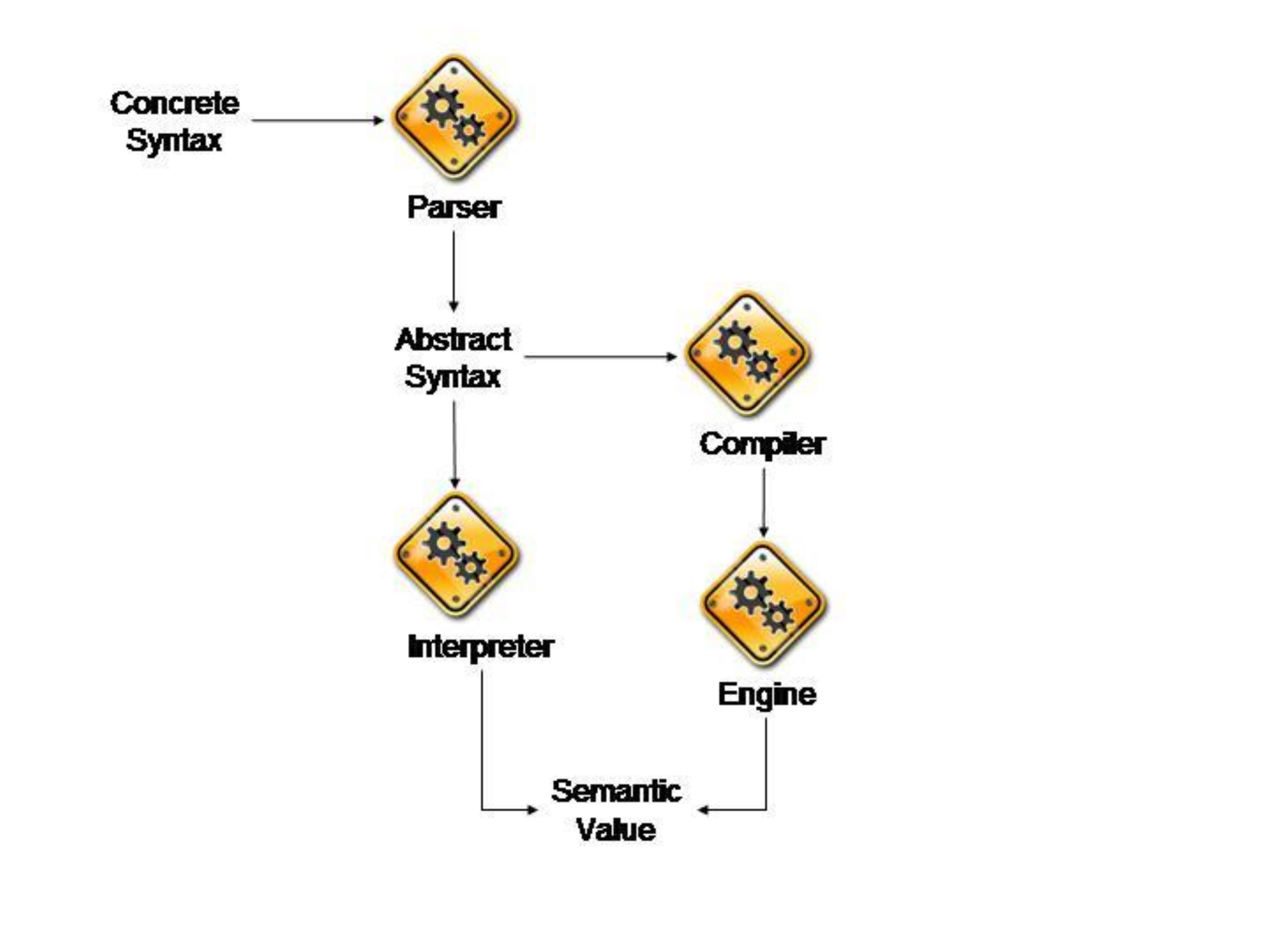}

\caption{Evaluation Process\label{fig:Evaluation-Process}}

\end{center} 
\end{figure}

An element is performable if it can be requested to do two tasks:
compile itself to a sequence of machine instructions and evaluate
itself to produce an element. In many ways these two tasks are similar
since compilation is translation from an element in one language to
an element in another; the instructions in the machine language can
each be asked to evaluate themselves. In order to design and develop
language constructs it is important to understand how to process syntax;
this section describes how XMF represents syntax and the various ways
of transforming and evaluating syntax.

Figure \ref{fig:Evaluation-Process} is an overview of the key syntax
processes. Processing starts with concrete syntax which is represented
as a sequence of characters. The characters may originate from a variety
of sources including files, strings or typed at a keyboard. A parser
is a processing engine that transforms a sequence of characters into
data elements; the data elements are referred to as abstract syntax.

Once syntax is represented in data, it can be processed in many different
ways. It can be transformed from one representation to another. It
can be analysed to see if it contains any errors. It can be executed
directly. The figure shows two different ways of evaluating syntax:
interpretation and compilation.

An interpreter is a program that takes another program as input and
runs it. An interpreter typically requires extra input to describe
the context of evaluation (for example the values for gobal variables
in the program). An interpreter is a program that itself must be written
in a language. The XOCL interpreter is written in XOCL and is easy
to extend.

A compiler is a program that translates a supplied program from a
source language to a target language. Typically the target language
has an interpreter which is somehow better than any interpreter for
the source language. For example the source language may not have
an interpreter or the target interpreter is much faster than any source
interpreter. In this case the XOCL compiler translates XOCL source
into XMF VM instructions for which there is an interpreter written
in Java. The XOCL compiler is written in XOCL and is easy to extend.

Every element that is to be used as syntax must implement the Performable
interface. Performable, requires that a syntax element knows how to
evaluate itself and how to compile itself. The interface is defined
as follows:

\begin{lstlisting}
Performable::compile(env,frame,isLast,saveSource)
// Produces a sequence of machine instructions.
// The env binds variable names that are currently
// in scope to type descriptors. The frame is an
// integer describing how many enclosing operations
// there are. isLast is true when there is nothing 
// further to perform. saveSurce is true when operations
// are to make theor source code available at run-time.
Performable::eval(target,env,imports)
// Evauates the element to produce a value. The target
// is the value of 'self' during evaluation. The 
// environment associates variable names with values.
// The imports are a sequence of imported name spaces.
Performable::FV()
// Returns a set of the free variables in the
// receiver.
Performable::maxLocals()
// Returns the maximum number of loal variables that
// is required in order to evaluate the receiver.
Performable::pprint(out,indent)
// Writes a textual version of the syntax to the
// output channel using indent as the current level
// of indentation after newlines.
\end{lstlisting}
Any new syntax should implement the Performable interface. Fortunately,
it is unusual to have to define all of this interface in practice
because sub-classes of Performable such as XOCL::Sugar shield the
user from the details. However, the following implementation of XOCL::While
shows how to implement a complete interface.

The instructions produced by compiling a while-loop involve skip-instructions.
The class While is a sub-class of Performable with performable attributes
for the while-test and the while-body:

\begin{lstlisting}
context While
  @Operation compile(env,frame,isLast,saveSource)
    let testCode = test.compile(env,frame,false,saveSource);
        bodyCode = body.compile(env,frame,false,saveSource);
        testLabel = Compiler::newLabel();
        endLabel = Compiler::newLabel() then
        returnValue = Compiler::labelInstrs(Seq{PushTrue()},endLabel)
    in Compiler::labelInstrs(testCode,testLabel) + 
       Seq{SkipFalse(endLabel)} + 
       bodyCode + 
       Seq{Pop()} + 
       Seq{SkipBack(testLabel)} + 
       returnValue
    end
  end
\end{lstlisting}
Given the above definition of compile for while-loops, XMF is able
to compile any file that contains a while-loop as part of the source
code. When the source file is compiled, the text is parsed and each
occurrence of concrete syntax for a while-loop is synthesized into
an instance of the class XOCL::While. The resulting syntax for the
complete file is then compiled. As part of the compilation, the while-loop
will perform the operation defined above and return the sequence of
instructions. The complete sequence of instructions for the whole
file is then written out as a binary file. The binary file is read
in at a later stage and the instructions are evaluated (either directly
when the file is read in or via the call of an operation defined in
the file).

Not all occurrences of a while-loop need to be compiled before they
can be performed. Expressions typed at the top-level of XMF are not
compiled, they are parsed to produce syntax objects and then evaluated.
It is also possible to evaluate the source code in a file directly
without having to compile it. The definition of eval for a while-loop
is given below:

\begin{lstlisting}
context While
  @Operation eval(target,env,imports)
    @While test.eval(target,env,imports) do
      body.eval(target,env,imports)
    end
  end
\end{lstlisting}
Although this definition seems at first to be self-referential (while
defined in terms of while), it is important to state that it is a
compiled operation and therefore the use of while in the body of the
operation is compiled using the definition of compile given above.
The key feature to note about the definition of eval is that the test
and body of the while-loop are evaluated using their own definitions
for eval.

A free variable is one that is used by a perfomable element, but which
is not locally defined by that element. In order to support compilation,
each performable elment must define FV. Since a while-loop does not
make direct reference to variables, the free variables are the union
of those for the test and body:

\begin{lstlisting}
context While
  @Operation FV():Set(String)
    test.FV() + body.FV()
  end
\end{lstlisting}
The opposite of free variables are bound variables. Again, in order
to support compilation it is necessary for the compiler to work out
how much local storage a performable element requires. Local variables
are introduced (for example) by a let-expression. A while-loop does
not directly introduce any local variables, but the test and body
might:

\begin{lstlisting}
context While
  @Operation maxLocals():Integer
    test.maxLocals().max(body.maxLocals())
  end
\end{lstlisting}
Finally, if the compiler is instructed to associate the source code
with each compiled operation, each element in the body of an operation
must be able to translate itself into a textual representation. This
is implemented for While as follows:

\begin{lstlisting}
context While
  @Operation pprint(out,indent)
    format(out,"@While ");
    test.pprint(out,indent);
    format(out," do~%~V",Seq{indent + 2});
    body.pprint(out,indent + 2);
    format(out,"~%~Vend",Seq{indent})
  end
\end{lstlisting}
While is a fairly simple example of a performable element. Here is
the complete implementation of a let-expression to give a :

\begin{lstlisting}
context Let
  @Operation compile(env,frame,isLast,saveSource)
    let valueCode = bindings->reverse->collect(b |
          b.value.compile(env,frame,false,saveSource))->flatten;
        letEnv = env.allocateLocals(bindings.name),env.maxLocal())
    in valueCode + 
       // Generate SETLOC instructions...
       letEnv.setLocalsCode(bindings.name) + 
       body.compile(letEnv,frame,isLast,saveSource)
    end
  end
context Let
  @Operation eval(target,env,imports)
    let newEnv = bindings->iterate(b e = env |
      e.bind(b.name,b.value.eval(target,env,imports)))
    in body.eval(target,newEnv,imports)
    end
  end
context Let
  @Operation FV():Set(String)
    bindings->iterate(binding FV = body.FV() - bindings.name->asSet |
      FV + binding.value.FV())
   end
context Let
  @Operation maxLocals():Element
    let valueMaxLocals = bindings->collect(b |
          b.value.maxLocals())->max;
        bindingMaxLocals = bindings->size;
        bodyMaxLocals = body.maxLocals()
    in valueMaxLocals.max(bindingMaxLocals + bodyMaxLocals)
    end
  end
context Let
  @Operation pprint(out,indent)
    format(out,"let ");
    if bindings.isKindOf(Seq(Element))
    then
      @For b in bindings do
        b.pprint(out,indent + 4);
        if not isLast
        then format(out,";~%~V",Seq{indent + 4})
        else format(out,"~%~V",Seq{indent})
        end
      end
    else bindings.pprint(out)
    end;
    format(out,"in ");
    body.pprint(out,indent + 3);
    format(out,"~%~Vend",Seq{indent})
  end
\end{lstlisting}
Defining new syntax classes in XMF requires two things to be associated
with the class: it must be a sub-class of Performable and therefore
implement the Performable interface as described in this section; it
must have a concrete syntax. Concrete syntaxes are defined in terms
of a grammar that synthesizes an instance of the syntax class. Concrete
syntaxes and grammars are the subject of the next chapter. 

The class defining the Object Command Language are defined in section
\ref{sec:OCL-Syntax-Classes}.

Fortunately, it is not usually necessary to define the complete Performable
interface for each new syntax class. This is because a new syntax
class can be defined in terms of a translation to existing syntax
classes that already implement the interface. This process is called
desugaring and is covered in section \ref{sec:Sugar-and-Expressions}.
Desugaring involves trnslating from one abstract syntax to another.
A useful technology to do this is quasi-quotes and this is the subject
of the next section.

\section{Quasi-Quotes and Lifting}

Suppose that we have a requirement for a new syntax for an until-loop.
An until-loop has the usual semantics and the syntax class has a body
and a test, both of which are performable. The syntax class Until
can fully implement the Performable interface, however, there is sufficient
syntax constructs in XOCL to support until-loops in terms of while-loops.
A traslation can be defined as follows:

\begin{lstlisting}
context Until
  @Operation translateToWhile(body,test)
    Order(body,While(test,body))
  end
\end{lstlisting}
The translateToWhile operation simply constructs instances of the
classes Order and While, supplying the appropriate body and test.
With larger syntax translations it becomes difficult to work with
target elements when they are directly constructed as instances of
the syntax classes. The reason is that we are used to seeing syntax
in concrete terms, not as instances of abstract classes. Worse still,
there are actually two uses of syntax in the body of translateToWhile:

\begin{enumerate}
\item The concrete syntax of object construction.
\item The abstract syntax of a command followed by a while-loop.
\end{enumerate}
When examples get a little larger than the one above, these issues
become very confusing. XMF provides technology to address this issue:
quasi-quotes. Quasi-quotes provide a means to construct syntax templates.
The body of transateToWhile can be viewed as a syntax template: the
fixed parts of the template are the ordering and the while-loop; the
variable parts are the supplied body and test. Quasi-quotes allow
the variable parts to be dropped into the fixed parts whih are expressed
using \emph{concrete syntax}. Here is the same operation using quasi-quotes:

\begin{lstlisting}
context Until
  @Operation translateToWhile(body,test)
    [| <body>; 
       @While <test> do 
         <body> 
       end
    |]
  end
\end{lstlisting}
The quotes are {\tt [|} and {\tt |]} surrounding concrete syntax. Within the
concrete syntax, code can be dropped in by surrounding the expressions
producing the code using {\tt <} and {\tt >}. In this case the expressions producing
the code are variables: body and test, but in principle the expressions
inside {\tt <} and {\tt >} can be anything that returns a performable element.

A typical pattern that occurs when working with syntax involves a
sequence of performable elements. Consider translating a sequence
of elements to produce a single expression that performs each expression
in order:
\begin{lstlisting}
@Operation orderCommands(commands:Seq(Performable))
  commands->iterate(command ordered = [| null |] |
    [| <ordered>; <command> |])
end
\end{lstlisting}
Translating syntax often involves the introduction and subsequent
reference of local variables. Introduction of local variables is typically
done via let-expressions and via operation arguments. This is easily
done using quasi-quotes, so long as you remember that variable names
are strings in let-bindings and operation arguments but variable references
are always done using the syntax class OCL::Var. Consider translating
a variable name, a classifier expression and a body expression into
an operation that checks the type of a supplied value, something like:

\begin{lstlisting}
@GuardedOp name(arg:type)
  body
end
\end{lstlisting}
is just like a normal operation except that the type of the value
supplied as arg is dynamically checked. The following translation
produces a normal operation and inserts the check:

\begin{lstlisting}
@Operation guardedOp(name:String,
                     arg:String,
                     type:Performable,
                     body:Performable)
  [| @Operation <name>(<arg>)
       if <Var(arg)>.isKindOf(<type>)
       then <body>
       else <Var(arg)>.error("Not of type " + <type>.toString())
       end
     end
  |]
end
\end{lstlisting}
Suppose that each a syntax construct allows the slots of an object
to be implicitly bound to variables, equivalent to a with-statement
in Pascal. Each variable is introduced using a let-expression:

\begin{lstlisting}
@Operation translateWith(object:Performable,
                         class:Class,
                         body:Performable)
  class.allAttributes()->iterate(a exp = body |
    let name = a.name().toString()
    in [| <name> = <object>.<name> in <exp> end |]
    end
  )
end
\end{lstlisting}
A problem with the translateWith translation is that the object expression
is performed many times. It is usual to introduce a local variable
for the value so that it is performed once. Also, suppose that the
with-translation is to be extended to put the values of the let-bound
variables back into the object at the end of the body:

\begin{lstlisting}
@Operation translateWith(object:Performable,
                         class:Class,
                         body:Performable)
  let A = class.allAttributes() then
      updateSlots = A->iterate(a exp = [| withResult |] |
        let name = a.name().toString()
        in [| withObject.<name> := <Var(name)>; <exp> |]
        end
      ) then
      completeBody = [| let withResult = <body> 
                        in <updateSlots> 
                        end |]
  in  [| let withObject = <object>
         in <A->iterate(a exp = completeBody |
               let name = a.name().toString()
               in [| let <name> = withObject.<name> 
                     in <exp> 
                     end |]
               end
            )>
         end
       |]
  end
end
\end{lstlisting}
Support that Point is a class with slots for the x and y position
in a diagram. A node is to be moved by dx and dy and will return the
new position as a pair:

\begin{lstlisting}
 translateWith(
   [| node.position |],
   Point,
   [| x := x + 1; y := y + 1; Seq{x,y} |])
\end{lstlisting}
The resulting performable element is pprinted as follows:

\begin{lstlisting}
let withObject = node.position
in let x = node.position.x
   in let y = node.position.y
      in let withResult = x := x + 1;
                          y := y + 1;
                          Seq{x,y}
         in withObject.x := x;
            withObject.y := y;
            withResult
         end
      end
   end
end
\end{lstlisting}

\chapter{Grammars}

Computer based languages are used as communication media; either system
to system or human to system. In either case, a computer must be able
to recognise phrases expressed in the language. When phrases are presented
to a computer they must be interpreted in some way. Typically, phrases
may be interpreted directly, as they are recognized, or some internal
representation for the phrases may be synthesized (abstract syntax)
and subsequently passed on to an internal module for processing.

There are a variety of formats by which language may be conveyed to
a computer; a common mechanism is by text. A standard way of defining
how to recognize and process text-based languages is by supplying
a \emph{grammar}. A grammar is a collection of rules that define the
legal sequences of text strings in a language. In addition to recognising
language phrases, a grammar may include actions that are to be performed
as sub-phrases are processed.

In order to use grammars to process language phrases we must supply
the grammar to a program called a \emph{parser}. The medium used to
express the grammar is usually text. BNF (and the extended version
EBNF) is a text-based collection of grammar formation rules. BNF defines
how to express a grammar and what the grammar means in terms of processing
languages. 

This chapter describes the XMF version of EBNF, called XBNF. XBNF
integrates EBNF with XOCL (and any language defined by XBNF). XMF
provides an XBNF parser that is supplied with an XBNF grammar and
will then parse and synthesize XMF-defined languages. XMF is a super-language
because (amongst other features) it supports the definition of new
language constructs that can be weaved into XOCL or used stand-alone.

Grammars (and therefore languages) in XMF are associated with classes.
Typically an XBNF grammar defines how to recognize legal sequences
of characters in a language and to syntheisize an instance of the
associated class (and populate the instance with instances of the
class's attribute types).

\section{Parsing and Syntheisizing}

This section describes the basics of using XBNF to define simple grammars
that recognize languages and synthesize XMF data values. A grammar
is an instance of the XMF class Parser::BNF::Grammar. A grammar consists
of a collection of \emph{clauses} each of which is a rule that defines
a non-terminal of the grammar. A non-terminal is a name (by convention
a non-terminal name starts with an upper case letter).

A clause has the following form:
\begin{lstlisting}
NAME ::= RULE .
\end{lstlisting}
where the RULE part defines how to recognize a sequence of input characters.
To illustrate the essential elements of a grammar definition we will
build a simple caculator that recognises arithmetic expressions and
executes (synthesizes) the expressions (integers) as the parse proceeds.

The grammar is constructed incrementally, the clauses are defined
in the context of an @Grammar:
\begin{lstlisting}
@Grammar
  CLAUSES
end
\end{lstlisting}Typically, a grammar has a starting non-terminal; this is the clause
that is used as the starting point of a parse. There is nothing special
about the starting non-terminal in the definition. In the case of
the calculator the starting non-terminal is Calc:

\begin{lstlisting}
Calc ::= Mult '=' 'end'.
\end{lstlisting}The clause for Calc defines that to recognize this non-terminal a
parse must recognize an Mult (defined below) followed by a \emph{terminal}
'='. A terminal is a sequences of characters in quotes; a terminal
successfully matches an input when the input corresponds to exactly
the characters, possibly preceded by whitespace. Therefore, a Calc
is recognized when a Mult is recognized followed by a =.

A Mult is a multiplicative expression, possibly involving the operators
{*} and /. We use a standard method of defining operator precedence
in the grammar, where addition operators bind tighter than multiplicative
operators. This is achieved using two different non-terminals:

\begin{lstlisting}
Mult ::= n1 = Add ( 
    '*' n2 = Mult { n1 * n2 } 
  | '/' n2 = Mult { n1 / n2 } 
  | { n1 } 
). 
\end{lstlisting}The clause for Mult shows a number of typical grammar features. A
Mult is successfully recognized when a Add is recognized followed
by an optional {*} or / operator.

Each non-terminal synthesizes a value when it successfully recognizes
some input. When one non-terminal \emph{calls} another (for example
Add is called by Mult) the value synthesized by the called non-terminal
may be optionally named in the definition of the calling non-terminal.
Once named, the syntheized value may be referenced in the rest of
the rule. Naming occurs a number of times in the definition of Mult
where the names n1 and n2 are used to refer to syntheisized numbers.

A clause may include optional parts separated by |. In general, if
a clause contains an optional component of the form A | B where A
and B are arbitrary components, then a parse is successful when either
A or B is successful. 

The clause for Mult contains three optional components that occur
after the initial Add is successful (hence the need for parentheses
around the optional components to force the Add to occur). Once the
Add is successful then the parse is successful when one of the following
occurs:

\begin{itemize}
\item a {*} is encountered followed by a Mult.
\item a + is encountered followed by a Mult
\item no further input is processed (this option must occur last since it
subsumes the previous two options -- options are tried in turn).
\end{itemize}
Clauses may contain \emph{actions}. A action is used to synthesize
arbitrary values and can refer to values that have been named previously
in the clause. An action is an XOCL expression enclosed in \{ and
\}. Just like calls to non-terminals in the body of a clause, an action
returns a value that may be named. If the last component performed
by the successful execution of a clause is an action then the value
produced by the action is the value returned by the clause.

The clause for Mult contains a number of actions that are used to
synthesize values (in this case evaluate numeric expressions). The
clause for Add is the same as that for Mult except that the clause
recognizes and synthesises addition expressions:

\begin{lstlisting}
Add ::= n1 = Int ( 
    '+' n2 = Add { n1 + n2 }
  | '-' n2 = Add { n1 - n2 }
  | { n1 } 
). 
\end{lstlisting}The complete grammar definition is given below:

\begin{lstlisting}
@Grammar
  Calc ::= Mult '=' 'end'.
  Mult ::= n1 = Add (
      '*' n2 = Mult { n1 * n2 }
    | '/' n2 = Mult { n1 / n2 }
    | { n1 }
  ).
   Add ::= n1 = Int (
       '+' n2 = Add { n1 + n2 }
    |  '-' n2 = Add { n1 - n2 }
    | { n1 }
  ). 
end
\end{lstlisting}
\section{Sequences of Elements}

A grammar rule may want to specify that, at a given position in the
parse, a sequence of character strings can occur. For example, this
occurs when parsing XML elements, where a composite element can have
any number of children elements each of which is another element.

XBNF provides the postfix operators {*} and + that apply to a clause
component X and define that the parse may recognize sequences of occurrences
of X. In the case of {*}, there may be 0 or more occurrences of X
and in the case of + there may be 1 or more occurrences of X. If X
synthesizes a value each time it is used in a parse then the use of
{*} and + synthesizes sequences of values, each value in the sequence
is a value syntheisized by X.

The following gramar shows an example of the use of {*}. XML elements
are trees; each node in the tree has a label. The following grammar
recognizes XML (without attributes and text elements) trees and synthesizes
an XMF sequence representation for the XML tree. The grammar shows
the use of the builtin non-terminal Name that parses XMF names and
synthesizes the name as a string:

\begin{lstlisting}
@Grammar
  Element ::= 
      SingleElement 
    | CompositeElement. 
  SingleElement ::= 
    '<' tag = Name '/>' 
    { Seq{tag} }.
  CompositeElement ::= 
    '<' tag = Name '>' 
      children = Element* 
    '<' Name '/>' 
    { Seq{tag | children} }. 
end
\end{lstlisting}
\section{Specializing Grammars}

Grammars may be specialized by adding extra clauses or extending existing
clauses. A grammar has an \emph{extends} clause that is followed by
comma separated references to parent grammars. The newly defined grammar
is the merge of the parent grammars and the newly defined clauses.
Any clauses in the parents or in the body of the new definition that
have the same names are merged into a single clause using the | operator.

For example, suppose we want to extend the grammar for XML given in
the previous section with the ability to recognize text elements. A
text element is supplied as a string within string-quotes. XBNF provides
a built-in non-terminal Str for strings. The new grammar is defined
as follows:

\begin{lstlisting}
@Grammar extends XML 
  Element ::= Str. 
end
\end{lstlisting}This is equivalent to the following grammar definition:

\begin{lstlisting}
@Grammar
  Element ::= 
      SingleElement 
    | CompositeElement
    | Str. 
  SingleElement ::= 
    '<' tag = Name '/>' 
    { Seq{tag} }.
  CompositeElement ::= 
    '<' tag = Name '>' 
      children = Element* 
    '<' Name '/>' 
    { Seq{tag | children} }. 
end
\end{lstlisting}
\section{Synthesizing Syntax}

Super-languages must support extensibility through language definition.
When a grammar defines a new language we will often want to synthesize
syntax values that are then passed to a compiler or an interpreter.
An XMF language definition often consists of a grammar that synthesizes
XOCL. Suppose we want to modify the calculator language defined above
to produce the arithmetic expression rather than execute the expressions
directly. Assuming that the OCL package is imported:

\begin{lstlisting}
@Grammar
  Calc ::= Mult '=' 'end'.
  Mult ::= n1 = Add (
      '*' n2 = Mult { BinExp(n1,"*",n2) }
    | '/' n2 = Mult { BinExp(n1,"/",n2) }
    | { n1 }
  ).
  Add ::= n1 = IntExp (
      '+' n2 = Add { BinExp(n1,"+",n2) }
    | '-' n2 = Add { BinExp(n1,"-",n2) }
    | { n1 }
  ). 
  IntExp :: n = Int { IntExp(n) }.
end
\end{lstlisting}Alternatively we might choose to use quasi-quotes:

\begin{lstlisting}
@Grammar
  Calc ::= Mult '=' 'end'.
  Mult ::= n1 = Add (
      '*' n2 = Mult { [| <n1> * <n2> |] }
    | '/' n2 = Mult { [| <n1> / <n2> |] }
    | { n1 }
  ).
  Add ::= n1 = IntExp (
      '+' n2 = Add { [| <n1> + <n2> |] }
    | '-' n2 = Add { [| <n1> - <n2> |] }
    | { n1 }
  ). 
  IntExp ::= n = Int { n.lift() }.
end
\end{lstlisting}
\section{Simple Language Constructs}

Language driven development involves working with and developing high-level
languages that allow us to focus on the the \emph{what}, rather than
the \emph{how} of applications. Languages can be defined in terms
of abstract syntax models or concrete syntax models. Grammars support
the definition of concrete syntax models and are used extensively
throughout the rest of this book.

A grammar for a language construct synthesizes instances of an abstract
syntax model. So long as the model has an execution engine then the
synthesized data can be executed. The syntax model and engine may
be provided, such as XOCL, or may be one that has been defined specially
for the application; the latter case is referred to as a \emph{domain
specific language}. 

This section provides many examples of new language constructs that
are defined using grammars. All of the syntax constructs are provided
as part of the XMF system and implemented as extensions to the basic
XOCL language. In most cases the grammars synthesize instances of
the XOCL model and can therefore be viewed as \emph{syntactic sugar}.
All of the language constructs are general purpose and should not
really be viewed as domain specific; however the techniques can be
used when defining domain specific constructs.

\subsection{When}

When is a simple language construct that behaves as sugar for an if-expression:

\begin{lstlisting}
@When guard do
  action
end
\end{lstlisting}It emphasizes that there is a guarded action and there is no alternative
action. Although there is no semantic difference between a when and
an if, it can be useful when reading code to see what was in the mind
of the developer. The grammar is as follows:

\begin{lstlisting}
@Grammar extends OCL::OCL.grammar
  When ::= guard = Exp 'do' action = Exp {
    [| if <guard> 
       then <action> 
       else "GUARD FAILS" 
       end 
    |]
  }.
end

\end{lstlisting}
\subsection{Cond}

Sometimes, if-expressions can become deeply nested and difficult to
read. A cond-expression is equivalent to a nested if-then-else, but
promotes all of the if-then parts to the top level:

\begin{lstlisting}
@Cond
  test1 do exp1 end
  test2 do exp2 end
  ...
  testn do expn end
  else exp
end
\end{lstlisting}The implementation of this construct provides an example of using
syntactic sugar. A class that inherits from XOCL::Sugar must implement
an operation desugar that is used to transform the receiver into a
performable element. These classes are often useful when the transformation
is complex and is better done as a separate step after the abstract
syntax synthesis. The grammar for the cond-expression is as follows:

\begin{lstlisting}
@Grammar extends OCL::OCL.grammar
  Cond ::= clauses = Clause* otherwise = Otherwise 'end' {
    Cond(clauses,otherwise)
  }.
  Clause ::= guard = Exp 'do' body = Exp 'end' {
    CondClause(guard,body)
  }.
  Otherwise ::= 'else' Exp | 
    { [| self.error("No else clause.") |] }.
end
\end{lstlisting}Cond inherits from XOCL::Sugar. A Cond has attributes, clauses (of
type Seq(CondClause)) and otherwise (of type Performable) and has
a desugar operation as follows:

\begin{lstlisting}
@Operation desugar():Performable
  clauses->reverse->iterate(clause exp = otherwise |
    [| if <clause.guard()>
       then <clause.body()>
       else <exp>
       end
    |])
end
\end{lstlisting}
\subsection{Classes and Packages}

Classes and packages are name spaces: they both contain named elements.
Although name spaces can contain any type of named element, packages
and classes are examples of name spaces that know about particular
types of named element. Packages know about sub-packages and classes.
Classes know about attributes, operations and constraints. 

To allow extensibility, name space definitions contain sequences of
element definitions. An element is added to a name space through the
add operation. Packages and classes are examples of name spaces that
hijack the add operation in order to manage particular types of known
named elements. The rest of this section defined package and class
definition language features. 

A second feature of name spaces is mutual-reference. Definitions within
a name space may refer to each other and the references must be resolved.
For example, this occurs when attribute types refer to classes in
the same package and when classes inherit from each other in a package
definition. One way to achieve resolution is via two-passes. The named
elements are added to the name space but all references are left unresolved.
A second pass then replaces all references with direct references
to the appropriate named elements. This section shows how this two-pass
system can be implemented.

Firstly, consider a package definition:

\begin{lstlisting}
@Package P
  @Class A
    @Attribute b : B end
  end
  @Class B extends A
    @Attribute a : A end
  end
end
\end{lstlisting}This definition is equivalent to the following expression (ignoring
the translation of the classes):

\begin{lstlisting}
  Package("P")
    .add(@Class A ... end)
    .add(@Class B ... end)
\end{lstlisting}The grammar for package definition is as follows:

\begin{lstlisting}
@Grammar extends OCL::OCL.grammar
  Package ::= n = Name defs = Exp* 'end' {
    defs->iterate(def package = [| Package(<n.lift()>) |] |
      [| <package>.add(<def>) |])
  }.
end
\end{lstlisting}Both class and attribute definitions contain references to named elements
that may be defined at the same time. For example in the package definition
P above, class A refers to B and vice versa. References are delayed
until all the definitions are created. Since they have been delayed,
the references must be resolved. But references cannot be resolved
until the outermost definition has been completed. An outermost definition
is introduced by a 'context' clause, therefore this is the point at
which resolution can be initiated. All elements have an init operation
that is used to initialise elements; name spaces hijack the init operation
so that delayed references can be resolved.

For example, attributes are defined by the following grammar (simplified
so that types are just names):

\begin{lstlisting}
@Grammar
  Attribute::= n = Name ':' t = Name 'end' {
    [| Attribute(<n.lift()>).type := <t.lift()> |]
  }.
end
\end{lstlisting}
The init operation defined for attributes is:

\begin{lstlisting}
@Operation init()
  if type.isKindOf(String)
  then self.type := owner.resolveRef(type)
  end;
  super()
end
\end{lstlisting}
Classes are created as follows:

\begin{lstlisting}
@Grammar extends OCL::OCL.grammar
  Class ::= n = Name ps = Parents defs = Exp* 'end' {
    ps->iterate(p classDef = 
      defs->iterate(def classDef = [| Class(<n.lift>) |] 
      | [| <classDef>.add(<def>) |]) 
    | [| <classDef>.addParent(<p.lift>) |])
  }.
  Parents ::=
    'extends' n = Name ns = (',' Name)* { Seq{n|ns} }
  | { Seq{"Object"}.
end
\end{lstlisting}
The context rule is as follows:

\begin{lstlisting}
  Context ::= 'context' nameSpace = Exp def = Exp {
    [| let namedElement = <def>;
           nameSpace = <nameSpace>
       in nameSpace.add(namedElement);
          namedElement.init()
       end |]
  }.
\end{lstlisting}Attributes in XOCL, usually occor inside class definitions and have
the following syntax:

\begin{lstlisting}
@Attribute <NAME> : <TYPE> [ = <INITEXP> ] [ <MODIFIERS> ] end
\end{lstlisting}The type of an attribute is an expression that references a classifier.
Usually this is done by name (if the classifier is in scope) or by
supplying a name-space path to the classifier. If the type is a collection
(set or a sequence) then it has the form Set(T) or Seq(T) for some
type T. When the attribute is instantiated to produce a slot, the
slot must be initialised. The initial value for classes is null. Each
basic type (String, Integer, Boolean, Float) has a predefined default
value ({}``'', 0, false, 0.0). The default value for a collection
type is the empty collection of the appropriate type. A specific attribute
may override the default value by supplying an initialisation expression
after the type.

An attribute within a class definition may supply modifiers. The modifiers
cause the slot to have properties and create operations within the
class. Attribute modifiers have the following form:

\begin{lstlisting}
'(' <ATTMOD> (',' <ATTMOD>)* ')'
\end{lstlisting}where an attribute modifier may be one of: ?, !, +, -, \textasciicircum{}.
The meaning of these modifiers is given below:

\begin{itemize}
\item ? defines an accesor operation for the attribute. For an attribute
named n, an operation named n() is created that returns the value
of the slot in an instance of the class.
\item ! defines an updater operation for the attribute. For an attribute
named n, an operation named setN(value) is created that sets the value
of the slot named n to value in an instance of the class.
\item + defines an extender operation for the attribute providing that the
type is a collection. For an attribute named n, an operation named
addToN(value) is created that extends the collection in the slot of
an instance of the class. If the type of the attribute is a sequence
type then the new value is added to the end of the sequence.
\item - defines a reducer operation for the attribute providing that the
type is a collection. For an attribute named n, an operation named
deleteFromN(value) is created that extends the collection in the slot
of an instance of the class.
\item \textasciicircum{} makes the slot a container providing that the type
inherits from XCore::Contained. In this case values in the slot will
be contained objects with a slot named owner. Using the updater and
extender operations to update the slot defined by this attribute will
modify the owner slot when the value is added. It is often desirable
to maintain linkage between a contained value and its owner, this
attribute modifier manages this relationship automatically providing
that the slot is always modified through the updater and extender.
\end{itemize}
The grammar for attribute definitions is given below. The grammar
synthesizes an instance of the syntax class XOCL::Attribute that is
then expanded to produce an expression that creates a new attribute
in addition to being used by a class definition to add in attribute
modifier operations:

\begin{lstlisting}
@Grammar extends OCL::OCL.grammar
  Attribute ::= 
    name = AttName       // Name may be a string.
    meta = MetaClass     // Defaults to XCore::Attribute
    ':' type = AttType   // An expression.
    init = AttInit       // Optional init value.
    mods = AttMods       // Optional modifiers.
   'end' { Attribute(name,mult,type,init,mods,meta) }.
  AttInit ::= 
    // Init values may be dynamic or static. A
    // dynamic init value is evaluated each time
    // a slot is created. A static init value is
    // evaluated once when the attribute is
    // created and used for all slots.
    '=' Exp 
  | '=' 'static' e = Exp { [| static(<e>) |] } 
  | { null }.
  AttMods ::= 
    mods = { AttributeModifiers() } 
    [ '(' AttModifier^(mods) (',' AttModifier^(mods))* ')' ] 
    { mods }.
  AttModifier(mods) ::= 
    mod = AttMod { mods.defineModifier(mod) }.
  AttMod ::= 
    '?' { "?" } 
  | '!' { "!" } 
  | '+' { "+" } 
  | '-' { "-" } 
  | '^' { "^" }.
  AttType ::= n = AttTypeName AttTypeTail^(n).
  AttTypeName ::= 
    n = Name ns = ('::' Name)* { Path(Var(n),ns) }.
  AttTypeTail(n) ::= 
    '(' args = CommaSepExps ')' { Apply(n,args) } 
  | { n }.
  AttName ::= Name | Str.
  MetaClass ::= 'metaclass' Exp | { null }.
end 
\end{lstlisting}
\subsection{Operations}

An operation has a basic form of:

\begin{lstlisting}
@Operation [ <NAME> ] <ARGS> <BODY> end
\end{lstlisting}The operation name may be optional and defaults to anonymous. The
following extensions to the form of basic operation definition are
noted:

\begin{itemize}
\item Operations have properties (allowing them to behave like objects with
slots). The properties are optional and specified between the operation
named and its arguments.
\item Arguments have optional types, specified as : <EXP> following the
argument name.
\item Argument names may be specified as XOCL patterns. The pattern matches
the supplied argument value.
\item Arguments are positional except in the case of an optional multiargs
specification as the last argument position. A multiargs specification
follows the special designator '.'. This is a mechanism that allows
operations to take any number of arguments. Arguments are matched
positionally and it is an error to supply less than the number of
declared positional arguments to an operation. If more than the expected
positional arguments are supplied and if the operation has a multiargs
designator then all the remaining supplied argument values are supplied
as a sequence and matched to the multiargs pattern.
\end{itemize}
The grammar for operation definition is shown below. the grammar synthesizes
an instance of XOCL::Operation (via mkOpDef) that is then analysed
and expanded to an operation creation expression:

\begin{lstlisting}
@Grammar extends OCL::OCL.grammar     
  Operation ::=        
    name = OpName 
    properties = Properties        
    '(' args = OpArgs multi = OpMulti ')'       
    type = ReturnType       
    body = Exp+        
    'end'       
    { mkOpDef(name,properties,args,multi,type,body) } .
  OpName ::= 
    Name 
  | { Symbol("anonymous") }.     
  OpArgs ::= 
    arg = Pattern args = (',' Pattern)* { Seq{arg | args } } 
  | { Seq{} }.     
  OpMulti ::= 
    '.' multi = Pattern { Seq{multi} } 
  | { Seq{} }.     
  ReturnType ::= 
    ':' TypeExp 
  | { NamedType() }.     
  Properties ::= 
    '[' p = Property ps = (',' Property)* ']' { Seq{p|ps} } 
  | { Seq{} }.     
  Property ::= n = Name '=' e = Exp { Seq{n,e} }.
\end{lstlisting}
\subsection{Enum}

Enumerated types occur frequently in models. An enumerated type consists
of a collection of predefined values; each value is classified by
the enumerated type and is different from all other values of the
type and different from all values of any other type. Examples include
Colour with values Red, Green and Blue. Another example is Directions
with values North, East, West and South. Here is an example of an
enumerated type and its usage:

\begin{lstlisting}
context Root
  @Enum Colour(Red,Green,Amber) end

context Root
  @Enum Position(North,South,East,West) end

context Root
  @Class TrafficLight
    @Attribute colour : Colour (?,!) end
    @Attribute position : Position (?,!) end
    @Constructor(colour,position) ! end
  end

context Root
  @Class RoadCrossing
    @Attribute light1 : TrafficLight 
      = TrafficLight(Colour::Red,Position::North) 
    end
    @Attribute light2 : TrafficLight 
      = TrafficLight(Colour::Red,Position::South) 
    end
    @Attribute light3 : TrafficLight 
      = TrafficLight(Colour::Green,Position::East) 
    end
    @Attribute light4 : TrafficLight 
      = TrafficLight(Colour::Green,Position::West) 
    end
    @Operation cycle()
      // Cycle the traffic lights...
    end
  end
\end{lstlisting}Many systems implement enumerated types as strings or integers. This
is an approximation to the above specification for enumerated types
since, if Red = 0 and West = 0 then both Colour and Direction are
really Integer and Red can easily be confused with West. An enumerated
type should be a classifier, and values of the type should be classified
by it. 

This is a good example of the use of meta-classes to implement new
types. The following class definition declares a class Enum; its definition
shows a number of useful features when declaring new meta-classes.
Firstly the definition:

\begin{lstlisting}
context XCore
  @Class Enum extends Class
    @Attribute names : Seq(String) (?) end
    @Constructor(name,names) ! 
      @For name in names do
        self.addName(name)
      end
    end
    @Operation default()
      self.getElement(names->head)
    end
    @Operation defaultParents():Set(Classifier)
      Set{NamedElement}
    end
    @Operation add(n)
      if n.isKindOf(String)
      then self.addName(n)
      else super(n)
      end
    end
    @Operation addName(name:String)
      self.names := names->including(name);
      self.add(self(name))
    end
  end
\end{lstlisting}Enum is a sub-class of XCore::Class. this means that instances of
Enum are classes that have their own instances - the values of the
enumerated type. Therefore, Colour is an instance of Enum and Red
is an instance of Colour. The class Enum has an attribute names that
is used to hold the declared names of its instances. When a new instance
of Enum is created, it is supplied with a name (Colour) and a collection
of names (Red,Green,Amber). 

In addition to the names being recorded as the value of the slot names,
each name gives rise to an instance of the enumerated type. Each instance
of the enumerated type is a new value that has a name, therefore Colour
must inherit from NamedElement. When a meta-class is defined it may
redefine the operation inherited from Class called defaultParents.
This must return a set of classifiers that instances of the meta-class
will inherit from by default. In the case of Enum, any instance of
Enum is declared to inherit from NamedElement, therefore the following
creates two named elements that are instances of the class Colour:
Colour({}``Red'') and Colour({}``Green'').

The operation Enum::addName is called for each of the names declared
for the enumerated type. An instance of Enum is a class therefore
it is a name-space and can contain named elements. the operation addName
created an instance of the enumerated type self(name) and then adds
it to itself. Each name declared for the enumerated type gives rise
to a new instance which is named and is guaranteed to be different
(when compared with =) to any other value.

A grammar for Enum completes the definition of enumerated types. Notice
that, since an enumerated type is a class it can contain attributes
and operations in its body:
\begin{lstlisting}
@Grammar extends OCL::OCL.grammar
  Enum ::= n = Name ns = Names es = Exp* 'end' {
    exps->iterate(e x = [| Enum(<name.lift()>,<names.lift()>) |] | 
      [| <x>.add(<e>) |])
  }.
  Names ::= '(' n = EnumName ns = (',' EnumName)* ')' { 
    Seq{n | ns} 
  }.
  EnumName ::= 
    Name
  | Str.
end
\end{lstlisting}

\section{Looping Constructs}

A @For-loop has a syntax construct defined by XOCL::For. 
When you define a @For-loop, the system creates an instance 
of XOCL::For and then desugars it. The @For-loop is a good 
example of a fairly extensive syntax transformation since 
the construct has many different forms controlled by a number 
of keywords. This section defines the syntax of a @For-loop 
and provides the definition of one of the syntax transformations.

The class XOCL::For has the following attribites:
\begin{lstlisting}
// The controlled variables...
@Attribute names     : Seq(String)      end  
// The type of for loop...
@Attribute directive : String           end  
// A guard on the bound element (null = true)...
@Attribute guard     : Performable      end  
// The collections...
@Attribute colls     : Seq(Performable) end  
// The body performed in the scope of 'name'...
@Attribute body      : Performable      end  
// Returns a value...
@Attribute isExp     : Boolean          end  
\end{lstlisting}
The @For-expression has the following grammar:
\begin{lstlisting}
@Grammar extends OCL::OCL.grammar 
   For ::=  
     // A for-loop may bind a sequence of names...
     names = ForNames     
     // The directive defines how to interpret the
     // collections...    
     dir = ForDirective     
     // A collection for each of the names...    
     colls = ForColls     
     // An optional boolean expression that
     // must be true in order to process
     // the names...    
     guard = ('when' Exp | { null })    
     // We are either performing commands or
     // producing a sequence of values...     
     isExp = ForType     
     // The body - either a command or an 
     // expression..   
     body = Exp    
     'end'     
     { For(names,dir,colls,guard,body,isExp) }. 
       
   ForColls ::= exp = Exp exps = (',' Exp)* { Seq{exp | exps} }.
    
   ForDirective ::= 
     // The directive controls how the collections
     // are processed...  
     // A walker produces instances of named types...   
     'classifiedBy'  { "classifiedBy" }     
     // Select elements from the collections...       
   | 'in'            { "in" }           
     // Reverse the collections...                 
   | 'inReverse'     { "inReverse" }         
     // Get keys from a table...           
   | 'inTableKeys'   { "inTableKeys" }         
     // Get values from a table...         
   | 'inTableValues' { "inTableValues" }       
     // Call ->asSeq...        
   | 'inSeq'         { "inSeq" }.
         
   ForNames ::=      
     name = AName       
     names = (',' AName)*        
     { Seq{name | names} }.
         
   ForType ::=      
     'do' { false }        
   | 'produce' { true }.
    
 end 
\end{lstlisting}
Each of the different @For-directives defines a different desugar 
operation for the components of a @For-expression abstract syntax 
construct. The following shows how a basic @For-loop using the 'in' 
directive is desugared:
\begin{lstlisting}
context XOCL::For
  @Operation desugarInAction():Performable
     [| let <self.collBindings()>
        in let isFirst = true
           in @While not <self.emptyCollCheck()> do
                let <self.bindNames()>
                in <self.takeTails()>;
                   let isLast = <self.emptyCollCheck()>
                   in <self.protect(body)>;
                      isFirst := false
                   end
                end
              end
           end
        end
     |]
  end
\end{lstlisting}
The operations used by desugarInAction are defined below:
\begin{lstlisting}
context XOCL::For
  @Operation bindNames():Seq(ValueBinding)       
     // Bind each of the controlled variables to the head of the
     // corresponding collection.      
     0.to(names->size-1)->collect(i |
       ValueBinding(names->at(i),[| <Var("forColl" + i)> ->head |]))        
   end

context XOCL::For
  @Operation collBindings():Seq(ValueBinding)     
     // Returns a sequence of bindings that guarantee the collections
     // are all sequences...       
     0.to(colls->size-1)->collect(i |
       ValueBinding("forColl" + i,[| <colls->at(i)> ->asSeq |]))
   end

context XOCL::For
  @Operation emptyCollCheck():Performable     
    // Check that any of the collection variables is empty...     
    1.to(colls->size-1)->iterate(i test = [| <Var("forColl0")> ->isEmpty |] |
      [| <test> or <Var("forColl" + i)> ->isEmpty |])       
  end

context XOCL::For
  @Operation protect(exp:Performable)  
    // If the guard is defined then check it before performing
    // the expression. Otherwise perform the expression. If
    // we are an expression then add the result of the expression
    // to the for results; these will be returned at the end of
    // the evaluation.    
    if guard = null
    then 
      if isExp
      then [| forResults := Seq{<exp> | forResults} |]
      else exp 
      end
    else 
      if isExp
      then [| if <guard> 
              then forResults := Seq{<exp> | forResults} 
              end |]
      else [| if <guard> 
              then <exp> 
              end |]
      end
    end
  end

context XOCL::For
  @Operation takeTails():Performable
    1.to(colls->size-1)->iterate(i updates = [| <Var("forColl0")> := <Var("forColl0")> ->tail |] |
      [| <updates>; <Var("forColl" + i)> := <Var("forColl" + i)> ->tail |])
  end
\end{lstlisting}

\section{XBNF}

This section describes the syntax for defining XBNF grammars. Since
XMF is a super-language, it can be used to define languages. XBNF
is just a language and can therefore be defined in XMF. One of the
goals of a super-language is meta-circularity. If you look at the
source code of XMF you will find a definition of the XBNF grammar
written in the XBNF language as follows (start reading at the clause
called Grammar):

\begin{lstlisting}
@Grammar extends OCL::OCL.grammar
    XBNF_Action ::=   
      // An action is either an expression in { and } 
      // which synthesizes a value or is a predicate that
      // must be true for the parse to proceed...     
      '{' exp = Exp '}' 
      { PreAction(exp) } 
    | '?' boolExp = Exp
      { PrePredicate(boolExp) }.
    
    XBNF_Atom ::=    
      // An atom is the basic unit of a clause...     
      XBNF_Action    
    | XBNF_Literal   
    | XBNF_Call      
    | '(' XBNF_Disjunction ')'.
  
    XBNF_Binding ::=    
      // A clause binding performs a grammar action and
      // associates the value produced with a named 
      // local...     
      name = Name '=' atom = XBNF_Sequence { 
        And(atom,Bind(name)) 
      }.
    
    XBNF_Call ::=    
      // Call a clause. The arguments are optional...     
      name = Name args = XBNF_CallArgs { Call(name,args) }.
      
    XBNF_CallArgs ::=     
      // Arguments supplied to a clause are optional.
      // The args must be preceded by a ^ to distinguish
      // the args from a pair of calls with 0 args...   
      '^' '(' n = Name ns = (',' Name)* ')' { Seq{n|ns} } 
    | { Seq{} }.
  
    XBNF_Clause ::= 
      // A clause is a named rule for parsing...   
      name = Name args = XBNF_ClauseArgs '::=' 
      body = XBNF_Disjunction '.' 
      { Clause(name,args,body) }. 
  
    XBNF_ClauseArgs ::=    
      '(' n = Name ns = (',' Name)* ')' { Seq{n|ns} } 
    | { Seq{} }.
    
    XBNF_Conjunction ::=    
      // Conjunction is just a sequence of 1 or more
      // clause elements...    
      elements = XBNF_Element+ { 
       elements->tail->iterate(e conj = elements->head | 
         And(conj,e)) 
    }. 
  
    XBNF_Disjunction ::=    
      // A disjunction is a sequence of elements
      // separated by | ...   
      element = XBNF_Conjunction (
        '|' rest = XBNF_Disjunction { Or(element,rest) } 
      | { element }).
   
    XBNF_Element ::=     
      XBNF_Optional    
    | XBNF_Binding   
    | XBNF_Sequence.
  
    Grammar ::=     
      parents = XBNF_GrammarParents
      imports = XBNF_GrammarImports
      clauses = XBNF_Clause* 
      'end'
      { Grammar(parents,clauses->asSet,"",imports) }.
  
    XBNF_GrammarImports ::=    
      // The imports of a grammar affect the grammars that are
      // available via @...    
      'import' class = Exp classes = (',' Exp)* { Seq{class | classes} } 
    | { Seq{} }.
  
    XBNF_GrammarParents ::=     
      // A grammar may inherit from 0 or more parent grammars.
      // The parent clauses are added to the child...   
      'extends' parent = Exp parents = (',' Exp)* 
      { parents->asSet->including(parent) } 
    | { Set{} }.
 
    XBNF_Literal ::=     
      // The following literals are built-in non-terminals of a
      // grammar. The action uses getElement to reference the
      // classes (and therefore the constructors) because a grammar
      // cannot reference a variable with the same name as a terminal
      // in an action...     
      // Get the next character...
    
      'Char'       { (Parser::BNF.getElement("Char"))() }          
      // Get the next line...       
    | 'Line'       { (Parser::BNF.getElement("Line"))() }         
      // Get a string...      
    | 'Str'        { (Parser::BNF.getElement("Str"))() }          
      // Get a terminal (in ' and ')...      
    | 'Terminal'   { (Parser::BNF.getElement("Term"))() }         
      // Return the current token...     
    | 'Token'      { (Parser::BNF.getElement("Tok"))() }     
      // Get an integer...           
    | 'Int'        { (Parser::BNF.getElement("Int"))() }         
      // Get a float...      
    | 'Float'      { (Parser::BNF.getElement("Float"))() }       
      // Get a name...     
    | 'Name'       { (Parser::BNF.getElement("Name"))() }    
      // Expect end-of-file...         
    | 'EOF'        { (Parser::BNF.getElement("EOF"))() }         
      // Throw away all choice points created since starting
      // the current clause...      
    | '!'          { (Parser::BNF.getElement("Cut"))() }       
      // Dispatch to the grammar on the most recently
      // synthesized value which should be a sequence of
      // names the represent a path to a classifier with
      // respect to the currently imported name-spaces...         
    | '@'          { (Parser::BNF.getElement("At"))() }          
      // Add a name-space to the currently imported 
      // name-spaces...       
    | 'ImportAt'   { (Parser::BNF.getElement("ImportAt"))() }     
      // Get the current state of the parsing engine...      
    | 'pState'     { (Parser::BNF.getElement("PState"))() }      
      // Get the current line position...     
    | 'LinePos'    { (Parser::BNF.getElement("LinePos"))() }     
      // Define a new terminal in the form NewToken(NAME,ID)...     
    | XBNF_NewToken                                   
      // Get a terminal name...                      
    | terminal = Terminal { (Parser::BNF.getElement("Terminal"))(terminal) }.
    
    XBNF_NewToken ::=    
      // A new token is defined as a name and an integer id.
      // The tokenizer used to parse the grammar is responsible
      // for returning a token with the type set to the id...   
      'NewToken' '(' n = Name ',' t = Int ')' {
        (Parser::BNF.getElement("NewToken"))(n,t) 
    }.
 
    XBNF_Optional ::=     
      // An optional clause element is tried and ignored if it fails...      
      '[' opt = XBNF_Disjunction ']'
      { Opt(opt) }.
  
    XBNF_Path ::= name = Name names = ('::' Name)* { Seq{name | names} }.
    
    XBNF_TypeCheck ::=    
      // An element that checks the type of the synthesized value...    
      element = XBNF_Atom (':' type = XBNF_Path { And(element,TypeCheck(type)) } 
    | { element }).
  
    XBNF_Sequence ::=    
      // An element an be followed by a * or a + for 0
      // or more and 1 or more repetitions...    
      element = XBNF_TypeCheck ( 
        '*' { StarCons(element) } 
      | '+' { PlusCons(element) } 
      | { element }
      ).
    
end
\end{lstlisting}

\part{Programming Examples}

\chapter{Executable Models and Simulation}

Super-languages must be expressive. Apart from language engineering,
super-languages play an important role in the design of systems by
providing a technology for building prototypes or executable models.
This chapter shows how XMF can be used to construct an executable
model of a job-shop system and then be used to perform some analysis
of the modelled system.

The simplest sketch of a system on the back of a napkin is a model
that can tell us something about the intended structure and behaviour
of the system. Models tend to fall into distinct categories that reflect
their intent. Models of usage (such as use-case models) try to capture
the client view of the system, i.e. what does the system offer? Archtecture
models try to capture the major components of the system and how they
will communicate. Data models try to capture the information that
the system processes. Dynamic models try to produce descriptions of
how the system behaves under certain conditions. In all cases, these
models may be constructed as lightweight sketches that ignore implementation
detail or constructed as detailed blueprints including key implementation
features.

To be effective, modelling must be integrated with a project development
method. There are many such methods; one significant method is to
use modelling to produce a complete executable simulation of the system.
The model is representative of the real system and is used to run
test cases. The model does not need to deal with complex implementation
details. Once the executable model has been developed and used to
analyse the behaviour of the required system, it can be used as an
executable blueprint for the real system.

There are a number of advantages to this approach: results can be
produced in a short time; requirements engineering can involve presenting
stakeholders with a simulation of the required system; the languages
used for executable modelling support good debugging and instrumentation;
the languages provide high level support for manipulating data and
representing key application components; using the executable model
as a blueprint significantly reduces the gap between design and implementation.

In order to simulate a system, a model must include the actions that
the system performs. There are a variety of ways of modelling actions
including state machines and model based action languages. Modelling
actions falls broadly into two categories that differ depending on
whether the model is to be executable or whether it is to be used
to analyse system executions. The former allows the model to be run.
The latter statically describes the properties of executions.

Section \ref{sec:A-Job-Shop-Simulation} specifies a simple job-shop
application and shows how the simulator will behave. Section \ref{sec:Job-Shop-Implementation}
describes how the simulation is implemented. Finally, section \ref{sec:Conflict-Analysis}
describes how dynamic execution behaviours can be modelled and subsequently
analysed.

\section{A Job-Shop Simulation}

\begin{figure} \begin{center}
\includegraphics[width=12cm]{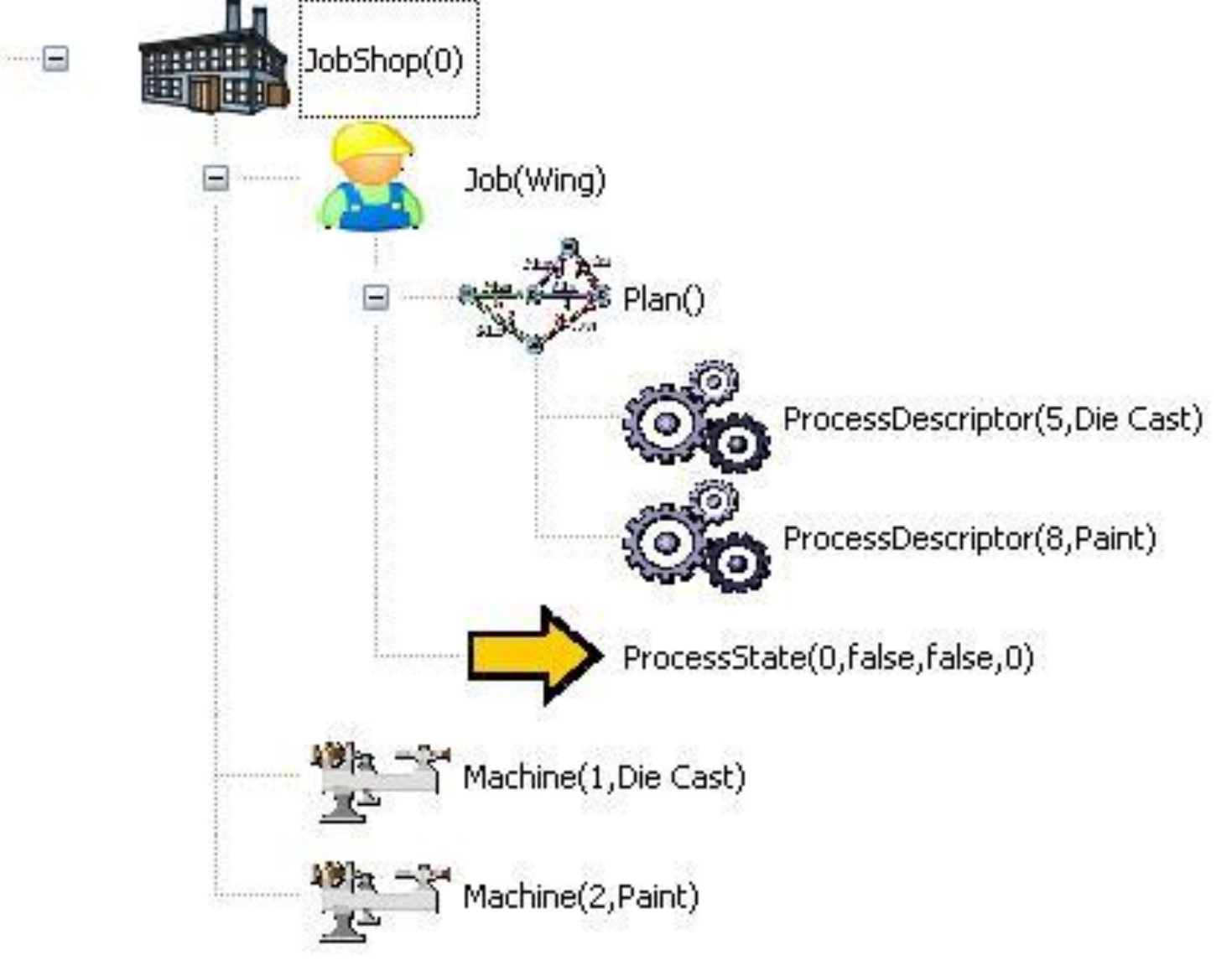}

\caption{A Job Shop Application\label{fig:A-Job-Shop-Application}}

\end{center} \end{figure}

\begin{figure}
\begin{center}

\subfigure[Initial State of the Job Shop]{\includegraphics[width=4cm]{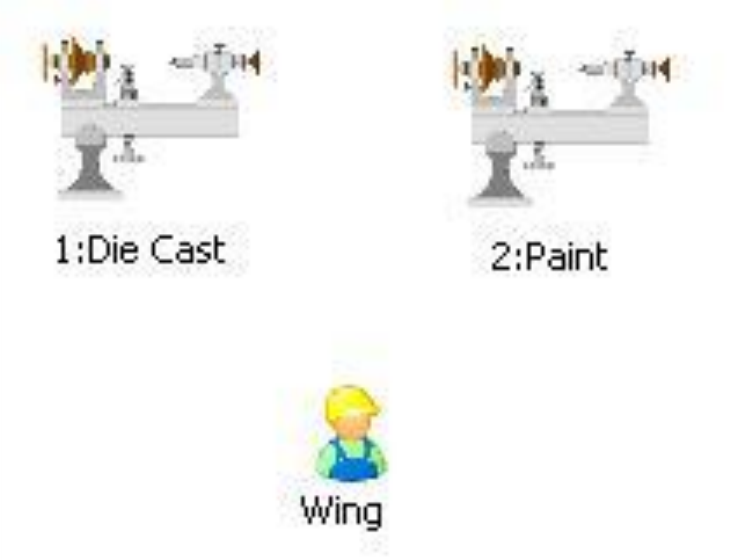}}

\subfigure[Allocated to the Die Cast Machine]{\includegraphics[width=4cm]{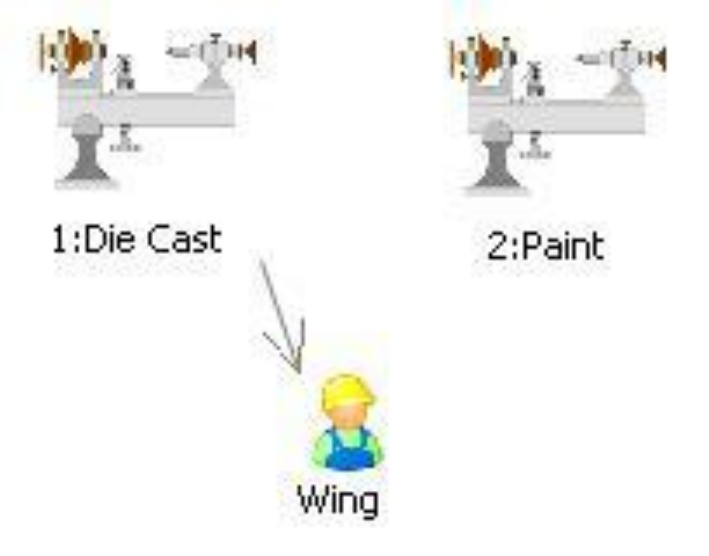}}

\subfigure[Allocation to the Paint Machine]{\includegraphics[width=4cm]{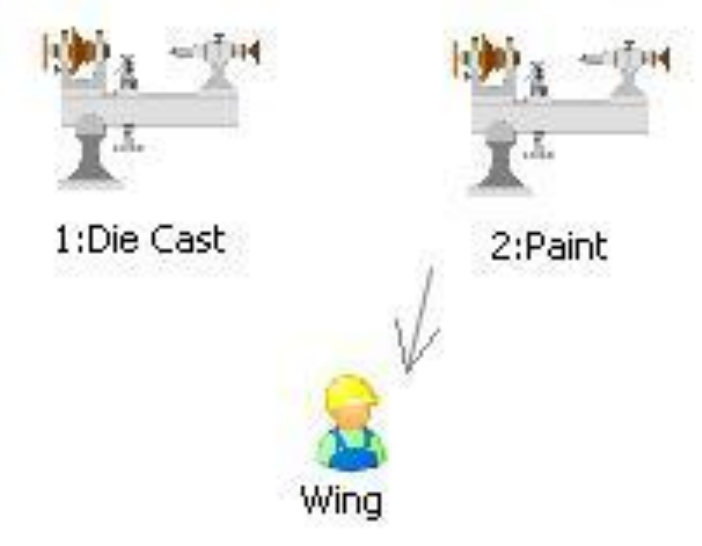}}

\subfigure[Final State]{\includegraphics[width=4cm]{Programming/Simulation/Images/Wing1.pdf}}

\caption{Job Shop Simulation\label{fig:Job-Shop-Simulation-Steps}}

\end{center}
\end{figure}

Scheduling jobs is a typical example of an application that can benefit
from simulation. There are many ways in which jobs can be scheduled
to machines that process the jobs and the behaviour of a scheduling
application can change depending on many different features of the
application. Opportunistic scheduling is a simple type of scheduling
whereby jobs are allocated to machines as soon as the machine becomes
available without any fancy analysis of the best use of resources.
This section describes an executable model for opportunistic scheduling.

Figure \ref{fig:A-Job-Shop-Application} shows a simple jobs shop
as a tree of components. The job shop contains jobs that must be processed
and machines that can process the jobs. In this case the application
is to process an aircraft wing. The single job named Wing has a plan
that specifies a sequence of processes that must be performed in order.
The wing must be case by a die casting machine and then painted by
a painting machine. The job-shop has exactly one machine of each type
(therefore the task is do-able).

Figure \ref{fig:Job-Shop-Simulation-Steps} shows the steps in performing
this simple simulation. The initial state of the jobs shop shows that
the Wing job is not allocated to any machine. The job is then allocated
to the Die Case machine for 5 units of time after which is switches
to the Paint machine for 8 units of time and finally the job is complete
and not allocated to any machines.

\begin{figure}
\begin{center}

\includegraphics[width=12cm]{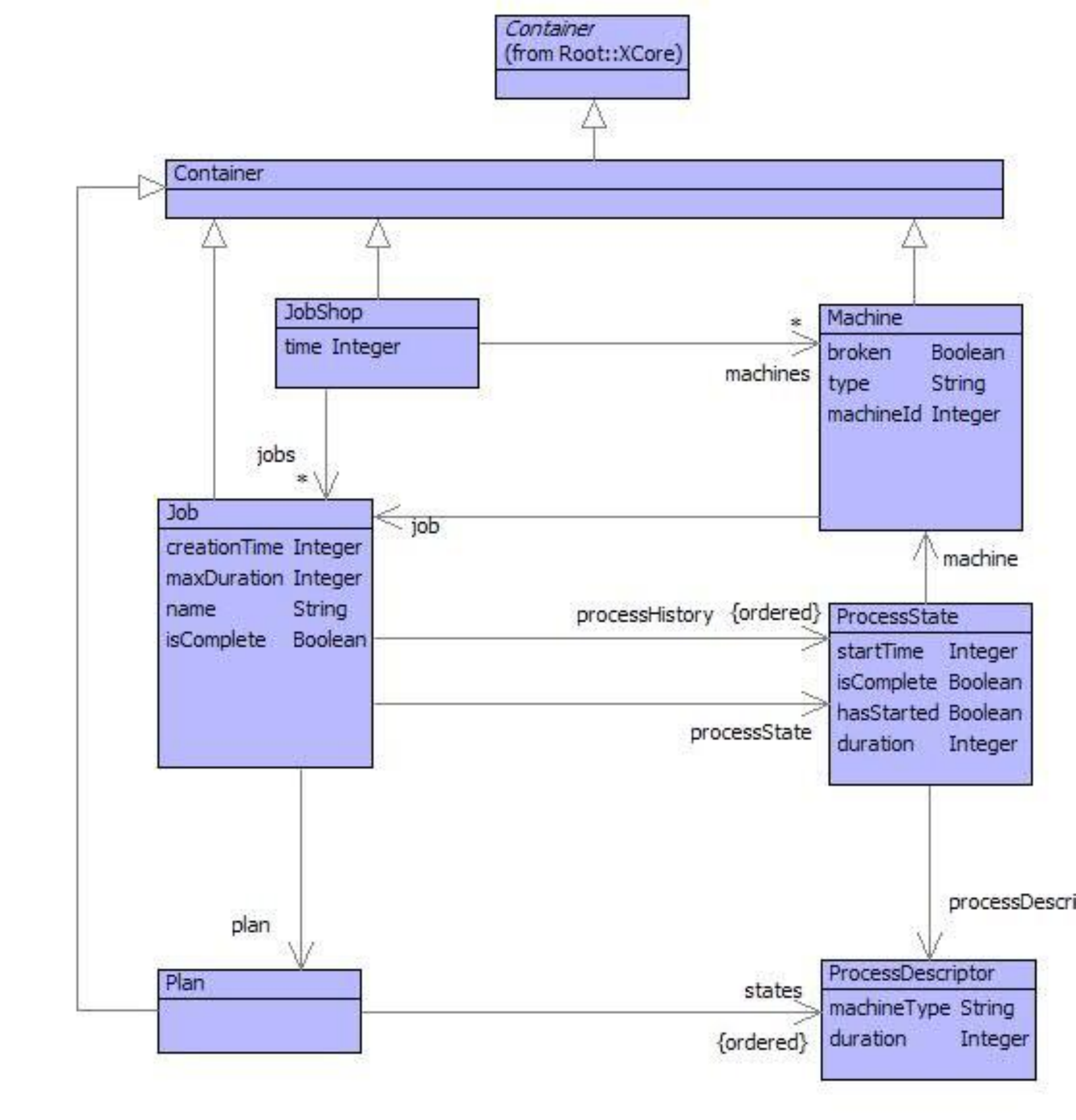}

\caption{JobShop Model\label{fig:JobShop-Model}}

\end{center}
\end{figure}

Figure \ref{fig:JobShop-Model} shows a model of a job-shop consisting
of jobs to be processed and machines that can process jobs. The required
application is an opportunistic scheduler that will allocate jobs
to appropriate machines as soon as they become available. Each job
has a plan that is a sequence of process descriptors: at any given
time, a job is either completed or is waiting to be allocated to a
machine or is being processed by a machine. A machine performs processes
of a given type and can only process one job at a time. 

The model allows various configurations of machines and jobs to be
created. By attaching actions to the classes in the model it is then
possible to simulate the execution of the configuration. The results
of the simulation can be used to analyse both the configuration and
the impact of modifications to the configuration. These include: determining
the overall execution time, whether there are any bottlenecks in the
system; the effect of adding and removing machines of various types;
and, the amount of time machines are left idle.

The rest of this section describes the classes in the model in detail
and gives some example executions of the simulation. The following
section describes the implementation of the executable model.

Class JobShop records the current time and has a number of jobs and
machines. At the end of the simulation, all of the jobs are complete.
Each machine has a unique id, a type and may be broken. The type of
the machine determines the tasks that it can perform. A machine has
a current job that may be empty. If it is non-empty then the machine
is currently processing the job and the type of the machine must match
the process at the head of the job's plan. 

Class Job has a name, a creation time (allowing for jobs to be dynamically
added to the system) and a maximum allowed duration (allowing for
the simulation to determine whether a job is taking too long). A job
becomes complete when the last process in its plan is performed by
a machine. 

Each job has a plan consisting of a sequence of process descriptors.
Class ProcessDescriptor defines the machine that must perform the
processing and the length of time that the process takes. 

A job is in a current state, which may be null if it is between tasks.
If the job is being processed by a machine then the current state
has a process descriptor and a machine: the type of the machine must
match the type in the prcess descriptor.

class ProcessState records the time at which it is created. It knows
the duration that the task must take from the process descriptor and
can therefore describe when it is complete. When a task is complete,
the current state of a job is set to null, however the job has a process
history that can be used for analysis (in a real implementation a
job may not require a process history). The process history is the
sequence of all process states that the job has performed.

\begin{figure}
\begin{center}
\includegraphics[width=14cm]{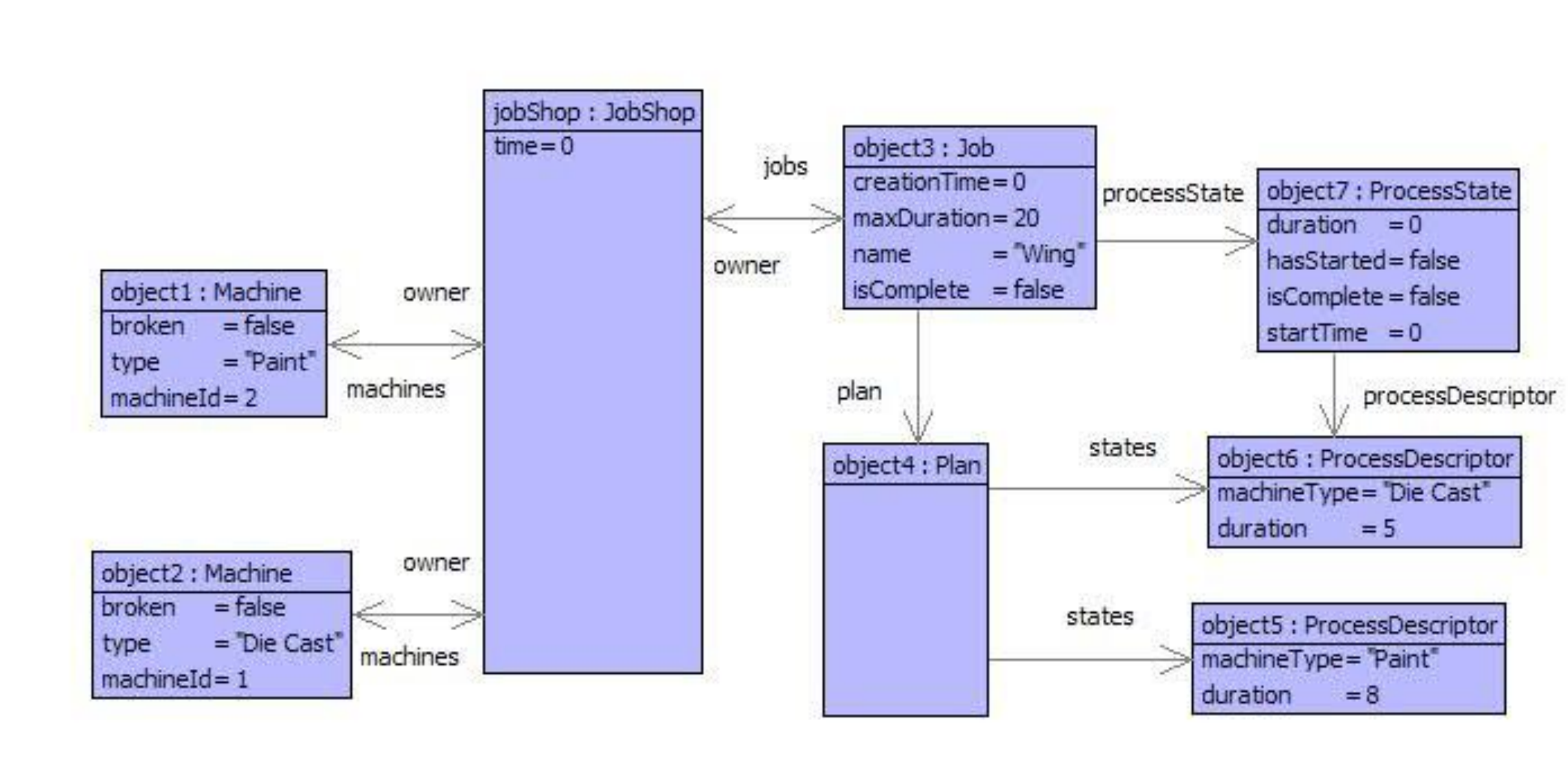}

\caption{A Job Shop Scenario. \label{fig:Process-Wing}}

\end{center}
\end{figure}

Figure \ref{fig:Process-Wing} shows a simple initial state for a
job shop that processes aircraft wings. The job is started with an
initial process state set up for the initial process descriptor in
the plan; however, the process state shows that the process has not
started and is not yet allocated to a machine.

\begin{figure}
\begin{center}

\includegraphics[width=14cm]{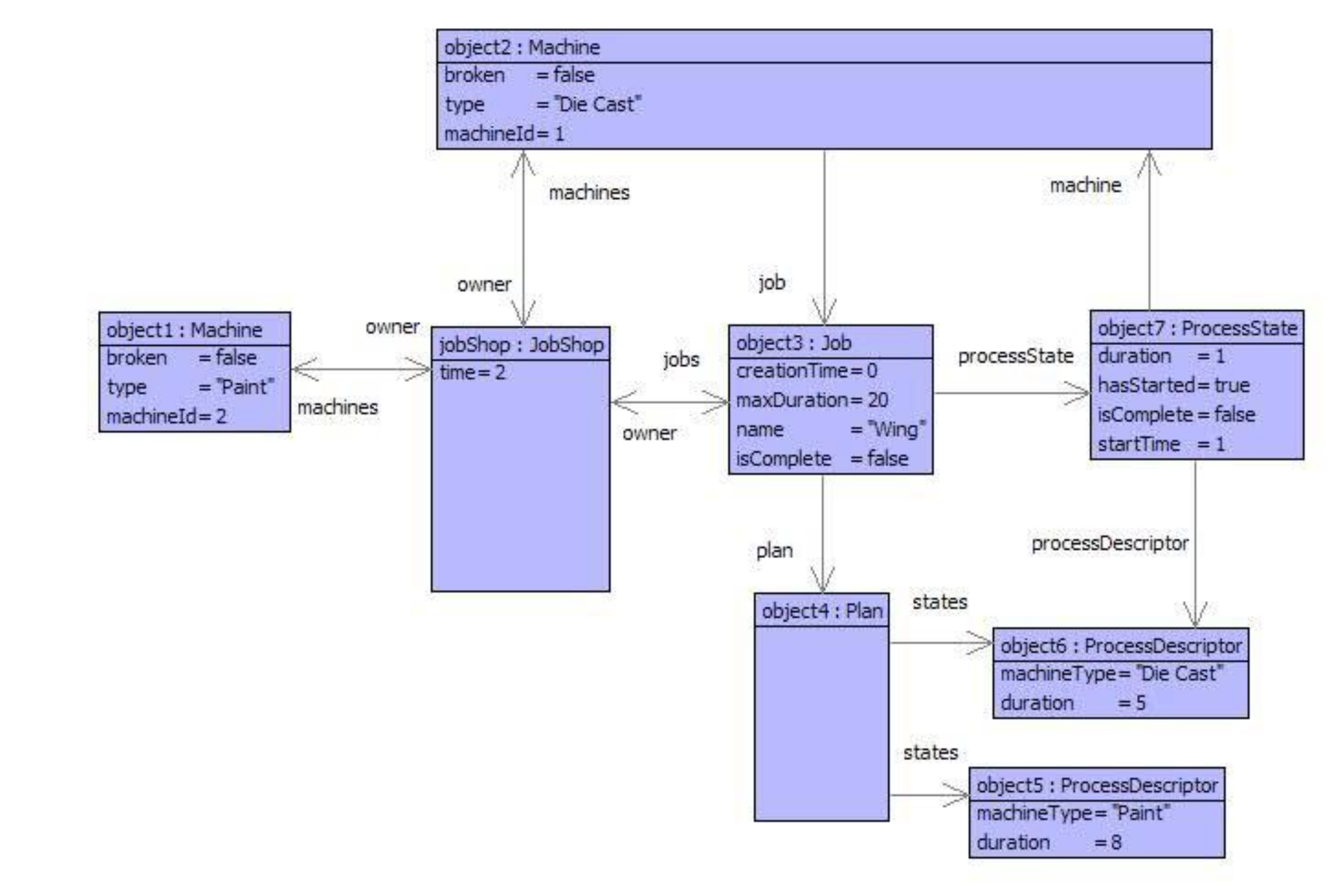}

\caption{At time 2 \label{fig:At-time-2}}

\end{center}
\end{figure}

Figure \ref{fig:At-time-2} shows the state of the simulator at time
2 when the initial process state has started. The process state is
allocated to a machine of the appropriate type, but is not yet complete.

\begin{figure}
\begin{center}

\includegraphics[width=14cm]{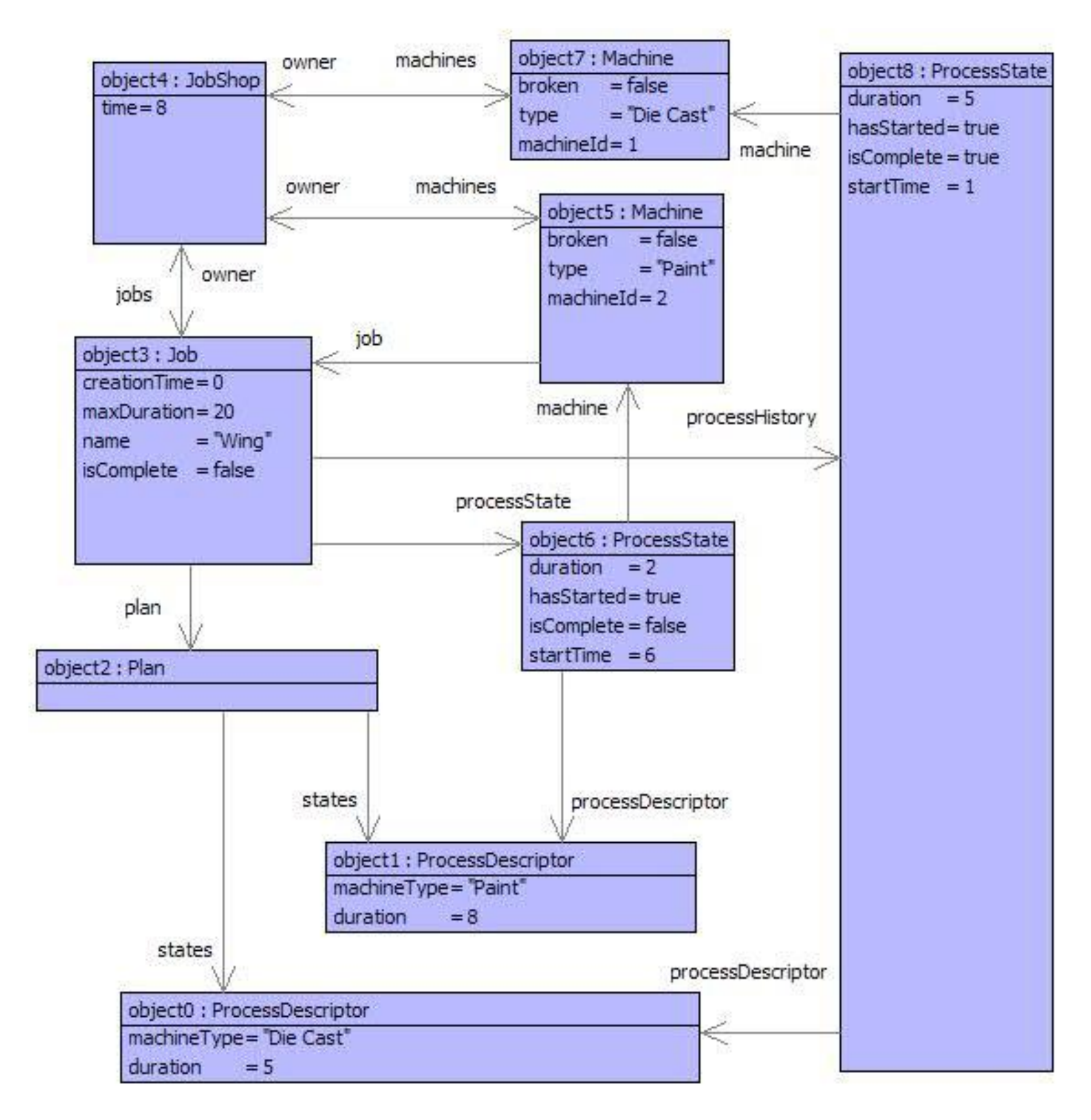}

\caption{After 8 time units.\label{fig:After-8-time}}

\end{center}
\end{figure}

Figure \ref{fig:After-8-time} shows the state of the simulation at
time 8. Machining at the Die Cast has completed and is shown on the
process history of the job. The process state shows that the job is
now scheduled at the paining machine. The simulation continues at
the painting machine until the time reaches 16 when the job becomes
complete.

A trace of the simulation is shown below:

\begin{lstlisting}
[1] XMF> ProcessWing::jobShop.run();
[1  ] Job Wing becomes scheduled at machine Die Cast:1.
[2  ] Machine Die Cast:1 processing job Wing (1 of 5 units)
[3  ] Machine Die Cast:1 processing job Wing (2 of 5 units)
[4  ] Machine Die Cast:1 processing job Wing (3 of 5 units)
[5  ] Machine Die Cast:1 processing job Wing (4 of 5 units)
[6  ] Machine Die Cast:1 completes job Wing.
[6  ] Job Wing becomes scheduled at machine Paint:2.
[7  ] Machine Paint:2 processing job Wing (1 of 8 units)
[8  ] Machine Paint:2 processing job Wing (2 of 8 units)
[9  ] Machine Paint:2 processing job Wing (3 of 8 units)
[10 ] Machine Paint:2 processing job Wing (4 of 8 units)
[11 ] Machine Paint:2 processing job Wing (5 of 8 units)
[12 ] Machine Paint:2 processing job Wing (6 of 8 units)
[13 ] Machine Paint:2 processing job Wing (7 of 8 units)
[14 ] Machine Paint:2 completes job Wing.
true
[1] XMF>
\end{lstlisting}Once the job shop simulation has completed, it is possible to perform
analysis on the job histories. One useful analysis is to determine
when any of the machines have been idle during the simulation. This
is shown below:

\begin{lstlisting}
[1] XMF> ProcessWing::jobShop.idleTimes();
Machine Paint:2 was idle at time 1.
Machine Paint:2 was idle at time 2.
Machine Paint:2 was idle at time 3.
Machine Paint:2 was idle at time 4.
Machine Paint:2 was idle at time 5.
Machine Die Cast:1 was idle at time 7.
Machine Die Cast:1 was idle at time 8.
Machine Die Cast:1 was idle at time 9.
Machine Die Cast:1 was idle at time 10.
Machine Die Cast:1 was idle at time 11.
Machine Die Cast:1 was idle at time 12.
Machine Die Cast:1 was idle at time 13.
true
[1] XMF>
\end{lstlisting}The following shows a slightly more interesting example of a job-shop
that processes many more jobs with more machines. In this example
there are several jobs competing for a single Die Cast machine; this
causes a bottleneck in the processing leading to 36 units of time
for the simulation. If another Die Cast machine is added then the
simulation shows the time is reduced to 27 units:

\begin{lstlisting}
[1] XMF> ProcessAircraft::jobShop.run();
[1  ] Job Wing1 becomes scheduled at machine Die Cast:3.
[2  ] Machine Die Cast:3 processing job Wing1 (1 of 5 units)
[3  ] Machine Die Cast:3 processing job Wing1 (2 of 5 units)
[4  ] Machine Die Cast:3 processing job Wing1 (3 of 5 units)
[5  ] Machine Die Cast:3 processing job Wing1 (4 of 5 units)
[6  ] Machine Die Cast:3 completes job Wing1.
[6  ] Job Wing1 becomes scheduled at machine Paint:4.
[7  ] Job Tail1 becomes scheduled at machine Die Cast:3.
[7  ] Machine Paint:4 processing job Wing1 (1 of 8 units)
[8  ] Machine Die Cast:3 processing job Tail1 (1 of 5 units)
[8  ] Machine Paint:4 processing job Wing1 (2 of 8 units)
[9  ] Machine Die Cast:3 processing job Tail1 (2 of 5 units)
[9  ] Machine Paint:4 processing job Wing1 (3 of 8 units)
[10 ] Machine Die Cast:3 processing job Tail1 (3 of 5 units)
[10 ] Machine Paint:4 processing job Wing1 (4 of 8 units)
[11 ] Machine Die Cast:3 processing job Tail1 (4 of 5 units)
[11 ] Machine Paint:4 processing job Wing1 (5 of 8 units)
[12 ] Machine Die Cast:3 completes job Tail1.
[12 ] Machine Paint:4 processing job Wing1 (6 of 8 units)
[13 ] Job Tail1 becomes scheduled at machine Paint:2.
[13 ] Job Tail2 becomes scheduled at machine Die Cast:3.
[13 ] Machine Paint:4 processing job Wing1 (7 of 8 units)
[14 ] Machine Paint:2 processing job Tail1 (1 of 8 units)
[14 ] Machine Die Cast:3 processing job Tail2 (1 of 5 units)
[14 ] Machine Paint:4 completes job Wing1.
[14 ] Job Wing1 becomes scheduled at machine Assemble:6.
[15 ] Machine Paint:2 processing job Tail1 (2 of 8 units)
[15 ] Machine Die Cast:3 processing job Tail2 (2 of 5 units)
[15 ] Machine Assemble:6 processing job Wing1 (1 of 3 units)
[16 ] Machine Paint:2 processing job Tail1 (3 of 8 units)
[16 ] Machine Die Cast:3 processing job Tail2 (3 of 5 units)
[16 ] Machine Assemble:6 processing job Wing1 (2 of 3 units)
[17 ] Machine Paint:2 processing job Tail1 (4 of 8 units)
[17 ] Machine Die Cast:3 processing job Tail2 (4 of 5 units)
[17 ] Machine Assemble:6 completes job Wing1.
[18 ] Machine Paint:2 processing job Tail1 (5 of 8 units)
[18 ] Machine Die Cast:3 completes job Tail2.
[18 ] Job Tail2 becomes scheduled at machine Paint:4.
[19 ] Machine Paint:2 processing job Tail1 (6 of 8 units)
[19 ] Job Nose becomes scheduled at machine Die Cast:3.
[19 ] Machine Paint:4 processing job Tail2 (1 of 8 units)
[20 ] Machine Paint:2 processing job Tail1 (7 of 8 units)
[20 ] Machine Die Cast:3 processing job Nose (1 of 5 units)
[20 ] Machine Paint:4 processing job Tail2 (2 of 8 units)
[21 ] Machine Paint:2 completes job Tail1.
[21 ] Machine Die Cast:3 processing job Nose (2 of 5 units)
[21 ] Machine Paint:4 processing job Tail2 (3 of 8 units)
[21 ] Job Tail1 becomes scheduled at machine Assemble:6.
[22 ] Machine Die Cast:3 processing job Nose (3 of 5 units)
[22 ] Machine Paint:4 processing job Tail2 (4 of 8 units)
[22 ] Machine Assemble:6 processing job Tail1 (1 of 3 units)
[23 ] Machine Die Cast:3 processing job Nose (4 of 5 units)
[23 ] Machine Paint:4 processing job Tail2 (5 of 8 units)
[23 ] Machine Assemble:6 processing job Tail1 (2 of 3 units)
[24 ] Machine Die Cast:3 completes job Nose.
[24 ] Machine Paint:4 processing job Tail2 (6 of 8 units)
[24 ] Machine Assemble:6 completes job Tail1.
[25 ] Job Nose becomes scheduled at machine Paint:2.
[25 ] Machine Paint:4 processing job Tail2 (7 of 8 units)
[26 ] Machine Paint:2 processing job Nose (1 of 8 units)
[26 ] Machine Paint:4 completes job Tail2.
[26 ] Job Tail2 becomes scheduled at machine Assemble:6.
[27 ] Machine Paint:2 processing job Nose (2 of 8 units)
[27 ] Machine Assemble:6 processing job Tail2 (1 of 3 units)
[28 ] Machine Paint:2 processing job Nose (3 of 8 units)
[28 ] Machine Assemble:6 processing job Tail2 (2 of 3 units)
[29 ] Machine Paint:2 processing job Nose (4 of 8 units)
[29 ] Machine Assemble:6 completes job Tail2.
[30 ] Machine Paint:2 processing job Nose (5 of 8 units)
[31 ] Machine Paint:2 processing job Nose (6 of 8 units)
[32 ] Machine Paint:2 processing job Nose (7 of 8 units)
[33 ] Machine Paint:2 completes job Nose.
[33 ] Job Nose becomes scheduled at machine Assemble:6.
[34 ] Machine Assemble:6 processing job Nose (1 of 3 units)
[35 ] Machine Assemble:6 processing job Nose (2 of 3 units)
[36 ] Machine Assemble:6 completes job Nose.
true
\end{lstlisting}
\section{Job-Shop Implementation}

\label{sec:Job-Shop-Implementation}

The previous section has specified a job shop scheduling simulation
in terms of a general model consisting of jobs, plans and machines.
Non-executable modelling would have to stop at this point: the execution
would be left implicit, perhaps specified as pre and post-conditions
on various model operations. Executable modelling allows the models
to be brought to life in terms of simulation. Depending on the application
requirements, it is possible for the executable model to \emph{be}
the application, however that is not the issue here. This section
describes the implementation of the key features of the simulation
by defining operations on the classes defined in figure \ref{fig:JobShop-Model}.

A job shop simulation is initiated by the run operation:

\begin{lstlisting}
context JobShop
  @Operation runJobs()
    @While not jobs->forAll(job | job.isComplete()) do
      self.time := time + 1
      @For machine in machines do
        machine.process(self)
      end
    end
  end
\end{lstlisting}Each machine processes the current job:

\begin{lstlisting}
context Machine
  @Operation process(jobShop:JobShop)
    if not broken
    then
      if job <> null
      then self.processJob(jobShop)
      else self.findJob(jobShop)
      end
    else format(stdout,"~S:~S is broken at the moment.",Seq{type,machineId})
    end
  end
\end{lstlisting}The current job being processed by a machine is handled by processJob.
Ticking a job causes the process state to be modified by one time
unit. If the job is complete then it is removed from the machine.
The operations for printing messages are not included in the code,
but conform to the sample runs given in the previous section:

\begin{lstlisting}
context Machine
  @Operation processJob(jobShop:JobShop)
    job.tick();
    if job.isComplete() 
    then 
      self.printCompleted(jobShop.time(),job);
      self.setJob(null)
    else self.printProcessing(jobShop.time(),job)
    end
  end
\end{lstlisting}A machine uses findJob to schedule the next job if the machine is
not currently processing a job. If a pending job is found then it
becomes the current job of the machine and the process state of the
job is updated using start():

\begin{lstlisting}
context Machine
  @Operation findJob(jobShop)
    @Find(job,jobShop.jobs())
      when 
        not job.processState().hasStarted() andthen
        job.processState().processDescriptor().machineType = type
      do self.setJob(job);
         job.processState().start(jobShop.time(),self);
         self.printJobScheduled(jobShop.time(),job)
    end
  end

context ProcessState
  @Operation start(time:Integer,machine:Machine)
    self.setHasStarted(true);
    self.setStartTime(time);
    self.setMachine(machine)
  end
\end{lstlisting}When a job is processed by a machine, its tick() operation is performed.
This updates the current process state:

\begin{lstlisting}
context Job
  @Operation tick()
    processState.tick();
    if processState.isComplete()
    then self.nextProcessState()
    end
  end

context ProcessState
  @Operation tick()
    if hasStarted and not isComplete
    then
      self.setDuration(duration + 1);
      if duration >= processDescriptor.duration()
      then self.setIsComplete(true)
      end
    end
  end
\end{lstlisting}If the process state of a job indicates that the current process is
complete, then the job moves to the next process descriptor in the
plan. The current process state is added to the process history for
the job so that analysis can be performed after the cimulation has
completed. The plan is asked for a new process state. If the plan
is complete then the job becomes complete, otherwise the process state
of the job contains the next process descriptor in the plan awaiting
scheduling on an appropriate machine:

\begin{lstlisting}
context Job
  @Operation nextProcessState()
    self.addToProcessHistory(processState);
    self.setProcessState(plan.nextProcessState(processState));
    if processState = null
    then self.setIsComplete(true)
    end
  end  

context Plan
  @Operation nextProcessState(current:ProcessState):ProcessState
    let index = states->indexOf(current.processDescriptor())
    in if index + 1 >= states->size
       then null
       else 
         let next = states->at(index + 1)
         in ProcessState(next)
         end
       end
    end
  end
\end{lstlisting}The operations above complete the basic simulator. It remains to perform
machine idle time analysis on a completed simulation, this is implemented
as an operation idleTimes on JobShop:

\begin{lstlisting}
context JobShop
  @Operation idleTimes()
    @Count t from 1 to time do
      @For machine in machines do
        if not jobs->exists(job | job.processedBy(machine,t))
        then self.printIdle(machine,t)
        end
      end
    end
  end

context Job
  @Operation processedBy(machine,time):Boolean
    processHistory->exists(state |
      state.machine() = machine and
      state.startTime() <= time and
      (state.startTime() + state.duration()) >= time)
  end
\end{lstlisting}The following is another analysis operation that calculates those
jobs that took longer than their alotted time to be processed:

\begin{lstlisting}
context JobShop
  @Operation lateJobs():Set(Job)
    jobs->select(job | job.maxDuration() < job.duration())
  end

context Job
  @Operation duration():Integer
    let state = processHistory->last
    in (state.startTime() + state.duration()) - creationTime
    end
  end
\end{lstlisting}
\section{Conflict Analysis}

\label{sec:Conflict-Analysis}

\begin{figure}
\begin{center}
\includegraphics[width=8cm]{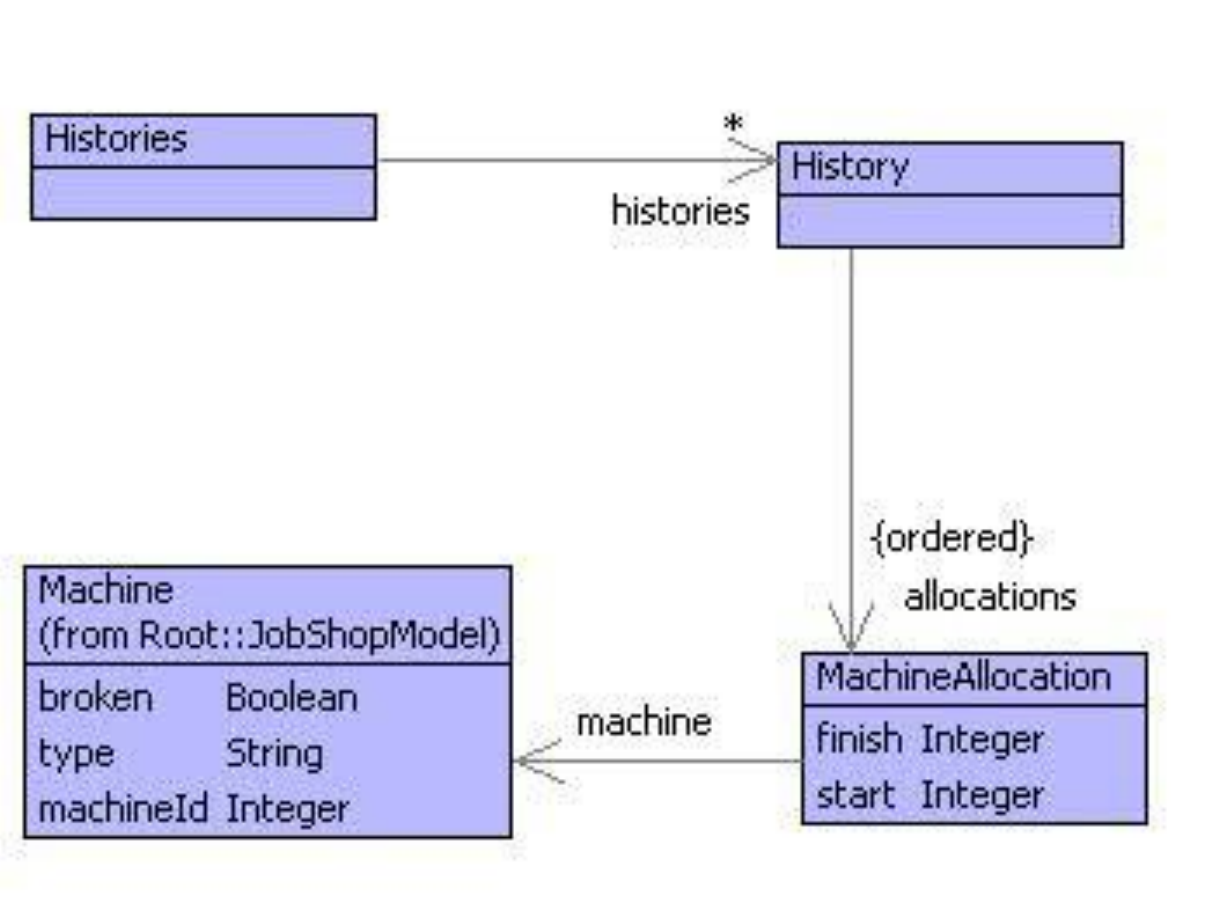}

\caption{History Models\label{fig:History-Models}}

\end{center}
\end{figure}

An interesting analysis dertmines whether or not there are two jobs
that ever require the same machine at the same time. This analysis
requires a slightly diffferent type of simulation from that described
so far. In order to determine whether there may be a conflict, the
job shop is \emph{symbolically executed,} creating histories of possible
allocations of jobs to machines. These allocations can then be analysed
to determine whether any of them overlap.

Given the initial state of the job-shop it is possible to calculate
the scheduling options by analysing the plan for each job and allocating
the job to the machines it requires. The result of doing this for
each job is a collection of possible \emph{histories} for the simulation.
An execution of the simulation is represented by one possible history
for each of the jobs. If there is a machine that is required by two
different jobs in all possible histories then it is not possible to
optimally schedule the jobs, i.e. there is a bottleneck arising from
a conflict.

Consider the ProcessAircraft example given in a previous section.
Here are all the histories calculated for the jobs:

\begin{lstlisting}
Histories calculated for job Wing1:
  Option:
     Machine 3 is used from 0 to 5
     Machine 4 is used from 5 to 13
     Machine 5 is used from 13 to 16
  Option:
     Machine 3 is used from 0 to 5
     Machine 4 is used from 5 to 13
     Machine 6 is used from 13 to 16
  Option:
     Machine 3 is used from 0 to 5
     Machine 2 is used from 5 to 13
     Machine 6 is used from 13 to 16
  Option:
     Machine 3 is used from 0 to 5
     Machine 2 is used from 5 to 13
     Machine 5 is used from 13 to 16
Histories calculated for job Tail1:
  Option:
     Machine 3 is used from 0 to 5
     Machine 4 is used from 5 to 13
     Machine 5 is used from 13 to 16
  Option:
     Machine 3 is used from 0 to 5
     Machine 4 is used from 5 to 13
     Machine 6 is used from 13 to 16
  Option:
     Machine 3 is used from 0 to 5
     Machine 2 is used from 5 to 13
     Machine 6 is used from 13 to 16
  Option:
     Machine 3 is used from 0 to 5
     Machine 2 is used from 5 to 13
     Machine 5 is used from 13 to 16
Histories calculated for job Tail2:
  Option:
     Machine 3 is used from 0 to 5
     Machine 4 is used from 5 to 13
     Machine 5 is used from 13 to 16
  Option:
     Machine 3 is used from 0 to 5
     Machine 4 is used from 5 to 13
     Machine 6 is used from 13 to 16
  Option:
     Machine 3 is used from 0 to 5
     Machine 2 is used from 5 to 13
     Machine 6 is used from 13 to 16
  Option:
     Machine 3 is used from 0 to 5
     Machine 2 is used from 5 to 13
     Machine 5 is used from 13 to 16
Histories calculated for job Nose:
  Option:
     Machine 3 is used from 0 to 5
     Machine 4 is used from 5 to 13
     Machine 5 is used from 13 to 16
  Option:
     Machine 3 is used from 0 to 5
     Machine 4 is used from 5 to 13
     Machine 6 is used from 13 to 16
  Option:
     Machine 3 is used from 0 to 5
     Machine 2 is used from 5 to 13
     Machine 6 is used from 13 to 16
  Option:
     Machine 3 is used from 0 to 5
     Machine 2 is used from 5 to 13
     Machine 5 is used from 13 to 16
\end{lstlisting}Figure \ref{fig:History-Models} shows a model of job-shop histories.
A history records when a machine is busy in terms of when a job is
allocated to the machine (the start time) and when a job is removed
from the machine (the finish time). A job-shop is translated to an
instance of the class Histories, by calculating all the possible sets
of histories; each history is an ordered collection of consistent
machine allocations.

Each plan produces n instance of Histories:

\begin{lstlisting}
context Plan
  @Operation histories(jobShop)
    let h = Histories()
    in @For state in states do
         let M = jobShop.machines()->select(machine | 
                   machine.type() = state.machineType())
         in h.addMachines(state.duration(),M)
         end
       end;
       h
    end
  end
\end{lstlisting}A histories, adds all the machines that could occur during a given
time duration. The histories are built up incrementally by adding
machines that might be used by process descriptors. Since scheduling
is opportunistic each machine is allocated at random:

\begin{lstlisting}
context Histories
  @Operation addMachines(duration,M)
    // Add all the machines for the given duration at the
    // end of the current histories.
    @For history in histories do
      self.deleteFromHistories(history);
      @For machine in M do
        self.addToHistories(history.extend(duration,machine))
      end
    end
  end
\end{lstlisting}\begin{lstlisting}
context History
  @Operation extend(duration,machine)
    // Add a new allocation. Returns a copy so that
    // other additions don't interfere...
    let newHistory = self.copy();
        d = allocations->iterate(a d = 0 | d + (a.finish() - a.start()))
    in newHistory.addToAllocations(MachineAllocation(d,d + duration,machine))
    end
  end
\end{lstlisting}Two machine allocations conflict when they overlap:

\begin{lstlisting}
context MachineAllocation
  @Operation conflict(other:MachineAllocation):Boolean
    machine = other.machine and
    ((start >= other.start and start <= other.finish) or
     (other.start >= start and other.start <= finish))
  end
\end{lstlisting}Finally, the job-shop configuration has a conflict when there are
two jobs that require the same machine during the same time interval
in all possible histories:

\begin{lstlisting}
context JobShop
  @Operation conflict():Boolean
    jobs->exists(j1 |
      jobs->exists(j2 |
        j1 <> j2 and
        j1.plan().histories(self).histories()->forAll(h1 |
          j2.plan().histories(self).histories()->forAll(h2 |
            h1.allocations()->exists(a1 |
              h2.allocations()->exists(a2 | a1.conflict(a2)))))))
  end
\end{lstlisting}

\chapter{Pretty Printing}

Super-languages deal with a great deal of structured data. How is
the information to be managed? Model developers need a way of interacting
with the information in a flexible way. They should be able to look
at the data in a variety of ways, slicing it up and presenting it
as appropriate.

One approach to presenting model data in a flexible way is to use
an extensible pretty-printer. A pretty-printer is a system that can
be used to print out data in a convenient form; it provides hooks
to change the appearance of data items and also provides a means of
controlling the layout of large quantities of data. This chapter shows
how a pretty-printer can be modelled and implemented as an execution
engine that runs the models. It is an example of how XMF can be used
to develop a language engine. The pretty-printer language has no concrete
syntax, however it is still a language and has an engine that executes
it. The super-language features of XMF make it idea for defining a
pretty-printer language and its execution engine.

\section{Example Printing}

Consider a database of bank records. Each record describes a bank
customer in terms of their name, age and the unique account identifiers.
If the database is printed out without any attempt at layout then
the following might be the result:

\begin{lstlisting}
DataBase[records = Seq{Record[accounts = Seq{A101,B204,
A108},age= 22,name = Fred Jones],Record[accounts = 
Seq{C202},age = 32,name = Sally Brown],Record[accounts = 
Seq{A102,B222},age = 24,name = Edward Clark]}]
\end{lstlisting}Obviously no attempt has been made at structuring the information;
newlines occur at arbitrary points in the data where the right-hand
edge of the output window (or \textit{page}) is reached. The following
are significant factors in controlling the presentation of data:

\begin{itemize}
\item The type of an element determines how it should be presented. For
example, an atomic data item such as a string or a number is just
printed out as is. An object contains slots and the slot values are
possibly large elements in themselves. A basic object might be displayed
as a sequence of slots where there are many choices governing the
layout of the slots. Particular types of object might be displayed
in specific ways; for example, internal features of an object may
be considered as hidden and not printed out. Other types of element
that require specific display include tables, collections and vectors.
\item The width of the page is an important factor when displaying elements.
Often there may be a choice regarding how an element is to be laid
out on the page, for example an object may list its slots one after
the other on the same line or list the slots one above the other on
different lines. The width of the page can be used in deciding amongst
the alternatives.
\item Independent of the width of the page, is the maximum width of output
on any given line. Call this the ribbon width; it differs from the
page width because it dones not take into consideration leading whitespace
on a line. The ribbon width can be used to decide amongst alternative
layout strategies.
\item Complex data is usually deeply nested, i.e. one element contains another,
and that contains more data etc. Sometimes a broad picture of the
data is required; other times it is necessary to drill deeply into
an element. 
\end{itemize}
Using the example data given above, the following shows a borad view
with a wide page width and ribbon:

\begin{lstlisting}
DataBase[records = Seq{
                     Record[accounts = ...,age = 22,name = Fred Jones],
                     Record[accounts = ...,age = 32,name = Sally Brown],
                     Record[accounts = ...,age = 24,name = Edward Clark]}]
\end{lstlisting}The ... shows that the level of nesting limit has been reached. If
the level of nesting is decreased by 1 then the display would show
that the database had 3 records, but not show the records themselves:

\begin{lstlisting}
DataBase[records = Seq{...,...,...}]
\end{lstlisting}Alternatively, consider a short ribbon width and a deep nesting:

\begin{lstlisting}
DataBase[records = Seq{
                     Record[
                       accounts =
                         Seq{A101,B204,A108}
                       age = 22,
                       name = Fred Jones],
                     Record[
                       accounts = Seq{C202}
                       age = 32,
                       name = Sally Brown],
                     Record[
                       accounts = Seq{A102,B222}
                       age = 24,
                       name = Edward Clark]}]
\end{lstlisting}Finally taking this to its limit:

\begin{lstlisting}
DataBase[
  records =
    Seq{
      Record[
        accounts =
          Seq{
            A101,
            B204,
            A108}
        age =
          22,
        name =
          Fred Jones],
      Record[
        accounts =
          Seq{C202}
        age =
          32,
        name =
          Sally Brown],
      Record[
        accounts =
          Seq{
            A102,
            B222}
        age =
          24,
        name =
          Edward Clark]}]
\end{lstlisting}
\section{A Document Model}

\begin{figure}
\begin{center}

\includegraphics[width=12cm]{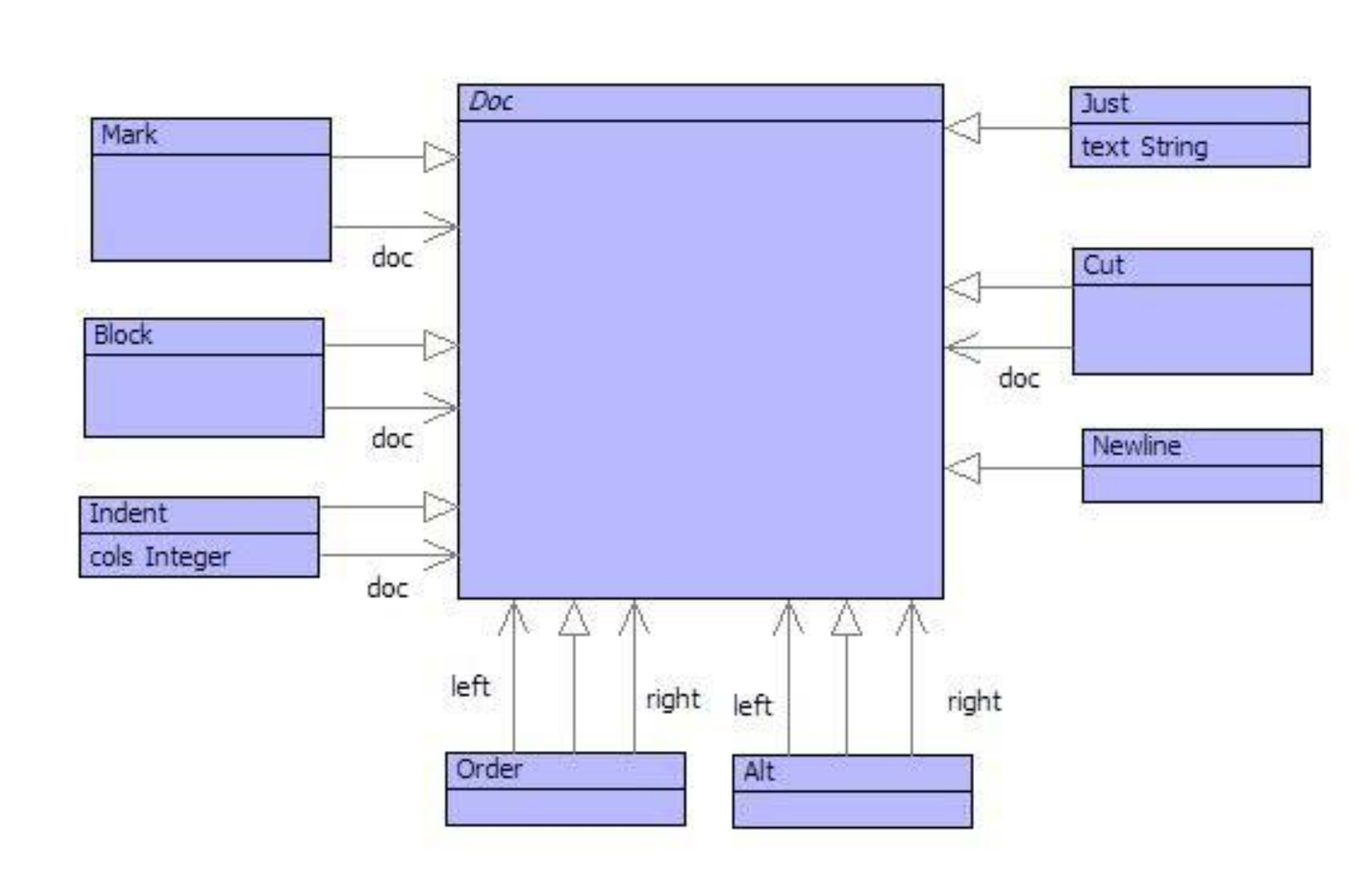}
\caption{A Document Model\label{fig:A-Document-Model}}

\end{center}
\end{figure}

Pretty printing can be achieved by writing an output operation for
each new model element. In the example above, the database and record
classes would each define an output operation. The output operations
need to keep track of the page and ribbon widths, and also deal with
manging options for output. 

The general output machinery is more or less the same for each new
class; just the details of what is being printed differs from class
to class. Therefore it makes sense to abstract the machinery into
a general model of documents. 

A document consists of output directives including the text to be
printed. A mapping translates instances of classes to the document
model. Therefore, the machinery for pretty-printing is defined once
in the document model and domain models are not polluted with details
of printing.

The document model is shown in figure \ref{fig:A-Document-Model}.
The rest of this section describes the classes in the document model
and shows how they affect the output when printed prettily.

A document Just(t) represents just the text t. When it is printed
with a page and ribbon width of 80, the text is displayed:

\begin{lstlisting}
XMF> Just("some text").pprint(80,80);
some text
\end{lstlisting}By itself, Just does not appear to do anything; however, as described
later, Just may cause an alternative formatting to be chosen because
the text does not fit onto the current line or the current ribbon.

Several documents can be joined together using Order. Viewing the
pretty printer as sending text to an output ribbon, Order(d1,d2) just
causes the output from d1 to be printed to the ribbon followed by
the output from d2:

\begin{lstlisting}
XMF> Order(Just("some text"),
       Order(Just(" more text"),
             Just(" the end."))).pprint(80,80);
some text more text the end.
\end{lstlisting}Newline can be used to insert breaks into the output:

\begin{lstlisting}
XMF> Order(Just("some text"),
       Order(Newline(),
         Order(Just("more text"),
           Order(Newline(),
             Just("the end."))))).pprint(80,80);
some text
more text
the end.
\end{lstlisting}Indentation is important in output, it is often used to represent
ownership or containment. For example, collections own their elements
and objects own their slots. The display of the elements and slots
can be indented in order to represent ownership.

The engine that prints a document maintains a context including the
current level of indentation. Each time a new line is printed, the
engine automatically tabs to the current indent by adding spaces to
the output ribbon. The value of the current indent can be increased
by a given value using the class Indent:

\begin{lstlisting}
XMF> Order(Just("some text"),
      Indent(2,
        Order(Newline(),         
          Order(Just("then more text"),
            Indent(2,
              Order(Newline(),
                Just("the end."))))))).pprint(80,80);
some text
  then more text
    the end.
\end{lstlisting}The current level of indentation can be set using the class Block.
The document Block(d) uses the position on the current line as the
level of indentation for processing d. Once d has been processes,
the level of indentation reverts back:

\begin{lstlisting}
XMF> Order(Just("some text"),       
       Order(
         Block(
           Order(Newline(),
             Just("more text"))),
         Indent(2,
           Order(Newline(),
            Just("the end."))))).pprint(80,80);
some text
         more text
  the end.
\end{lstlisting}The arguments to the pprint operation are the page width and ribbon
width respectively. The docment classes described so far provide a
fixed output format: the values of page and ribbon width make no difference
to the output. As described earlier, it is desirable to be able to
print an element in different formats depending on the context: a
wide page width will allow an object to list its slots on the same
line ehwreas a narrow page width forces the slots to be listed one
above the other. 

The document Alt(d1,d2) defines two alternative layouts: d1 and d2.
When pretty printing the document Alt(d1,d2), the printer first tries
d1, if it finds that any of the output of d1 exceeds the line or ribbon
width then the printer discards the output from d1 and proceeds with
d2. Consider the following operations:

\begin{lstlisting}
@Operation line(.ss:Seq(String)):Doc
  ss->tail->iterate(s d = Just(ss->head) | Order(d,Just(" " + s)))
end
  
@Operation stack(.ss:seq(string)):Doc
  ss->tail->iterate(s d = Just(ss->head) | Order(d,Order(Newline(),Just(s))))
end
\end{lstlisting}The operation line transforms a sequence of strings into an ordered
document. The operation stack transforms a sequence of strings into
a document that displays the strings one above the other. The following
code shows two uses of the same document, the first prints on a page
width of 80 with a ribbon width of 80 and the second on the same page
with a ribbon width of 10. In the first case the text can fit on the
same line. In the second case the text cannot fit onto a ribbon with
of 10 and therefore the second option is used:

\begin{lstlisting}
XMF> let strings = Seq{"some text","more text","end."}
     in Alt(line(strings),stack(strings)).pprint(80,80)
     end;
some text more text end.
XMF> let strings = Seq{"some text","more text","end."}
     in Alt(line(strings),stack(strings)).pprint(80,10)
     end;
some text
more text
end.
\end{lstlisting}
\section{A Pretty Printing Machine}

\begin{figure}
\begin{center}

\includegraphics[width=12cm]{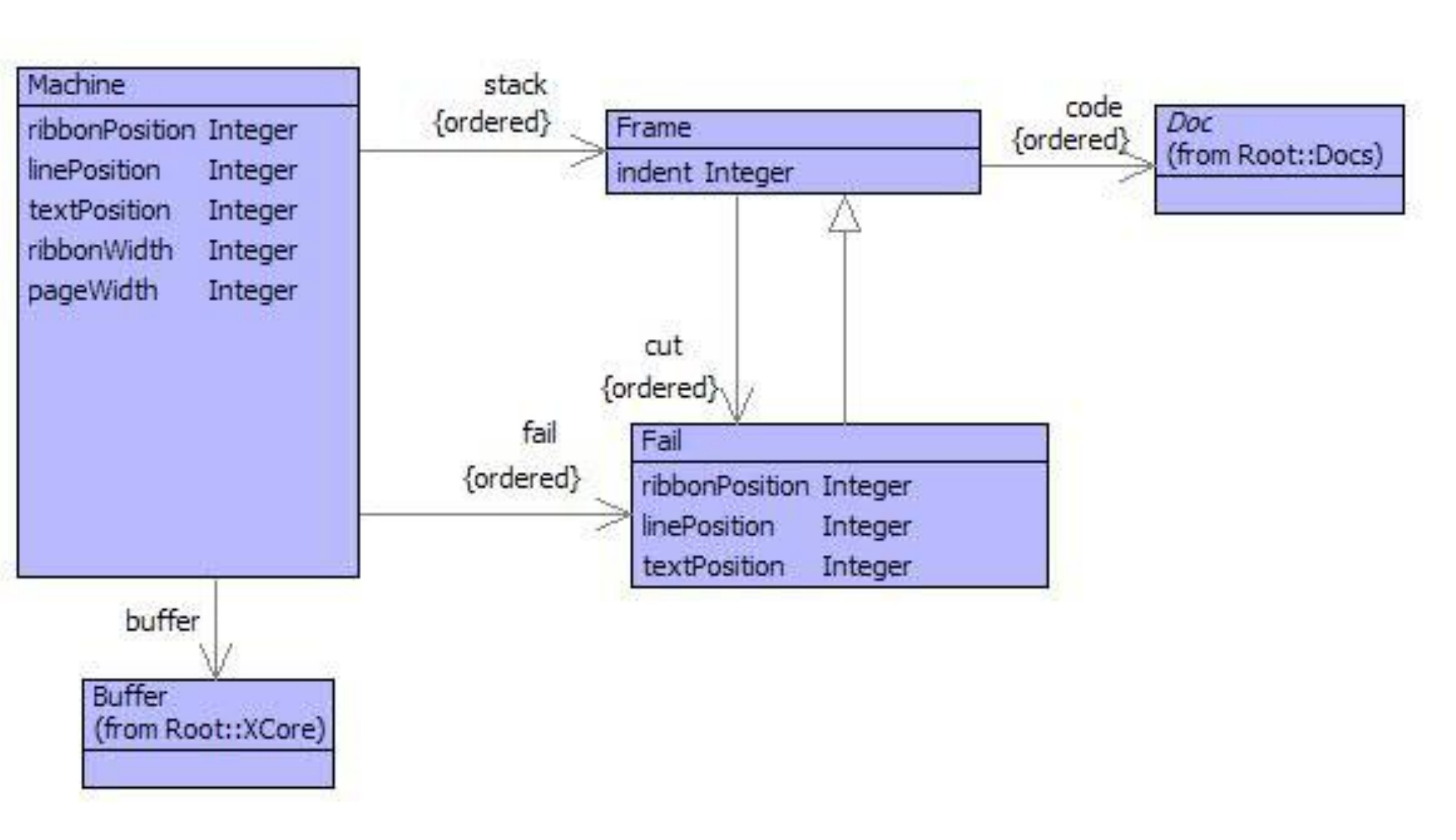}

\caption{The Pretty Print Engine\label{fig:The-Pretty-Print-Engine}}

\end{center}
\end{figure}

A document is pretty-printed by a pretty-printing machine. The machine
\textit{runs} the document and the state of the machine includes the
page and ribbon width and the current text position on the page. The
machine model is shown in figure \ref{fig:The-Pretty-Print-Engine},
the rest of this section describes how the machine works.

A pretty-printer is an instance of the class Machine. The state of
a machine consists of the following:

\begin{itemize}
\item ribbonPosition is the next output position on the current line ignoring
any leading indentation;
\item linePosition is the next output position on the current line;
\item textPosition is the next output position on the page;
\item ribbonWidth is the maximum number of characters on a line ignoring
any leading indentation;
\item pageWidth is the maximum number of character on a line;
\item a buffer containing the output (next buffer position is given by textPosition);
\item stack is a sequence of frames;
\item fail is a sequence of fail frames.
\end{itemize}
The stack contains a sequence of frames. Each frame contains a sequence
of documents to be printed along with the indentation used for their
output. Frames allow the indentiation in a document to change: when
the machine runs it prints out all the documents in the frame at the
head of the stack, when those documents are completed then the machine
pops the stack and continues with the next frame. If a document wants
to change the indentation for a given document then it pushes a new
frame at the head of the stack.

A machine has a sequence of fail frames. When a Just(t) document is
executed at the head of the stack, if the text t cannot be printed
out in the current page and ribbon width then the machine attempts
to \textit{fail}. Failure involves returning to a choice point defined
by an instance of the class Fail containing a saved machine state.
To fail, the machine resets all the saved state in the Fail instance
and then pops the fail stack.

\begin{lstlisting}
@Operation run()
  @While not stack->isEmpty do
    if stack->head.code()->isEmpty
    then self.stack := stack->tail
    else
      let frame = stack->head then
          instr = frame.code()->head 
      in stack->head.setCode(frame.code()->tail);
         @Case instr of
           Just(text) do
             if self.canPrint(text) or fail->isEmpty
             then self.write(text)
             else self.fail() 
             end
           end
           Order(left,right) do
             frame.setCode(Seq{left,right} + frame.code()) 
           end
           Indent(cols,doc) do
             self.pushFrame(frame.indent() + cols,frame.cut(),Seq{doc})
           end
           Block(doc) do
             self.pushFrame(linePosition,frame.cut(),Seq{doc})
           end
           Cut(doc) do 
             self.fail := frame.cut()
           end
           Mark(doc) do
             self.pushFrame(frame.indent(),fail,Seq{doc})
           end
           NewLine() do
             self.newline()
           end
           Alt(left,right) do
             frame.setCode(Seq{left | frame.code()});
             self.pushFail(Seq{right | frame.code()})
           end
         end
      end
    end
  end
end
\end{lstlisting}%
\begin{figure}
\begin{center}
\begin{tabular}{|c|c|}
\hline
\begin{minipage}{2in}
\begin{lstlisting}

(1)Machine[80,50,0,0,0,
  [],
  Seq{Frame[0,
    code = Seq{
      some text; 
       more text; 
       the end.},
    cut = Seq{}
    ]},
  Seq{}]

\end{lstlisting}
\end{minipage}
&
\begin{minipage}{2in}
\begin{lstlisting}
(2)Machine[80,50,0,0,0,
  [],
  Seq{Frame[0,
    code = Seq{
      some text, 
       more text; 
       the end.},
    cut = Seq{}
    ]},
  Seq{}]
\end{lstlisting}
\end{minipage}
\\\hline
\begin{minipage}{2in}
\begin{lstlisting}

(3)Machine[80,50,9,9,9,
  [some text],
  Seq{Frame[0,
    code = Seq{
      more text; 
      the end.},
    cut = Seq{}
    ]},
  Seq{}]

\end{lstlisting}
\end{minipage}
&
\begin{minipage}{2in}
\begin{lstlisting}
(4)Machine[80,50,9,9,9,
  [some text],
  Seq{Frame[0,
    code = Seq{
      more text, 
      the end.},
    cut = Seq{}
    ]},
  Seq{}]
\end{lstlisting}
\end{minipage}
\\\hline
\begin{minipage}{2.3in}
\begin{lstlisting}

(5)Machine[80,50,19,19,19,
  [some text more text],
  Seq{Frame[0,
    code = Seq{ the end.},
    cut = Seq{}
    ]},
  Seq{}]

\end{lstlisting}
\end{minipage}
&
\begin{minipage}{2.7in}
\begin{lstlisting}
(6)Machine[80,50,28,28,28,
  [some text more text the end.],
  Seq{Frame[0,
    code = Seq{},
    cut = Seq{}
    ]},
  Seq{}]
\end{lstlisting}
\end{minipage}
\\\hline
\end{tabular}

\caption{Execution of Ordered Text\label{fig:Execution-of-Ordered}}

\end{center}
\end{figure}

Figure \ref{fig:Execution-of-Ordered} shows the steps taken by the
machine when it processes a document consisting of 3 ordered strings.
The machine is printed (using the pretty printer) as Machine{[} followed
by the page width, ribbon width, text position, line position and
ribbon position. The contents of the output buffer is on the following
line inside {[} and ]. The stack frames are printed next, followed
by the fail frames. Each frame is printed as Frame{[} followed by
the value used for indentation. The code and cut values follow on
the next lines. To save space, Order(d1.d2) is printed as d1;d2.

State (1) is the starting state. State (2) shows that the ordered
document at the head of the stack is processed first. State (3) has
output the first string, notice how the values of the positions in
the machine state have been updated. State (4) handles the ordered
document at the head of the stack. States (5) and (6) output the second
and third strings. State (6) is a terminal state because there is
a single frame with empty code.

Consider pretty-printing the following:

\begin{lstlisting}
Order(Just("some text"),
  Order(
    Block(
      Order(Newline(),
        Just("more text"))),
    Indent(2,
      Order(Newline(),
        Just("the end.")))))
\end{lstlisting}The machine starts in the following state:

\begin{lstlisting}
Machine[80,50,0,0,0,
  [],
  Seq{
    Frame[0,
      code = Seq{
               some text;
               ...;
               Indent[2
                 ...;
                 ...]},
      cut = Seq{}
      ]},
  Seq{}]
\end{lstlisting}The machine prints some text on the output and then encounters the
block:

\begin{lstlisting}
Machine[80,50,9,9,9,
  [some text],
  Seq{
    Frame[0,
      code = Seq{
               Block[
                 doc = 
                   ...;
                   ...];
               Indent[2
                 ...;
                 ...]},
      cut = Seq{}
      ]},
  Seq{}]
\end{lstlisting}A new frame is pushed that records the current indent position:

\begin{lstlisting}
Machine[80,50,9,9,9,
  [some text],
  Seq{
    Frame[9,
      code = Seq{
               Newline[];
               more text},
      cut = Seq{}
      ],
    Frame[0,
      code = Seq{
               Indent[2
                 ...;
                 ...]},
      cut = Seq{}
      ]},
  Seq{}]
\end{lstlisting}The newline and more text is printed with respect to the frame at
the head of the stack:

\begin{lstlisting}
Machine[80,50,28,18,9,
  [some text
         more text],
  Seq{
    Frame[9,
      code = Seq{},
      cut = Seq{}
      ],
    Frame[0,
      code = Seq{
               Indent[2
                 ...;
                 ...]},
      cut = Seq{}
      ]},
  Seq{}]
\end{lstlisting}The empty frame is popped, processing resumes with the Indent causing
a new frame to be pushed with an indent of 2:

\begin{lstlisting}
Machine[80,50,28,18,9,
  [some text
         more text],
  Seq{
    Frame[2,
      code = Seq{
               Newline[];
               the end.},
      cut = Seq{}
      ],
    Frame[0,
      code = Seq{},
      cut = Seq{}
      ]},
  Seq{}]
\end{lstlisting}Finally, the newline and text is processed with respect to the frame:

\begin{lstlisting}
Machine[80,50,39,10,8,
  [some text
         more text
  the end.],
  Seq{
    Frame[2,
      code = Seq{},
      cut = Seq{}
      ],
    Frame[0,
      code = Seq{},
      cut = Seq{}
      ]},
  Seq{}]
\end{lstlisting}All frames are empty and are then popped.

\section{Machine Implementation}

The machine shown in the previous section relines on a number of operations
in order to run. This section defines the operations and a machine
loader that creates an initial state.

When the machine encounters literal text to output, it must decide
whether the output will fit onto the current line given the page and
ribbon width. If the text cannot fit then the machine backtracks (if
possible) to an alternative layout. The following operation calaulcted
whether the text will fit: 

\begin{lstlisting}
context Docs
  @Operation canPrint(text,w,r,pl,pr)
    pl + text->size < w and
    pr + text->size < r
  end
\end{lstlisting}When the machine writes some text, the current state must be updated:

\begin{lstlisting}
context Machine
  @Operation write(text)
    emit(text,buffer,textPosition);
    self.textPosition := textPosition + text->size;
    self.linePosition := linePosition + text->size;
    self.ribbonPosition := ribbonPosition + text->size
  end
\end{lstlisting}The text output from the machine is recorded in a buffer:

\begin{lstlisting}
context Docs
  @Operation emit(s,b,i)
    @Count x from i to i + s->size do
      b.put(x,s->at(x - i))
    end
  end
\end{lstlisting}When a context switch takes place, the machine pushes a frame:

\begin{lstlisting}
context Machine
  @Operation pushFrame(indent,code,cut)
    self.stack := Seq{Frame(indent,code,cut) | stack}
  end
\end{lstlisting}A new choice point is pushed onto the fail stack:

\begin{lstlisting}
context Machine
  @Operation pushFail(code)
    self.fail := Seq{
      Fail(stack->head.indent(),
           textPosition,
           linePosition,
           ribbonPosition,
           code,
           stack->head.cut()) | 
      fail}
  end
\end{lstlisting}When the machine fails, a fail frame is popped from the fail stack
and the state of the machine is reset to the point at which the choice
point was recorded:

\begin{lstlisting}
ontext Machine
  @Operation fail()
    self.stack := fail->head.stack();
    self.stack := Seq{fail->head | stack};
    self.textPosition := fail->head.textPosition();
    self.linePosition := fail->head.linePosition();
    self.ribbonPosition := fail->head.ribbonPosition();
    self.fail := fail->tail
  end
\end{lstlisting}Writing a new-line updates the state of the machine:

\begin{lstlisting}
context Machine
  @Operation newline()
    let frame = stack->head
    in emit("\n" + spaces(frame.indent()),buffer,textPosition);
       self.textPosition := textPosition + frame.indent() + 1;
       self.linePosition := frame.indent();
       self.ribbonPosition := 0
    end
  end
\end{lstlisting}Finally, given a document to pretty-print, it is loaded onto a machine
as follows:

\begin{lstlisting}
context Machine
  @Operation load(code)
    self.stack := Seq{};
    self.pushFrame(0,code,Seq{});
    self.textPosition := 0;
    self.linePosition := 0;
    self.ribbonPosition := 0;
    self.fail := Seq{};
    buffer.setSize(0)
  end
\end{lstlisting}
\section{A Document Mapping}

We have defined a language for representing documents and a machine
that performs the document language. The document language can be
the target of any number of translations from source models where
we want to pretty-print the data. This section describes a translation
from XCore to the document language.

The translation from XCore to the documentation language is defined
as a dispatcher. This allows new handlers to be defined for user-defined
element types and allows the existing handlers to easily be redefined.
The dispatching class is defined as follows:

\begin{lstlisting}
context Walkers

  @Class PPrint metaclass Dispatcher 
  
    // The following constants define the default
    // pretty-printing parameters...
    
    // The depth limit...
    @Bind PRINTDEPTH   = 5              end
    
    // The limit on the number of elements 
    // in a sequence...
    @Bind PRINTLENGTH  = 10             end
    
    // If possible everything should fit 
    // into the page width...
    @Bind PAGEWIDTH    = 100            end
    
    // If possible everything should fit into 
    // the ribbon width on a page...
    @Bind RIBBONWIDTH  = 40             end
    
    // The following controls whether or not operations
    // are printed in all their glory. Should be true 
    // only when debugging compiled operations...
    @Bind PPRINTOPS   = false           end
   
    // Controls whether quasi-quotes are used to
    // print expressions...
    @Bind PPRINTEXP   = true            end
    
    // The following constants are used throughout
    // the Element pretty-printer...
    @Bind EQUALS      = Just(" = ")     end
    @Bind COMMA       = Just(",")       end
    @Bind LSQUARE     = Just("[")       end
    @Bind RSQUARE     = Just("]")       end
    @Bind NOTHING     = Just("")        end
    @Bind LCURL       = Just("{")       end
    @Bind RCURL       = Just("}")       end
    @Bind SEQ         = Just("Seq{")    end
    @Bind SET         = Just("Set{")    end
    @Bind BAR         = Just(" | ")     end
    @Bind DOTS        = Just("...")     end
    @Bind HASH        = Just("###")     end
    @Bind VECTOR      = Just("Vector[") end
    @Bind TABLE       = Just("Table[")  end
    
    // Controls the depth in each dispatcher...
    @Attribute depth     : Integer = PPrint::PRINTDEPTH   end
    
    // Controls the length in each dispatcher...
    @Attribute length    : Integer = PPrint::PRINTLENGTH  end
    
    // Shqring is handled by associating elements with labels
    // and using the labels for subsequent occurrences of an 
    // element. The jlabels table contains an empty string
    // literal that is updated with the label if the associated
    // element is ever encountered twice. If the label is included
    // in the first generation then it will be "" until the element
    // is encountered again at which point it is replaced with
    // a #(n)= which defines the label n t be the element...
    @Attribute jlabels   : Table   = Table(10)            end
    
    // A table associating elements with tags. Each element is
    // allocated a tag that can be used in subsequent references...
    @Attribute tlabels   : Table   = Table(10)            end
    
    // A counter that is used to generate labels...
    @Attribute labelc    : Integer = 0                    end
    
    // Whether operations are pretty-printed in long form...
    @Attribute pprintOps : Boolean = PPrint::PPRINTOPS    end
    
    // Whether expressions are pretty-printed using quasi-
    // quotes...
    @Attribute pprintExp : Boolean = PPrint::PPRINTEXP    end
    
    @Constructor() end
    
    @Constructor(depth,length) ! end
    
    @Operation dispatch(element:Element,depth:Element)
    
      // Call dispatch to translate an element to a document.
      // The current depath-level is supplied. If the max depth
      // level is reached then generate a short version of the 
      // element...
      if depth > self.depth
      then Just("<a " + element.of().name() + ">")
      else super(element,depth)
      end
    end
    
    @Operation label(e)
    
      // Record a label for e...
      labels.put(e,self.nextLabel())
    end
    
    @Operation indent()
    
      // Increase the indentation...
      self.indent := indent + 2
    end
    
    @Operation mark(element)
    
      // An element is marked by associating it with an
      // empty label. This empty label is generated along
      // with the document for the element. If the element
      // is subsequently encountered then the empty label 
      // is replaced by a unique label and the generated
      // occurrence becomes the defining occurrence...
      if jlabels.hasKey(element)
      then jlabels.get(element)
      else
        let just = Just("");
            tag = self.nextLabel()
        in jlabels.put(element,just);
           tlabels.put(element,tag);
           just
        end
      end
    end
    
    @Operation nextLabel()
    
      // Get a new label...
      self.labelc := labelc + 1;
      labelc
    end
    
    @Operation ref(element)
    
      // A subsequent occurrence of an element. Generates
      // a reference to the label. Note that the label is
      // replaced if it is empty. Therefore the original
      // occurrence of the label will become the defining
      // occurrence...
      if jlabels.hasKey(element)
      then 
        let just = jlabels.get(element);
            tag = tlabels.get(element)
        in just.setText("#(" + tag + ")=");
           Just("#(" + tag + ")")
        end
      else self.error("Cannot reference " + element)
      end
    end
    
  end
\end{lstlisting}Every element in XMF is to be pretty-printable, therefore we define
a pprint operation on Element that uses the dispatcher to translate
the receiver into a document element and then uses the pretty-printing
engine to translate the document into a string. The definition of
pprint for Element is as follows:

\begin{lstlisting}
import Walkers;
import PPrint;

context Element
  @Operation pprint()
  
    // Uses defaults defined in PPrint...
    self.pprint(PAGEWIDTH,RIBBONWIDTH,PPRINTDEPTH,PPRINTLENGTH)
  end

context Element
  @Operation pprint(width:Integer,
                    ribbon:Integer,
                    depth:Integer,
                    length:Integer,
                    linePosition:Integer):String
    // Returns a string after pretty-printing the supplied
    // value.
    let printer = Walkers::PPrint(depth,length) then
        doc = printer.dispatch(self,0);
        machine = Machine(width,ribbon)
    in machine.load(Seq{doc},linePosition);
       machine.run();
       machine.text()  
    end
  end
\end{lstlisting}Now, Element and appropriate sub-classes of Element must define handlers
for the dispatcher. Each handler produces a document that describes
how to prettily display the element. Here is the handler for Element:

\begin{lstlisting}
import Doc;

@Handler XCore::Element in PPrint(element,depth,encountered)
  // If there is no handler defined for the type of the receiver then
  // the toString() operation is used to produce the pretty output...
  Just(element.toString())
end;
\end{lstlisting}When a set is pretty-printed, we will try to print out all the elements
on a single line. If they do not fit then the elements are displayed
on separate lines.

\begin{lstlisting}
@Handler XCore::SetOfElement in PPrint(set,depth,encountered) 

  // Try to print the elements of the set on a single line.
  // If that fails then indent and print the elements on
  // separate lines.
  
  // Sets do not have state therefore we can ignore encountered...
  
  let seq = set->asSeq then
      // Get the documents for each of the elements, truncate to the 
      // current value of length if necessary...   
      docs = seq->take(length.min(seq->size))
                ->collect(e | self.dispatch(e,depth+1)) then    
      // Add a comma after each of the documents as a separator...  
      docs = if docs->isEmpty 
             then Seq{} 
             else docs->butLast->collect(d | 
                    Order(d,COMMA)) + 
                  Seq{docs->last} 
             end then  
      // The single line option..
      singleLine = docs->iterate(d l = NOTHING | Order(l,d));
      // The separate lines option - add newlines 
      // after each element (and comma)...
      nestedLine = 
        Indent(2,docs->iterate(d l = Just("") | 
                   Order(l,Order(Newline(),d))))
  in 
     // Record the current level of indentation (newlines will auto-tab)...
     Mark(  
      // Record the choice points for cut...
      Block(
        Order(
          // Display the Set{ token...
          SET,
           Order(
            // Try a single line, and fail to separate lines if necessary...
            Alt(singleLine,nestedLine),
             // However we get here, accept the output and throw away any
             // choice points up to the recent Block...
             Cut(
              // If we truncated then print out '...' ...
              Just(if seq->size <> docs->size 
                   then ",...}" 
                   else "}" 
                   end))))))
  end     
end;
\end{lstlisting}

\chapter{Teamworking}

Super-languages are often used to process structured data in flexible
ways. The features provided by super-languages, such as pattern matching,
make it easy to process structured data. In addition, super-languages
are meta-aware and can reason about information in terms of its type.
This chapter shows how XMF can be used to process data that represents
information from different teams and how the high-level features can
process the information to detect inconsistencies in the data.

Modelling can be used to address large-scale business problems, where
teams of individuals collaborate to produce different parts of a model.
Teamworking raises the issue of model management: managing different
versions of a model and merging different versions into a single coherent
whole. Each team works independently on their version of the model.
The different versions must be compared for consistency. Inconsistencies
are identified and can be offered up for conflict resolution. Once
the versions are consistent then the approach allows them to be merged
into a single model.

The approach uses a meta-description of modelling data and therefore
does not depend on the language that has been used for modelling.
This is attractive because model comparison engines can be constructed
once and then used throughout an organisation.

\section{Overview}

\begin{figure}
\begin{center}

\includegraphics[width=12cm]{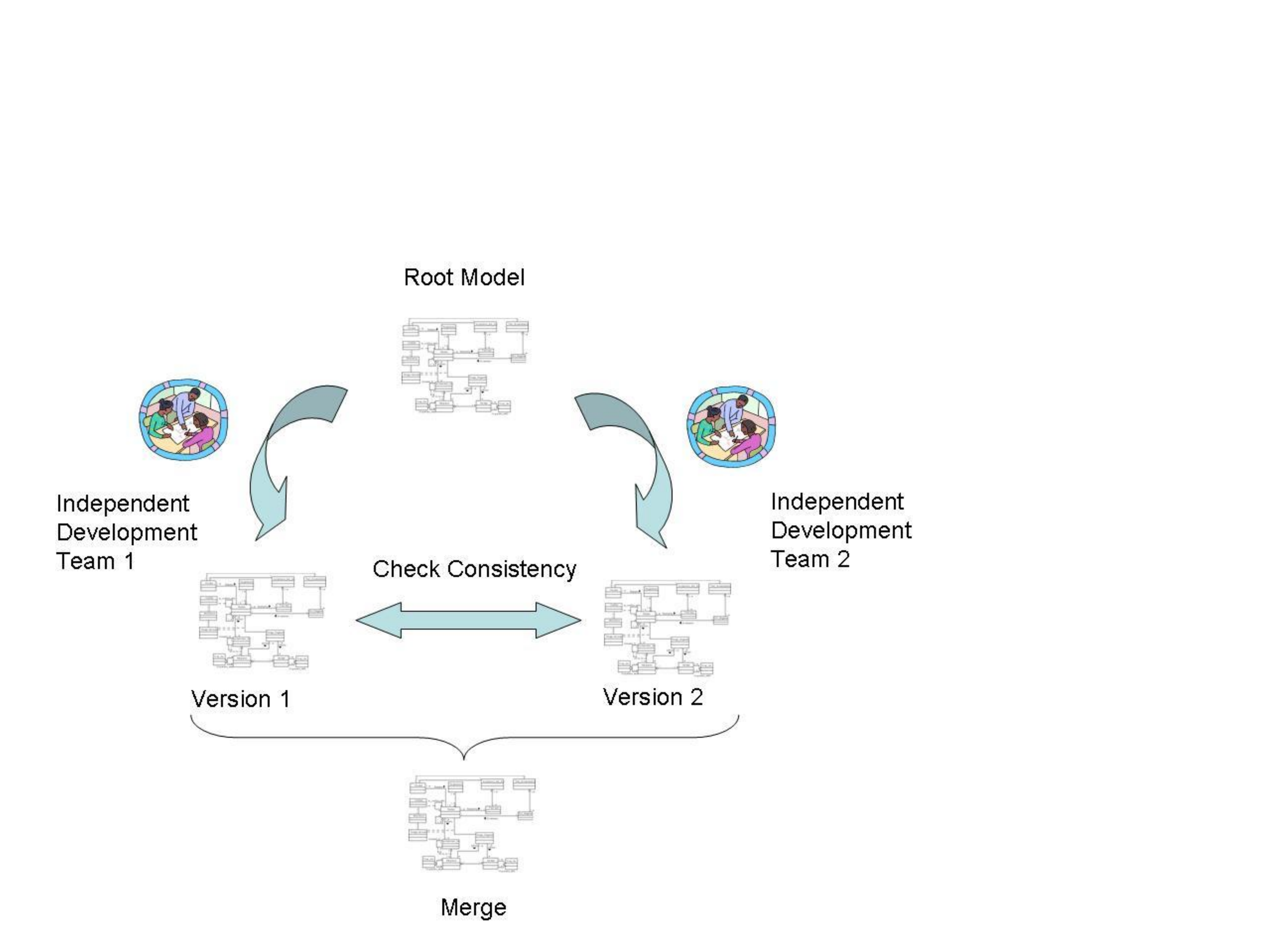}

\caption{Overview Of Teamworking\label{fig:Overview-Of-Teamworking}}

\end{center}
\end{figure}

Systems that support teamworking, such as CVS and SVN provide functionality
that is shown in figure \ref{fig:Overview-Of-Teamworking}. A root
system is modified independently by two or more teams. The modifications
are referred to as \textit{deltas} leading to two or more new \textit{versions}. 

Once satisfied that their independent versions are OK, the teams want
to merge. To do this the versions must be compared for inconsistencies
which must be resolved. Inconsistencies may occur because each team
has modified an element in the root in a different way. Resolution
may involve one team changing their version so that it is consistent
with the other or may involve both teams removing the conflicting
parts.

Once the conflicts have been resolved, the versions can be merged
to provide a new baseline of the system. This process can be performed
any number of times and allows multiple teams to work concurrently
on a large system.

Conventional technologies for teamworking, such as CVS and SVN, work
with files. They compare the contents of the files on a character-and-line
basis. Conflicts arise when two versions of the same file have differing
changes at the same position.

Modelling technologies have a similar requirement to teamworking as
a file-content based approach. It is worth comparing the basic features
of the technologies, since the key aspects are similar whilst the
implementation details are different. When comparing files the following
features are important:

\begin{enumerate}
\item Files have unique identifiers: their names within the file-system.
These identifiers are used by the teamworking technology when comparing
files for potential conflicts, and when merging files.
\item Files have an internal structure: the lines and characters. The internal
structure of a file is used by the teamworking technology when comparing
files for potential conflicts. The internal structure is also used
when merging files.
\end{enumerate}
When comparing models similar features arise:

\begin{enumerate}
\item Model elements must have unique identifiers. These may be automatically
imposed by the system or may be supplied by the user. For example,
when a model element is created it may be tagged with the unique combination
of the user's details and the current time and date. Alternatively,
a naming scheme within the model may be used to uniquely identify
elements. In practice it may be necessary to use a combination of
system and user supplied identifiers.
\item Models have internal structure. At the meta-level, all model elements
are just objects with named slots. Conflicts arise when the same slot
of a uniquely tagged object is changed by both development teams.
Also, when merging models, changes to different slots of the same
object can be merged providing that it is uniquely tagged.
\end{enumerate}
\begin{figure}
\hfill{}\begin{lstlisting}
let counter = 0
in @Operation newTag()
     counter := counter + 1;
     counter
   end
end
  
@Class TaggedObject 
  @Attribute tag : Integer = newTag() (?,!) end
end
\end{lstlisting}\hfill{}

\caption{A Class for Uniquely Tagging Instances\label{fig:A-Class-for}}

\end{figure}

For the purposes of this technical note the tags will be automatically
generated for model elements. Figure \ref{fig:A-Class-for} shows
a class TaggedObject whose instances are uniquely tagged through the
use of an auxiliary operation newTag with local state 'counter'.

\begin{figure}
\begin{center}

\includegraphics[width=12cm]{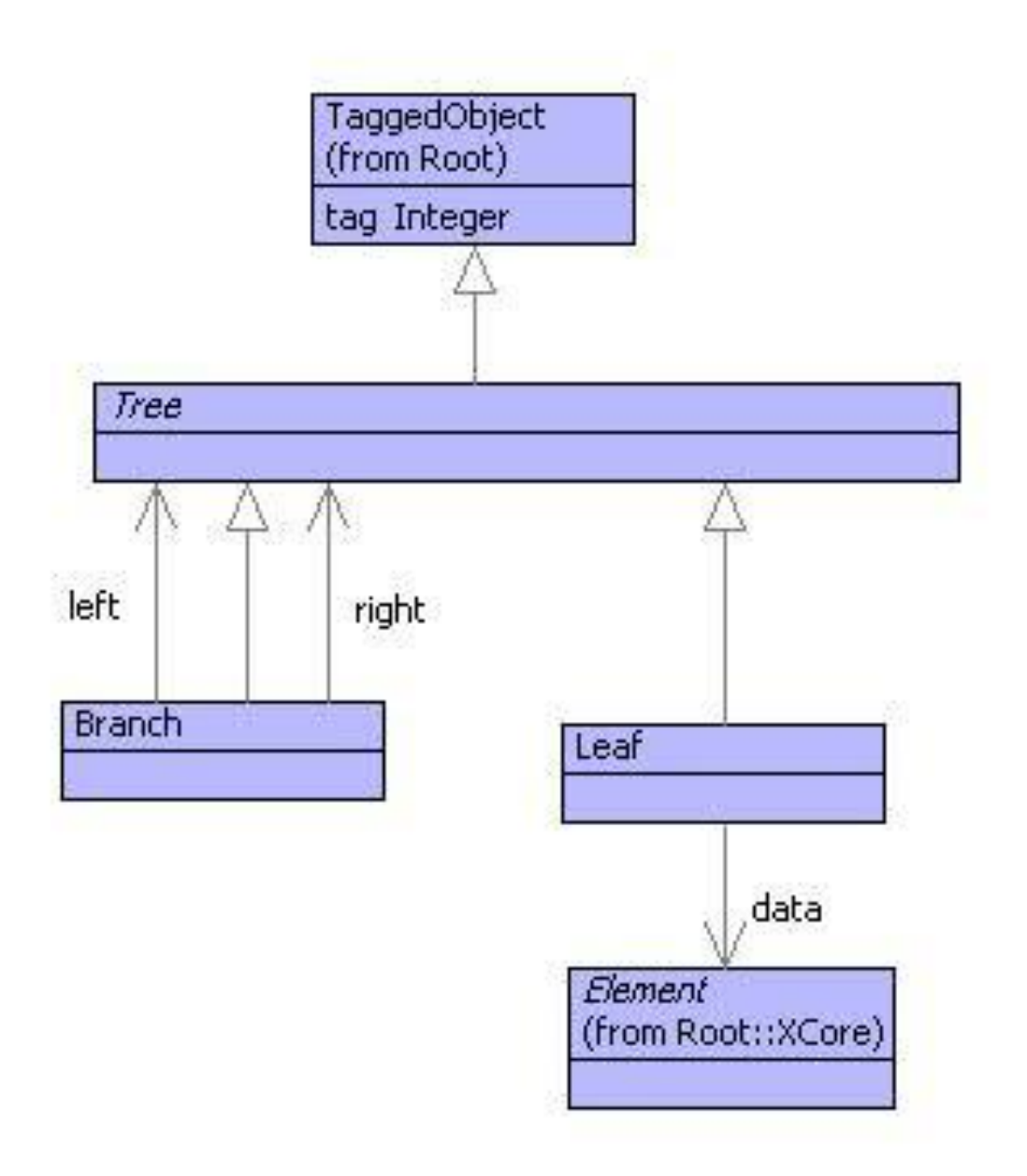}

\caption{A Model of Binary Trees\label{fig:Binary Trees}}
\end{center}
\end{figure}

Example elements in this technical note are taken from the model shown
in figure \ref{fig:Binary Trees}. As noted above, model elements
are just objects. Provided that they are uniquely tagged and the structure
of the data is known, model-based teamworking technology can be applied
to any model. The example data forms binary trees where the elements
are all uniquely tagged and the leaves of the tree have arbitrary
data elements attached to them.

\begin{figure}
\begin{center}

\includegraphics[width=12cm]{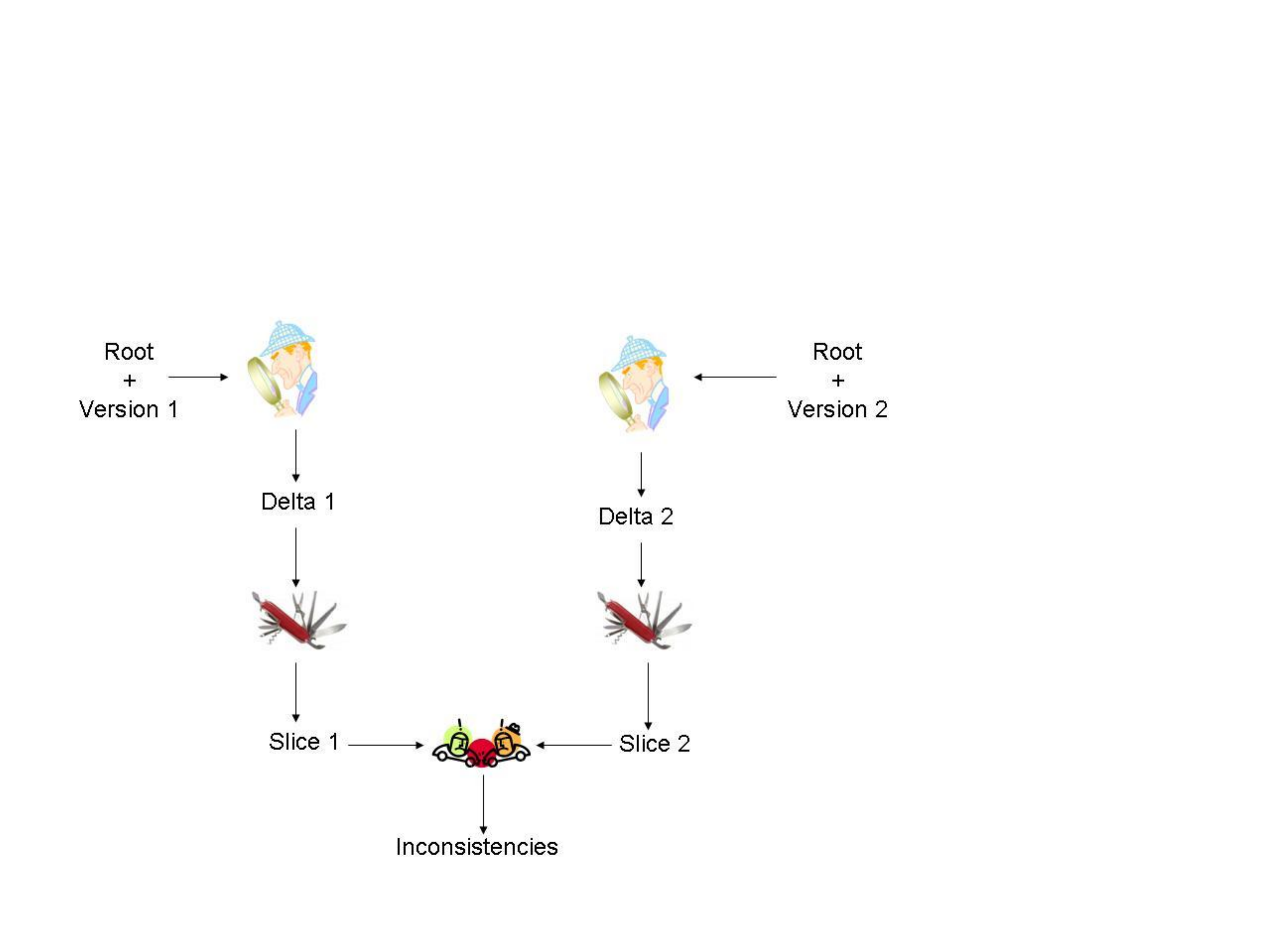}

\caption{Overview of Teamworking Implementation\label{fig:Overview-of-Teamworking}}

\end{center}
\end{figure}

\section{Overview of the Proposed Approach}

Figure \ref{fig:Overview-of-Teamworking} shows an overview of the
proposed tool support for teamworking. Both teams produce a new version
from the root. The root and version are compared and produce deltas.
Applying the delta to the root produces the version. The root can
be sliced with respect to the delta to produce just those model elements
that have changed. Two slices can be compared to see if the changes
are compatible; any incompatible changes are produced as inconsistencies.

If the two versions produce inconsistencies then they must be rationalized.
An inconsistency involves an element from each version and the slot
that has been modified. An inconsistency is rationalized when one
of the modifications (or both) is removed from the corresponding delta.
Once removed, the process is performed again, producing fewer inconsistencies.

The rationalization process continues until no inconsistencies arise.
In this case the deltas are independent and can be performed in any
order to produce a final merged model.

\section{The Delta Machine}

\begin{figure}
\begin{center}

\includegraphics[width=12cm]{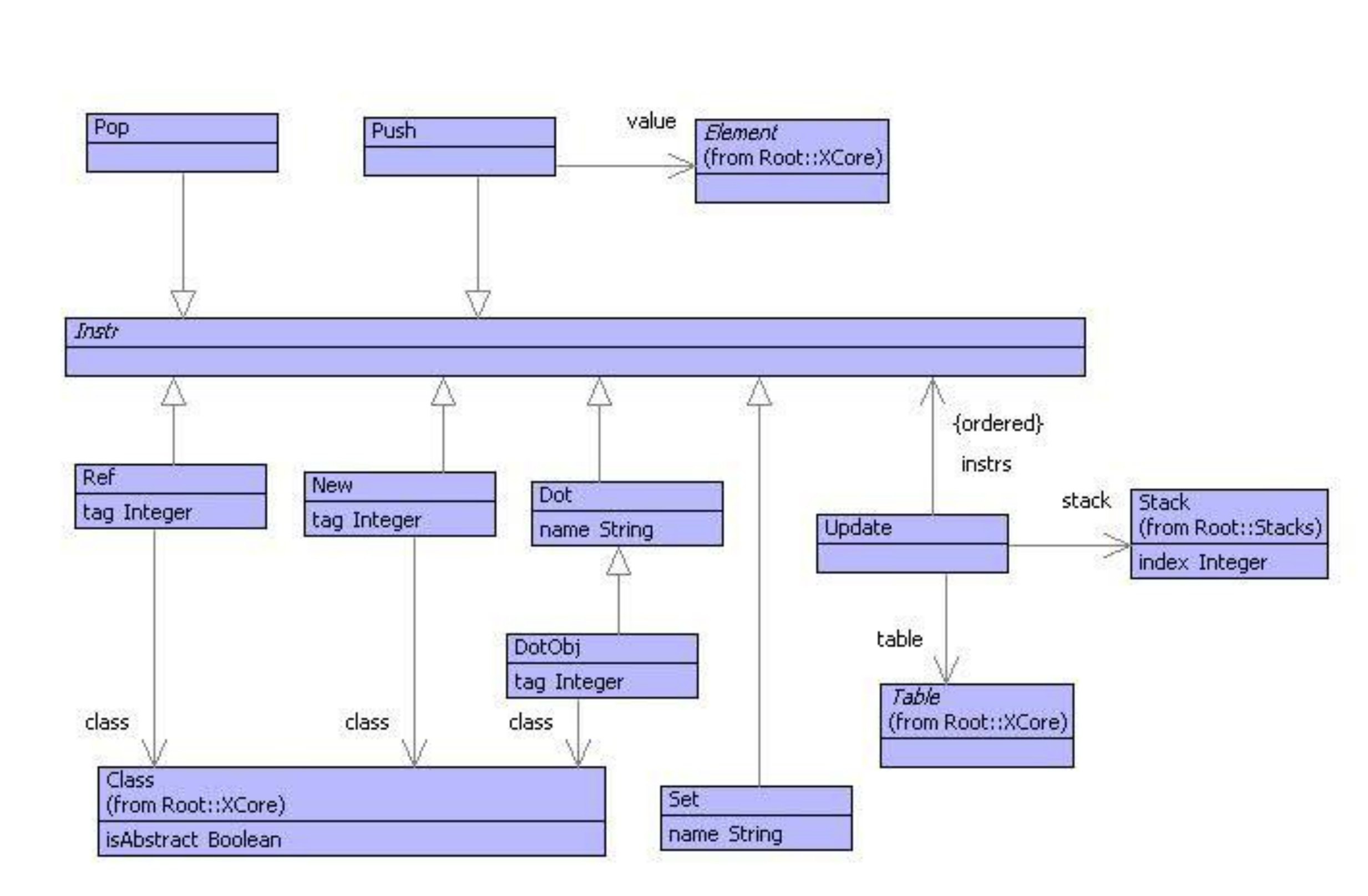}

\caption{The Delta Machine\label{fig:The-Delta-Model}}

\end{center}
\end{figure}

Teamworking produces two or more versions based on a shared root.
Each version adds new elements and makes modifications to the root.
Large parts of the root will be left unchanged and can be ignored
when the versions are compared and merged. It is therefore useful
to be able to isolate the changes or \textit{deltas} that have been
applied by each team to the shared root. This section describes a
language and its execution engine used to model deltas.

Model elements are uniquely tagged objects. An object consists of
slots; each slot has a name and a value. A value is either atomic
(such as an integer or a string) or is an object. Models are arranged
as graphs where the nodes in the graph are values and the edges are
slots. A model-graph may have cycles, i.e. an object can refer directly
or indirectly to itself.

A delta is a modification to a model. A delta can change a slot value.
The new value may be atomic, may introduce a reference to an existing
object or may create a completely new object. Imagine a machine that
takes as input a model and a delta and \textit{runs} the delta producing
a new model. The delta machine is shown in figure \ref{fig:The-Delta-Model}.
A delta is a sequence of instructions, the machine is defined by a
class called Update.

\begin{figure}
\hfill{}\begin{lstlisting}
@Operation run1(instr)
    @Case instr of
(1)   Push(v) do
        stack.push(v)
      end
(2)   Pop() do
        stack.pop()
      end
(3)   Dot(n) do
        stack.push(stack.top().get(n))
      end
(4)   Set(n) do
        stack.top().set(n,stack.pop())
      end
(5)   Ref(c,i) do
        stack.push(table.get(i))
      end
(6)   New(c,i) do
        let o = c()
        in o.setTag(i);
           table.put(i,o);
           stack.push(o)
        end
      end
      else self.error("Unknown instr: " + instr.toString())
    end
end
\end{lstlisting}\hfill{}

\caption{The Delta Machine Interpreter\label{fig:The-Delta-Machine}}

\end{figure}

The delta machine execution engine is defined in figure \ref{fig:The-Delta-Machine}.
The machine has a stack, used to hold for intermediate results and
the model being updated. The machine also has a table associating
the tags in the model with the model elements, used to move elements
around the model. To run the machine, the root model element is pushed
onto the stack, the table is populated with all associations in the
model and sequence of instructions is loaded onto the machine.

The machine executes by case analysis on the instructions. Each instruction
is popped from the instruction stack and performed by the operation
run1. Case (1) pushes an atomic value v onto the stack. Case (2) pops
the value at the head of the stack. Case (3) expects an object at
the head of the stack and pushes the value of slot n onto the stack.
Case (4) expects a value above an object on the stack. The value is
popped and the slot named n is updated. Case (5) pushes an existing
element in the model (held in the table) onto the stack. Case (6)
creates a new object of type c with tag i and pushes it onto the stack.

\begin{figure}
\subfigure[Tree Before Update]{

\includegraphics[width=12cm]{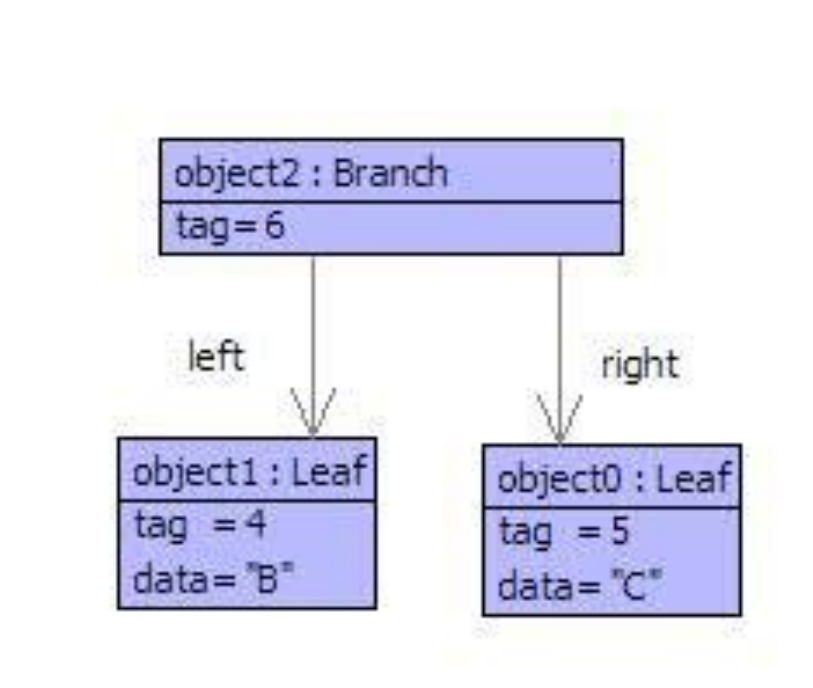}}

\subfigure[Tree After Update]{

\includegraphics[width=12cm]{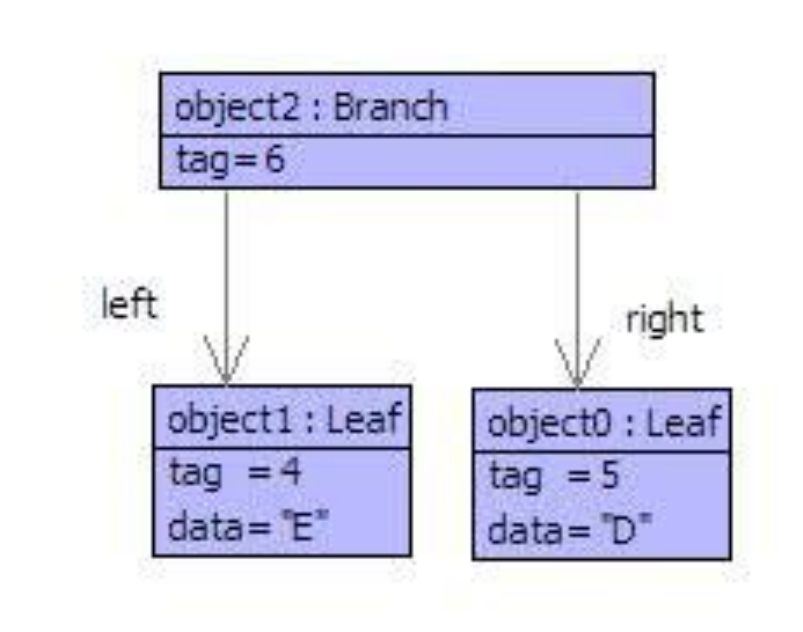}}

\caption{Tree Update\label{fig:Tree-Update}}

\end{figure}

Figure \ref{fig:Tree-Update} shows a tree before and after modification.
The delta machine instructions that perform the change are as follows:

\begin{lstlisting}
Dot(left)
Dot(data)
Pop()
Push(E)
Set(data)
Set(left)
Dot(right)
Dot(data)
Pop()
Push(D)
Set(data)
Set(right)
\end{lstlisting}The delta machine program above runs by navigating from the root using
the Dot instructions. Each slot is referenced, its value is pushed
and then the slot is set. The delta machine program is produced by
a model comparator as described in the next section. The instructions
can be optimized by omitting redundant navigations:

\begin{lstlisting}
Dot(left)
Push(E)
Set(data)
Set(left)
Dot(right)
Push(D)
Set(data)
Set(right)
\end{lstlisting}
\section{Model Comparison}

Teamworking involves a root model and two or more versions. Given
a root and a version, a delta is a machine program that is produced
by comparing the root and version. The result is a machine program
that captures exactly the steps necessary to transform the root into
the version. This section describes a model comparator that is used
to produce a delta.

\begin{figure}
\hfill{}\begin{lstlisting}
@Operation compareObjects(new,old,oldDB,compared)
  if tag(new) = tag(old)
  then 
    if compared.hasKey(tag(new)) 
    then Seq{}
    else 
      compared.put(tag(new),tag(old));
      compareSlots(slots(new),slots(old),oldDB,compared)
    end
  elseif oldDB.hasKey(tag(new))
  then 
    let oldSlots = slots(oldDB.get(tag(new)))
    in Seq{Pop(),Ref(new.of(),tag(new))} + 
         compareSlots(slots(new),oldSlots,oldDB,compared)
    end
  else Seq{Pop()} + newObject(new,oldDB,compared)
  end
end
\end{lstlisting}\hfill{}

\caption{Object Comparison\label{fig:Object-Comparison}}

\end{figure}

Consider the models shown in figure \ref{fig:Tree-Update}. Suppose
that the before model is the root and the after model is a version
produced in a teamworking scenario. What steps are necessary to produce
the delta that transforms the root into the increment?

Starting with the root element of both models, compare the objects.
They both have the same tag and therefore at this stage no changes
have taken place. Now compare slots from each object. Start with the
left slot: traverse the slot, emit a Dot instruction, compare the
values found there, and emit a Set instruction. At this stage the
delta is:

\begin{lstlisting}
Dot(left),...,Set(left),...
\end{lstlisting}The values are both Leaf objects with the same tag. Therefore, compare
the data slot values: emit a Dot and a Set as before:

\begin{lstlisting}
Dot(left),Dot(data),...,Set(data),Set(left),...
\end{lstlisting}When the data slot values are compared, it is found that the value
has changed to E. Therefore emit a Pop instruction and Push the new
value:

\begin{lstlisting}
Dot(left),Dot(data),Pop(),Push(E),Set(data),Set(left),...
\end{lstlisting}If this is repeated for the right slot then the complete delta is
produced as shown in the previous section.

Figure \ref{fig:Object-Comparison} shows the operation which is the
entry point for the object comparison. The four arguments are: the
new object (initially the root element of the new version); the old
object (initially the root element of the root model) ; a table (oldDB)
associating tags with elements from the root model; and, a table (compared)
that records comparisons as they are performed so that the any comparison
happens once (otherwise cycles in the model would lead to cycles in
the comparator).

The second part of the operation shown in figure \ref{fig:Object-Comparison}
has not been covered by the running example. If the tags of the new
and old objects are not the same then this means that the old object
has been replaced with the new object. In this case, either some existing
element from the root model has been linked into the model, or the
new model contains a completely new object. This is determined using
the oldDB which contains associations for all the tags in the root
model.

\begin{figure}
\hfill{}\begin{lstlisting}
@Operation compareSlots(newSlots,oldSlots,oldDB,compared)
  if newSlots->isEmpty
  then Seq{}
  else
    let newSlot = newSlots->head then
        name = slotName(newSlot)
    in @Find(oldSlot,oldSlots)
         when slotName(oldSlot) = name
         do let newSlots = newSlots->tail;
                oldSlots = oldSlots->excluding(oldSlot)
            in compareSlot(newSlot,oldSlot,oldDB,compared) +
               compareSlots(newSlots,oldSlots,oldDB,compared)
       end
    end
  end
end
\end{lstlisting}\hfill{}

\caption{Slot Comparison\label{fig:Slot-Comparison}}

\end{figure}

Slot comparison is defined by the operation in figure \ref{fig:Slot-Comparison}.
The compareSlots operation simply runs through the new slots, finding
old slots with the same name and comparing them using the operation
compareSlot.

\begin{figure}
\hfill{}\begin{lstlisting}
@Operation compareSlot(newSlot,oldSlot,oldDB,compared)
  let name = slotName(newSlot);
      newValue = slotValue(newSlot);
      oldValue = slotValue(oldSlot)
  in if newValue.isReallyKindOf(Object)
     then 
       Seq{DotObj(name,value.of(),tag(newValue))} + 
         compareObjects(newValue,oldValue,oldDB,compared) + 
       Seq{Set(name)}
     else 
       Seq{Dot(name)} + 
         compareValues(value,oldValue,oldDB,compared) + 
       Seq{Set(name)}
     end
  end
end
\end{lstlisting}\hfill{}

\caption{Comparing Slot Values\label{fig:Comparing-Slot-Values}}

\end{figure}

Comparing slot values is shown in figure \ref{fig:Comparing-Slot-Values}.
This operation produces the Dot ... Set machine instructions. For
reasons to be explained later, if the slot values are objects then
a DotObj instruction is used.

\begin{figure}
\hfill{}\begin{lstlisting}
@Operation newObject(new,oldDB,compared)
  if compared.hasKey(tag(new))
  then Seq{Ref(new.of(),tag(new))}
  else 
    compared.put(tag(new),null);
    Seq{New(new.of(),tag(new))} +
      newSlots(slots(new),oldDB,compared)
  end
end
\end{lstlisting}\hfill{}

\caption{Creation of New Object\label{fig:Creation-of-New}}

\end{figure}

Figure \ref{fig:Creation-of-New} shows the instructions produced
when a slot in the new model contains an object that does not match
that in the old model. If the object has been encountered before then
a Ref instruction is produced (the machine will push a reference to
an existing object). Otherwise, the new object has not been seen before,
it is recorded and newSlots is used to process the slot of the new
object.

\begin{figure}
\hfill{}\begin{lstlisting}
@Operation newSlots(slots,oldDB,compared)      
  if slots->isEmpty
  then Seq{}
  else 
    let slot = slots->head
    in newSlot(slot,oldDB,compared) +
       newSlots(slots->tail,oldDB,compared)
    end
  end
end
  
@Operation newSlot(slot,oldDB,compared)
  if value.isReallyKindOf(Object)
  then 
    if oldDB.hasKey(tag(slotValue(slot)))
    then 
     Seq{Ref(slotValue(slot).of(),tag(slotValue(slot)))} +
      compareObjects(new,oldDB.get(tag(new)),oldDB,compared)
    else newObject(slotValue(slot),oldDB,compared)
    end
  else Seq{Push(slotValue(slot))}
  end +
  Seq{Set(slotName(slot))}
end
\end{lstlisting}\hfill{}

\caption{Checking Slots of a New Object\label{fig:Checking-Slots-of}}

\end{figure}

Checking new slots is defined in figure \ref{fig:Checking-Slots-of}.
Each of the new sots is inspected in turn; the job of newSlots is
to emit instructions that will initialize the slots of the newly created
object at the head of the machine stack.

If the slot value is an object that also occurs in the old model then
a Ref instruction is emitted and the two objects are compared. Otherwise,
if the value is an object it must be new, so further instructions
are produced by newObject. Otherwise the value must be atomic so it
is simply pushed. In all cases the last instruction emitted is a Set
instruction to modify the newly created object at the head of the
machine stack.

\begin{figure}
\hfill{}\subfigure[Tree Before Update]{

\includegraphics[width=12cm]{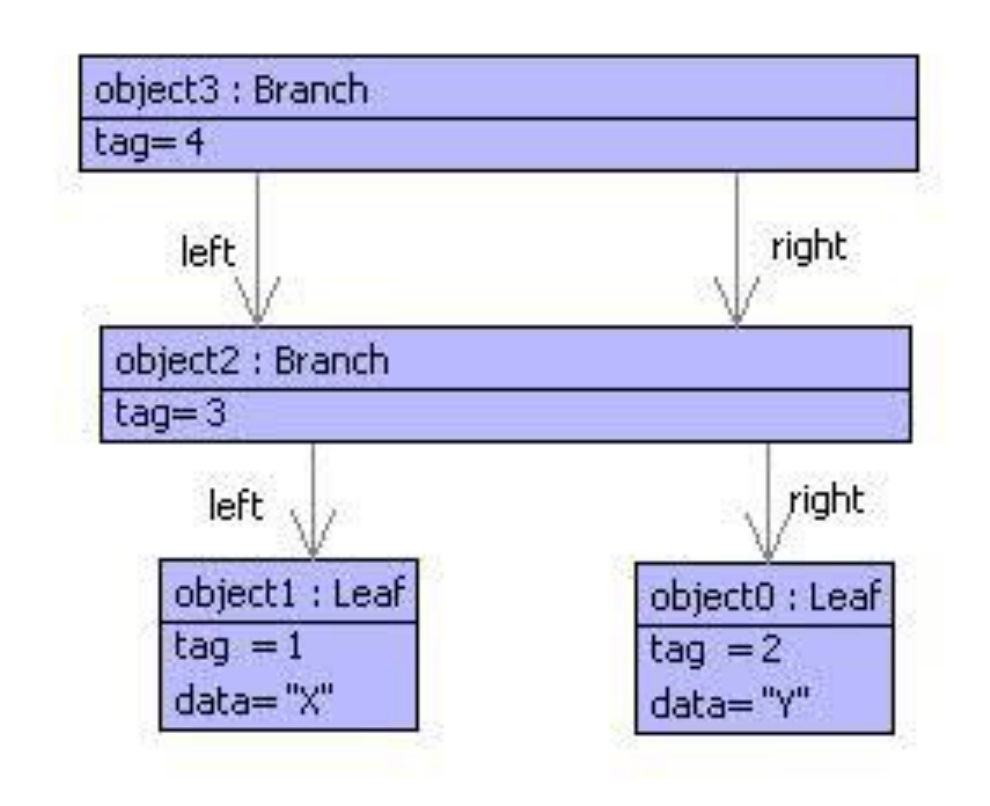}}

\subfigure[Tree After Update]{

\includegraphics[width=12cm]{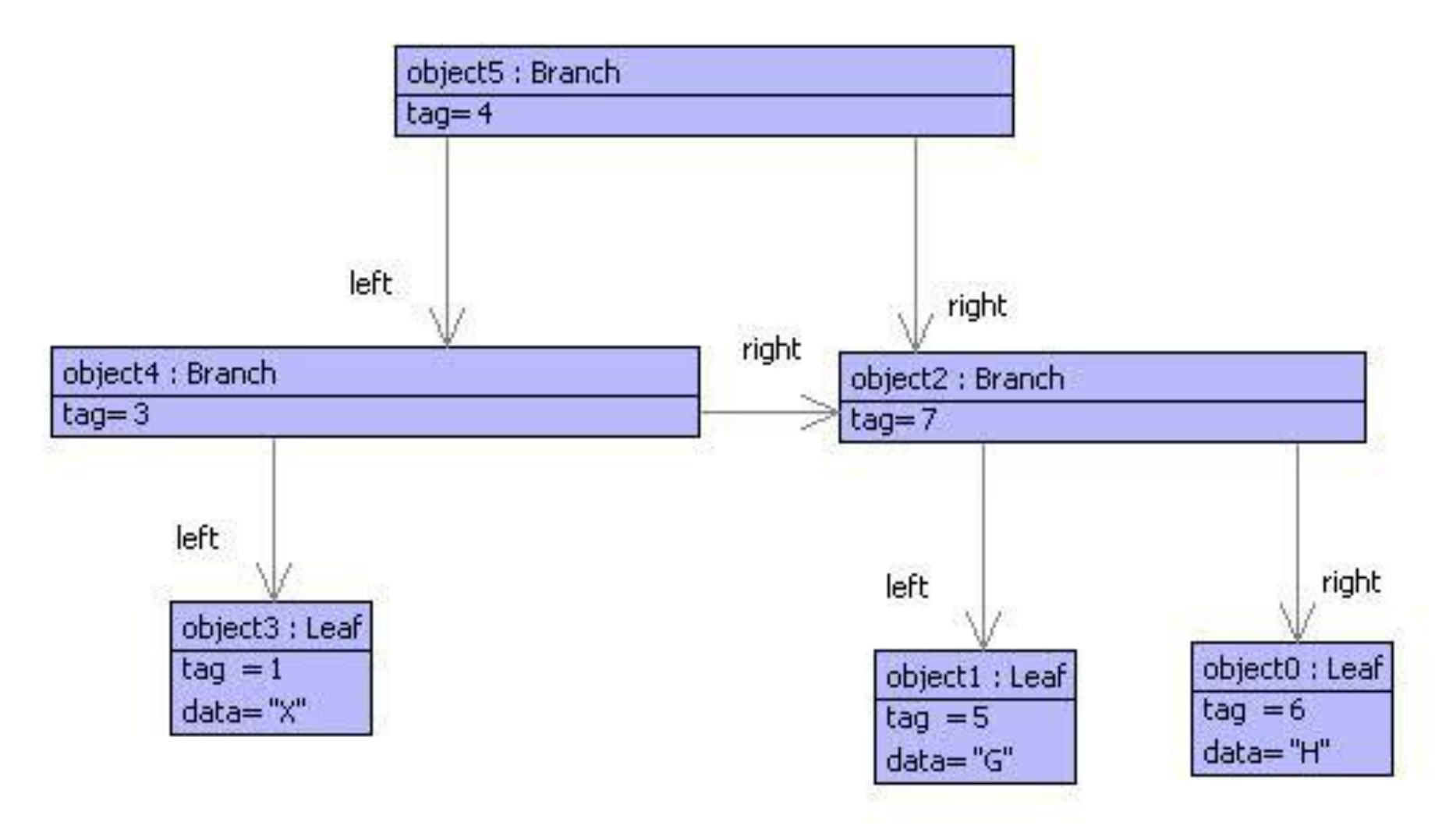}}

\caption{A More Complex Tree Modification\label{fig:A-More-Complex}}

\end{figure}

Figure \ref{fig:A-More-Complex} shows a slightly more complex tree
modification in which a new branch is created and shared between two
different places in the tree. Comparison of the before and after states
produces the following delta:

\begin{lstlisting}
Dot(left)
New(<Class Branch>,14)
New(<Class Leaf>,12)
Push(G)
Set(data)
Set(left)
New(<Class Leaf>,13)
Push(H)
Set(data)
Set(right)
Set(right)
Set(left)
Ref(<Class Branch>,14)
Set(right)
\end{lstlisting}
\section{Model Slices}

Teamworking involves a root model and two or more version models.
The version models are produced by independent teams making modifications
to the same root model. Many of the elements in the root models may
be unchanged in the increments. When comparing the two versions for
consistency it is useful to be able to ignore the unchanged elements.
A model that describes just the changed elements in a version is called
a \textit{slice.} Two slices can be easily compared since they just
contain the elements that have changed. Two models are inconsistent
if the same elements have been changed in different ways; slices make
consistency checking easy. This section describes how to slice a model
with respect to a delta.

\begin{figure}
\begin{center}

\hfill{}\subfigure[Sliced Elements]{

\includegraphics[width=12cm]{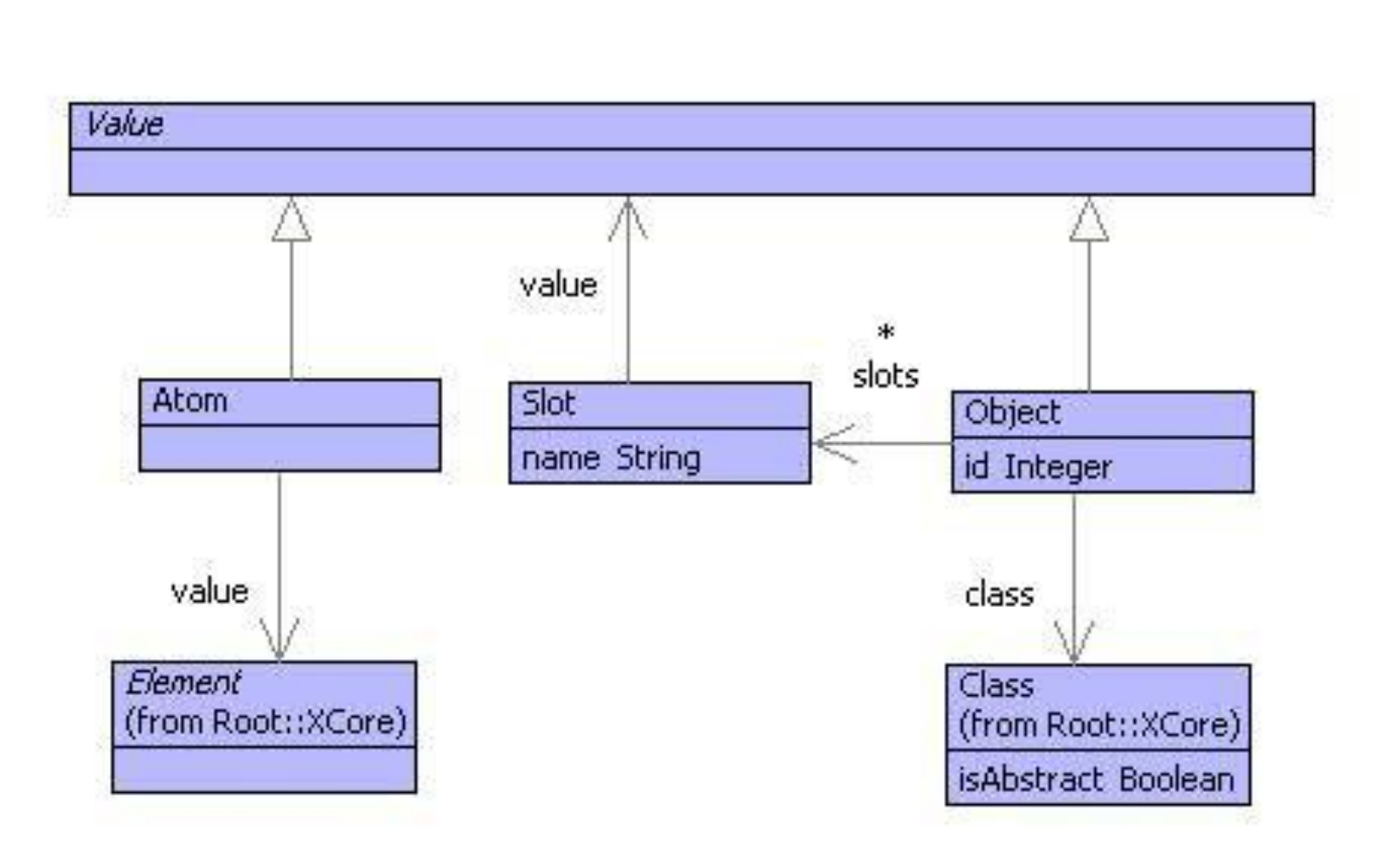}}

\subfigure[Slicing Engine]{

\includegraphics[width=12cm]{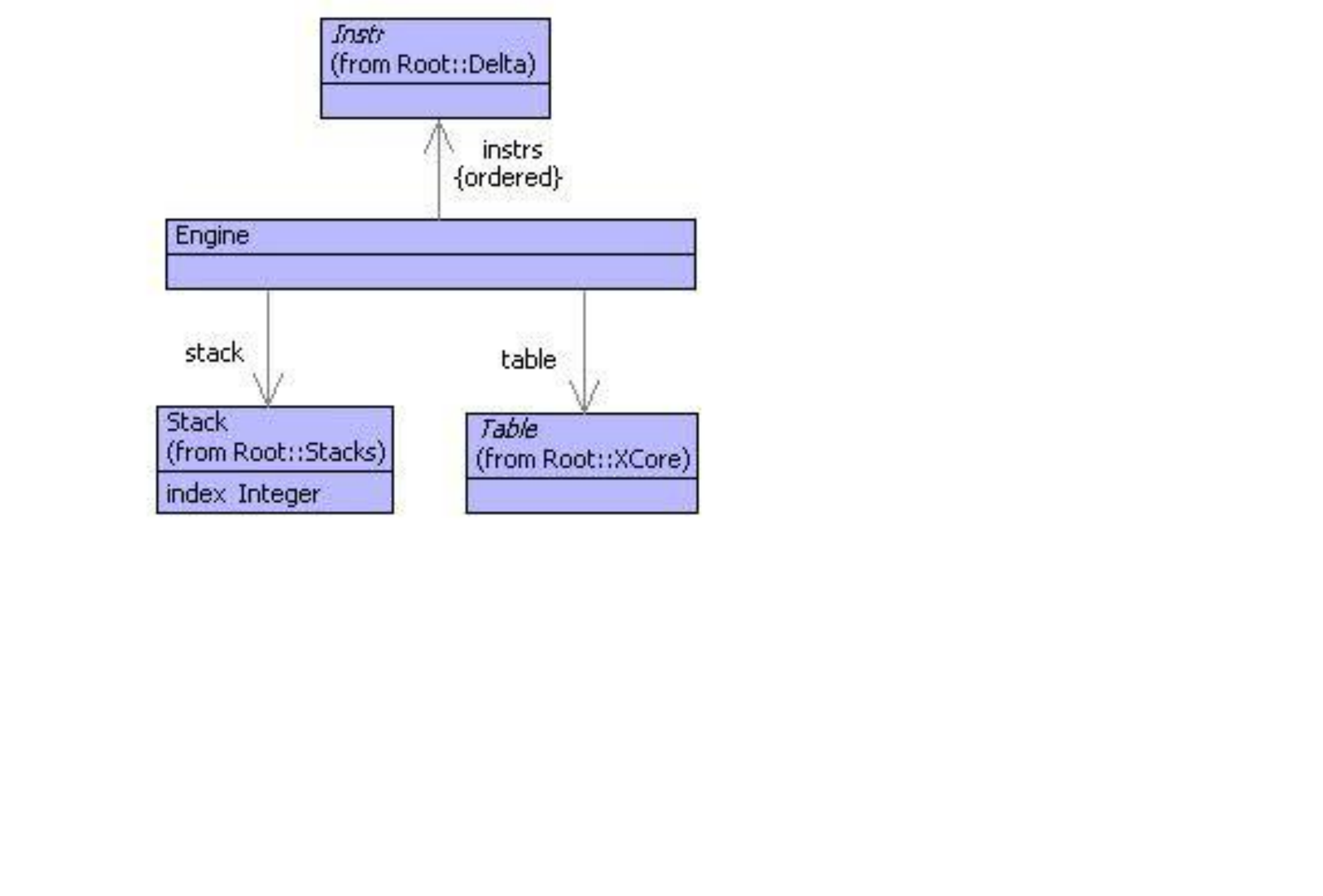}}

\caption{Model Slice\label{fig:Model-Slice}}

\end{center}
\end{figure}

Figure \ref{fig:Model-Slice} shows a model used to represent a slice.
Running a delta against a model using the slicing engine produces
a slice; the engine is like the delta machine except that it interprets
the delta instructions differently. To slice a model with respect
to a delta, the model is pushed onto the slicing engine, the delta
is loaded onto the machine and the instructions are performed. The
result at the head of the stack is an instance of the sliced object
model; only those things that are changed are contained in the slice.

\begin{figure}
\begin{center}

\hfill{}\begin{lstlisting}
@Operation slice(instr)
    @Case instr of
(1)   DotObj(n,c,i) do
        if table.hasKey(i)
        then stack.push(table.get(i))
        else
          let o = Object(c,i)
          in table.put(i,o);
             stack.push(o)
          end
        end
      end
(2)   New(c,i) do
        let o = Object(c,i)
        in stack.push(o);
           table.put(i,o)
        end
      end
(3)   Set(n) do
        let value = stack.pop() then
            obj = stack.top();
            slot = Slot(n,value)
        in obj.addToSlots(slot)
        end
      end
(4)   Push(v) do
        stack.push(Atom(v))
      end
      else self.error("Unknown instr: " + instr.toString())
    end
end
\end{lstlisting}\hfill{}

\caption{The Slicing Interpreter\label{fig:The-Slicing-Interpreter}}

\end{center}
\end{figure}

Figure \ref{fig:The-Slicing-Interpreter} defines the interpreter
that performs a single delta instruction on the slicing engine. If
the delta performs an update then the slicing engine records this
in the data model that is constructed at the head of the stack. It
is driven by case analysis on the instruction. Case (1) occurs when
an object slot is being referenced. In an optimized delta, slots that
contain objects are the only slots that need referencing (otherwise
they are being updated with atomic data and do not need referencing).
If the slot value has been seen before then the object data is in
the table, otherwise an object is created (recording that the object
is going to be updated) and recorded in the table.

Case (2) records the creation of a new object. Case (3) records the
update of a slot; the value is at the head of the stack. Case (4)
pushes some atomic data onto the stack ready for updating a slot.

\begin{figure}
\begin{center}

\includegraphics[width=12cm]{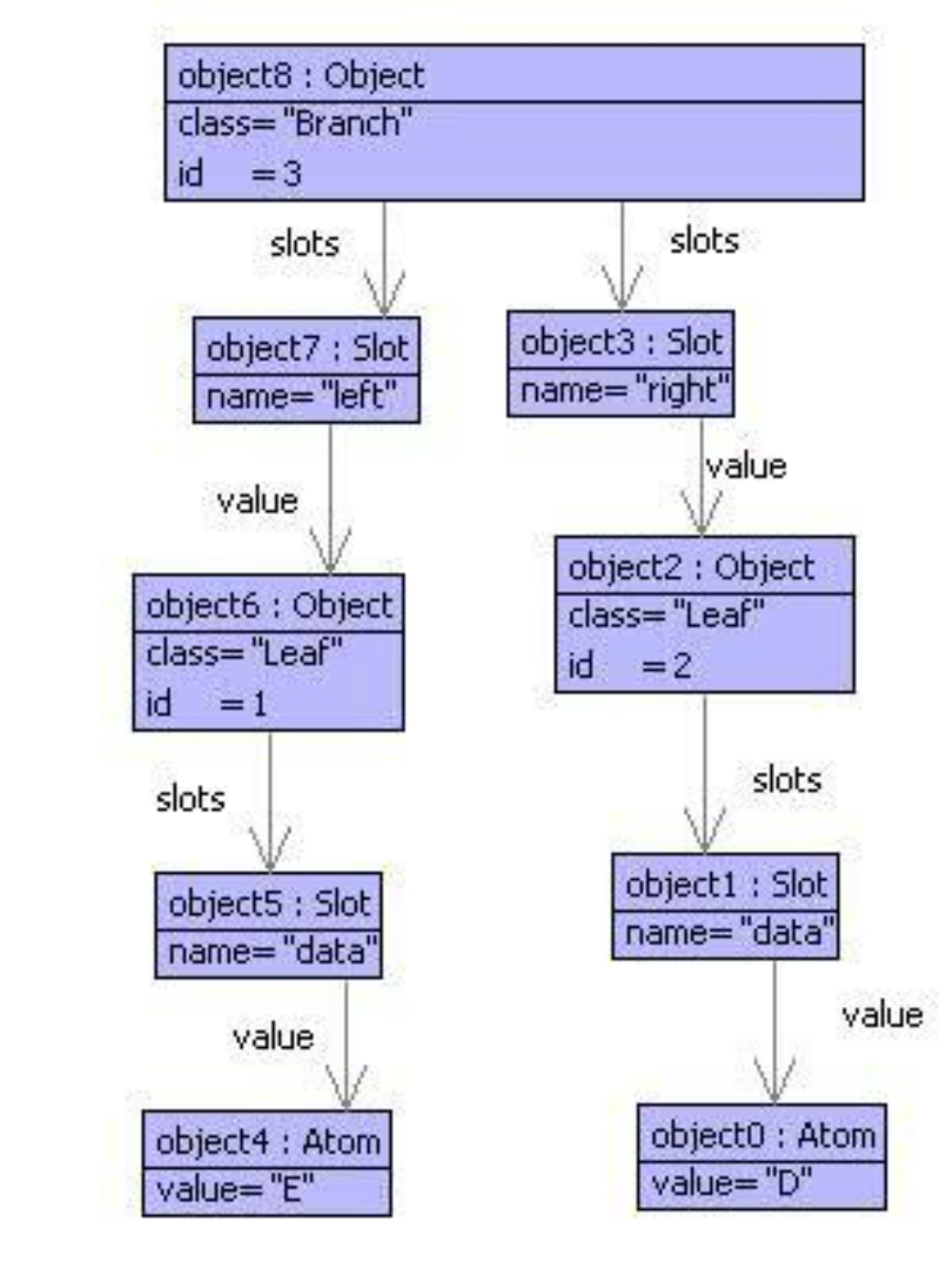}

\caption{Model Slice\label{fig:SlicedModel}}

\end{center}
\end{figure}

Figure \ref{fig:SlicedModel} shows a slice created by running the
delta against the starting snapshot shown in figure \ref{fig:Tree-Update}.
It clearly shows that the changed slots are the data of the left hand
branch and the data of the right hand branch.

\section{Conflicts}

Given two slices that arise from the same root model, it remains to
compare them to see if the changes are compatible. Since slices just
contain the changes that have occurred, conflict detection involves
determining whether the same slots have changed differently in the
two slices. This section defines an operation that takes two slices
and produces a set of conflicts; a conflict records two objects, a
slot and two incompatible values.

\begin{figure}
\hfill{}\begin{lstlisting}
@Operation objectConflicts(o1,db1,o2,db2)
  let S1 = o1.slots().name;
      S2 = o2.slots().name then
      common = S1->intersection(S2);
      C1 = slotChanges(o1,S1 - S2,db1,db2);
      C2 = slotChanges(o2,S2 - S1,db1,db2) then
      Conflicts = C1 + C2
  in @For name in common do
       let s1 = o1.slot(name);
           s2 = o2.slot(name) then
           C = conflictingSlots(o1,s1,db1,o2,s2,db2)
       in Conflicts := Conflicts + C
       end
     end;
     Conflicts
  end
end
\end{lstlisting}\hfill{}

\caption{Object Conflicts\label{fig:Object-Conflicts}}

\end{figure}

Figure \ref{fig:Object-Conflicts} shows the main conflict detection
operation. It is supplied with two objects and two tables associating
tags with objects. The common slot names are calculated and used to
check for any slot value conflicts (using conflictingSlots). Slots
in slice o1 that do not occur in o2 cannot conflict, and vice versa.
However, such slots may contain values that conflict so these are
checked using slotChanges.

\begin{figure}
\hfill{}\begin{lstlisting}
@Operation slotChanges(o1,slots,db1,db2)
  let C = Set{};
      v = s.value()
  in @For name in slots do
       let s = o1.slot(name)
       in if v.isKindOf(Object)
          then 
            if db2.hasKey(v.id())
            then
              let o2 = db2.get(v.id())
              in C := C + objectConflicts(v,db1,o2,db2)
              end
            else C := C + slotChanges(v,slots(v),db1,db2)
            end
          end
       end;
       C
     end
  end
end
\end{lstlisting}\hfill{}

\caption{Slot Changes\label{fig:Slot-Changes}}

\end{figure}

Figure \ref{fig:Slot-Changes} shows the definition of an operation
that calculates slot conflicts arising from an object o1. If any of
the slots contain an object then either the object is referenced in
the other slice via db2, or not. If the other slice contains a reference
to the object then objectConflicts is used to calculate any conflicts,
otherwise slot changes is used on each slot of v.

\begin{figure}
\hfill{}\begin{lstlisting}
@Operation conflictingSlots(o1,s1,db1,o2,s2,db2)
  let n = s1.name();
      v1 = s1.value();
      v2 = s2.value()
  in if v1.equal(v2)
  then 
    if v1.isKindOf(Object)
    then objectConflicts(v1,db1,v2,db2)
    elese Set{}
    end
  else Set{Conflict(n,o1,v1,o2,v2)}
  end
end
\end{lstlisting}\hfill{}

\caption{Conflicting Slots\label{fig:Conflicting-Slots}}

\end{figure}

Figure \ref{fig:Conflicting-Slots} calculates conflicts between slots
with the same name. If the values are the same and the values are
objects then the slots are compared. If the values are different then
a conflict is produced. Otherwise the values are the same and no conflict
arises.

\part{Language Engineering}

\chapter{Domain Specific Languages}

Software Frameworks and Domain Specific Languages are related in the sense 
that both aim to support the construction of specific types of software application. 
In many ways you can view the essence of a DSL as a software framework. 
Super-languages must support he definition of DSLs through language extension
facilities.

This chapter describes the relationship between a Software Framework and
a DSL. It gives an example of a DSL based on that of Martin Fowler from 
his presentation at JAOO 06.

\section{Software Frameworks and Domain Specific Languages}

Domain Specific Languages are increasingly popular as a technology 
and an approach to constructing applications. The idea is that a specific 
aspect of an application can be supported through the design and use of a 
language. The language may be used to configure parts of the application, 
to describe the GUI, to encode processing rules etc. Proponents of DSLs 
claim that, since the language abstracts away from the implementation 
details of the code otherwise necessary to achieve the same effect, the 
overall quality of the process and product is increased. The aspect 
supported by the DSL is easier to implement, the development is less 
error prone (since the language is specifically designed to support the task), 
the debugging and maintenance of the application is easier.

A Software Framework is a collection of classes that capture a design 
pattern or an aspect of an application. The key feature of a software 
framework is that it represents a reusable component where some of the 
features of the framework are fixed in structure and behaviour and some 
are left open to extension by the user. A framework differs from a more 
traditional software library in that the user provides the framework with 
the implementation of services that the framework calls. In a software 
library this works the other way round where the user calls services 
provided by the library (the so-called {\em Inversion of Control}). There 
are many examples of large scale general purpose software frameworks, 
such as Eclipse, but many packages provided for OO languages can be 
thought of as frameworks. Frameworks are also an architectural design pattern.

Domain specific languages usually (but not always) {\em do something}. To 
achieve this, the language has a semantics (a description of how it 
should operate) and some means of implementing the semantics - a DSL 
engine. The DSL engine is supplied with a representation of programs in 
the DSL. This program-as-data is then processed by the engine to achieve 
the required task.

A DSL engine is exactly the same as a Software Framework. The engine 
expects to be supplied with an implementation of a specific collection of 
services {\em implemented as data}. The task of the framework is to perform some 
standard, albeit domain specific, processing which uses the services 
supplied as DSL programs when appropriate.

It is a fact of Software Engineering that anything you can implement 
in program code, you can also implement in data with an associated 
engine that processes the data. This is exactly the relationship between 
Software Frameworks and DSLs. But why would you implement the services 
as data rather than implement them directly as program code (for example 
as an implementation of an abstract method or an interface)?

The answer is fundamentally an issue of {\em quality}. By pushing more of the 
framework services into data, the framework increases its level of control  
- nothing can escape! Therefore the more the framework matures, the higher 
the quality of the applications that use it. Data can also be processed in 
many ways: checking it, repairing it, translating it. This level of control 
over the services is not available to frameworks that represent their information 
as program code.

There are differences between a Software Framework and a DSL. Typically, a 
DSL will have some {\em concrete syntax} which is used to express the data to be 
supplied to the engine. Frameworks do not tend to have specific language 
syntax - they use a general purpose programming language to construct the 
services. However, these differences are not as significant as they might 
appear at first.

A DSL is usually implemented in terms of a parser that translates the 
concrete syntax into the abstract syntax (the data structures) and then 
an engine that processes the abstract syntax (the framework). The essence 
of the DSL is the abstract syntax and the engine - after all that is the 
part which is doing all the real work. The concrete syntax just makes the 
DSL easy to use and there may be many concrete syntaxes for the same DSL. 
Therefore, the {\em essence} of a DSL is the data and the engine - a Software 
Framework!

The rest of this document drills into the view of DSL as Software Framework. 
We use a simple example application to illustrate some of the points. The 
example (in addition to the link to Software Frameworks) is taken from a talk 
given by Martin Fowler at JAOO 06.

\section{Example Application}

Suppose we have data that contains information about customer events. 
A customer event might be a service call including the name of the customer, 
the type of the call (failed hardware, billing query etc) and the date. This 
information may be provided in real-time or in a log file as text:
\begin{lstlisting}
SVCLFOWLER    10101MS0120050313
SVCLHOHPE    10201DX0320050315
SVCLTWO     x10301MRP220050329
USGE10301TWO     x50214..7050329
\end{lstlisting}
The requirement is to have a Java program that processes the information. 
Obviously the first task of the Java application is to split the input 
strings up in terms of their fields. If there are a very small number of 
types of call then it would be OK to just write the appropriate string 
manipulation calls on the input.

However, this will rapidly become tedious and error prone as the number 
of different types of call grows. This leads to the idea of replacing 
the repeating patterns in the code with a software framework. Service 
plug-points in the framework allow the developer to register different 
types of call and the associated string manipulation that goes with it. 
This might be done using an abstract class:
\begin{lstlisting}
framework.registerStrategy(new ReaderStrategy() {
  
  public boolean recognizes(String s) {
    return hasPrefix(s,"SVCL");
  }

  public Record extract(String s) {
    return new ServiceCall(s.subString(4,18),...);
  }
});
\end{lstlisting}
The framework is responsible for processing the raw input text and calling 
the various registered recognizers in order to process the lines.

In developing a framework such as that shown above, we observe that the 
services are written in Java. There is little the framework can do to 
analyze each registered service in order to make sure that it is legal. 
Rules governing a well-formed service might include checking that the 
indices used to chop up the string are non-conflicting and that the prefix 
labels for all recognizers are unique.

In addition to the quality control objections raised above, the service-as-code 
implementation uses a great deal of Java which is standard for each of the 
recognizers. Essentially, there is a pattern to defining a recognizer where 
only the prefix name, the type of the record and indices change. It would be 
much better from the point of quality to be able to abstract away from this 
{\em implementation noise}.

It would be much better to represent the framework services as data. This is 
because the framework can perform analysis on the well-formedness of the 
services and also because data representations of services provides scope 
for concrete syntax which will address the problems that arise from repeating 
boilerplate code.

Fowler gives a service representation something like the following:
\begin{lstlisting}
public void ConfigureCallReader(Framework framework) {
  framework.registerStrategy(ConfigureServiceCall());
  framework.registerStrategy(ConfigureUsage());
}
 
private ReaderStrategy ConfigureServiceCall() {
  ReaderStrategy result = new ReaderStrategy("SVCL",typeof(ServiceCall));
  result.addFieldExtractor(4,18,"CustomerName"));
  result.addFieldExtractor(19,23,"CustomerID"));
  result.addFieldExtractor(24,27,"CalltypeCode"));
  result.addFieldExtractor(28,35,"DataOfCallString"));
}
 
private ReaderStrategy ConfigureUsage() {
  ReaderStrategy result = new ReaderStrategy("USGE",typeof(Usage));
  result.addFieldExtractor(4,8,"CustomerID"));
  result.addFieldExtractor(9,22,"CustomerName"));
  result.addFieldExtractor(30,30,"Cycle"));
  result.addFieldExtractor(31,36,"ReadDate"));
}
\end{lstlisting}
Notice how everything about the service is now represented as data that 
is supplied to the framework. The framework has a model of this language 
(the classes for ReaderStrategy, FieldExtractor etc) that it can manipulate 
the 'program' against. Therefore it can perform analysis on the registered 
services and translate them into something else if it wishes. Neither of 
these activities would be possible of the services were supplied as code.

Finally, note that the services-as-data example above involves further 
boilerplate code. In some ways this is worse than the boilerplate in the 
services-as-code example since there is more of it! Furthermore, we are 
dealing with two languages here: Java is used to represent a language for 
reader strategies. It is therefore difficult to see the wood for the trees.

The solution is to design a concrete syntax for the call language and to 
have a parser translate the concrete syntax into instances of the abstract 
syntax (classes ReaderStrategy, FieldExtractor etc). There can be any number 
of concrete syntaxes for a language. Fowler proposes both XML and a bespoke 
syntax.

XML is popular because it is standard. Whilst XML is widely used for 
configuring tools and frameworks, it is in many ways a lowest common 
denominator and lacks the variety of syntax features that is the essential 
reason for using a concrete syntax in the first place. Having said that, 
XML might be attractive because it can be processed easily by several tools. 
A suitable representation for the DSL described above is:
\begin{lstlisting}
<ReaderConfiguration>
  <Mapping code='SVCL' targetClass='ServiceCall'>
    <Field name='CustomerName' startPos='4' endPos='18'/>
    ...
  </Mapping>
  <Mapping code='USGE' targetClass='Usage'>
    ...
  </Mapping>
  <Strategy>
    <ref name='SVCL'/>
    ...
  </Strategy>
</ReaderConfiguration>
\end{lstlisting}
A bespoke syntax for a DSL should be designed for readability and therefore 
to aid development and maintainability. Here is a possible syntax:
\begin{lstlisting}
@Reader CallReader

  map(SVCL,ServiceCall)
    4-18:CustomerName
    19-23:CustomerID
    24-27:CallTypeCode
    28-35:DataOfCallString
  end

  map(USGE,Usage)
    4-8:CustomerID
    9-22:CustomerName
    30-30:Cycle
    31-36:ReadDate
  end

  do 
    ServiceCall
    Usage
end
\end{lstlisting}

\section{Abstract Syntax}

\begin{figure}
\begin{center}
\includegraphics[width=12cm]{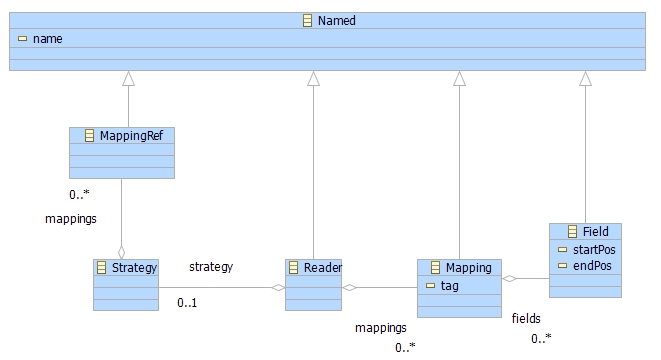}
\caption{DSL Abstract Syntax\label{fig:DSLAbstractSyntax}}
\end{center}
\end{figure}

Our DSL requires an abstract syntax to represent the information 
that is supplied to the framework. Figure \ref{fig:DSLAbstractSyntax} shows 
classes and relationships defined by the abstract syntax.The rest of this 
section discusses how these classes are used in the definition of a reader.

Named is an abstract class that defines a name of type string. A Reader 
is a container of mappings and a strategy. The mappings define the tagged 
and formatted fields that are received as input from the log file. Each 
Mapping is named and collectively they define a library available to the 
containing reader reader. The Strategy is used to reference mapping by name. 
The strategy is used to register mappings with the framework. In a more 
sophisticated language, the strategy could impose constraints on how the 
mappings are used, for example by requiring them to be used in a particular 
order.

\section{Implementation}

There are a number of ways in which we can implement the relationship 
between the abstract syntax defined in the previous section and the 
software framework. Two standard ways of achieving this are: to embed 
the language into the same platform as the framework; or, to translate 
the syntax onto a different platform. There is no real reason why you 
would do one rather than the other. In many cases it will be determined 
by other factors, for example when the DSL implementation technology is 
different from the framework technology.

\subsection{Embedded}

Suppose that the technology platform used to represent the DSL is the 
same as that for the framework. In this case we may choose to dynamically 
link the syntax to the framework which will allow new reader definitions 
to be added while the framework is running. Suppose that the framework is 
implemented in XMF and that the class Framework implements a register 
operation:
\begin{lstlisting}
context Framework
  @Operation register(reader:Reader) 
    reader.strategy().register(self,reader)
  end

context Strategy
  @Operation register(framework:Framework,reader:Reader)
    @For ref in mappings do
      ref.register(framework,reader)
    end
  end

context MappingRef
  @Operation register(framework:Framework,reader:Reader)
    reader.getMapping(name).register(framework)
  end

context Mapping
  @Operation register(framework:Framework)
    let r:ReaderStrategy = ReaderStrategy(tag,typeof(name)
    in framework.registerStrategy(r);
       @For field in fields do
         field.register(r)
       end
    end
  end

context Field
  @Operation register(readerStrategy:ReaderStrategy)
    readerStrategy.addFieldExtractor(startPos,endPos,name)
  end
\end{lstlisting}

\subsection{Translational}

If the DSL is implemented in a different technology to the DSL 
or if you wish to impose a phase distinction between the definition-time 
(processing the DSL) and run-time (running the framework) then you may 
choose the translate the DSL to the source code of a target language. 
To do this you need to write an exported that transforms instances of 
the abstract syntax classes to text. Again, support hat the abstract 
syntax classes are implemented in XMF and that we want to attach an 
operation to Reader that produces the Java source code defined above.

XMF provides code templates that can be used to embed the source code 
of one language in that of another. This is defined in the package CodeGen. 
CodeGen allows you to define a class which will be used to represent code 
templates in terms of drop and lift delimiters. If we define a code 
generation template called X using CodeGen then we might decide to use 
drop delimiters {\tt <} and {\tt >} and to use lift delimiters {\tt [} and {\tt ]}. 
Then we can write code templates of the form:
\begin{lstlisting}
// We are writing in the language of the
// surrounding context (e.g. XOCL)
@X
  // Write syntax in language X
  <
    // In here we have reverted back to XOCL
    [ 
      // Now we are back with language X again.
      < 
        // Back to XOCL
      >
      // Finished with XOCL, back to X
    ]
    // Finished with X back to XOCL
  >
  // Finished with XOCL back to X
end
// The end of the code template - back to XOCL.
\end{lstlisting}
The ability to switch back and forth between languages is very powerful. 
The templates observe variable scoping. The only special thing to be aware 
of is that within a template you must use 'e\_nd' instead of 'end' since the 
only time the keyword 'end' can be used is at the very end of the template.

Here is the code generator for Reader. The operation uses the template for Java:
\begin{lstlisting}
context Reader
  @Operation toJava(out:OutputChannel)
    @Java(out,9)
      public void Configure<name>(Framework framework) { 
        <@Loop ref in strategy.mappings() do 
           [framework.registerStrategy(Configure<ref.name()>()); 
           ] 
         e_nd>
      }

    <@Loop mapping in mappings do
        private ReaderStrategy Configure<mapping.name()>() {
          ReaderStrategy result = new ReaderStrategy("<mapping.tag()>",typeof(<mapping.name()>));
          <@Loop field in mapping.fields() do
             [result.addFieldExtractor(<field.startPos()>,<field.termPos()>,"<field.name()>"));
             ]
           e_nd>
        }
       ]
     e_nd>
    end
  end
\end{lstlisting}

\section{Concrete Syntax}

A DSL may have one or more concrete syntaxes. The syntax provides a way 
of easily constructing the service descriptions and a way of easily 
maintaining them. Typically a syntax for a DSL would be designed so 
that it abstracts away from the details of how the abstract syntax 
structures are created. The concrete syntax also reduces to 1 the 
number of languages that a develop has to deal with. For example, 
the developer is not using Java in order to construct programs in 
another language.

This section shows how two concrete syntaxes can be added on top 
of the abstract syntax for the call-data processing language. XMF 
is used to define the syntaxes in both cases. The first syntax is 
an embedded language within XOCL, therefore call-data processing 
readers can be defined in XOCL code. The second syntax is XML.

\subsection{Embedded}

XMF allows classes to have grammars. The grammar marks the class as 
a new syntax construct that can be used as part of XOCL or as a stand-alone 
language. A new syntax class C can be used as part of XOCL using the 
special '@' character before the name of the class:
\begin{lstlisting}
@C
  // Syntax for the new construct here...
end
\end{lstlisting}
The syntax construct for new reader definition shown above can be 
defined as follows:
\begin{lstlisting}
context Reader
  @Grammar
    // Start with the syntax rule named Reader...
    Reader ::= 
      // A reader is a name...
      n = Name 
      // ... followed by some mappings...
      M = Mapping* 
      // ... and then a strategy...
      s = Strategy 
      'end' 
      // The rule creates and returns a new reader...
      { Reader(n,s,M) }.
    Mapping ::= 
     // A mapping is the tag and the name...
     'map' '(' t = Name ',' n = Name ')' 
     // ... followed by some field specifications...
     F = Field* 
     'end' 
     // This rule creates and returns a new mapping...
     { Mapping(t,n,F) }.
    Field ::= 
      // A field specification is a start and end index
      // followed by a record name...
      s = Int '-' e = Int ':' n = Name 
     { Field(s,e,n) }.
    Strategy ::= 
      // A strategy is a sequence of names...
     'do' N = Name* 
     // Each name is turned into a mapping ref...
     { Strategy(N->collect(n | MappingRef(n))) }.
  end
\end{lstlisting}
Once the new syntax construct is defined, we can use it anywhere 
that an XOCL expression is expected. For example, we can define 
an operation that returns two readers:
\begin{lstlisting}
context Root
  @Operation test()
    let reader1 =
          @Reader CallReader
            map(SVCL,ServiceCall)
              4-18:CustomerName
              19-23:CustomerID
              24-27:CallTypeCode
              28-35:DataOfCallString
            end
            map(USGE,Usage)
              4-8:CustomerID
              9-22:CustomerName
              30-30:Cycle
              31-36:ReadDate
            end
            do 
              ServiceCall
              Usage
          end;
        reader2 =
          @Reader BankAccountReader
            map(DEPOSIT,Deposit)
              0-10:Name
              11-20:Account
              21-30:Amount
            end
            map(WITHDRAW,Withdraw)
              0-10:Name
              11-20:Account
              21-30:Amount
            end
          do
            Deposit
            Withdraw
          end
    in Seq{reader1,reader2}
    end    
  end
\end{lstlisting}

\subsection{XML}

An alternative syntax for our language is XML. This is possibly 
less attractive than the embedded syntax defined above because 
the XML concrete syntax is universal and therefore could be viewed 
as being the lowest common denominator. In many ways, XML is in a 
one-to-one correspondence with the abstract syntax and therefore 
really does not add much to the readability of a language. However, 
there may be other reasons why XML is attractive for the concrete 
syntax of a language. For example, you may have other XML-based tools 
that need access to the reader definitions.

XMF provides many different ways of processing XML documents. A useful 
approach is to define XML syntax rules very similar to those defined 
in the previous section. This can be done as follows:
\begin{lstlisting}
context Reader
  @Grammar ReaderXMLGrammar
     Reader ::=
       // An XML document containing a reader has a root
       // element with a ReaderConfiguration tag. The element
       // has an attribute called 'name' and children that
       // are described by the Mapping and Strategy rules...
       <ReaderConfiguration n=name>
         M = Mapping*
         s = Strategy
       </ReaderConfiguration>
       // Once a ReaderConfiguration element has been
       // successfully consumed according to the rules,
       // A reader is created and returned...
       { Reader(n,s,M) }.
    Mapping ::=
      <Mapping c=code t=targetClass>    
        F = Field*
      </Mapping>
      { Mapping(c,t,F) }.
    Field ::=
      <Field n=name s=startPos e=endPos/>
      { Field(n,s,e) }.
    Strategy ::=
      <Strategy>
        // The children of a strategy element are all
        // elements with tag 'Ref'. Each is processed in
        // turn by translating it to a MappingRef and then
        // all the children can be references as N...
        N = (<Ref n=name/> { MappingRef(n) })*
      </Strategy>
      { Strategy(N) }.
   end
\end{lstlisting}

\section{Conclusion}

This chapter has described the relationship between Software 
Frameworks and Domain Specific Languages, noting that DSLs are 
a special case of frameworks where the services are all represented 
in data. DSLs add an extra dimension to Software Frameworks by 
allowing the services to be defined using a specially designed 
concrete syntax. The relationship has been described using an 
example DSL due to Martin Fowler which has been implemented using 
XMF.

\chapter{Data-Centric Language Constructs}

Super-languages must provide mechanisms that allow new language constructs
to be defined for representing data. The new constructs must make
it easy to represent the required data structures. This chapter shows
how new language constructs can be defined for records and trees.

\section{Simple Records\label{sub:Simple-Records}}

This section provides an introduction to the definition of a new language
construct. All aspects of a new construct are covered, but the construct
is rather simple since it supports data records consisting of fields
with names and values. The record construct is an example of an integrated
language construct since the field values can be any language construct
supported by XOCL. Since XOCL is itself open-ended, this allows record
field values to be expressed as any language construct including records
and those constructs we have not even thought of yet!

Our requirement is for a database containing records. Each record
has a collection of fields with a name and a value. A database may
be filtered by supplying it with a predicate that expects a record.
The result is a new database containing all those fields that satisfied
the predicate.

\begin{figure}
\begin{center}

\includegraphics[width=12cm]{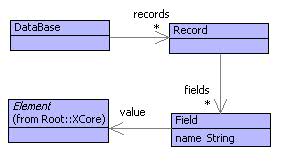}

\caption{\label{Data Records}A Simple Model Of Data Records}

\end{center}
\end{figure}

It is usual to start off with a data definition for the elements that
are required. Figure \ref{Data Records} shows the contents of a package
called Records. A database consists of a set of records. Each record
is a set of fields. Each field has a value that can be of any data
type (since Element is the super-class of everything).

The next step is to define a concrete syntax for the Record construct
that will synthesize an expresstion whose evaluation produces an instance
of the class Record. The complete grammar for Record is defined below,
followed by a step-by-step description:

\begin{lstlisting}
context Record
  @Grammar extends OCL::OCL.grammar
    Record ::= r = { [| Record() |] } Fields^(r) 'end'.
    Fields(r) ::= r = Field^(r) FieldTail^(r) | { r }.
    Field(r) ::= n = Name '=' e = Exp { 
      [| <r>.addToFields(Field(<n.lift()>,<e>)) |] 
    }.
    FieldTail(r) ::= ',' Fields^(r) | { r }.
  end
\end{lstlisting}When a syntax construct starts with the token @Record, XMF looks for
a grammar associated with the class Record. If it finds one then the
grammar should have a rule with the name Record; this rule is the
starting point for the parse.

The rule named Record defined above, creates an expression named r
that will create a new record when evaluated. The expression r is
passed as an argument to the rule named Fields. Once fields has completed
its parse, the keyword 'end' is expected. If successful, the value
returned by Fields is the result of the parse. So far we have dealt
with the following concrete syntax:

\begin{lstlisting}
@Record
  ???
end
\end{lstlisting}The rule Fields deals with the part labelled ??? above. Fields will
either recognize nothing (an empty record) or will recognize a sequence
of fields separated by commas. If there are fields, the rule uses
Field followed by FieldTail, otherwise the rule returns the supplied
expression r. Note the repeated use of the variable r in the rule
Fields. It is used as a formal argument, a modified variable, an actual
argument and in the body of an action.

A field is defined by the rule Field, it is a name n, followed by
= followed by an expression e. If successful, it synthesizes an expression
that adds a field to the supplied record r. So far, the following
can be recognized:

\begin{lstlisting}
@Record
  name = "Fred"
end
\end{lstlisting}FieldTail is used to allow more than one field to be recognized. A
FieldTail is either a comma followed by some fields (notice how the
record expression is threaded through so that it is continually built
up), or is nothing. 

The grammar is complete and the following records can be recognized
and the appropriate expressions synthesized:

\begin{lstlisting}
@Record
  name = "Fred",
  age = 35
end
\end{lstlisting}which synthesizes the following expression:

\begin{lstlisting}
Record()
  .addToFields(Field("name","Fred"))
  .addToFields(Field("age",35))
\end{lstlisting}
\section{Trees}

Information often naturally forms a tree structure where each component
of information is related to several children. Think of company structures:
responsibilities from CEO downwards or groupings from the board down.
Trees are often used to model physical entities such as cars where
a car consists of a body, an engine, electrical system, interior etc.
and where a body consists of doors, a windscreen etc.

Trees are useful when we want to organise a model in terms of parent-child
relationships which may be physical containment or may be logical
groupings (think of nested sets). This section gives an example of
a language for expressing trees for classification.

Consider a collection of bank records that are to be classified with
respect to awarding a loan. The first step to awarding the loan is
to determine an income threshold. If the applicant does not have an
annual salary of four times the loan amount then the loan is denied.

The next step is to consider the current debt position of the applicant.
If the applicant has debts of more than £10K then the loan is denied.
If they have debts of £5 - £10K then the loan application is referred
otherwise the application might be awarded.

Finally, if the applicant has an account with the bank, then the loan
is awarded otherwise it is referred.

\begin{figure}
\begin{center}

\includegraphics[width=12cm]{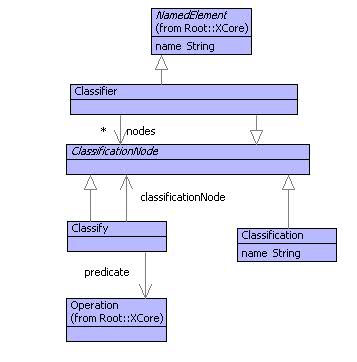}

\caption{\label{Classification-Tree-Model}Classification Trees}

\end{center}
\end{figure}

Figure \ref{Classification-Tree-Model} shows a simple model of a
classification tree. A classifier is a named element (so that it can
be put in name spaces and then referred to from elsewhere). A classification
is a property of some data (for example Award the loan, Deny the loan
or Refer the loan). Finally, a classify, is something that applies
a predicate to some data to determine whether the classification given
by the classificationNode applies.

To construct a classification tree, proceed as follows. Start with
a classifier node. Its child nodes, will be classify nodes with disjoint
predicates. The child node of a classifier node will be either a classification
or another classifier. Proceed with the tree until all categories
are classified.

The following example shows a classifier for a loan:

\begin{lstlisting}
@Operation loan(bank,amount)
  @Classifier Loan
    @Classify(applicant)
      applicant.lookup("salary") * 4 > amount
      @Classifier
        @Classify(applicant)
          applicant.lookup("debt") >= 10000
          @Classification Deny end
        end
        @Classify(applicant)
          applicant.lookup("debt") < 10000 and
          applicant.lookup("debt") >= 5000
          @Classification Refer end
        end
        @Classify(applicant)
          applicant.lookup("debt") < 5000
          @Classifier
            @Classify(applicant)
              applicant.lookup("bank") = bank
              @Classification Award end
            end 
            @Classify(applicant)
              applicant.lookup("bank") <> bank
              @Classification Refer end
            end
          end
        end
      end
    end
    @Classify(applicant)
      applicant.lookup("salary") * 4 <= amount 
      @Classification Deny end
    end
  end
end
\end{lstlisting}The operation loan accepts a bank and a loan amount. It returns a
classifier for the amount. The classifier first checks that the applicant
has an adequate annual salary for the required loan. If this condition
is not met then the application is denied. Otherwise, the amount of
debt is analysed. There are three outcomes: the first denies the application;
the second refers the application and the third analyses the bank
of the applicant. If the bank of the applicant is the same as that
receiving the application then the load is awarded otherwise the application
is referred.

The operation loan may be applied to any number of banks and loan
amounts. In each case it returns a different classifier.

The classification language is defined by adding a grammar to each
of the appropriate classes in the classification model. Each class
gets its own grammar so that the classification language is not fixed.
The classification language is integrated into XOCL and the components
of the language are extensible (new types of classifier node can be
defined later and they can be integrated with no changes to the existing
components).

Classification is the simplest grammar:

\begin{lstlisting}
@Grammar
  Classification ::= n = Name 'end' { 
    [| Classification(<n.lift()>) |] 
  }. 
end

\end{lstlisting}since it recognizes a name and synthesizes an instance of the appropriate
class. Classify, involves capturing XOCL expressions and is defined
as follows:

\begin{lstlisting}
@Grammar extends OCL::OCL.grammar
  Classify ::= '(' n = Name ')' e = Exp c = Exp 'end' { 
    [| Classify(@Operation(<n>) <e> end,<c>) |] 
  }.
end
\end{lstlisting}Notice that Classify synthesizes an operation definition whose argument
name is supplied in tthe classify definition. This is a standard way
of capturing executable code (in this case e) within synthesized elements.
The body of the operation, e, may refer to any variables that are
in scope. The argument n is deliberately placed in scope and allows
the body to refer to data that will be supplied by the instance of
Classify.

Classification involves sequences of XOCL expressions (an arbitrary
number of child classification nodes); it is defined as follows:

\begin{lstlisting}
@Grammar extends OCL::OCL.grammar
  Classifier ::= n = OptName c = Classifications 'end' {
    [| Classifier(<n.lift()>,<c>) |] 
  }.
  Classifications ::= e = Exp cs = Classifications { 
    [| <cs> ->including(<e>) |] 
  } | { [| Set{} |] }.
  OptName ::= Name | { "anon" }.
end
\end{lstlisting}The key aspect of the definition above is how the set of child classification
nodes is constructed using the Classifications rule. This recognizes
nothing, in which case it synthsizes an expression that creates the
empty set; or, it recognizes an expression e (a child node definition)
followed by more classifications cs and then synthesizes an expression
that adds e to cs. This is an example of a recursive rule%
\footnote{The rule is right recursive because the recursive use of Classifications
occurs after some input must be consumed. XMF does not support left
recursion where Classifications could occur on the left since this
would lead to Classifications calling itself without consuming any
input.%
}

with a base case that creates an empty set.

\begin{itemize}
\item Types and simple type checking (Dynamic - type check the classifiers.
vs Static)
\item Rendering configuration data.
\end{itemize}

\chapter{Graphs}

Unfortunately, information is not always convniently organised in
a tree structure. A tree structure does not make allowances for relationships
that span the tree or where cycles occur in the data. For example,
what happens when a company employee fills two roles within the company
in different departments? It would be approprate for the employee
to occur underneath both departments in the tree; the employee is
\textit{shared} between the departments or equivalently there are
two different \textit{paths} from the root of the tree to the employee.

Trees do not represent sharing and multiple paths very well. There
are strategies; for example, XML is a tree structured data format
where labels are used to mark up the elements in order to represent
sharing. When data representation requires sharing, it is usually
because the data naturally forms a \textit{graph}. Graphs can be encoded
in trees (and other forms of data representation), but if the data
requires a graph then it is probably best to use one and be done with
it.

\begin{figure}
\begin{center}

\includegraphics[width=12cm]{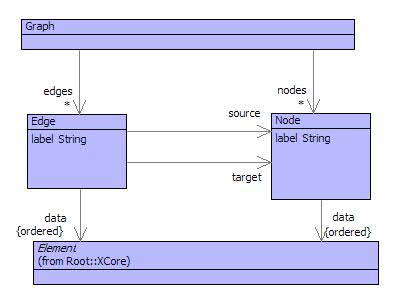}
\caption{Graphs\label{fig:Graphs}}

\end{center}
\end{figure}

Figure \ref{fig:Graphs} shows a model of graphs. A graph consists
of a set of nodes and a set of edges. Each node has a label and a
sequence of data. The label is used to identify the node and the data
is used by applications for whatever purpose they require.

Each edge has a label and data, and also has a source and target node.
Edges go between nodes and are directed. The diagram shown in figure
\ref{fig:Graphs} is itself an example of a graph where the nodes
are displayed as class boxes and the edges are shown as attribute
edges. Notice that the node labelled Element is shared (parent-wise)
between Edge and Node; equivalently there are two paths from the rot
node (labelled Graph) to the node labelled Element: Graph, Edge, Element
and Graph, Node, Element. Such sharing is difficult to represent using
trees.

Graphs are a very rich form of data representation. There is a wealth
of material written about graphs and how to process them. Here are
some useful operations on graphs defined in XOCL:

\begin{lstlisting}
context Graph
  @Operation nodeFor(label:String):Node
    @Find(node,nodes)
      when node.label() = label
      else null
    end
  end
\end{lstlisting}The operation nodeFor returns the node with the supplied label, or
null if no such node exists. The operation edgesFrom returns the set
of edges from a given node:

\begin{lstlisting}
context Graph
  @Operation edgesFrom(n:Node):Set(Edge)
    edges->select(e | e.source() = n)
  end
\end{lstlisting}Graphs are generally useful and therefore it is appropriate to have
a general purpose language construct to define a graph. As mentioned
above, each use of a graph structure will attach different amounts
of data to the nodes and labels. The data is used to support the application
specific processing of the graph. Therefore, a general purpose language
construct for graph construction should support:

\begin{enumerate}
\item Arbitrary node and edge data.
\item Plug-points for the sub-classes of Graph, Node and Edge that are used
to represent the graph.
\end{enumerate}
Here are two examples of different graph applications:

\begin{lstlisting}
@Graph(Routes,Location,Road)
  London()
    M1(200) -> Leeds
    A1(50)  -> Cambridge
end

@Graph(Plan,Task,Dependency)
  Start("January")
    -> Contractors
    -> Plans
  Contractors("March")
  Plans("April")
end
\end{lstlisting}The first graph is represented as an instance of the class Routes
where the nodes and edges are instances of the classes Location and
Road. These classes are sub-classes of Graph, Node and Edge repsectively.
Locations have no data; the three locations have labels London, Leeds
and Cambridge. 

An edge is listed below the source node. In the first example graph,
there are two edges with labels M1 and A1. The edges have data 100
and 50 (being the distance in miles) and the label of the edge target
is givn after ->.

The second example is a plan graph. The nodes have data that represents
the month at which the task is completed. Edges have no labels or
data (they just represent dependencies).

The proposed structure for a graph definition has plug-points for
the graph, node and edge classes and a body consisting of a sequence
of node definitions. A node definition n is a node label, followed
by node data in parentheses followed by a series of edge definitions
for which the source of the node is n. An edge definition is an optional
edge label, optional edge data in parentheses, an arrow and then the
label of the target node. Here is an example:

\begin{lstlisting}
@Graph(G,N,E)
  n1(a,b,c)
    e1(d) -> n2
    e2() -> n3
  n2()
    e3() -> n1
  n3()
\end{lstlisting}When the parser encounters a graph definition it will synthesize program
code that, when evaluated, produces the required graph. Are there
any rules that need to be observed when this synthesis takes place?
Given the model of graphs in figure \ref{fig:Graphs}, a graph contains
nodes and edges, and edges link nodes. Here is a possible program
that produces the graph above\begin{lstlisting}
(1) let g = G()
(2) in g.addToNodes(N("n1",Seq{a,b,c}));
       g.addToNodes(N("n2"));
(3)    g.addToNodes(N("n3"));
(4)    g.addToEdges(E("e1",Seq{d},g.nodeFor("n1"),g.nodeFor("n2")));
       g.addToEdges(E("e2",g.nodeFor("n1"),g.nodeFor("n3")));
(5)    g.addToEdges(E("e3",g.nodeFor("n2"),g.nodeFor("n1")));
(6)    g
    end
\end{lstlisting}Line (1) creates the graph using the supplied class G (a sub-class
of Graph). Each node must be added first in lines (2-3) so that edges
can then be created \textit{between} the nodes in lines (4-5). Note
that the supplied classes N and E are used to create the nodes and
edges. Finally the graph is returned in line (6).

The rules for graph construction are: create the graph, add the nodes
and then add the edges. Unfortunately, the graph definition construct
does not follow this pattern; it interleaves node and edge definitions.
A strategy is required to untangle this interleaving.

One way to address the interleaving is to have the parser synthesize
an intermediate graph definition that is processes using two or more
passes. This is perfectly respectable, and often a sensible way forward
when the required processing is fairly complex. 

In this case, the processing is not that complex, so another strategy
is used. To see how this works, a few definitions are required. An
\textit{edge constructor} expects a graph and an edge class; it adds
some edges to the supplied graph. A \textit{node constructor} expects
a graph, a node class, an edge class and a collection of edge constructors;
it adds some nodes to the supplied graph and then uses the edge constructor
to add some edges.

Node definitions are synthesized into node constructors and edge definitions
into edge constructors. The trick is to build up the edge constuctors
so that they are performed after all the node constructors. Since
the edge constructors are supplied to the node constructors, this
should be easy. Using the running example from above:

\begin{lstlisting}
n3 =
  @Operation(nodeConstructor)
    @Operation(g,N,E,edgeConstructor)
      g.addToNodes(N("n3"));
      nodeConstructor(g,N,E,edgeConstructor)
    end
  end
\end{lstlisting}The node definition for n3 is transformed into an operation that is
supplied with a node constructor and returns a node constructor. This
construction allows n3 to be linked with other noe constuctors without
knowing any details -- i.e. n3 can be defined \textit{in isolation}. 

The node constructor for n2 is similar, but involves the addition
of an edge constructor:

\begin{lstlisting}
n2 =
  @Operation(nodeConstructor)
    @Operation(g,N,E,edgeConstructor)
      let e3 = 
        @Operation(g,E)
          g.addToEdges(E("e3",g.nodeFor("n2"),g.nodeFor("n1")))
        end
      in g.addToNodes(N("n2"));
         nodeConstructor(g,N,E,addEdges(edgeConstructor,e3))
      end
    end
  end
\end{lstlisting}Note how the edge constructor for e3 is added to the supplied edge
contructor (using the yet-to-be-defined addEdges) when the supplid
node constructor is activated. This is the key to deferring the construction
of edges until all the nodes have been defined.

What should addEdges do? It is used to link all the edge constructors
together so that they all get activated. It takes two edge constructors
and returns an edge constructor:

\begin{lstlisting}
@Operation addEdges(ec1,ec2)
  @Operation(g,E)
    ec1(g,E);
    ec2(g,E)
  end
end
\end{lstlisting}The noe constructor for n1 is similar, but two edge constructors are
required:

\begin{lstlisting}
n1 =
  @Operation(nodeConstructor)
    @Operation(g,N,E,edgeConstructor)
      let e1 = 
        @Operation(g,E)
          g.addToEdges(E("e1",g.nodeFor("n1")Seq{d},g.nodeFor("n2")))
        end;
          e2 = 
        @Operation(g,E)
          g.addToEdges(E("e1",g.nodeFor("n1"),g.nodeFor("n3")))
        end then
          edges = addEdges(e1,e2)
      in g.addToNodes(N("n2"));
         nodeConstructor(g,N,E,addEdges(edgeConstructor,edges))
      end
    end
  end
\end{lstlisting}The complete graph can now be defined by linking the node constructors
together and supplying a graph:

\begin{lstlisting}
let nc = addNodes(n1,addNodes(n2,addNodes(n3,noNodes)))
in nc(G(),N,E,@Operation(g,E) g end)
end
\end{lstlisting}Each of the node constructors are linked via an operation addNodes.
The left-hand argument of addNodes is an operation that maps a node
constructor to a node constructor. The right-hand argument is a node
constructor. It is easier to see how this works from the definition:

\begin{lstlisting}
@Operation addNodes(nodeCnstrCnstr,nodeCnstr2)
  @Operation(g,N,E,edgeConstructor)
    let nodeCnstr1 = nodeCnstrCnstr(nodeCnstr2)
    in nodeCnstr1(g,N,E,edgeConstructor)
    end
  end
end
\end{lstlisting}The mechanism used by addNodes is an example of a typical pattern
that threads sequences of operations together. It allows the node
constuctor encoded in nodeCnstrCnstr to occur before that encoded
in nodeCnstr2 while also allowing the edge constructors produced by
the first to be handed on to the second (because they are to be deferrred
until all the nodes are added to the graph).

There are two types of constructor, each of which can occur repeatedly
in a sequence: nodes and edges. When this occurs, it is usual to have
some way to encode an empty sequence; in this case there are noNodes
and noEdges. Both of these are constructors:

\begin{lstlisting}
@Operation noNodes(g,N,E,edgeConstuctor)
  edgeConstructor(g,E)
end

@Operation noEdges(g,E)
  null
end
\end{lstlisting}
noNodes is a node constructor that starts edge construction. Therefore,
noNodes should be the right-most node constuctor in a sequence that
is combined using addNodes. noEdges does nothing, and can occur anywhere
in a sequence.

The grammar for graph definition synthesizes node and edge constructors
combined usin addNodes and addEdge. When an empty sequence is encountered,
the gramar synthesizes noNodes and noEdges respectively. The grammar
is defined below:

\begin{lstlisting}
  @Grammar extends OCL::OCL.grammar
    Data ::= '(' es = CommaSepExps ')' { SetExp("Seq",es) }.
    Edges(s) ::= e = Edge^(s) es = Edges^(s) 
      { [| addEdges(<e>,<es>) |] } 
    | { [| noEdges |] }.
    Edge(s) ::= l = Label d = OptData '->' t = Label { [| 
      @Operation(g,E)
        g.addToEdges(E(<l>,<d>,g.nodeFor(<s>),g.nodeFor(<t>)))
      end 
    |] }.
    Graph ::= '(' mkGraph = Exp ',' mkNode = Exp ',' mkEdge = Exp')' 
      GraphBody^(mkGraph,mkNode,mkEdge).
    GraphBody(mkGraph,mkNode,mkEdge) ::= ns = Nodes 'end' { [| 
      <ns>(<mkGraph>(),<mkNode>,<mkEdge>,@Operation(g,E) g end) 
    |] }.
    Label ::= NameExp | { "".lift() }.
    NameExp ::= n = Name { n.lift() }.
    Nodes ::= n = Node ns = Nodes 
      { [| addNodes(<n>,<ns>) |] } 
    | { [| noNodes |] }.
    Node ::= l = Label d = Data e = Edges^(l) { [|
      @Operation(Cn)
        @Operation(g,N,E,Ce)
          g.addToNodes(N(<l>,<d>));
          Cn(g,N,E,addEdges(<e>,Ce))
        end
      end
    |] }.
    OptData ::= Data | { [| Seq{} |] }.
  end
\end{lstlisting}

\chapter{Collections and Comprehensions}

It is usual for data to form collections: employees; customer transactions;
connected clients. When modelling collections of data it is very convenient
to have language structures that make it easy to create, transform
and filter the collections. This chapter describes how such a language
construct can be defined. 

\begin{figure}
\begin{center}

\includegraphics[width=12cm]{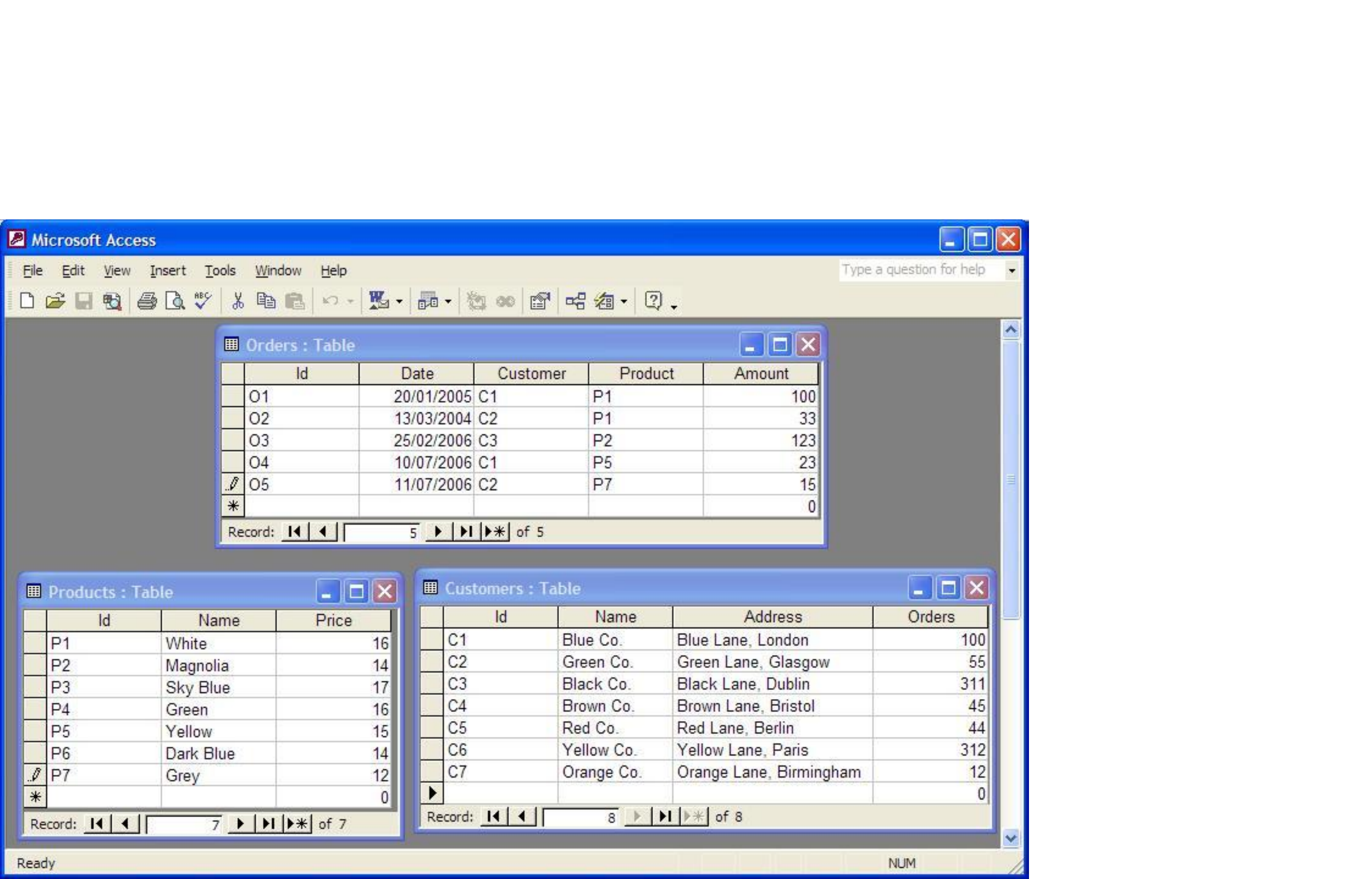}

\caption{\label{fig:A-Customer-Order}A Customer Order Database}

\end{center}
\end{figure}

A typical example of data collections occurs in a relational database.
Figure \ref{fig:A-Customer-Order} shows a portion of a MS Access
database containing three tables relating to a pain manufacuring company.
The Customers table shows all the customers who have bought paint
from the company; each customer has a unique id, a name, an address
and a running tally of the number of orders that have ben placed.

The Products table shows all the types of paint that the company sells
along with the current price per unit. Finally, the Orders table shows
the orders that have been placed, linking customers, products and
amount.

\begin{figure}
\begin{center}

\includegraphics[width=12cm]{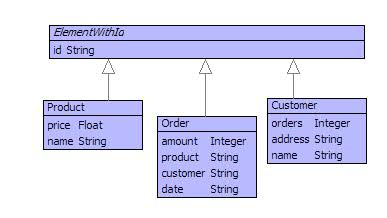}

\caption{Database Tables as Classes\label{fig:Database-Tables-as}}

\end{center}
\end{figure}

Database tables contain records expressed as rows. Each table is modelled
as a class and the records modelled as instances of the class grouped
together as a set. The diagam in figure \ref{fig:Database-Tables-as}
shows three classes corresponding to the three database tables above.

The model above has chosen to represent the database tables directly:
each table uses the id field as a primary key and each class has modelled
this directly. Unlike database records, objects have a unique identity,
so this is not strictly necessary. It is useful to retain the id field
to show how database style queries and operations are supported by
the collection language construct below.

The database records are modelled as instances of the classes as shown
below:

\begin{lstlisting}
Root::Customers := Set{
  Customer("C1","Blue Co.","Blue Lane, London",100),
  Customer("C2","Green Co.","Green Lane, Glasgow",55),
  Customer("C3","Black Co.","Black Lane, Dublin",311),
  Customer("C4","Brown Co.","Brown Lane, Bristol",45),
  Customer("C5","Red Co.","Red Lane, Berlin",44),
  Customer("C6","Yellow Co.","Yellow Lane, Paris",312),
  Customer("C7","Orange Co.","Orange Lane, Birmingham",122)
};
Root::Products := Set{
  Product("P1","White",16),
  Product("P2","Magnolia",14),
  Product("P3","Sky Blue",17),
  Product("P4","Green",16),
  Product("P5","Yellow",15),
  Product("P6","Dark Blue",14),
  Product("P7","Grey",12)
};
Root::Orders := Set{
  Order("O1","20/01/2005","C1","P1",100),
  Order("O2","13/03/2004","C2","P1",33),
  Order("O3","25/25/2006","C3","P2",123),
  Order("O4","10/07/2006","C1","P5",23),
  Order("O5","11/07/2006","C2","P7",15)
};
\end{lstlisting}Each table is modelled as a global variable whose value is a set of
objects. New records can be added by updating the variable:

\begin{lstlisting}
@Operation newId(E:Set(ElementWithId)):String
  let i = 0
  in @While (not E->isEmpty) and 
            (not E->exists(e | e.id = "X" + i)) do
       i := i + 1
     end;
     i
  end
end
@Operation addCustomer(name:String,address:String)
  let id = newId(Customers) then
      newCustomer = Customer(id,name,address,0)
  in Root::Customers := Customers->including(newCustomer)
  end
end
\end{lstlisting}The operation newId is defined to automatically calculate a new identifier
for a collection of elements. The addCustomer operation modifies the
collection of customers by adding a newly constucted customer. Removing
customers can be achieved by a similar update to the global variable.

A query on one or more database tables involves performing a test
on each of the records in the table and producing a new table as a
result. In the model, a table is represented as a collection of elements.
The new language construct Cmp (short for \textit{comprehension})
allows us to do a query as follows:

\begin{lstlisting}
(1) @Operation ordersGreaterThan(x:Integer):Seq(Customer)
(2)   @Cmp customer where
(3)     customer <- Customers,
(4)     ? customer.orders > x
      end
    end
\end{lstlisting}The operation ordersGreaterThan (1) takes an integer x and returns
a collection of customers. Line (2) introduces the query as a Cmp
expression that returns all those customers (2) where each customer
is an element of the Customers table (3) and for which the orders
of each customer are grater than x (4).

A Cmp expression has the following general form:

\begin{lstlisting}
@Cmp body where clauses end
\end{lstlisting}A clause is either a binding (name <- collection) or a filter (? test).
The Cmp expression performs each clause in turn. A binding selects
each element from the collection in turn and, for each selection,
evaluates the rest ot the Cmp expression. A filter evaluates the test
(referencing any names bound tfrom its left); if the test succeeds
then evluation proceeds, otherwise evaluation retuns to the last selction
and tried with another element.

Consider the definition of ordersGreaterThan again. Line (3) causes
each sucessive customer to be selected from the Customers table. Line
4 allows only those customers to proceed whose orders value passes
the test. Finally, a sequence is returned containing all those customers
(2) that get through the test.

The following is an example of a query that involves multiple binding
and filters:

\begin{lstlisting}
@Operation customersWhoBuy(product:String):Seq(String)
  @Cmp c.name where
    c <- Customers,
    o <- Orders,
    p <- Products,
    ? p.name = product,
    ? o.customer = c.id,
    ? o.product = p.id
  end
end
\end{lstlisting}The operation customersWhoBuy is supplied with the name of a product
and returns a sequence of customer names for those customers who have
bought the supplied product.

A typical operation that occurs in a database is a \textit{join} where
two different tables are joined on two fields with the same value.
The records that have been defined so far have ben instances of specific
classes. A join does not necessary produce a record with a meaningful
type (although it might); section \ref{sub:Simple-Records} defines
a language construct for representing arbitrary records. The following
expression performs a join on the three tables producing a new table
of records showing the amount of each product for each customer:

\begin{lstlisting}
@Cmp 
  @Record
    customerName = c.name,
    productName = p.name,
    amount = o.amount
  end
where
  c <- Customers,
  p <- Products,
  o <- Orders,
  ? o.customer = c.id,
  ? o.product = p.id 
end
\end{lstlisting}%
\begin{figure}
\begin{center}

\includegraphics[width=12cm]{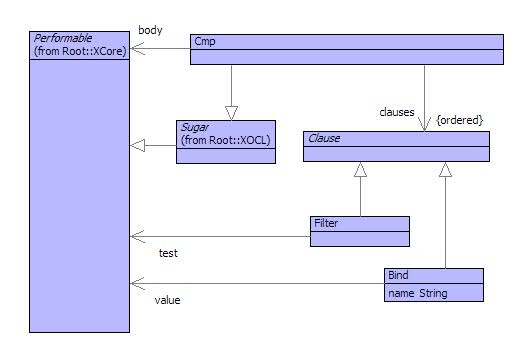}

\caption{Comprehensions\label{fig:Comprehensions}}

\end{center}
\end{figure}

A Cmp expression is defined in terms of a package of syntax classes
shown in figure \ref{fig:Comprehensions}. The class Cmp is defined
as sugar which means that it must define an operation called desugar
that turns an instance of Cmp into a Performable thing (usually an
instance of OCL classes). 

A Cmp has a performable body and a sequence of clauses. Each clause
is either a binding or a filer. A Cmp is desugared as follows:

\begin{lstlisting}
   @Operation desugar()
    @Case self of
(1)   Cmp(body,Seq{}) do
        [| Seq{<body>} |]
      end
(2)   Cmp(body,Seq{Bind(n,v)}) do
        [| <v> ->asSeq ->collect(<n> | <body>) |]
      end
(3)   Cmp(body,Seq{Bind(n,v)|clauses}) do
        [| <Cmp(Cmp(body,clauses),Seq{Bind(n,v)})>.flatten() |]
      end
(4)   Cmp(body,Seq{Filter(e)|clauses}) do
        [| if <e> then <Cmp(body,clauses)> else Seq{} end |]
      end
    end
  end
\end{lstlisting}The translation of a Cmp expression into basic XOCL is defined by
case analysis. Case (1) occurs when there are no clauses. In this
case the body is transformed into a single element sequence. Case
(2) occurs when there is a single binding. In this case the value
of the binding produces a collection and the body is performed for
each element of the collection producing a sequence. 

Case (3) occurs when there is a binding and some clauses. In this
case the following equivalence is used:

\begin{lstlisting}
@Cmp b where                            @Cmp 
  x <- S,         is equivalent to         @Cmp b where Cs end
  Cs                                    where
end                                       x <- S
                                        end->flatten
\end{lstlisting}Case (4) tests the expression e. If the result is true then the Cmp
expression proceeds otherwise it produces no elements. This has the
effect of filtering out the current variables that have been bound
so far.

Finally, the Cmp expression has a grammar that synthesizes an instance
of Cmp that, since it is an instance of Sugar, calls desugar automatically:

\begin{lstlisting}
@Grammar extends OCL::OCL.grammar
  Cmp ::= body = Exp cs = Clauses 'end' { Cmp(body,cs) }.
  Clauses ::= Where | { Seq{} }.
  Clause ::= Bind | Filter.
  Bind ::= n = Name '<-' e = Exp { Bind(n,e) }.
  Filter ::= '?' e = Exp { Filter(e) }.
  Where ::= 'where' c = Clause cs = (',' Clause)* { Seq{c|cs} }
end
\end{lstlisting}Applications for business intelligence and KPIs.

\section{Transforming Comprehensions}

The previous section has shown how a comprehension expression can
be transformed into an XOCL expression, thereby enriching XOCL with
a new language construct. Comprehensions have often been proposed
as a construct suitable for performing database queries since they
abstract away from the details of how the queries are performed and
how the result sets are built up.

A comprehension can be viewed as a query over a database if the tables
are viewed as sets of records. In this way the queries are performed
by selecting records, filtering them with predicates and joining them
back together. However, whilst this is an attractive way to construct
database queries, it does not solve the problems of how the queries
are performed and how the comprehensions can be embedded into a standard
programming language.

This section sketches out how comprehensions can be viewed as database
queries by showing how to desugar a comprehension into a standard
imperative language that provides a database interface. The translation
is deliberately simple; once it has been described the conclusion
outlines how the translation can be made more realistic.

Instead of the Cmp::desugar operation defined in the previous section,
a new operation is defined. The operation is called toSQL and takes
two arguments: the an expression to be transformed into an imperative
program and a variable. At run-time, the variable will be bound to
a set of values; after performing the comprehension program, the set
will contain the result. 

The toSQL operation proceeds by case analysis on the expression, each
case is addressed in turn. The first case deals with the aituation
where there are no qualifiers:

\begin{lstlisting}
@Operation toSQL(cmp,var)
  @Case cmp of
    Cmp(body,Seq{}) do
     [| let S = Set{}
        in <toSQL(body,[| S |])>;
           <var> := <var> ->including(S)
        end |]
    end
\end{lstlisting}The resulting program code creates a new set S and then uses this
as the container for the body of the comprehension. The resulting
set S is added to the supplied set (value of var). The next case deals
with a single filter:

\begin{lstlisting}
    Cmp(body,Seq{Filter(p)}) do
      [| if <p> 
         then <self.toSQL(Cmp(body,Seq{}),var)>
         else <var> := <var> ->including(Set{}) 
         end |]
    end
\end{lstlisting}If the predicate expression p is true then the body of the comprehension
is performed with the supplied var as the target set. Otherwise the
empty set is added to the target var. The next case deals with the
situation where there are multiple qualifiers:

\begin{lstlisting}
    Cmp(body,Seq{q|qs}) do
      [| <self.toSQL(Cmp(Cmp(body,qs),Seq{q}),var)>;
         <var> := <var> ->flatten;
         <var>
      |]
    end
\end{lstlisting}The same transformation as before is applied to the comprehension
whereby the multiple qualifiers are reduces by nesting the comprehension.
In this case the transformation on the nested comprehension will produce
an impreative program that updates the supplied set. Therefore, after
the supplied set is updated it is flattened before returning it. Finally,
the case where variables are bound is dealt with as follows:

\begin{lstlisting}
    Cmp(body,Seq{Bind(name,OCL::Var(tableName))}) do
      [| let S = DB.SQL("SELECT * FROM " + <tableName.lift()>);
             P = Set{}
         in @For <name> in S do
              <self.toSQL(body,[| P |])>
            end;
            <var> := <var> ->including(P)
         end |]
    end
  end
end
\end{lstlisting}Here it is assumed that all bindings select their elements from named
database tables. In this case the result set S is constructed by calling
a builtin procedure DB.SQL that takes an SQL query as a string. A
for-loop is used to iterate through the result set and the body of
the loop is generated by translating the body of the comprehension.
Notice that the set supplied to the body translation is P - the new
set created as part of the binding. The final action is to add the
set P to the supplied set var.

\begin{lstlisting}
let S = DB.SQL("SELECT * FROM " + "V");
    P = Set{}
in @For x in S do
     let S = DB.SQL("SELECT * FROM " + "W");
         P = Set{}
     in @For y in S do
          if x < y
          then
            let S = Set{}
            in S := S->including(x);
               P := P->including(S)
            end
          else
            P := P->including(Set{})
          end
        end;
        P := P->including(P)
     end;
     P := P->flatten
   end;
   S := S->including(P)
end;
S := S->flatten
\end{lstlisting}

\chapter{Language Emulation}

Super-languages provide support for syntax extension and for weaving new language
constructs into the base super-language. In addition they must allow new 
external languages to be defined so that applications can be constructed
by combining components that are written in mixtures of languages.

In addition to inventing new language constructs that capture programming
idioms and that provide domain specific abstractions, a super-language
should allow constructs from existing conventional languages to be used.
Ultimately, a super-language should provide support for complete emulation
of an existing conventional language or some sub-set of it.

A super-language must be able to assimilate conventional languages because
most projects do not start with a clean sheet. Existing code must be weaved
into a new application and a super-language should be able to accommodate
legacy code whilst providing scope for DSLs and new language features.

This chapter describes how XMF can be used to support features from existing
conventional languages.

\section{MicroJava}

XMF is shipped with an implementation of Java. The definition is found in 
the package Languages::MicroJava. MicroJava is a sub-set of the Java language
that supports classes, fields and methods. The MicroJava grammar supports
most Java features and is a useful starting point for designing Java-like
languages.

The MicroJava language does not support direct execution of programs. There
is no MicroJava compiler or interpreter. MicroJava is intended as a starting 
point for experimentation with Java or as a mechanism for exporting Java
code from XMF. MicroJava supports:
\begin{itemize}
\item A grammar that translates MicroJava concrete syntax into an instance
of the MicroJava model.
\item A pretty-printer that exports a textual representation of MicroJava
source code (a Java export mechanism).
\item A translation from MicroJava abstract syntax to XOCL abstract syntax.
\end{itemize}
Therefore, to execute MicroJava you can: pretty-print it and use a Java
compiler on the result; translate it to XOCL and execute that; write your
own MicroJava compiler or interpreter.

The grammar for MicroJava is defined by the class Java in the Languages::MicroJava
package. The following shows an example MicroJava program defined in a standard
XMF source file:
\begin{lstlisting}
parserImport XOCL;
parserImport Languages::MicroJava;
 
context Root
 
@Java
 
  public class Printer {
 
    private int copies;
    private String string;
 
    public Test(int copies,String string) {
    }
 
    private boolean printCount() {
      while(copies > 0) {
        this.write(string);
	    this.copies = this.copies - 1;
      }
      return true;
    }
 
    // Embedded XOCL
 
    with JOCL {
      @Operation write(string:String)
        format(stdout,"~S~%",Seq{string})
      end 
    }
  }
 
end
\end{lstlisting}
The @Java construct will cause the MicroJava program text to be parsed,
an instance of the MicroJava abstract syntax model to be created and then 
the translator to XOCL to be performed. The result of the parse is XOCL
abstract syntax which is then compiled or interpreted by the XOCL
execution system as normal.

You could use MicroJava to experiment with interesting language
extensions to Java. Each new construct can be added to the MicroJava
abstract syntax model. If you add a toXOCL operation to the new
abstract syntax class then the new MicroJava construct can be executed
via a translation to XOCL. See the source code for MicroJava for more
details about how the toXOCL operation works for each abstract syntax
class.

Notice in the MicroJava program shown above the use of a non-Java 
construct: {\tt with}. The {\tt with} keyword is followed by the
name of a class (that must be in-scope) that defines (via a grammar) 
a language. The source code between {\tt \{} and {\tt \}} in the {\tt with}
is written in the language.

The {\tt with} construct allows arbitrary languages to be nested
in MicroJava. In the example above, XOCL has been nested in MicroJava using
the following class definition for JOCL:
\begin{lstlisting}
context Root
  @Class JOCL
    @Grammar extends OCL::OCL.grammar
      // A grammar invoked using @X must be defined
      // by a class named X and have a clause named X.
      // The JOCL clause just proxies the SimpleExp
      // clause defined by OCL...
      JOCL ::= SimpleExp.
    end
  end
\end{lstlisting}

\section{Lisp}

Lisp is an interesting language because its syntax is very unconventional.
Lisp has a very small meta-syntax (just lists and atoms). The language is
built around this syntax in two ways:
\begin{itemize}
\item By adding builtin operators. The more orthogonal builtin operators are
added to a Lisp engine, the more applications can be built. 
\item By adding special forms. A special form is a Lisp list that
is recognized by the Lisp interpreter or compiler and treated specially.
\end{itemize}
This section provides a complete mini-Lisp system including a package of
builtin operators that you can extend. Execution of the Lisp system is
performed by a translation to XOCL. The translator is called from the grammar
rules - so there is no abstract syntax representation of Lisp in this example
(but you could add one in if you wanted to manipulate the Lisp expressions).
The translator, shows how Lisp special forms are handled.

Here is an example Lisp program:
\begin{lstlisting}
parserImport XOCL;

import LispOps;

context Root
  @Operation testFact(n:Integer):Integer
    // Define a factorial function and supply the 
    // integer n...
    @Lisp
      (let ((fact (Y (lambda(fact)
                       (lambda(n)
                         (if (eql n 0)
                             1
                             (mult n (fact (sub n 1)))))))))
           (fact n))
    end
  end
\end{lstlisting}
The builtin operators are defined in the package LispOps. A Lisp
program is executed in the current scope so that Lisp variables 
are the same as XOCL variables.

Note that recursive definitions are handled using a Y operator
that is defined in LispOps. It is beyond the scope of this book
to go into the detailed semantics of Lisp and Y - you'll just have
to trust us that Y is the fixed point operator for which:
\begin{lstlisting}
((Y f) x) = x
\end{lstlisting}
The grammar for Lisp parses the text and construct a sequence
containing the Lisp program. The program is then passed to an
operator eval that is defined in the package LispAux which
translates the sequence to XOCL code. The grammar is defined
below:
\begin{lstlisting}
parserImport XOCL;
parserImport Parser::BNF;
 
import OCL;
import LispAux;
 
context Root 
  @Class Lisp 
    @Grammar  
      Lisp ::=  e = Exp  'end' { eval(e) }.
        
      Exp ::= 
        Atom      // A basic syntax element.
      | List      // A sequence of expressions.
      | Const     // A (back)quoted element.
      | Comma.    // An unquoted element.
      
      Atom ::= 
        Int       // Integers.
      | Str       // Strings.
      | Float     // Floats
      | n = Name  // Names are turned into variables for syntax.
        { Var(n) }.
      
      List ::= '(' es = Exp* ListTail^(es).
        
      ListTail(es) ::=       
        // A list tail may be a '.' followed by an element then ')'
        // or may just be a ')'...        
        '.' e = Exp ')' 
        { es->butLast + Seq{es->last | e} }
        
      | ')' { es }.
      
      Const ::=        
        // A constant is a quoted expression. Use backquote
        // which allows comma to be used to unquote within
        // the scope of a constant...       
        '`' e = Exp 
        { Seq{Var("BackQuote"),e} }.
        
      Comma ::=        
        // A comma can be used within the scope of a
        // backquote to unquote the next expression...       
        ',' e = Exp 
        { Seq{Var("Comma"),e} }.        
    end
  end
\end{lstlisting}
The auxiliary definition {\tt eval} and {\tt quote} perform opposite
translation duties. The {\tt eval} operation translates Lisp source
code that is to be evaluated into XOCL code. The {\tt quote} operation
translates Lisp source code that is protected by a quote (actually
a backquote character here which allows comma to be used). The 
function of these two operations is given below:
\begin{lstlisting}
parserImport XOCL;
parserImport Parser::BNF;

import OCL;
  
context Root
  @Package LispAux  
    // Operations that are used by the Lisp language when
    // translating the parsed syntax into XOCL...    
    @Operation eval(e):Performable    
      // This operation takes a single value produced by the Lisp
      // parser and produces an XOCL performable element...      
      @TypeCase(e)      
        // Atoms all produce the XOCL equivalent syntax...       
        Integer do
          e.lift()
        end        
        Float do
          e.lift()
        end        
        Var do        
          // Variables are evaluated in the surrounding lexical
          // environment...         
          e
        end        
        String do
          e.lift()
        end        
        Seq(Element) do        
          // Sequences may start with special names which are
          // evaluated specially...         
          @Case e of          
            Seq{Var("BackQuote"),e} do           
              // Suppress the evaluation of the expression e in the
              // XOCL code that is produced...             
              quote(e)
            end           
            Seq{Var("Comma"),e} do          
              // Only legal within the scope of a backquote and 
              // therefore can only be processed by the quote
              // operation defined below...              
              self.error("Comma found without backquote")
            end            
            Seq{Var("lambda"),args,body} do             
              // Create a lexical closure....             
              lambda(args,eval(body))
            end           
            Seq{Var("if"),test,e1,e2} do           
              // Choose which of the two expressions e1 and e2
              // to evaluate based on the result of the test...             
              [| if <eval(test)> then <eval(e1)> else <eval(e2)> end |]
            end           
            Seq{Var("let"),bindings,body} do           
              // A let-expression introduces identifier bindings into
              // the current lexical environment. This can be viewed
              // as simple sugar for the equivalent lambda-application...              
              let formals = bindings->collect(b | b->head);
                  actuals = bindings->collect(b | b->at(1))
              in eval(Seq{Seq{Var("lambda"),formals,body} | actuals})
              end
            end           
            Seq{op | args} do           
              // Evaluate the argument expressions...             
              [| let A = <SetExp("Seq",args->collect(arg | eval(arg)))>
                 in 
                    // Then invoke the operation...                   
                    <eval(op)>.invoke(self,A)
                 end
              |]
            end           
            Seq{} do             
              // nil is self evaluating...             
              [| Seq{} |]
            end
          end
        end
      end
    end
    
    @Operation quote(e):Performable   
      // This operation takes a singlelisp expression produced by the
      // grammar and returns an XOCL expression that protects the
      // evaluation (similar to lifting the value represented by the
      // expression)...      
      @Case e of     
        Seq{Var("Comma"),e} do       
          // Comma undoes the effect of backquote...          
          eval(e)
        end        
        Seq{h | t} do        
          // Create a pair...          
          [| Seq{ <quote(h)> | <quote(t)> } |]
        end        
        Seq{} do         
          // The empty sequence (nil in Lisp)...         
          [| Seq{} |] 
        end        
        Var(n) do        
          // Protected names are just strings (symbols in Lisp)...          
          n.lift()
        end        
        else         
          // Otherwise just return an XOCL expression
          // that recreates the atomic value...          
          e.lift()
      end
    end
    
    @Operation lambda(A,body:Performable):Operation     
      // Returns an XOCL operation expression that
      // expects the arguments represented vy the sequence
      // A...     
      Operation("lambda",args(A),NamedType(),body)
    end
    
    @Operation args(A)     
      // Turn a sequence of OCL::Var into a sequence of
      // OCL::Varp (because operation expressions in XOCL use
      // Varp as the class of simple named arguments)...      
      @Case A of      
        Seq{} do
          A
        end        
        Seq{Var(n) | A} do
          Seq{Varp(n) | args(A)} 
        end       
      end
    end
  end
\end{lstlisting}
You can see from the definition of {\tt eval} that the translation must
handle a number of special forms. These are all sequences whose head is
a variable with a special name. These are:
\begin{description}
\item{\tt lambda} A function. This must create a closure and can be
directly translated to an XOCL operation.
\item{\tt Quote} A protected sequence. The rest of the sequence is
handed to {\tt quote} which translates the elements to the equivalent
XOCl constant builting expressions. Note that {\tt quote} must detect
the occurrence of {\tt comma} which un-quotes the a Lisp expression.
Unquoting is handled by calling {\tt eval} again.
\item{\tt if} An is-expression just translates directly onto an XOCL
if-expression.
\item{\tt let} A let-expression performs parallel binding. This can be
achieved by a translation to a Lisp expression that creates local
variables by applying a lambda-function.
\end{description}
The builtin operators of Lisp are all defined in the LispOps package
which is imported for the scope of any Lisp program. You can add any
operation definitions you like into this package. Here are some
examples including the infamous Y operator:
\begin{lstlisting}
parserImport XOCL;
parserImport Parser::BNF;
 
import OCL;
    
context Root 
  @Package LispOps  
    // A package of basic Lisp operations. Think of these
    // as the builtin operations and make sure you import
    // them before referencing them...  
    @Operation add():Integer    
      // Adding up integers...      
      args->iterate(arg n = 0 | n + arg)
    end
    
    @Operation mult(.args:Seq(Integer)):Integer    
      // Multiplying integers...      
      args->iterate(arg n = 1 | n * arg)
    end
    
    @Operation sub(x:Integer,y:Integer):Integer    
      // Subtracting...      
      x - y
    end
    
    @Operation list(.args):Seq(Element)    
      // Creating a list from some elements...     
      args
    end
    
    @Operation Y(f)    
      // The infamous Y combinator for finding a fixed point...      
      let op = @Operation(x) f(@Operation(.y) (x(x)).invoke(null,y) end) end
      in op(op)
      end
    end
    
    @Operation eql(x,y):Boolean   
      // Comparing two elements...      
      x = y
    end    
  end
\end{lstlisting}

\chapter{Business Rules}

Businesses are run in terms of various types of information. This
includes resource information, marketing strategies, customer information,
etc. A key part of the information underlying a business are the policies
by which the business operates. Policies are the guidelines by which
all components of the business are co-ordinated. Policies are used
to determine what happens when various events take place within the
business. Events include day-to-day occurrences such as new customer
contact, and also include longer term strategic events such as a company
merger or implementing a new product.

Often the rules that drive a business are left implicit within the
structure of the organisation. Some of the rules may be written down
in company literature (such as an employee handbook) and others may
simply be in the heads of individuals. There is significant advantage
in producing a complete collection of the rules that drive a business.
Once the rules are brought together in a uniform representation they
can be analysed in terms of their impact on the business. 

Representing the rules as executable models has added benefit: the
rules can be used to simulate the business in different scenarios.
By executing the business in terms of its models, it is possible to
try out different strategies and to get a feel for the effect of changes
to how the business operates. Furthermore, it may be possible to run
certain parts of the business directly from the executable business
rules; changes to the strategy can then be implemented directly by
changing the rules.

Typically, business rule systems are event driven. The rules monitor
business data; changes to the business data are detected as events
by the rules which in turn fire actions that update the business data.
For example, rules may monitor an invoice database; an event is raised
when an invoice is received from a customer causing the rule to fire
and an email to the appropriate sales manager is sent.

A rules engine must manage large collections of rules that monitor
events across a very large amount of business information. A naive
implementation of a business rules engine might run all rules against
all the business data each time an event occurs. This is far too expensive
because of the speed of change in many businesses and the combinations
of data that must be checked. 

Fortunately, the problem has been solved by a special type of matching
algorithm called the Rete Algorithm that uses caching of partial matches
to produce huge reductions in the amount of retesting required by
a rule engine each time the data changes. This chapter describes how
to develop language constructs for business rules and to implement
a business rule engine using XOCL.

\section{Example Business Rules}

\begin{figure}
\begin{center}

\includegraphics[width=12cm]{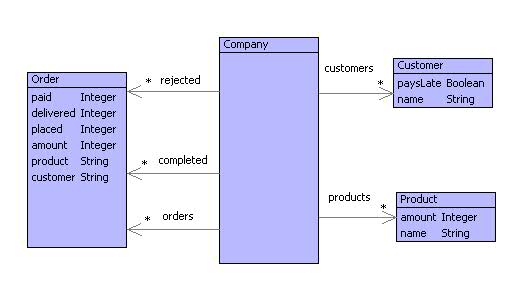}

\caption{A Sales Order Information Model\label{fig:A-Sales-Order}}

\end{center}
\end{figure}

Consider the sales order information model shown in figure \ref{fig:A-Sales-Order}.
This is part of a larger information model that is used to represent
the state of a company at any moment in time. Customers initially
contact the company with a sales order for a named product. If the
customer does not already exist then they are added to the customers
database. The order is then analysed to determine whether it can be
met. An order can only be processed if the company has sufficient
amount of the required product. If the order cannot be met then it
is placed in the rejected orders database and the customer is informed
by letter. 

The order may also be rejected if the customer pais late. The company
has a policy that requires payment for an order to be received within
30 days of the order being received. If the customer fail to satisfy
this policy then they become marked and all subsequent orders are
rejected until appropriate representations are made to senior management.

Once an order has been received and accepted, the product is packaged
up and sent to the customer. On dispatching the goods, the order is
moved to the completed orders database.

The order system described above includes an number of policies that
can be expressed using business rules. Business rules are grouped
into business aspects, the sales order processing system is such an
aspect:

\begin{lstlisting}
@RuleBase OrderProcessing
  // Business Rules...
end
\end{lstlisting}Each major policy that is used to drive the business is expressed
as a rule. The following rule describes what happens when an order
is received:

\begin{lstlisting}
context OrderProcessing
(1)  @Rule ReceiveOrder
(2)    Company[c]
(3)    Order[o]
(4)    ? not c.hasOrder(o) end
(5)  -> 
(6)    c.addToOrders(o)
     end
\end{lstlisting}The rule named ReceiveOrder is defined in line (1). It has a head
(lines (2) - (4)) and a body (line(6)). The head of the rule describes
a situation that may occur in the business. ReceieveOrder describes
a situation where an order is received by the business but has not
yet been entered into the orders database. Line (2) detects any company
and associates the variable c with the company. Line (3) detects any
order and associates variable o with it. Line (4) is a general condition
that can be any XOCL expression using any of the variables in scope.

The body of a rule is an action that updates the current state of
the business. ReceiveOrder has a body that updates the orders database.
Line (6) is the action performed by the rule and may be any XOCL action.
In this case the order is added to the company.

An order cannot be satisfied when there is insufficient product:

\begin{lstlisting}
context OrderProcessing
  @Rule UnableToSatisfyOrder
    Company[c]
(1) Order[o,product=pName,customer=cName,amount=pRequested,delivered=0]
(2) Product[p,name=pName,amount=pAvailable]
(3) Customer[k,name=cName,paysLate=false]
(4) ? c.orders()->includes(o) end
(5) ? c.products()->includes(p) end
(6) ? c.customers()->includes(k) end
(7) ? pRequested > pAvailable end
  -> c.deleteFromOrders(o);
     c.addToRejected(o)
  end
\end{lstlisting}The UnableToSatisfyorder rule matches a company, an order, a product
and a customer and places consitions on the relationships between
them. Line (1) matches any order whose delivered date is 0 (signifying
that the order has yet to be delivered). The slots product, customer
and amount are matahced to variables that are referenced elsewhere
in the rule.

Line (2) matches a product whose name is referenced in the prodict
in line (1). Line (3) matches a customer whose name matches that used
in the order matched in line (1). The customer must pay on time for
this business rule to apply. Lines (4 - 7) are conditions on the values
of the variables matched in lines (1 - 3). In particular the condition
on line (7) states that the requested amount of product exceeds the
available amount and therefore the order cannot be satisfied.

The body of the UnableToSatisfyOrder removes the order from the orders
database and adds it to the rejected orders database.

The company has a policy of rejecting orders from customers who pay
late:

\begin{lstlisting}
context OrderProcessing
  @Rule RejectOrder
    Company[c]
    Order[o,customer=cName]
    Customer[k,name=cName,paysLate=true]
    ?c.orders()->includes(o) end
  -> c.deleteFromOrders(o);
     c.addToRejected(o)
  end
\end{lstlisting}An order is completed when the sales staff update the database with
the data of dispatch. At this point the order is removed from the
orders database and added to the completed orders database. At this
point the amount of product is reduced by the required amount:

\begin{lstlisting}
context OrderProcessing
  @Rule CompleteOrder
    Company[c]
    Order[o,delivered=dTime,product=pName,amount=pRequested]
    Product[p,name=pName]
    ? dTime > 0 end
    ? c.orders()->includes(o) end
    ? c.products()->includes(p) end
  -> c.deleteFromOrders(o);
     c.addToCompleted(o);
     p.setAmount(p.amount() - pRequested)
  end
\end{lstlisting}Finally, the company has a policy of marking customers who pay late.
The payment period is set at 30 days:

\begin{lstlisting}
context OrderProcessing
  @Rule OrderPaid
    Company[c]
    Order[o,customer=cName,placed=placedDate,paid=paidDate]
    Customer[k,name=cName,paysLate=false]
    ? paidDate > placedDate + 30 end
  -> k.setPaysLate(true)
  end
\end{lstlisting}Suppose a company has a customer and a product. The following shows
an instance of the class Company:

\begin{lstlisting}
Company[
  rejected = Set{},
  completed = Set{},
  orders = Set{},
  customers = Set{Customer[
                    paysLate = false,
                    name = C1]},
  products = Set{Product[
                   amount = 100,
                   name = P1]}]
\end{lstlisting}An order is received and is processed by business rule ReceiveOrder.
The following shows how the company object is updated when the order
is received:

\begin{lstlisting}
Company[
  rejected = Set{},
  completed = Set{},
  orders = Set{Order[
                 paid = 0,
                 delivered = 0,
                 placed = 0,
                 amount = 34,
                 product = P1,
                 customer = C1]},
  customers = Set{Customer[
                    paysLate = false,
                    name = C1]},
  products = Set{Product[
                   amount = 100,
                   name = P1]}]
\end{lstlisting}The order is delivered at time 10. By rule CompleteOrder, the state
is as follows:

\begin{lstlisting}
Company[
  rejected = Set{},
  completed = Set{Order[
                    paid = 0,
                    delivered = 10,
                    placed = 0,
                    amount = 34,
                    product = P1,
                    customer = C1]},
  orders = Set{},
  customers = Set{Customer[
                    paysLate = false,
                    name = C1]},
  products = Set{Product[
                   amount = 66,
                   name = P1]}]
\end{lstlisting}Suppose the system is reset, and run again, but this time the payment
is received at time 50. The delay in payment is greater than 30 days
which is the business policy of the company,. Therefore, by rule OrderPaid
the new state is:

\begin{lstlisting}
Company[
  rejected = Set{},
  completed = Set{Order[
                    paid = 50,
                    delivered = 10,
                    placed = 0,
                    amount = 34,
                    product = P1,
                    customer = C1]},
  orders = Set{},
  customers = Set{Customer[
                    paysLate = true,
                    name = C1]},
  products = Set{Product[
                   amount = 66,
                   name = P1]}]
\end{lstlisting}Now, if a new order is received from the same customer then by rule
RejectOrder:

\begin{lstlisting}
Company[
  rejected = Set{Order[
                   paid = 0,
                   delivered = 0,
                   placed = 0,
                   amount = 20,
                   product = P1,
                   customer = C1]},
  completed = Set{Order[
                    paid = 50,
                    delivered = 10,
                    placed = 0,
                    amount = 34,
                    product = P1,
                    customer = C1]},
  orders = Set{},
  customers = Set{Customer[
                    paysLate = true,
                    name = C1]},
  products = Set{Product[
                   amount = 66,
                   name = P1]}]
\end{lstlisting}
\section{Business Rules Implementation}

\begin{figure}
\begin{center}

\includegraphics[width=12cm]{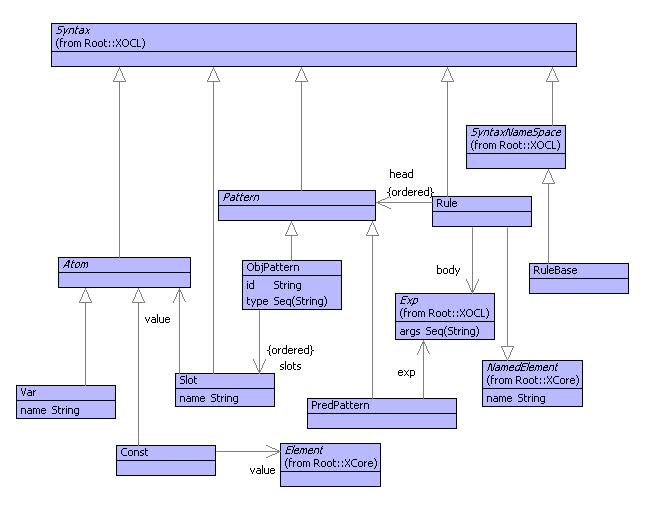}

\caption{Business Rules Model\label{fig:Business-Rules-Model}}

\end{center}
\end{figure}

The first step in implementing the business rules engine describes
in the previous section is to construct some syntax classes and a
grammar that synthesizes the rules. Figure \ref{fig:Business-Rules-Model}
shows the model of the business rules syntax. All the classes in the
model inherit from XOCL::Syntax so that instances of the classes evaluate
to themselves (i.e. there is no evaluation phase-shift). The class
RuleBase is a name-space for rules; name-spaces are special and inherit
from SyntaxNameSpace. Each rule is a named element and has a head
and a body. Each element of the rule-head is a pattern: either an
object pattern or a predicate pattern. An object pattern references
a type by giving its absolute path as a sequence of strings. An object
pattern has an id that names an oject that matches the pattern. Each
slot of an object pattern has a name and an atomic value; the atomic
value is either a variable or a constant. 

A predicate pattern is an expression. The free variables of the predicate
are encoded as arguments of the expression. This makes it easy to
supply the values of the free variables to the expression when it
is evaluated. The body of a rule is also an expression. 

The rest of this section describes the grammar that translates from
concrete rule-base syntax to instances of the rule-base classes described
above.

A rule-base is a name-space and as such contains a collection of sub-expressions
each of which are added to the rule-base:

\begin{lstlisting}
@Grammar extends OCL::OCL.grammar
  RuleBase ::= name = Name rules = Exp* 'end' {
    rules->iterate(r rb = RuleBase(name) | rb.add(r))
  }.
end
\end{lstlisting}The syntax for rules is defined separately from that of rule-bases.
A rule is defined in isolation and is then added to the rule-base
either as part of the rule-base definition or via a context definition.
The grammar definition for Rule is straightforward except for the
use of XOCL::Exp to encode the predicate patterns and the rule-bodies.
Note how the FV operation of Performable is used to calcuate the free
variables referenced in an expression and set them as the arguments
of the operation that implements the expression:

\begin{lstlisting}
@Grammar extends OCL::OCL.grammar
  Rule ::= 
    name = Name 
    patterns = RulePattern* '->' exp = Exp 'end' {
      Rule(name,patterns,Exp(exp,exp.FV()->asSeq,null))
  }.
  RulePattern ::= 
    ObjPattern 
  | PredPattern.
  PredPattern ::= '?' exp = Exp 'end' { 
    PredPattern(Exp(exp,exp.FV()->asSeq,null)) 
  }.
  ObjPattern ::= 
    type = PType '[' id = Name slots = (',' Slot)* ']' {
      ObjPattern(type,id,slots)
  }.
  PType ::= n = Name ns = ('::' Name)* { Seq{n | ns} }.
  Slot ::= name = Name '=' a = SlotValue { Slot(name,a) }.
  SlotValue ::= 
    n = Name { Var(n) } 
  | s = Str { Const(s) } 
  | i = Int { Const(i) } 
  | 'true' { Const(true) } 
  | 'false' { Const(false) }.
end
\end{lstlisting}
\section{Rete Networks}

\begin{figure}
\begin{center}

\includegraphics[width=12cm]{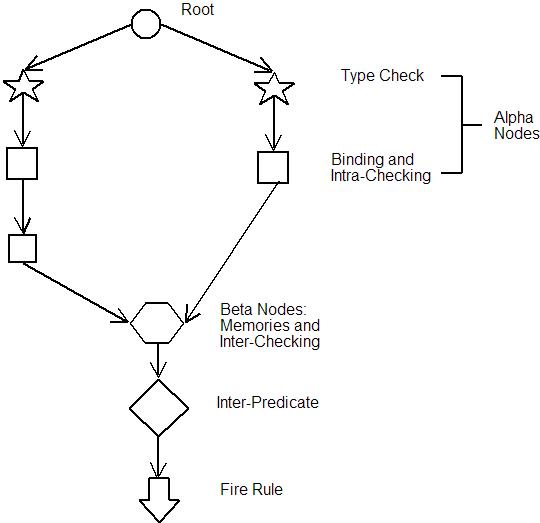}

\caption{A Rete-Network Fragment\label{fig:A-Rete-Network-Fragment}}

\end{center}
\end{figure}

A naive implementation of business rules would use an engine that
applies all business rules to all business data each time an event
occurs. Unfortunately, the performance of such an engine would not
be acceptable. Each rule typically involves matching and combining
multiple data elements. Each head pattern may contain multiple object
patterns. Each object pattern may match many different data elements
and a rule will become enabled for each combination of matching elements.
As the numbers of data elements increase, the number of combinations
that cause a rule to become enabled rises dramatically. If all combinations
must be checked and re-checked for each change in the business data
then the performance of rule matching will quickly become unacceptable.

The Rete Network matching algorithm was developed to address this
problem. It caches partial rule matches so that when a business event
occurs, only rule matching activities involving the changed data are
necessary. No matching involving unchanged data needs to take place.
This leads to a rule matching engine that can accommodate very large
number of rules and business data.

To see how the matching algorithm works, consider a simple rule:

\begin{lstlisting}
@Rule Example
  C1[c1,s1=v]
  C2[c2,s2=s1,s3=w]
  ? v > w end
-> e
end
\end{lstlisting}
The example rule matches any number of C1 and C2 instances such that
c2.s2 = c1.s1 and where c1.s1 > c2.s3. Rete proceeds as follows: all
new instances of C1 match the first pattern and are stored in the
rule memory. All new instance of C2 match the second pattern (ignoring
the s2 consistency issue), these are also stored in the rule memory.
The two parts of the rule memory are referred to as left and right
(C1 and C2 instances). 

As soon as a change occurs that leads to elements in both left and
right memories, the rule then matches the most recent entry with all
elements in the other memory. If a c1 is added to the left memory
then it is compared with element elements in the right memory; if
a c2 is added to the right memory then it is compared to all elements
in the left memory. Comparison checks for slot consistency: c2.s2
= c1.s1. For each pair that passes the consistency test, the algorithm
contines with the match.

Combined pairs of C1 and C2 instances are then filtered by the predicate
pattern v > w where v is a slot from the left-memory element and w
is from the right-memory element. If the predicate is astified then
the pair passes through and, since there is nothing left to check
, the rule is enabled. An enabled rule is added to the \emph{conflict
set}. If there is a non-empty conflict set then a rule is selected
(using a suitable choice algorithm) and its body is performed with
respect to the pair of C1 and C2 instances.

The structure of a Rete Network for Example is given in figure \ref{fig:A-Rete-Network-Fragment}.
A new or changed data element is supplied to the network root node.
The root node feeds the element to test nodes that check the type
of the element. In the case of Example, the types are C1 and C2 respectively.
If the type checks pass then the elements are passed on to \emph{alpha-nodes}
that check the internal consistency of the data and bind any variables.
In the case of Example, C1 requires a single alpha-node to match s1,
C2 requires two alpha nodes to match s2 and s3. 

Alpha nodes then feed into the left and right memories of a \emph{beta-node}.
A beta-node is responsible for pairing up matching data from the left
and right and performing inter-data checks. Once the beta-nodes are
satisfied they pass the data on to inter-element predicate nodes and
then on to nodes that fire rules into the conflict set.

Rules may have more patterns than Example given above. In this case,
for each object pattern pairing, the rule will have a pair of left
and right memories to store matching elements. The rest of this section
describes an executable model for Rete Networks.

\begin{figure}
\begin{center}

\includegraphics[width=12cm]{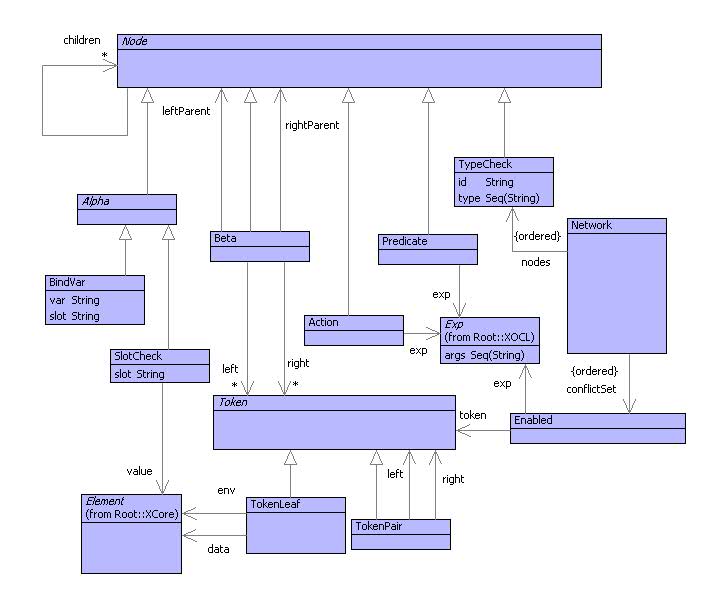}

\caption{Rete Network Model\label{fig:Rete-Network-Model}}

\end{center}
\end{figure}

Figure \ref{fig:Rete-Network-Model} shows the model for a Rete Network.
A network consists of a collection of type checking nodes and a collection
of enabled rules. The rest of the network consists of nodes of different
types, all of which have 0 or more children.

Alpha nodes are either binding nodes or slot checking nodes. Predicate
and action nodes use expressions to represent executable predicates
and rule bodies respectively. Beta nodes have a left and right memory.

Tokens contain data elements to be matched by the network. A token
is either a single element of data (TokenLeaf) together with an environment
that binds rule variables with slot values, or a combination of two
tokens (TokenPair). The idea is that a leaf is initially added to
the network; each time the token is combined via a beta-node, the
tokens from the left and right memories are conmbined to become token
pairs. After several combinations, a rule is fired into the conflict
set together with a tree of tokens.

\begin{lstlisting}
context Network
  @Operation add(value)
    @For node in nodes do
      node.add(TokenLeaf(value,Seq{}),self)
    end
  end
\end{lstlisting}\begin{lstlisting}
context TypeCheck
  @Operation add(token,network)
    if token.data().isKindOf(type->lookup)
    then 
      @For node in children do
        node.add(token.bind(id,token.data()),self,network)
      end
    end
  end  
\end{lstlisting}\begin{lstlisting}
context SlotCheck
  @Operation add(token,parent,network)
    if token.data().hasSlot(slot) andthen
       token.data().get(slot) = value
    then
      @For node in children do
        node.add(token,self,network)
      end
    end
  end
\end{lstlisting}\begin{lstlisting}
context BindVar
  @Operation add(token,parent,network)
    if token.data().hasSlot(slot)
    then
      @For node in children do
        node.add(token.bind(var,token.data().get(slot)),self,network)
      end
    end
  end
\end{lstlisting}\begin{lstlisting}
context Beta
  @Operation add(token,parent when parent = leftParent,network)
     self.addToLeft(token);
     @For rightToken in right do
       let token = token + rightToken
       in if token.consistent()
          then 
            @For node in children do
              node.add(token,self,network)
            end
          end 
       end
     end
   end
\end{lstlisting}\begin{lstlisting}
context Predicate
  @Operation add(token,parent,network)
    let args = exp.args->collect(a | self.lookup(a,token))
    in if exp.op.invoke(network,args)
       then super(token,parent,network)
       end
    end
  end
\end{lstlisting}\begin{lstlisting}
context Action
  @Operation add(token,parent,network)
    network.addToConflictSet(Enabled(token,exp))
  end
\end{lstlisting}
\section{Rule Compilation}

Rules are compiled into a network by translating each of the head
patterns into sequences of alpha nodes and then combining the patterns
using beta-nodes. Compilation of a rule-base is supplied with a network.
The compiler updates the network with all the rules:

\begin{lstlisting}
context RuleBase
  @Operation compile(network)
    @For rule in self.contentsOf(Rule) do
      rule.compile(network)
    end;
    network
  end
\end{lstlisting}A rule consists of a head and a body. The head consists of a sequence
of patterns; each pattern may be an object pattern or a predicate
pattern. The object patterns are compiled first. Each object pattern
is of the form C{[}s1,s2,...,sn] where C is a classifier and each
si is a slot. An object pattern is compiled into a sequence of nodes
that test properties of tokens that are added to the network. The
nodes must check that the token has an object of type C and then must
check each of the slots of the token. Each slot must match each of
the slots si in turn: of the slot requires a variable to be bound
then the token is extended with a binding for the variable. if the
slot requires the object to have a particular slot value then the
token is checked, if the token's object does not have the required
slot value then the token is discarded.

If a rule head contains a sequence of object patterns then they are
compiled and the resulting sequences of nodes are combined pair-wise
using beta-nodes. The beta-nodes merge the tokens that match the two
object patterns. Beta-nodes are also responsible for checking the
consistency of bindings for repeated variable occurrences.

Finally, after each of the object patterns have been added to the
network and merged using beta-nodes, the predicate patterns are compiled.
Each predicate pattern produces a node that is supplied with a token.
If the predicate is satisfied by the data and environment in the token
then the token is passed on otherwise it is discarded:

\begin{lstlisting}
context Rule
  @Operation compile(network)
    let oPatterns = head->select(p | p.isKindOf(ObjPattern));
        pPatterns = head->select(p | p.isKindOf(PredPattern))
    in pPatterns->iterate(p node = self.compilePatterns(oPatterns,network) |
         let newNode = p.compile()
         in node.addToChildren(newNode); 
            newNode
         end).addToChildren(Action(body))
    end
  end
context Rule
  @Operation compilePatterns(Seq{pattern},network)
    pattern.compile(network)
  end
context Rule
  @Operation compilePatterns(Seq{pattern | patterns},network)
    pattern.compile(network)
      .join(self.compilePatterns(patterns,network))
  end
\end{lstlisting}Nodes are joined together as follows:

\begin{lstlisting}
context Node
  @Operation join(node when node.isKindOf(Predicate))
    self.addToChildren(node);
    node
  end
context Node
  @Operation join(node)
    let beta = Beta(self,node)
    in self.addToChildren(beta);
       node.addToChildren(beta);
       beta
    end
  end
\end{lstlisting}A predicate pattern is compiled to become a predicate node:

\begin{lstlisting}
context PredPattern
  @Operation compile()
    Predicate(exp)
  end
\end{lstlisting}An object pattern is compiled into a type checking node followed by
a sequence of slot checking nodes:

\begin{lstlisting}
context ObjPattern
  @Operation compile(network)
    self.compileSlots(slots,network.typeCheck(type,id))
  end
context ObjPattern
  @Operation compileSlots(Seq{},node) 
    node 
  end
context ObjPattern
  @Operation compileSlots(Seq{slot | slots},node)
    let newNode = slot.compile()
    in node.addToChildren(newNode);
       self.compileSlots(slots,newNode)
    end
  end
\end{lstlisting}An object pattern requires that the type of the data in a token is
checked. A TypeCheck node is added to the network. When a token is
received by a type check node, the type of the token data is checked.
If the type matches the required classifier then the token passes
on otherwise it is discarded:

\begin{lstlisting}
context Network
  @Operation typeCheck(type,id)
    @Find(node,nodes)
      when node.type() = type
      else 
        let node = TypeCheck(type,id)
        in self.addToNodes(node);
           node
        end
    end
  end
\end{lstlisting}Slots are compiled to become nodes that bind variable or check the
values in slots:

\begin{lstlisting}
context Slot
  @Operation compile()
    value.compile(name)
  end
context Var
  @Operation compile(slot)
    BindVar(slot,name)
  end
context Const
  @Operation compile(slot)
    SlotCheck(slot,value)
  end
\end{lstlisting}The following rule-base:

\begin{lstlisting}
@RuleBase Test
  @Rule r1
    C[c,slot1=v,slot2=100]
    D[d,slot3=v]
    ? v > 10 end
  ->
    e
  end
end
\end{lstlisting}is compiled by Test.compile(Network()) to produce the following network:

\begin{lstlisting}
Network[
  conflictSet = Seq{},
  nodes =
    Seq{
      TypeCheck[
        type = Seq{D},
        id = d,
        children =
          Set{
            #(3)=BindVar[
              slot = slot3,
              var = v,
              children =
                Set{#(4)=Beta[
                      left = Set{},
                      right = Set{},
                      leftParent = #(5)=SlotCheck[
                                     slot = slot2,
                                     value = 100,
                                     children = Set{#(4)}],
                      rightParent = #(3),
                      children =
                        Set{Predicate[
                              exp = [| v > 10 |],
                              children =
                                Set{Action[
                                      exp =
                                        [| e |],
                                      children = Set{}]}]}]}]}],
      TypeCheck[
        type = Seq{C},
        id = c,
        children = Set{BindVar[
                         slot = slot1,
                         var = v,
                         children = Set{#(5)}]}]}]
\end{lstlisting}
\section{Network Execution}

Once compiled, the network is used to implement the rules. When an
object is created it is added to the network. The state of the object
is recorded in the network in order to maximise the speed of multipl
object matching. When an object changes, the object is removed from
the network and then re-introduced. This section describes how the
network executes in terms of adding and removing objects.

Objects are added to a network as tokens. The token allows the engine
to manage the object along with any variable bindings that have been
associated with it during the match:

\begin{lstlisting}
@Class Token isabstract 
  @Operation add(other:Token):Token
    TokenPair(self,other)
  end
  @Operation consistent()
    // Repeated use of variables have the same values...
    self.env()->forAll(b1 |
      self.env()->forAll(b2 |
        b1->head = b2->head implies b1->tail = b2->tail))
  end
  @AbstractOp env():Seq(Element)
  end
  @AbstractOp lookup(name:String)
    // Look up the supplied variable name
  end
end
\end{lstlisting}When an object is introduced to the network, a leaf-token is created
to contain and manage it. The class LeafToken is a sub-class of Token
that uses an environment to lookup the supplied variable name. Two
tokens are merged on the output of a beta-node. Two tokens are merged
to produce a single token as an instance of TokenPair which is a sub-class
of Token and just contains two component token instances. The token-pair
implements lookup by delegating to the two component tokens.

An object is added to the network as follows:

\begin{lstlisting}
context Network
  @Operation add(value)
    @For node in nodes do
      node.add(TokenLeaf(value,Seq{}),self)
    end;
    self
  end
\end{lstlisting}In general, a node just passes a token on to its children:

\begin{lstlisting}
context Node
  @Operation add(token,parent,network)
    @For node in children do
      node.add(token,self,network)
    end
  end
\end{lstlisting}One of the first inodes in a network is a type-check node that filters
out objects by type:

\begin{lstlisting}
context TypeCheck
  @Operation add(token,network)
    if token.data().isKindOf(type->lookup)
    then 
      @For node in children do
        node.add(token.bind(id,token.data()),self,network)
      end
    end
  end
\end{lstlisting}Once an object has been determined to be of the required type it must
pass all the alpha-node checks before joining beta-memories and possibly
firing rules into the conflict set. Each rule may involve slot-checks:

\begin{lstlisting}
context SlotCheck
  @Operation add(token,parent,network)
    if token.data().hasSlot(slot) andthen
       token.data().get(slot) = value
    then
      @For node in children do
        node.add(token,self,network)
      end
    end
  end
\end{lstlisting}If the rule specifies that the slot value is just a variable then
the following node is used to bind the variable:

\begin{lstlisting}
context BindVar
  @Operation add(token,parent,network)
    if token.data().hasSlot(slot)
    then
      @For node in children do
        node.add(token.bind(var,token.data().get(slot)),self,network)
      end
    end
  end
\end{lstlisting}A predicate node is used to check an arbitrary boolean expression
using the values of varaibles that have been bound at that point in
the match:

\begin{lstlisting}
context Predicate
  @Operation add(token,parent,network)
    let args = exp.args->collect(a | self.lookup(a,token))
    in if exp.op.invoke(network,args)
       then super(token,parent,network)
       end
    end
  end
\end{lstlisting}That concludes the alpha-nodes. Next we must deal with the inter-clause
and inter-rule matching. This is handles by beta-nodes. The behaviour
of a beta-nodes when a token is added depends on whether the token
arrives form the left or the right. This is tested by defining two
'add' operations in BetaNode that use guards on the operation arguments
comparing the supplied parent with the left and right parent of the
node:

\begin{lstlisting}
context BetaNode
  @Operation add(token,parent when parent = leftParent,network)
    self.addToLeft(token);
    @For rightToken in right do
      let token = token + rightToken
      in if token.consistent()
         then 
           @For node in children do
             node.add(token,self,network)
           end
         end 
      end
    end
  end
context BetaNode
  @Operation add(token,parent when parent = rightParent,network)
    self.addToRight(token);
    @For leftToken in left do
      let token = leftToken + token
      in if token.consistent()
         then 
           @For node in children do
            node.add(token,self,network)
           end
         end
      end
    end
  end
\end{lstlisting}If a token ever reaches an action node then it has enabled a rule
and the rule is added to the conflict set of the network:

\begin{lstlisting}
context Action
  @Operation add(token,parent,network)
    network.addToConflictSet(Enabled(token,exp))
  end
\end{lstlisting}The network implements a remove operation that performs the same function
as add defined above, except that the tokens are removed from the
beta-memories when they arrive at beta-nodes and also removed from
the conflict set of the rule becomes enabled.

\section{Monitoring Classes}

In order to complete the implementation of business rules we require
that newly created objects are added to the network and when the object's
state is updated, they are removed and then re-inserted into the network.
To do this automatically we implement a new-meta-class that manages
instances appropriately:

\begin{lstlisting}
import Daemon;

context Root
  @Class MonitoredClass extends Class
    // Classes whose instances are to be monitored by a
    // rule-based that has been compiled into a network
    // should be instances of the meta-class MonitoredClass.
    @Bind daemon =
      // The following daemon is added to all new instances
      // of monitored classes. The daemon ensures that all
      // changes to the object cause the network to be
      // informed via add and remove...
      Daemon("MonitorDaemon",ANY,
        @Operation(object,slot,new,old)
          object.of().network().remove(object);
          object.of().network().add(object)
        end)
    end
    
    // Each monitored-class must specify which network is
    // to be used for its instances...
    @Attribute network : Network (?,!) end   
    
    @Operation add(element)
      // The network for a monitored-class can be
      // specified in the definition of the class
      // and added to the class via this operation...
      @TypeCase(element)
        Network do
          self.network := element
        end
        else super(element)
      end
    end
    
    @Operation invoke(this,args)
      // By defining an invoke operation we hi-jack the
      // class instantiation process for a monitored-class.
      // The new instance is created in the usual way but
      // the daemon is added to the object before it is returned 
      // and the new object is added to the network...
      let newObj = super(this,args)
      in newObj.addDaemon(MonitoredClass::daemon);
         network.add(newObj);
         newObj
      end
    end
  end
\end{lstlisting}

\chapter{Model View Controllers}

A typical example of model synchronization occurs when a model is
visualized as a diagram. There are many software tools available that
provide access to a model via a diagram. You can create an instance
of the model by interacting with the diagram. Sometimes, the tool
also supports interacting with the model instance via other routes:
via property editors or a scripting language. Where the instance can
be modified in this way, it is important that the diagram view is
updated to be consistent. Consistency update may be automatic or may
be explicitly invoked via a refresh.

This chapter defines a modelling language, a diagram model and shows
how the two can be synchronized by modelling a mapping between them.
It turns out that the mappings conform to a small number of patterns
that can be reused to create model synchronizers.

\begin{figure}
\begin{center}

\includegraphics[width=12cm]{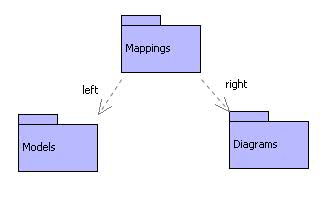}

\caption{\label{Mapping-Overview}Overview}

\end{center}
\end{figure}

Figure \ref{Mapping-Overview} provides an overview of the synchronization
architecture for the example. The left-hand package defines a simple
data modelling language consisting of the usualy components: packages;
classes; attributes and inheritance. The right-hand package defines
a simple model for diagrams including: diagrams; nodes; edges and
display elements. A diagram display element has a position and may
be either a box (container of display elements) or a text string.

The package in the centre contains a definition of mappings that are
used to synchronize models and diagrams. The idea is that a mapping
connects instances from the left with their corresponding instances
on the right such that when a change occurs on one side, the change
can be reflected in the corresponding element on the other side.

\begin{figure}
\begin{center}

\includegraphics[width=12cm]{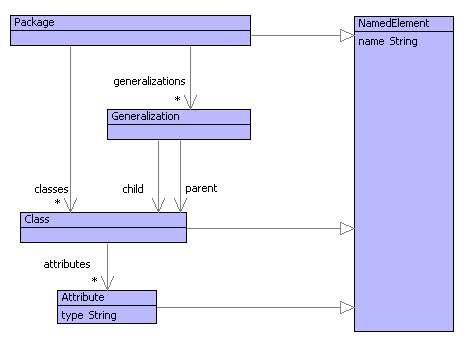}

\caption{\label{Data-Models}Data Models}

\end{center}
\end{figure}

\section{Model-Diagram Synchronization}

Here's an example. Support there is a class named C in a package.
Then there should be a diagram with a node for C. The node for C contains
a box which in turn contains a text item whose text is the string
{}``C''. Mappings are used to:

\begin{itemize}
\item connect the package to the diagram
\item connect the class to the node
\item connect the class name to the text item
\end{itemize}
Suppose that the name of C is changed on the diagram by editing the
text item. The change is propagated to the mapping between the class
name and the text item. Since the text item has changed, the class
is updated to be consistent. Alternatively, if the class name is changed
then the mapping allows the change to be propagated to the text item.

\begin{figure}
\begin{center}

\includegraphics[width=12cm]{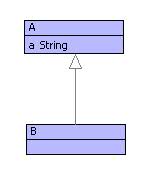}

\caption{Example\label{Example-Model}}

\end{center}
\end{figure}

Notice that in both the scenarios given above, it was not necessary
for the model element to know about the diagram element and vice versa.
This is an important issue: when model instances are synchronized,
the machinery must ensure a separation of concerns. There is no reason
why the model should know about the diagram -- to do so would compromise
reusability of the model. Similarly, if the diagram elements are directly
associated with the model elements then this makes it difficult to
use the diagram model in a variety of circumstances (domain specific
languages for example).

\section{Models}

\begin{figure}
\begin{center}

\includegraphics[width=12cm]{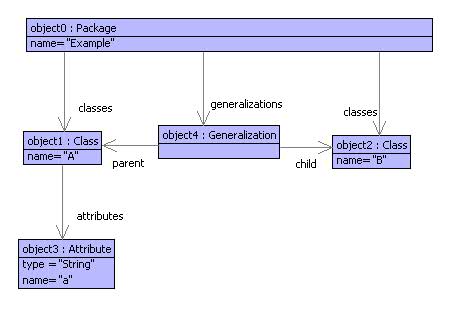}

\caption{Model Snapshot\label{Model-Snapshot}}

\end{center}
\end{figure}

Figure \ref{Data-Models} shows a simple data model consisting of
packages, classes, attributes and generalizations. This model is representative
of many data modelling notations. 

Figure \ref{Example-Model} shows an example model as it would be
seen in a diagram editor. The snapshot shown in figure \ref{Model-Snapshot}
shows the same model as an instance of figure \ref{Data-Models}.

\section{Diagrams}

\begin{figure}
\begin{center}

\includegraphics[width=12cm]{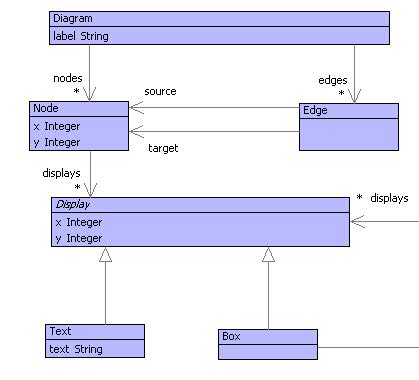}

\caption{\label{fig:Diagrams}Diagrams}

\end{center}
\end{figure}

Diagrams contain nodes with edges between them. Each node has a position
on the diagram and contains a collection of displays. A display has
a co-ordinate that is relative to its container; it may be a text
item or a box (bordered container of display elements). The model
for diagrams is shown in figure \ref{fig:Diagrams}.

\begin{figure}
\begin{center}

\includegraphics[width=12cm]{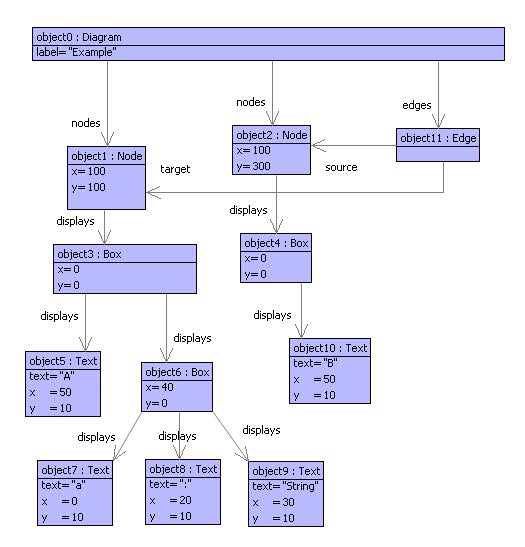}

\caption{Diagram Snapshot\label{fig:Diagram-Snapshot}}

\end{center}
\end{figure}

The diagram of the model shown in figure \ref{Example-Model} is shown
in figure \ref{fig:Diagram-Snapshot}. A diagram has a label that
is intended to be the same as the name of the corresponding package.
Each class is shown as a node and generalizations are shown as edges
between class nodes. A class is drawn as a box, with the name as a
text item at the top followed by attribute boxes. Each attribute box
contains text items for the name, {}``:'' and the type of the attribute.
Note that in the example, the positions of the nodes and displays
are illustrative.

Diagrams know nothing about classes, attributes and generalizations.
This is as it should be since we could choose to use the diagram model
to display a variety of information including class-models, snapshot-models,
CPM networks and decision trees.

\section{Mappings}

\begin{figure}
\begin{center}

\includegraphics[width=12cm]{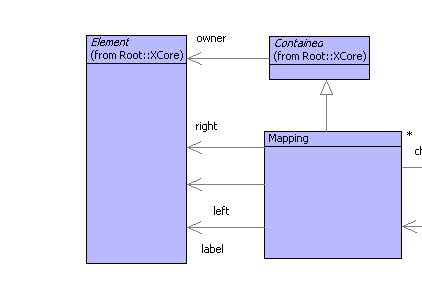}

\caption{\label{fig:Mappings}Mappings}

\end{center}
\end{figure}

Figure \ref{fig:Mappings} shows a model of mappings that can be used
to synchronize a class-model with a diagram. A mapping is a tree structure
that link elements from a left-hand model with elements from a right-hand
model. Each mapping has a label that can be used to distinguish between
different mappings between the same left and right elements.

Mappings sit between the left and right-hand elements in order to
record the fact that they should be the same in some sense. As such,
a mapping can have intimate knowledge of the left and right-hand model
structures. In the case of a class-model and a diagram, mappings can
be expected to know that a class has a name and the related diagram
node has a text item that is used to display the class name. Neither
the class-model or the diagram model know anything about each other,
so there is a clear division of concerns: cross-model information
is limited to the mapping.

Mappings can be used to relate any elements from the left and right-hand
models and a mapping is is a tree structured thing (made up of \textit{maplets}).
It is up to us as modellers to decide how to structure and use the
mapping. Each mapping has children that are other mapping components.
The idea is that if the left and right elements are composed of children
then the mapping will have maplets that appropriately relate those
children. For example, a package may be related to a diagram via a
mapping composed of maplets that relate the package's classes with
the diagram's nodes.

Mappings simply record the associations between left and right-hand
elements. There is no requirement for mappings to be used in any particular
way. However, it turns out that there are useful patterns of mapping
usage, both in terms of how they relate models that are themselves
structured using standard patterns.

\begin{figure}
\begin{center}

\includegraphics[width=12cm]{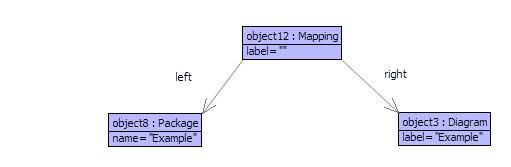}

\caption{\label{fig:A-Package-Mapping}A Package Mapping}
\end{center}
\end{figure}

Figure \ref{fig:A-Package-Mapping} shows how a package and a diagam
are associated using mappings. This is a root mapping and is responsible
for synchronizing the package with all its components against the
diagram with all its components. The mapping object12 is individually
responsible for keeping the name of the package synchronized with
the diagram label; it devolves responsibility for the class and generalization
synchronization to its child maplets.

\begin{figure}
\begin{center}

\includegraphics[width=12cm]{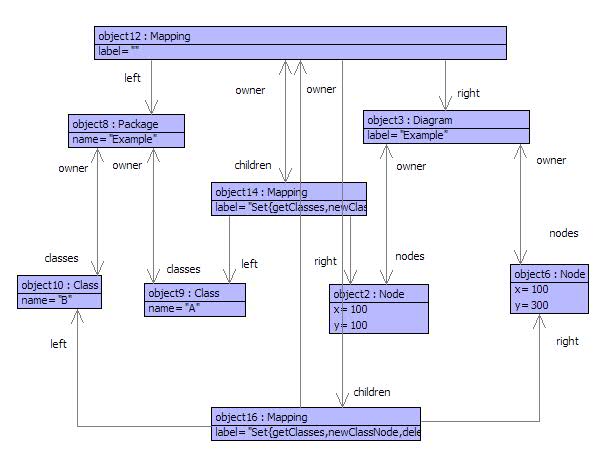}

\caption{\label{fig:Class-Mappings}Class Mappings}

\end{center}
\end{figure}

Figure \ref{fig:Class-Mappings} shows the child maplets of the root
package/diagram mapping. The root mapping has two children:; object14
and object16 that are responsible for associating classes A and B
with their corresponding diagram nodes. In each case the maplets have
labels (the detail of which is not shown) that identify them as being
\textit{class-associating} mappings.

It is worth considering how this mapping migh be used as part of a
tool. Suppose that the name of the package (object8) is modified,
to X. The resulting event causes a synchronization to occur (or the
user manually causes a synchronization to occur) which in turn asks
the mapping (object12) to synchronize its left and right elements.
At this stage, the name at object8 is X whereas the label at objct3
is Example. The mapping object12 detects this inconsistency and modifies
the label (causing a chage in the GUI). The story would be much the
same if the label was changed causing the packag name to be modified.

Now consider what happens if we add a class node to the diagram. Suppose
that objects 6, 16 and 10 have yet to be added (i.e. the model contains
only class A). The user adds a node to the diagram (object6) and the
synchronization is started, as before, at object12. 

Part of the synchronization task of object12 is to ensure that, if
the diagram has changed, every node has a corresponding class. The
mapping object12 can do this because it has access to (and knowledge
of) both packages and diagrams. When this check is performed, object12
detects that there is a node (object6) for which there is no child
maplet. Therefore, the model and diagram are inconsistent. The remedy
is to create a class (object10) and associate it via a new maplet
object16 with the diagam node. The story is much the same if a class
is added to the model: a new node is added to the diagram and a maplet
to the root mapping.

\section{Mapping Patterns}

Mappings are used to link left-hand instances to rght-hand instances
and mappings are structured into trees of maplets. So far, synchronization
has been described in rather hand-wavy terms. This is because, in
general, many different synchronization strategies and implementations
are possible ranging from those that are implemented by hand (and
guided by the mappings) to those that are fully automated.

In many circumstances, it is possible to define mapping patterns such
that mappings can be applied to instances of left and right models
in chunks. Moreover, the pattern chunks can be made to be self manging
in terms of synchronization. The next couple of sections show how
an analysis of the model-diagram example leads to patterns and then
shows how the patterns can be defined in XMF. The following section
then gives the full implementation of the model diagram application
in terms of the patterns.

\section{Equality Pattern}

Consider classes and attributes. They have a feature in common: they
both have names. Consider class diagrams and the rendering of classes
and attributes. They are each rendered as text items in boxes such
that the synchronization ties up the name with the string in the text
item.

Class and attribute name synchronization is an example of and \textit{equality
pattern}. Where a left-hand item (the name) is defined to be equal
to a right-hand item (the string). The difference between the two
examples, occurs in terms of how the name/string is accessed, updated
and created.

Suppose that an equality synchronization pattern is defined. What
are its different modes of operation:

\begin{enumerate}
\item If a left-hand instance exists but a right-hand instance does not
then the maping is being used to genrate the right from the left.
It must be possible to creat a new instance of the appopriate right-hand
element and link it to the left with a maplet.
\item If a left-hand instance changes then we must be able to update the
right-hand instance.
\item If a right-hand instance exists, but a left-hand instance does not
then proceed as for 1 (right to left).
\item If a right-hand instance chages then proceed as for 2 (right to left).
\end{enumerate}
There are clearly two major modes (left, right) each of which has
minor modes (in-sync, update, generate). Assuming that it is always
possible to tell which side has changed, the two major modes can be
defined separately. The following XMF operation definition gives the
laft-major mode definition for the equality pattern:

\begin{lstlisting}
@Operation syncEqualLeft(lab,getLeft,setRight,newRight,mapping)
(1)  let left = mapping.left();
         right = mapping.right()
(2)  in @Find(child,mapping.children())
(3)       when child.label() = lab
(4)       do if getLeft(left) <> child.left()
             then
(5)            child.setLeft(getLeft(left));
               child.setRight(getLeft(left));
               setRight(right,getLeft(left))
          end
         else
(6)        let new = newRight(right,getLeft(left)) then
(7)            child = Mapping(lab,getLeft(left),new)
(8)        in mapping.addToChildren(child)
           end
        end
    end
  end
\end{lstlisting}The operation syncEqualLeft takes 5 arguments: a label used to mark
the equality maplet; an accessor for the left-hand element; an updater
for the right-hand element; a creator for the right-hand element;
and a mapping. The mapping associates two elements that have sub-components
related by equality. For example, the mapping may associate a package
and a diagram such that the name and label are syncronized. For example,
the mapping may associate a class and a node such that the class-name
and the string in a text item are synchronized.

Line (1) extracts the left and right elements from the mapping. Lines
(2-3) select a maplet linking the synchronized elements. If this exists
then a check is made at (4) to see if the leftt-hand element has changed.
If it has changed then the element will be out of sync with the maplet
and lines (5-) update the appropriate components thereby synchronizing.

Line (6) occurs when the maplet for the synchronized elements dos
not exist. This occurs when the right-hand element is being generated
from the left. In this case, a new right-hand element is constructed
(6). Imagine class-names being synchronized with text items: the arguments
in (6) are the node and the class-name.

Line (7) creates a new maplet (assumed synchronized) and (8) adds
the maplet to the parent mapping.

\section{Containment Pattern}

Consider packages and classes (similarly classes and attributes) compared
with diagrams and nodes (similarly nodes and contained boxes). In
both cases one element contains a collection of sub-elements: packages
contain classes; diagrams contain nodes. Synchronization of such elements
can be captured as a containment pattern with the following modes:

\begin{enumerate}
\item Addition of a new left-hand instance (e.g. a class) causes addition
of a corresponding right-hand instance (e.g. a node).
\item Deletion of a left-hand instance causes deletion of the corresponding
right-hand instance.
\item Addition of a right-hand instance (as for 1 but right to left).
\item Deletion of a right-hand instance (as for 2 but right-to-left).
\end{enumerate}
The pattern can be applied to mny different types of containment structure.
The differences occur in terms of access to, creation of and deletion
of the contained elements. Also, as for equality, there are two modes
depending on whether the left hand element has chaged or the right.
The left-hand mode is defined by the following XMF operation:

\begin{lstlisting}
@Operation syncSetsLeft(lab,getLeft,newRight,deleteRight,mapping,subSync)
(1) let left = mapping.left();
        right = mapping.right()
(2) in @For x in getLeft(left) do
(3)      @Find(child,mapping.children())
(4)        when child.left() = x
(5)        do subSync(child)
           else
(6)          let y = newRight(mapping,x,right) then
(7)              child = Mapping(lab,x,y)
(8)          in mapping.addToChildren(child);
(9)             subSync(child)
             end
         end
       end;
(a)    @For child in mapping.children() 
(b)      when child.label() = lab 
(c)      do @Find(x,getLeft(left))
(d)           when child.left() = x
              else
(e)             mapping.deleteFromChildren(child);
(f)             deleteRight(right,child.right())
            end
       end
    end
  end
\end{lstlisting}The operation syncSetsLeft expects 6 arguments:; a label used to tag
maplets; an accessor for components of the left element; a constructor
for new right-hand elements; an operation that deletes right-hand
elements, a mapping and an element synchronizer.

Line (1) extracts the left and right-hand elements. Each sub-component
x is extracted at (2) and handled in turn. A maplet that associates
x is extracted at (3-4). If the maplet child exists then the synchronizer
is applied to it.

Otherwise (6) creates a new right-hand element, (7) creates a new
maplet, (8) adds the maplet to the parent mapping and (9) applies
the synchronizer.

Lines (a-f) check whether any children have been deleted from the
left-hand element. At (e) a maplet exists for which there is no longer
a sub-component x. In this case (e) removes the maplet and (f) deletes
the right-hand element.

\section{Model-Diagam Implementation}

The model-diagram synchronizer can now be implemented in terms of
the patterns defined in the previous sections. A synchronizer is an
operation that accepts a mapping and synchronizes the left and right-hand
elements. The implementation of the synchronizer that is used when
a model eleent changes is shown below:

\begin{small}
\begin{lstlisting}
@Operation syncPackage(mapping)
  syncEqualLeft("name",
    getName,
    setLabel,
    newLabel,
    mapping);
  syncSetsLeft("classes",
    getClasses,
    newClassNode,
    deleteClassNode,
    mapping,
    syncClass);
  syncSetsLeft("inherits",
    getGens,
    newGenEdge,
    deleteGenEdge,
    mapping,
    emptyMap)
end
  
@Operation syncClass(mapping)
  syncEqualLeft("name",
    getName,
    setClassNodeName,
    newClassNodeName,
    mapping);
  syncSetsLeft("atts",
    getAtts,
    newAttBox,
    deleteAttBox,
    mapping,
    syncAttribute)
end
  
@Operation syncAttribute(mapping)
  syncEqualLeft("att",
    getNameAndType,
    setNameAndType,
    newNameAndType,
    mapping)
end
\end{lstlisting}
\end{small}

The operation syncPackage causes changes in the package name to update
the diagram label, changes in the package classes to update the nodes
on the diagram, and chages in the generalization links to cause corresponding
changes in the diagram. Each class is synchronized using syncClass.

The operation syncClass causes chages in the class name to update
the text in the class node name box, and changes to the class attributes
to cause corresponding changes to the boxes on the diagram.

\section{A Simple Database Synchronizer}

The following shows how the patterns can be used to synchronise packages
and classes with database tables. A table is represented simply as
a sequence of records. Each record is just a sequence of field values: 

\begin{lstlisting}
Root::Packages := Seq{};

Root::ClassSets := Seq{};

Root::Classes := Seq{};
  
context Root
  @Operation deleteClass(table,id)
    Root::ClassSets := table->reject(r | r->head = id)
  end
  
context Root
  @Operation setPackageName(id,name)
    @Find(package,Packages)
      when package->head = id
      do Root::Packages := 
           Packages->excluding(package)
                   ->including(Seq{id,name,package->at(2)});
         name
    end
  end
  
context Root
  @Operation newClassRecord(map,class,pid)  
    @Find(package,Packages)
      when package->head = pid
      do let C = ClassSets->select(r | r->head = package->at(2));
             cid = "C" + Classes->size
         in Root::ClassSets := 
              ClassSets->including(Seq{package->at(2),cid});
            Root::Classes := 
              Classes->including(Seq{cid,class.name()});
            cid
         end
    end
  end
  
context Root
  @Operation deleteClassRecord(pid,cid)
    @Find(package,Packages)
      when package->head = pid
      do let Cid = package->at(2)
         in @Find(C,ClassSets)
              when C->head = Cid and C->at(1) = cid
              do Root::ClassSets := ClassSets->excluding(C);
                 Root::Classes := Classes->reject(r | r->head = cid)
            end
         end
    end
  end
  
context Root
  @Operation syncPackageTable(mapping)
    syncEqualLeft(getName,setPackageName,setPackageName,mapping);
    syncSetsLeft(getClasses,newClassRecord,deleteClassRecord,mapping,emptyMap)
  end
\end{lstlisting}

\chapter{Generating Text from Models}

\section{Introduction}

An important transformation involves turning data into text. The text
may be program code, HTML, XML, natural language, or some other format.
The output may be an end in itself (for example documentation) or
may be an intermediate format ready to be processed by a tool (for
example HTML or program code) or may be a save format (for example
XML). Many modelling tools provide support for \textit{code templates}
that allow models to be transformed to program code. These are a special
case of a model to text transformation.

Often, text is generated by running over the data producing standard
chunks of text (or \textit{boiler plate}) with data inserted at appropriate
points. For example, when producing HTML, there will usually be a
standard header and standard components of table definitions.

This chapter describes a language construct that allows text to be
conveniently produced from models. Furthermore, the construct makes
it possible to dip in-and-out of the model and the text interleaved
to any depth, making it easy to structure the transformation so that
it reflects the text as seen in the output.

\section{Examples}

Consider the following operation that expects a class as an argument
and produces a table definition:

\begin{lstlisting}
@Operation classToTable(c)
(1)  @Table(stdout,7)
(2)    table <c.name> {
(3)      // Table for <c.path()>
(4)      <@For a in c.attributes do
(5)        [ field <a.name> : <a.type.name>;
(6)        ]
(7)       e_nd> 
(8)    }
(9) end
end
\end{lstlisting}Line (1) introduces a text transformation of type Table. In general
a text transformation produces literal text interspersed with text
generated by arbitrary bits of XOCL program. Each text transformation
differs in terms of the delimiters used to start and end the literal
text and the XOCL programs. In the case of Table, the delimiters are
<, >, {[} and ].

The arguments used by Table in line (1) are an output channel (to
send the text to) and an integer that indicates where the left margin
should start. In the example, the left margin starts at position 7:
when generating text, whitespace characters up to position 7 are ignored;
or, put another way, everything is left-shifted by 7 characters.

Line (2) starts with some literal text (table) that is sent to the
output channel. Then the \textit{drop} delimiters < and > are encountered.
The drop delimiters surround XOCL code that produces some text to
be sent to the output channel. In this case, the name of the supplied
class is generated. Line (2) concludes with more literal text.

Line (3) is a mixture of literal text (a comment) and the path of
the supplied class. Line (4) introduces a dropped XOCL expression:
a for-loop. Each attribute of the class c produces some text in lines
(5-6) which are surrounded by \textit{lift} delimiters {[} and ].
The lift delimiters are used within a dropped expression to surround
literal text. It should be noted that lifts and drops can be arbitrarily
interleaved.

Lines (5-6) produce text output that describes each of the attributes
in turn. Note that the newline at the end of line (5) is part of the
output.

Line (7) completes the dropped for-loop. Note that normally an XOCL
for-loop is termined by an end keyword. Within a dropped expression,
end cannot be used since it will confuse the lifting mechanism (as
explained below), so the convention is to use e\_nd within lifted
text. The only end-keyword that is permitted must be that at the end
of the outermost lifted text -- line (9).

In review, the Table construct is an example of a model to text template.
The body of the template is literal text, up to an open drop delimiter.
The text inside the drop delimiter, up to the corresponding closing
drop delimiter, is XOCL code (containing e\_nd instead of end where
necessary). The dropped XOCL code can do anything, but should return
a string that is dropped into the surrounding literal text of the
template. Within dropped XOCL code, literal text can be used surrounded
by lift delimiters. The lift delimiters surround a model to text template
(just like Table), however, nested templates can use the delimiters
rather than the @TemplateName ... end notation. Drops and lifts can
be nested to any depth, and in general the actual delimiters used
to demote the lift and drop can be defined (so that they do not clash
with literal text used in the body of the template). Newlines within
lifted text are carried through to the output with the left hand margin
affected by the initial value supplied to the template (7 in the example
above).

Supplying Class to the Table template produces the following:

\begin{lstlisting}
table Class {
  // Table for Root::XCore::Class
  field attributes : Set(Attribute);
  field isAbstract : Boolean;
  field constructors : Seq(Constructor);
}
\end{lstlisting}Notice how the literal text in the lifted parts has been copied through
verbatim, whereas the name of the class and the names of the attributes
have been dropped into the text. The following example shows how drops
and lifts can be interleaved to good effect. The aim is to produce
an HTML page containing the description of a class. The attributes
of the class are to be rendered as a table; each row of the table
describes an attribute: the first row is the name of the attribute
and the second row is the name of the attribute's type.

\begin{lstlisting}
@Operation classToHTML(class:Class,out:OutputChannel)
  @HTML(out,4)
    <HTML>
      <HEAD>
        <TITLE> {class.name} </TITLE>
      </HEAD>
      <FONT SIZE="+2">
        <B> { class.name } </B>
      </FONT>
      <BR><BR><HR>
      <FONT SIZE="+1">
        <B>Overview</B>
      </FONT>
      <BR>
      <P> {c.doc().doc}
      <BR><HR><BR>
      <FONT SIZE="+1">
        <B>Parents:</B>
      </FONT>
      { @For p in class.allParents() do
          [ { p.name.toString() } ]
        e_nd }
      <BR><HR><BR>
     <TABLE WIDTH="100%" CELLPADDING="3" CELLSPACING="0">
        <TR BGCOLOR="#CCCCFF" CLASS="TableHeadingColor">
          <TD COLSPAN=3>
            <FONT SIZE="+2">
              <B>Attributes</B>
            </FONT>
          </TD>
        </TR>
        { @For a in class.attributes do
           [ <TR>
               <TD>
                 <B> {a.name.toString()} </B>
               </TD>
               <TD>
                 <B> { a.type.name().toString() } </B>
               </TD>
             </TR>
           ]
         e_nd
       }
     </TABLE>
   </HTML>
  end
end
\end{lstlisting}%
\begin{figure}
\begin{center}

\includegraphics[width=12cm]{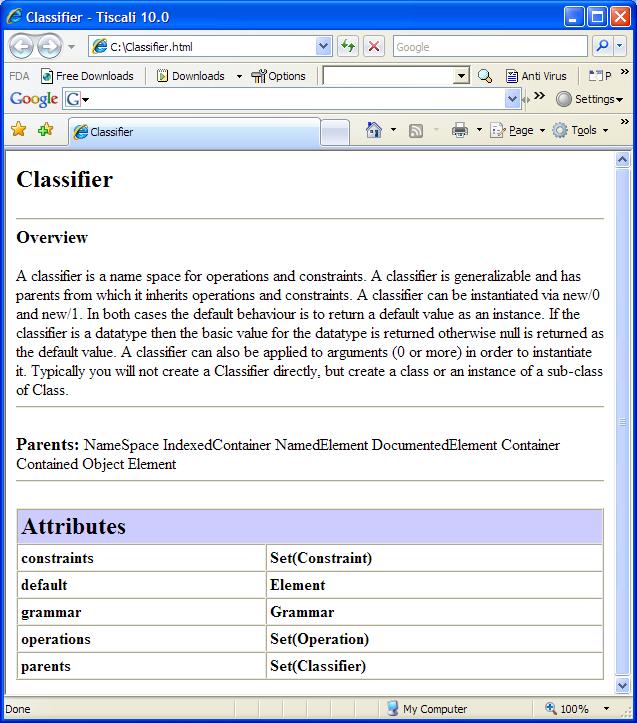}

\caption{G\label{fig:Generating-HTML}enerating HTML}

\end{center}
\end{figure}

In classToHTML notice how the table rows for each attribute are priduced
by a dropped for-loop. The body of the for-loop is a lifted table-row,
where the attribute name and type are dropped in. This interleaving
is typical of the template mechanism. Notice how the structure of
the required output is retained (i.e. an HTML page) where the model
data is dropped in at appropriate points. The HTML page for the class
Classifier is viewed in a web browser as shown in figure \ref{fig:Generating-HTML}.

\section{Specification}

Code templates can be implemented as a transformation into XOCL code.
Each element of the text in a template is turned into a format statement.
Literal text ets turned into a format statement that just prints out
the text. Dropped expressions are transformed into format statements
that evaluate the expressions and then print out their results. The
complications arise from keeping track of the indentation and from
the interleaving of lifts and drops.

To understand how the transformation works it is useful to look at
a few examples. Suppose we have a construct X that is defined to be
a template:

\begin{lstlisting}
@X(out,0)
  Some text.
end
\end{lstlisting}is translated to the following code:

\begin{lstlisting}
// Output spaces...
format(out,"~V",Seq{3});
// Output the text...
format(out,"Some text.")

If newlines are included in the text then they are faithfully recreated
in the output:

\begin{lstlisting}
@X(out,0)
  Some 
    text.
end
\end{lstlisting}produces the following code...

\begin{lstlisting}
// Output spaces...
format(out,"~V",Seq{3});
// Output the text...
format(out,"Some");
// Newline and pad to the appropriate column...
format(out,"~%~V",Seq{5});
format(out,"text.")
\end{lstlisting}Now, suppose that a simple expression is dropped into the middle...

\begin{lstlisting}
@X(out,0)
  Some { dropped } text.
end
\end{lstlisting}The output should include the value of the variable named 'dropped'
in between the literal text for {}``Some'' and {}``text.''. Therefore
the code that is produced looks like the following:

\begin{lstlisting}
// Pad to the appropriate column...
format(out,"~V",Seq{3});
// Output the literal...
format(out,"Some "});
// Get the value of dropped and check
// that it is a string. If so then
// print it...
let str = dropped
in if str.isReallyKindOf(String)
   then format(out,str)
   end
end;
// The rest of the literal output...
format(out," text.")
\end{lstlisting}Within a dropped expression it is possible to introduce a lifted literal:

\begin{lstlisting}
@X(out,0)
  Some { dropped1 [ then lifted ] dropped2 } text.
end
\end{lstlisting}What we would like to achieve for the nested lifting is the following:

\begin{lstlisting}
format(out,"~V",Seq{3});
format(out,"Some "});
let str1 = dropped1
in if str1.isReallyKindOf(String)
   then format(out,str1);
        format(out," then lifted ");
        let str2 = dropped2
        in if str2.isReallyKindOf(String)
           then format(out,"str2)
           end
        end
   end
end;
format(out," text.")
\end{lstlisting}Note that the nested lift, is equivalent to another occurrence of
@X ... end, therefore we could just produce the following code:

\begin{lstlisting}
format(out,"~V",Seq{3});
format(out,"Some "});
let str = dropped1
in if str.isReallyKindOf(String)
   then format(out,str)
   end
end;
@X(out,3) then lifted end;
let str = dropped2
in if str.isReallyKindOf(String)
   then format(out,str)
   end
end;
format(out," text.")
\end{lstlisting}Now we have a general rule for dealing with nested lifts: just wrap
the nested lift up in the appropriate template construct (passing
in the current margin for indentation). This will deal with any level
of nesting and interleaving:

\begin{lstlisting}
@X(out,0)
  Some { dropped1 [ then { nested } lifted ] dropped2 } text.
end
\end{lstlisting}Is translated to:

\begin{lstlisting}
format(out,"~V",Seq{3});
format(out,"Some "});
let str = dropped1
in if str.isReallyKindOf(String)
   then format(out,str)
   end
end;
@X(out,3) then { nested } lifted end;
let str = dropped2
in if str.isReallyKindOf(String)
   then format(out,str)
   end
end;
format(out," text.")
\end{lstlisting}which in turn becomes:

\begin{lstlisting}
format(out,"~V",Seq{3});
format(out,"Some "});
let str = dropped1
in if str.isReallyKindOf(String)
   then format(out,str)
   end
end;
format(out," then ");
let str = nested
in if str.isReallyKindOf(String)
   then format(out,str)
   end
end;
format(out," lifted ");
let str = dropped2
in if str.isReallyKindOf(String)
   then format(out,str)
   end
end;
format(out," text.")
\end{lstlisting}
\section{Design}

In order to perform the transformation from a template to code, we
must keep track of the level of nesting. This can be done using a
simple two-state transition machine. the machine is wither currently
lifting or is currently dropping. At each stage the machine consumes
input characters until it receives a lifting or dropping token. When
it receives a token it changes state. The machine builds literals,
lifts and drops. Each time a literal character is encountered it is
added to the literal currently being constructed. When the machine
encounters an end token, it constructs an appropriate element (either
a lift or a drop) and consumes the literal tokens. When the machine
encounters a start-token (either lift or drop) then it changes state
and starts a new collection of literals:

\begin{figure}
\begin{center}

\includegraphics[width=12cm]{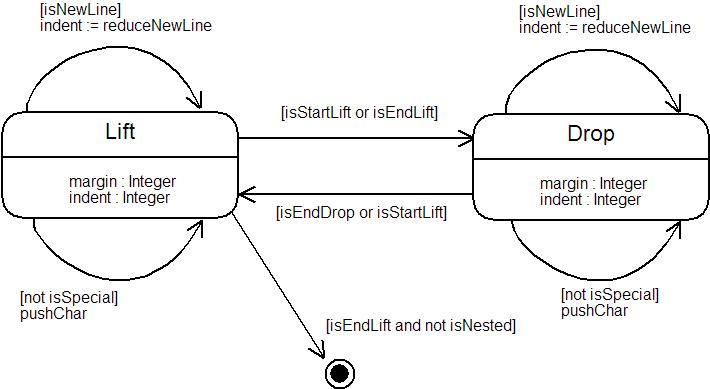}

\caption{L\label{fig:Lift-and-Drop}ift and Drop Mechanism}

\end{center}
\end{figure}

The state machine consumes an input string and produces a code element
that represents the structure of the template. The model of code elemnts
is shown below:

\begin{figure}
\begin{center}

\includegraphics[width=12cm]{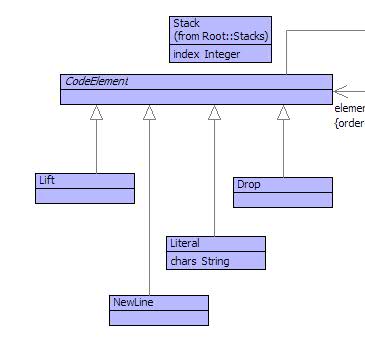}

\caption{Code Elements}

\end{center}
\end{figure}

A code element is a literal string containing text that is to be printed
on the output channel. Code elements retain newlines so that the indentation
can be controlled. The structure of lift and drop blocks in a template
is retained by constructing instances of Lift and Drop. An lift or
drop object contains a sequence of code elements produced by translating
the input string in between the start and end tokens.

For example, the template:

\begin{lstlisting}
@X(out,0)
  Some text.
end
\end{lstlisting}produces the following code elements when processed by the state machine:

\begin{lstlisting}
Lift(0,Seq{
  NewLine(0,3),          
  Literal(Some text.),          
  NewLine(0,1),          
  Literal()
})
\end{lstlisting}Note that the NewLine instances contain two values. The first value
is the amount to left-shift all input. This value is supplied to the
template as the argument after the output channel. In the example,
@X(out,0) ... end supplied 0 as the adjustment to the left-hand margin.

The following example includes further newlines:

\begin{lstlisting}
@X(out,0)
  Some 
    text.
end
\end{lstlisting}and produces the following code elements:

\begin{lstlisting}
Lift(0,Seq{          
  NewLine(0,3),          
  Literal(Some),          
  NewLine(0,5),          
  Literal(text.),          
  NewLine(0,1),          
  Literal()
})
\end{lstlisting}Dropped text becomes wrapped in an instance of Drop:

\begin{lstlisting}
@X(out,0)
  Some { dropped } text.
end
\end{lstlisting}is translated into the following instance by the state machine:

\begin{lstlisting}
Lift(0,Seq{       
  NewLine(0,3),
  Literal(Some ),
  Drop(Seq{
    Literal( dropped )
  }),
  Literal( text. )
})
\end{lstlisting}Finally, if we interleave drops and lifts then these are faithfully
reproduced in the code element structure:

\begin{lstlisting}
@X(out,0)
  Some { dropped1 [ then { nested } lifted ] dropped2 } text.
end
\end{lstlisting}becomes:

\begin{lstlisting}
Lift(0,Seq{
  NewLine(0,3),
  Literal(Some ),
  Drop(Seq{
    Literal( dropped1 ),
    Lift(2,Seq{
      Literal(then ),
      Drop(Seq{
        Literal( nested )
      }),
      Literal( lifted )
    }),
    Literal( dropped2 )
  }),
  Literal( text.)
})
\end{lstlisting}
\section{Implementation}

A code template is implemented as a language construct that is defined
in terms of tokens for the start and end of lift and drop respectively.
The language construct grammar simply consumes all the text in between
the start of the template and the keyword 'end'. The text is then
loaded onto the machine and the machine is executed to produce a code
element. The code element is then transformed into abstract syntax
that is the returned by the grammar. Here is an example template:

\begin{lstlisting}
@Class X extends CodeGen::Generator
    @Grammar extends OCL::OCL.grammar
      X ::= 
        '(' // Get the output channel... 
            out = Exp 
            // get the current level of indent...
            ',' indent = Int 
        ')'     
        // Get the raw text...   
        s = Char*   
        'end'            
         { let // The control tokens for the code generator...
               startLift = "[";
               endLift = "]";
               startDrop = "{";
               endDrop = "}";  
               // All references to 'end' are protected...   
               protectEnd = s.asString().subst("end","e_nd",true) then   
               // All references to newline characters are protected...   
               protectNewline = protectEnd.subst("\n","\\n",true) then
               newString = startLift + protectNewline + endLift then
               // Create the mapping to a code element...
               mapping = Mapping(newString,startLift,endLift,startDrop,endDrop) then
               // Perform the mapping (run the machine) ...
               lift = mapping.processLift(indent,indent)    
           in 
              // Translate the code element to abstract syntax and return...
              lift.desugar("CodeGen::X",
                startLift,endLift,startDrop,endDrop,startExtract,endExtract,out,0) 
           end
       }.       
    end
\end{lstlisting}In the example template X defined above, the start- and end-lift tokens
are {[} and ] respectively, and the start- and end-drop tokens are
\{ and \} respectively. Note how the keyword 'end' replaces occurrences
of the keyword 'e\_nd' before the machine is performed. Including
XOCL code in templates must use the keyword 'e\_nd' instead of 'end'
in order that the transation does not get confused.

The mapping is loaded with the token information and the input string.
Notice how the input text (protectNewline) is loaded onto the machine
by surrounding it with the start- and end-lift tokens. This is because
the main body of the template is equivalent to an outermost 'lift'. 

The machine is executed with respect to the following state:

\begin{lstlisting}
@Class Mapping
    // The input string...
    @Attribute chars        : String end
    // The stack of code elements being generated...
    @Attribute stack        : Stack = Stack() end
    // The current position in the string...
    @Attribute index        : Integer end
    // The token indicating a start lift...
    @Attribute startLift    : String end
    // The token indicating the end of a lift...
    @Attribute endLift      : String end
    // The token indicating the start of a drop...
    @Attribute startDrop    : String end
    // The token indicating the end of a drop...
    @Attribute endDrop      : String end
    // The constructor...
    @Constructor(chars,startLift,endLift,startDrop,endDrop) end
    ...
\end{lstlisting}The machine constructs values on a stack. At any given time there
is a sequence of code elements being built at the head of the stack.
To support this, the machine arranges for the head element of the
stack to be another stack onto which the current sequence of code
elements are pushed. The following operations manipulate the stack:

\begin{lstlisting}
context Mapping
    @Operation pushChar()
      // Elements are added to the top-stack. Generally
      // the head element is a buffer. The next literal
      // char is added to the top-buffer...
      let elements = stack.top() then
          b = elements.top()
      in b.add(chars->at(index));
         self.index := index + 1
      end
    end
context Mapping   
    @Operation pushElement(element)
      // Add an element to the top-stack...
      let elements = stack.top()
      in elements.push(element)
      end
    end
context Mapping
    @Operation reduceDrop()
      // An end-drop has been consumed. The top-stack
      // contains the elements to be dropped...
      let elements = stack.pop().asSeq()
      in Drop(elements)
      end
    end
context Mapping         
    @Operation reduceLift()
      // An end-lift has been encountered. The top-stack
      // contains the elements to be lifted...
      let elements = stack.pop().asSeq()
      in Lift(elements)
      end
    end
context Mapping         
    @Operation reduceLit()
      // The literal being constructed at the head of the
      // top-stack has terminated. The buffer is replaced
      // by a literal...
      let elements = stack.top() then
          b = elements.pop() then
          str = b.toString()
      in elements.push(Literal(str))
      end
    end
context Mapping    
    @Operation reduceNewLine()
      // When new-lines are encountered in the template
      // they are retained in the code-element so that
      // indentation can be manipulated...
      self.reduceLit();
      self.index := index + 1;
      let whiteSpace = self.skipWhiteSpace()
      in stack.top().push(NewLine(margin,whiteSpace));
         self.restart();
         whiteSpace
      end
    end
context Mapping   
    @Operation restart()
      // Create a new empty literal buffer in a
      // sequence of existing code-elements...
      let elements = stack.top()
      in elements.push(Buffer(10,true))
      end
    end
context Mapping
    @Operation start()
      // Start a completely new sequence of code-elements...
      let elements = Stack()
      in stack.push(elements);
         self.restart()
      end
    end
\end{lstlisting}The machine needs to recognize when the current position in the input
corresponds to one of the tokens. a number of token predicates are
implemented. All token recognition predicates follow the same basic
implementation which is shown as follows:

\begin{lstlisting}
context Mapping
  @Operation hasPrefix(p:String):Boolean
    // Returns true when the current input starts 
    // with the token p...
    let hasPrefix = true;
        i = index
    in @While hasPrefix and not i = chars->size and (i - index) < p->size do
         hasPrefix := chars->at(i) = p->at(i - index);
         i := i + 1
       end;
       hasPrefix
    end
  end
context Mapping
  @Operation isStartLift():Boolean
    // Returns true when the next token is start-lift...
    self.hasPrefix(startLift)
  end
\end{lstlisting}The machine is implemented using two main operations: processLift
and processDrop. Each operation defines the state-machine processing
that occurs when the machine is in the corresponding state. The first
operation that is called is processLift (since the complete template
corresponds to an outermost lift):

\begin{lstlisting}
@Operation processLift()
  // Advance past the start-lift token...
  self.consumeStartLift();
  // Until we get to the end of this lift...
  @While not self.isEndLift() do
    if self.isNewLine()
    then 
      // Record the newline...
      self.reduceNewLine()
    elseif self.isStartDrop()
    then 
      // A nested drop is to be processed.
      // Any literal text is consumed...
      self.reduceLit();
      // Change state...
      self.processDrop();
      // Prepare to carry on with the lift...
      self.restart()
    else 
      // Consume a literal character...
      self.pushChar()
    end
  end;
  // Advance past the end-lift token...
  self.consumeEndLift();
  // Consume the most recent literal...
  self.reduceLit();
  // Build a lift from the tokens on the
  // stack...
  self.reduceLift()
end
\end{lstlisting}The implementation of processdrop is very similar, however it uses
different tokens:

\begin{lstlisting}
@Operation processDrop(margin,indent0,indent)
   self.consumeStartDrop();
  @While not not self.isEndDrop() do
    if self.isNewLine()
    then self.reduceNewLine()
    elseif self.isStartLift()
    then
      self.reduceLit(); 
      self.processLift();
      self.restart()
    else self.pushChar()
    end
  end;
  self.consumeEndDrop();
  self.reduceLit();
  self.reduceDrop()
end
\end{lstlisting}The code elements produced by the machine are translated into XOCL
code. As we have seen in the design, outer-most literals are simply
formatted to the output channel. Nested drops are translated to XOCL
code that evaluates an expression to produce a string and then formats
the string to the output channel. Lifts that are nested within drops
are translated to templates are re-processed using the machine. The
re-processing is now at the outermost level and therefore produces
XOCL code. 

The processing is implemented using two code element operations: desugar
and dropString. The outermost Lift instance is sent a 'desugar' message
in order to produce the XOCL code:

\begin{lstlisting}
context Lift    
  @Operation desugar(path:String,
                     lstart:String,
                     lend:String,
                     dstart:String,        
                     dend:String,
                     out:Performable,
                     nesting:Integer):Performable
    // The arguments are the name-space path to the template
    // the tokens, the output channel and the level of nesting.
    // Desugar the lifted elements. Lifting an element produces code,
    // each code element will produce some string output when it is
    // executed...
    elements->iterate(e code = [| null |] |
      [| <code>; 
         // Add 1 to the level of nesting...
         <e.desugar(path,lstart,lend,dstart,dend,out,level+1)> 
      |])
  end
\end{lstlisting}Desugaring a literal or a new-line is straightforward:

\begin{lstlisting}
context Literal
  @Operation desugar(path:String,lstart,lend,dstart,dend,out,level):Performable
    [| format(<out>,"~S",Seq{<chars.lift()>}) |]
  end

context Newline
  @Operation desugar(path:String,lstart,lend,dstart,dend,out,level):Performable
    [| format(<out>,"~%~V",Seq{<(indent - base).lift()>}) |]
  end
\end{lstlisting}Desugaring an instance of Drop involves inserving the code for the
dropped expression and then printing the resulting string to the output
channel at run-time. The literals in between the start of the drop
and the end of the drop is to be viewed as XOCL code. Therefore we
must construct a string containing the code, then turn the string
into XOCL code by parsing it. Code elements can be transformed into
strings ready for the XOCL parser using the 'dropString' operation
as called in the following operation:

\begin{lstlisting}
context Drop
  @Operation desugar(path:String,lstart,lend,dstart,dend,out,level):Performable
      // Turn the elements into code. The elements are strings that
      // are concatenated to produce XOCL code that is parsed using 
      // the OCL grammar...
      let str = elements->iterate(e s = "" | 
                  s + e.dropString(path,out,lstart,lend,dstart,dend,0)) then
          code = OCL::OCL.grammar.parseString(str,"Exp1",Seq{XOCL})
      in [| let s = <code>
            in if s.isReallyKindOf(XCore::String)
               then format(<out>,s)
               end
            end
         |]
      end
    end
\end{lstlisting}Now we need to implement 'dropString' for each of the code element
classes. The result from 'dropString' must be a string that corresponds
to the XOCL code for the element. Literals and Newlines are straightforward:

\begin{lstlisting}
context Literal
  @Operation dropString(path:String,out,lstart,lend,dstart,dend,level):String
    // Careful to replace outermost ocurrences of e_nd in user code with
    // end...
    if level = 0
    then chars.subst("end","e_nd",true)
    else chars
    end
  end
context Newline
  @Operation dropString(path:String,out,lstart,lend,dstart,dend,level):String       
    "\n" + formats("~V",Seq{indent})     
  end
\end{lstlisting}A Lift is dropped by transforming it into a template construct:

\begin{lstlisting}
context Lift
  @Operation dropString(path:String,out,lstart,lend,dstart,dend,,level):String
    // Dropping a lift with respect to the level of nesting. If the
    // level is 0 then the lift and drop cancel out and we produce the
    // code as a string (ready to splice into surrounding code)...
    if level = 0  
    then
      "@"+path+"(" + out.pprint() + "," + margin + ") " + 
         elements->iterate(e s = "" | 
           s + e.dropString(path,out,lstart,lend,dstart,dend,level+1)) + 
      " end"
    else 
      lstart + 
        elements->iterate(e s = "" | 
          s + e.dropString(path,out,lstart,lend,dstart,dend,level+1)) + 
      lend
    end
\end{lstlisting}Finally, a Drop generates a drop string by wrapping itself with drop
start- and end-tokens:

\begin{lstlisting}
context Drop
  @Operation dropString(path,out,lstart,lend,dstart,dend,level):String
    
    // To create a string from a drop. Wrap it with the 
    // drop-start and drop-end tokens...
      
    dstart + elements->iterate(e s = "" | 
      s + e.dropString(path,out,lstart,lend,dstart,dend,level)) + 
    dend
  end
\end{lstlisting}

\chapter{A Problem of Composition}

Pattern matching is a useful technique when manipulating data. This
chapter describes how to implement pattern directed rules that transform
data and how to address a problem that arises with this approach.
The implementation also shows how to use operations to delay expressions
in rules and how to get the variables bound by the pattern matching
into the delayed expressions.

When mapping models to code (and other structured targets) it would
be nice of the structure of the target matched the structure of the
source. When this happens, individual elements in the source model
can be transformed into elements of the target model, and the resulting
target elements are then just grouped together to form the result.
For example, when translating UML model classes with attributes to
Java class definitions, each UML class turns into a Java class definition,
each attribute in the UML class produces a corresponding attribute
in the Java class. The structure is the same, but the detail of representation
is different.

Unfortunately this situation does not often occur. For example, UML
models consist of separate elements for classes, associations and
state machines; each of these model elements are distinct, but they
reference each other. A translation from UML to Java may choose to
produce a single class definition consisting of:

\begin{itemize}
\item Java classes for UML classes;
\item Java fields for UML class attributes and association role ends;
\item A state field.
\item An enumeration in each Java class for each UML message.
\item A single Java method in a class for handling messages;
\item A case for each transition within the message handler.
\end{itemize}
The information in the Java program is nested within the Java class.
The corresponding information in the UML model is spread out. The
structures are not equivalent.

It is certainly possible to pass tables around the source model and
build up the required parts of the target model. However, it is possible
to do better than that and to use a language driven approach to hide
away the details of how the target model is built up. The transformation
language constructs can be pattern directed and the output can be
declarative in such a way as to focus on mapping source elements to
target elements without having to say too much about the mechanisms
by which the output is produced.

This chapter is about the design of a language for producing code
for models that solves the composition problem. The structure of this
chapter is as follows: section \ref{sec:Translating-Executable-Models}
describes a simple executable model and its translation to Java; section
\ref{sec:Transformation-Architecture} describes the architecture
of the solution; section \ref{sec:A-Rule-Based-Language} describes
a language for rule based model transformation and shows how it can
be used to implement the transformation rules for the example; section
\ref{sec:Rules-Implementation} provides the implementation of the
language constructs.

\section{Translating Executable Models\label{sec:Translating-Executable-Models}}

This section describes a typical scenario where the translation from
models to program code requires the target code to pull together information
from different parts of the source model. A fully executable program
is produced by translating class models and state machines to Java.
The example is a car cruise controller and is very simple, but is
representative of a large number of strategies for producing programs
from models.

\begin{figure}
\begin{center}

\includegraphics[width=12cm]{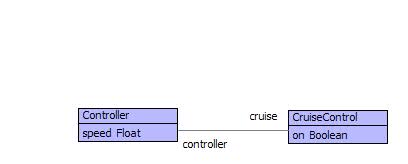}

\caption{Classes for a simple cruise control in a car engine system. Typical
of a system that is to be controlled via a state machine.\label{fig:Cruise-Control-Model}}

\end{center}
\end{figure}

Figure \ref{fig:Cruise-Control-Model} shows a simple model consisting
of classes, attributes and associations. A car engine controller manages
the speed of the car and communicates with the cruise control system.
The cruise control system is managed via a mechanical switch on the
dashboard.

\begin{figure}
\begin{center}

\includegraphics[width=12cm]{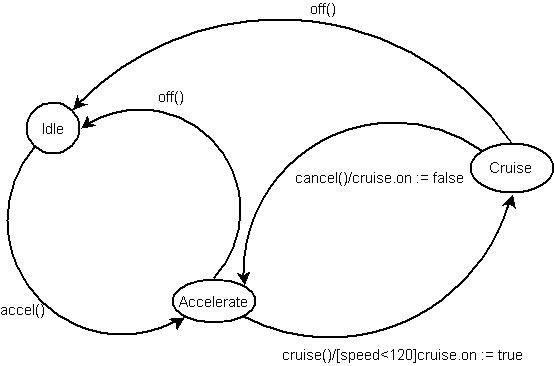}

\caption{A state machine describing the behaviour of the Controller class for
the car cruise control system. The controller exists in one of the
states (labelled ellipses). Transitions between states occur when
the controller receives the messages. Guards in {[} and ] determine
whether the transition is legal. Actions on the transitions describe
what happens when the transition takes place.\label{fig:Cruise Control}}

\end{center}
\end{figure}

Figure \ref{fig:Cruise Control} shows a state machine that defines
the behaviour of the engine controller. The engine starts in the Idle
state when it is switched on. When the controller detects the accelerator
being pressed, it switches to the Accelerate state. When the car is
accelerating, the driver can press the cruise control button on the
dashboard causing the controller to switch to the Cruise state providing
that the speed is less than a preset amount (120). In the cruising
state, the driver can switch the control off by pressing the accelerator.
When switching to and from the cruising state, the engine controller
communicates with the cruise controller, setting its 'on' state.

Given the behavioural models defined in figures \ref{fig:Cruise-Control-Model}
and \ref{fig:Cruise Control}, it is possible to produce a fully executable
Java program. The program consists of class definitions for the two
classes in the model. Each class includes fields, accessors, updaters
and a message handling operation; the definition of these components
comes from different aspects of the source model. 

The rest of this section shows the program code generated from the
Controller class. The example is representative of the code generated
from any class in a behavioural model. Firstly, each model class produces
a Java class definition (the missing definitions are given subsequently):

\begin{lstlisting}
class Controller {
  // Messages...
  // Fields...
  // Accessors...
  // Updaters...
  // Message passing...
}
\end{lstlisting}The state machine for a class defines a collection of messages on
the transitions. Each of the messages has a unique identifier that
is defined as a collection of constants in the class:

\begin{lstlisting}
  // Messages...
  public static final int accel  = 0;
  public static final int cruise = 1;
  public static final int off    = 2;
  public static final int cancel = 3;
\end{lstlisting}The state of an object is given by field definitions that come from
different aspects of the behavioural model:

\begin{itemize}
\item Each model class defines a collection of attributes with simple types
(such as String or Integer); for example the engine controller class
defines an attribute 'speed' of type 'Float'. 
\item The model requires that each object exists in one of a given number
of states; for example the engine controller has states for Idle,
Cruise and Accelerate.
\item The model contains a number of associations between classes. The associations
allow instances of the classes to communicate with each other. The
engine controller class has an association with the cruise controller
class with role end names cruise and controller.
\end{itemize}
The Java field definitions for the engine controller are given below:

\begin{lstlisting}
  // Fields...
  Float speed;          // Simple attribute.
  String state;         // Current state.
  CruiseControl cruise; // Association role end.
\end{lstlisting}Each of the Java fields given above are private to the class definition.
To provide access to the state of a Java instance, the following accessors
and updaters are defined:

\begin{lstlisting}
  // Accessors and updaters...
  public CruiseControl getcruise() { return cruise; }
  public Float getspeed() { return speed; }
  public void setcruise(CruiseControl cruise) { 
    this.cruise = cruise; 
  }
  public void setspeed(Float speed) { 
    this.speed = speed; 
  }
\end{lstlisting}State machines define the behaviour of objects in response to receiving
messages. This can be defined in Java using a method in each class
that handles messages:

\begin{lstlisting}
public void send(int message,Object[] args) {
  switch(message) {
    // Transitions...
  }
}
\end{lstlisting}Each transition consists of source and target states s and t, the
transition occurs when a message m is received and when a guard g
is true. When the transition fires, an action a is performed and the
receiver changes to the target state t. Each message is implemented
as a case in the Java switch statement:

\begin{lstlisting}
    case m:
      // Transition...
      if(state == s && g) {
        a;
        state = t;
      }
      break;
\end{lstlisting}
The state machine for the engine controller is shown below:

\begin{lstlisting}
  // Message passing...
  public void send(int message,Object[] args) {
    switch(message) {
      case Controller.accel:
        if(state.equals("Idle"))
          state = "Accelerate";
        break;
      case Controller.cruise:
        if(state.equals("Accelerate") && speed < 120)
          cruise.seton(true);
          state = "Cruise";
        break;
      case Controller.off:
        if(state.equals("Accelerate"))
          state = "Idle";
        break;
      case Controller.cancel:
        if(state.equals("Cruise"))
          cruise.seton(false);
          state = "Accelerate";
        break;
      case Controller.off:
        if(state.equals("Cruise"))
          state = "Idle";
        break;
      default: throw new Error("No message " + message);
    }
\end{lstlisting}
\section{Transformation Architecture\label{sec:Transformation-Architecture}}

Before looking at a language that supports the generation of code
for models, it is worth considering the steps taken by a mechanical
process. The idea is that the mechanical process is driven by code
generation rules and that the rules encode any domain specific information
about the models being transformed. The code generation rules are
pattern directed and follow the structure of the source model:

\begin{lstlisting}
for each class c
  emit a Java class definition.
  emit a state field.
  for each attribute a of c
    emit a field definition for a.
    emit an access and updater for a.
  emit a message handling method.
for each association A
  emit a field for end1
  emit a field for end2
for each state machine m for class c
  for each transition t of m
    emit a constant field to c for message(t)
    emit a message case for t to c
\end{lstlisting}The transformation defined above follows the structure of the source
model. However, the code that is produced (or \emph{emitted}) by the
transformation rules does not follow the same structure; for example
fields arising from associations must be added to class definitions
that are produced by other parts of the transformation. It is attractive
to have the transformation follow the structure of the source model
but some technology must be employed to tie up the various target
components. An effective way to do this is to use labels: emitted
code is labelled and may contain labels, once the labelled target
components are produced a subsequent phase is used to resolve label
references.

\begin{figure}
\begin{center}

\includegraphics[width=12cm]{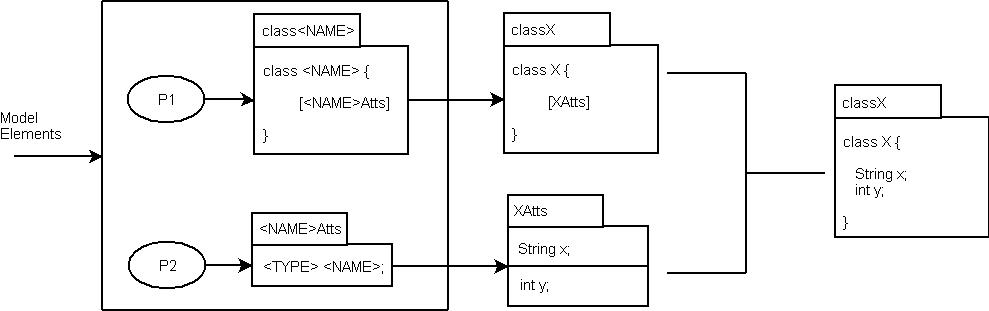}

\caption{An overview of rule firing. Elements match against patterns, document
templates are produced. The templates contain label references. A
resolution step replaces label references with the documents that
they refer to.\label{fig:Mapping-Cycle}}

\end{center}
\end{figure}

An overview of the architecture of the transformation process shown
in figure \ref{fig:Mapping-Cycle}. Model elements are supplied to
the rules in the rule base. Each rule consists of a source pattern
and a target pattern. Two rules are shown:

\begin{enumerate}
\item R1: Pattern P1 matches a class in the model. The output pattern describes
a Java class labelled with the class name.
\item R2: Pattern P2 matches an attribute in the model. The output pattern
describes a Java field labelled with the name of the owning class.
\end{enumerate}
When model elements are supplied to a rule-base consisting of a collection
of rules, each rule is tried in turn. If the input elements match
the patterns then the rule is \emph{enabled}. Patterns contain variable
names and an enabled rule has an \emph{environment} of variable bindings.
The first enabled rule is \emph{fired}. Firing a rule produces a \emph{document
template} by replacing all occurrences of pattern variables in the
output pattern. Turning an output pattern into a document template
is referred to as \emph{forcing} the output pattern with respect to
the environment. 

Figure \ref{fig:Mapping-Cycle} shows three document templates produced
from the rule-base. The first is labelled with classX and is the result
of mapping a model class named X to a Java class. The second and third
are both labelled with XAtts and are the result of mapping two class
attributes named x and y to Java fields. The classX template contains
a reference to a label XAtts that is a placeholder for the fields
of the class. 

A document template is transformed into a document by replacing all
label references with documents. This process is called \emph{displaying}
the document template. The display process is shown as the final step
in figure \ref{fig:Mapping-Cycle} where the occurrence of XAtts in
the class definition is replaced with all the field definitions labelled
XAtts.

Rule firing may involve multiple passes over the data; the output
pattern of each rule may refer to other rule-bases (including its
own) and supply data to those rule-bases to produce the final document
templates. For example, the rule R1 is supplied with a class, emits
a document template for the class but also calls the rule-base again
with the attributes of the class so that rule R2 fires on each attribute.
All rule firing is completed before document display takes place.

\section{A Rule Based Language\label{sec:A-Rule-Based-Language}}

The transformation from a model to code is performed by a rule-base
containing a collection of rules. The rule base is supplied with one
or more model elements as input. It tries the rules in turn until
one of the rules matches the input. The body of the matching rule
is forced to produce a document template. The template may be labelled
and may contain labelled placeholders. The rule-base may be performed
multiple times against different model elements. Once execution is
complete, the templates are displayed by matching placeholder labels
against labelled output. The result is an output document with no
unresolved labels that can be displayed as text. This section describes
the features of the rule-base language and concludes with the complete
behavioural modelling rule-base as described in the previous section.

A rule-base with the name N has the following format:

\begin{lstlisting}
@RuleBase N
  // Rules...
end
\end{lstlisting}A rule named R has the following format:

\begin{lstlisting}
@Rule R P1,P2,...,Pn -> 
  D1 D2 ... Dm 
end
\end{lstlisting}where each Pi is an element pattern and each Di is a document pattern.
The idea of a rule is that it must be supplied with n inputs. If each
of the n inputs match the patterns then the rule is \emph{enabled}.
An enabled rule is \emph{fired} by forcing the output defined by each
Di in turn. The result produced by the rule firing is the last document
Dm. Each pattern Pi may contain variables that match against the corresponding
input. If the rule is enabled then the collection of variables and
values provides an environment for the rule firing.

Here is a simple example of a rule:

\begin{lstlisting}
@Rule Anything x ->
  <x>
end
\end{lstlisting}The pattern 'x' matches anything as input. A document < exp > performs
the expression exp in the supplied rule-firing environment. The result
of x is either a document or an element that is transformed into a
string. In the case of Anything, the rule will produce the stringified
version of x as a document.

A pattern may match an object:

\begin{lstlisting}
@Rule ToJava Class[name=n] ->
  "public class " + <n> + "{"
end
\end{lstlisting}The pattern in the rule ToJava matches instances of Class and matches
the name of the class with the variable 'n'. Two documents are composed
using the '+' operator. Supplying the class Element to this rule (firing
and displaying the result) produces:

\begin{lstlisting}
public class Element {
\end{lstlisting}Indentation in a document is produced using the matching directives
->{[} and ]:

\begin{lstlisting}
@Rule ToJava Class[name=n] ->
  "public class " + <n> + "{" +
    ->[ nl +
        "public String state;"
    ] + nl +
   "}"
end
\end{lstlisting}Supplying Element to the above rule produces:

\begin{lstlisting}
public class Element {
  public String state;
}
\end{lstlisting}The nl directive produces a newline and tabs to the current level
of indentation. Collections are processed using the \{ and \} directives:

\begin{lstlisting}
@Rule ToJava Class[name=n,attributes=A] ->
  "public class " + <n> + "{" +
    ->[ nl +
        "public String state;" + nl
        { <A> <map> nl empty }
    ] + nl +
  "}"
end
\end{lstlisting}The \{ and \} contain four components \{ S M C D \} where: S defines
the collection; M defines a mapping that is applied to each element
of S in turn; C is a document combinator and D is a default document.
The easiest way to understand this construct is via an example. Given
a collection S = \{x,y,z\} then \{ S M C D\} produces C(M(x),C(M(y),M(z))).
Given a collection S = \{\} then the default D is produced.

The collection component S may be a label or an expression. In the
example above <A> is an expression where A is provided by the context
of the rule firing. The mapping component M is an expression. The
variable map always refers to the containing rule-base and therefore
allows the rule base to be applied as part of a rule firing. The combiner
nl causes the documents to be combined by joining them together with
newlines. The document empty is just that. Note that nl and empty
are builtin so we don't put < and > round them.

Suppose that C is a class with two attributes x and y of type String
and Integer respectively. Using the above definition to transform
C produces the following (assuming suitable rule definitions for attribute
to field transformation):

\begin{lstlisting}
public class C {
  public String state;
  String x;
  int y;
}
\end{lstlisting}The ToJava rule above is incomplete since there is no rule for mapping
attributes, here they are:

\begin{lstlisting}
@Rule MapStrAtt 
  Attribute[name=n,type=NamedElement[name="String"]] -> 
    "String " + <n> + ";" 
end
@Rule MapIntAtt 
  Attribute[name=n,type=NamedElement[name="Integer"]] -> 
    "int " + <n> + ";" 
end
// More cases...
\end{lstlisting}Labels can be used either by themselves in documents or as the source
of collection templates (as in <A> above). Documents can be tagged
with labels using 'emit'. Here are the class and attribute rules re-written
to use labels:

\begin{lstlisting}
@Rule ToJava Class[name=n,attributes=A] ->
  "public class " + <n> + "{" +
    ->[ nl +
        "public String state;" + nl
        { [n + "Atts"] <@Operation(a) map(a,n) end> nl empty }
    ] + nl +
  "}"
end
@Rule MapStrAtt 
  Attribute[name=n,type=NamedElement[name="String"]],
  className -> 
    emit[className + "Atts"]
      "String " + <n> + ";" 
end
\end{lstlisting}Notice that the mapping components of the collection expression is
modified to become an operation that supplies two elements to the
mapping (a,n). Each attribute mapping rule has two inputs: the attribute
and the name of the class. The attribute rule emits (and returns)
a Java field definition. When a document is emitted against a label
(in this case className + {}``Atts''), it is added to the collection
of documents for that label. When the resolution phase is performed,
the collection expression will use all of the documents registered
against the label.

This concludes the overview of the rule language for transforming
model elements. The key features of element patterns are: constants;
variables; object patterns with slots. The key features of the document
patterns are: literal strings; delayed expressions in < ... >; combination
with +; indentation and newlines with ->{[} ... ] and nl; labels with
{[} ... ]; tagged documents using 'emit'; combining collections with
\{ ...\}.

Finally, the rules for the Java mapping are defined below. There are
two rule bases. The first is used to map model types to Java types
and the second is used to map packages, classes, attributes and state
machines to Java classes. The first rule base is defined in its entirety
below and the second is defined on a rule-by-rule basis.

Model types are named elements. The names must be mapped to the appropriate
Java types. To simplify the example, sets and sequences are translated
to vectors:

\begin{lstlisting}
@RuleBase Types
  @Rule String NamedElement[name='String'] ->
    "String"
  end
  @Rule Integer NamedElement[name='Integer'] ->
    "int"
  end
  @Rule Integer NamedElement[name='Boolean'] ->
    "boolean"
  end
  @Rule Sequence Seq[elementType=t] ->
    "Vector<" + <map(t)> + ">"
  end
  @Rule Set Set[elementType=t] ->
    "Vector<" + <map(t)> + ">"
  end
  @Rule Default NamedElement[name=n] -> 
    <n>
  end
\end{lstlisting}A package is a collection of classes, associations and sub-packages.
Each package gives rise to a document labelled with the name of the
package. The contents of the document are produced by mapping the
contents:

\begin{lstlisting}
@Rule TranslatePackage 
  Package[name=n,classes=C,associations=A,packages=P] ->
    emit["Package-" + n] 
      { <C> <map> nl empty } +
      { <A> <map> nl empty } +
      { <P> <map> nl empty }
end
\end{lstlisting}A class is transformed into a Java class definition. Some of the components
of the Java class can be produced directly from the model, other components
are produced from model elements that originate elsewhere. The following
rule uses labels as placeholders for the program code that is generated
elsewhere. The comments in the rule describe each of the major components:

\begin{lstlisting}
@Rule TranslateClass 
 Class[name=n,attributes=A] ->
  // Emit a labelled class definition...
  emit["Class-" + n]
   "class " + <n> + " {" + 
   // Indent the body of the definition...
   ->[ nl +
     // Transitions define messages, 
     // each message defines a constant...
     { [n + "transitions"] id nl empty } + nl +
     // Each attribute defines a simple-typed Java field...
     { <A> < @Operation(a) map(a,n) end> nl empty } + nl +
     // Each instance has its own state...
     "String state;" + nl +
     // Attributes are defined by associations...
     { [n + "attributes"] id nl empty } + nl +
     // Accessors and updaters are defined elsewhere...
     { [n + "accessors"] id nl empty } + nl +
     { [n + "updaters"] id nl empty } + nl +
     // Each class has a message handler...
     "public void send(int message,Object[] args) {" +
     // Indent the body of the method...
     ->[ nl + 
       "switch(message) {" +
       // Indent the body of the switch
       ->[ nl +
         // Each transition produces a message case...
         { [n + "messages"] id nl empty } + nl +
         // In case the message is not handled...
         "default: throw new Error(\"No message \" + message);" 
       ] + nl + 
       "}"
     ]
   ] + nl +
  "}"
end
\end{lstlisting}Each attribute defined by a class in the model produces an accessor,
an updater and a Java field definition. The methods are labelled so
that all accessors and updaters for the class are defined in the same
place in the output. The field definition is returned:

\begin{lstlisting}
@Rule TranslateAttribute Attribute[name=n,type=t],c ->
  // Produce an accessor method for this field...
  emit[c + "accessors"]
    <Types.apply(t)> + " get" + <n> + "()" +
    "{ return " + <n> + "; }"
  // Produce an updater method for this field...
  emit[c + "updaters"]
    "void set" + <n> + "(" + <Types.apply(t)> + " " + <n> + ")" +
    "{ this." + <n> + " = " + <n> + "; }"
  // Produce the field definition...
  <Types.apply(t)> + " " + <n> + ";"
end
\end{lstlisting}An association consists of two ends; each end has a name and is attached
to a type. An association produces a pair of field definitions in
the Java classes and also adds accessors and updaters. The transformation
is performed by the following two rules:

\begin{lstlisting}
@Rule TranslateAssociation Association[end1=e1,end2=e2] ->
  // Produce definitions for the class at end1...
  <map(e1,e2)>
  // Produce definitions for the class at end2...
  <map(e2,e1)>
  // Just for side effect...
  empty
end

@Rule TranslateEnd 
  End[name=n1,type=t1],End[name=n2,type=t2] ->
  // produce a field definition for the class t1...
  emit[t1.name() + "attributes"]
    <Types.apply(t2)> + " " + <n2> + ";"
  // Produce an accessor method for t1...
  emit[t1.name() + "accessors"]
    "public " + <Types.apply(t2)> + " get" + <n2> + "() { " +
    ->[ nl + 
      "return " + <n2> + ";" + nl  
    ] + nl + "}"
  // produce an updater for t1...
  emit[t1.name() + "updaters"]
    "public void set" + <n2> + 
      "(" + <Types.apply(t2)> + " " + <n2> + ") { " +
    ->[ nl + 
      "this." + <n2> + " = " + <n2> + ";" + nl  
    ] + nl + "}"
end
\end{lstlisting}A state machine belongs to a class and defines states and transitions.
The state machine is implemented in Java as an enumerated type for
the messages and as a message handling method. Each transition defines
what happens when an instance of the class receives a message. A message
is defined as an enumerated type implemented as a bunch of constants
in the class. A transition is implemented as a case in the message
handling method. The rest of the rules define the transformation of
a state machine to its implementation.

\begin{lstlisting}
@Rule TranslateMachine StateMachine[class=c,states=S,trans=T] ->
  // Translate the messages to constants. Use ->asSet
  // to remove duplicate message names...
  <map(T.message->asSet->asSeq,T.message->asSet->asSeq,c)>
  // translate the transitions to message handling cases...
  <map(T,c)>
end
\end{lstlisting}The following rules produce constant definitions for the messages
used by the state machine:

\begin{lstlisting}
@Rule TranslateMessages allMessages,Seq{message|messages},class ->
  // Produce a constant (use the position 
  // of the name as its value)...
  <map(message,class,allMessages->indexOf(message))>
  // Map the rest of the messages...
  <map(allMessages,messages,class)>
end
@Rule NoMessagesLeft allMessages,Seq{},class ->
  // No messages, requires no constants...
  empty
end
@Rule TranslateMessage message,Class[name=className],index ->
  // Produce a constant with value index...
  emit[className + "transitions"]
    "public static int " + <message> + " = " + <index> + ";"
end
\end{lstlisting}Finally, each transition produces a case entry in the message handling
method of the class. The case checks that the receiver is in the correct
state and that the guard is true. If so then the action is performed
and the receiver changes to the target state:

\begin{lstlisting}
@Rule TranslateTrans Seq{
  Trans[source=s,        // Source state name
        target=t,        // Target state name
        message=m,       // Message name
        condition=p,     // Condition (Java code)
        action=a         // Action (Java code)
       ] | T},           // T is more transitions
  c ->                   // Class that owns the transition
  // Produce a case definition for the message handling
  // method for class c...
  emit[c.name() + "messages"]
   // This case handles the message m...
   "case " + <c.name()> + "." + <m> + ":" + 
   ->[ nl + 
     // check we are in the appropriate state s and
     // the predicate p is true...
     "if(state.equals(" + <str(s)> + ") && " + <p> + ")" + 
     ->[ nl +
       // Perform the action...
       <a> + ";" + nl +
       // Change to the target state...
       "state = " + <str(t)> + ";"
     // Break out of the switch...
     ] + nl + "break;"
   ]
  // Translate the rest of the transitions...
  <map(T,c)>
end

@Rule TranslateNoTrans Seq{},c ->
  // The base case...
  empty
end
\end{lstlisting}
\section{Implementation\label{sec:Rules-Implementation}}

The previous section has defined the syntax of a language for pattern
directed document generation rules. The language addresses the code
generation composition problem whereby the structure of the input
model elements does not match the structure of the output code. Labels
are used to tag the generated code (or \emph{document} \emph{templates})
and to define placeholders that are then resolved in a subsequent
phase. 

This section defines the rule-based document generation language.
There are several parts to the definition: the language of patterns,
the language of documents and the label resolution mechanism. Each
of these are explained in turn.

\subsection{Pattern Matching}

\begin{figure}
\begin{center}

\includegraphics[width=12cm]{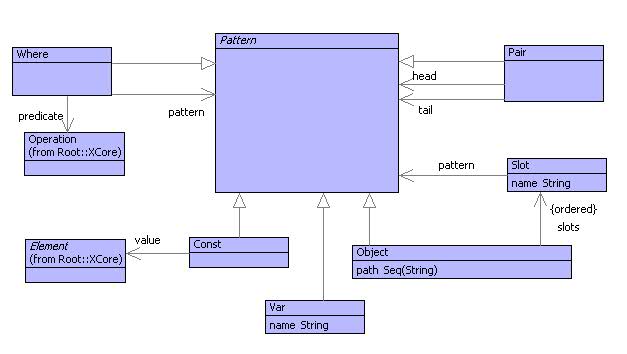}

\caption{Patterns\label{fig:Patterns}}

\end{center}
\end{figure}

Figure \ref{fig:Patterns} shows the definition of a pattern language.
Patterns occur frequently in language definitions where it is useful
to extract elements from data depending on the structure of the data.
When this is a requirement is it almost always cost effective to define
a pattern language and a pattern matching mechanism that it is to
write the corresponding program code that extracts the elements each
time they are required. For example, the pattern:

\begin{lstlisting}
  C[x=10,y=v]
\end{lstlisting}is equivalent to the code:

\begin{lstlisting}
if element.isKindOf(C)
then
  if element.x = 10
  then // bind "v" to 10
  else // fail match
  end
else // fail match
end
\end{lstlisting}Given that patterns can involve multiple elements and can be nested,
the saving in terms of code is quite significant. Furthermore, modelling
the pattern language provides other benefits. The language becomes
circumscribed and its processing can be very effectively controlled.
For example error handling can be reported in terms of the original
patterns and not the implementation of the patterns. Optimizations
for pattern matching can be universally and retrospectively applied.
The patterns can be translated to other implementation platforms,
for example by exporting to a programming language.

The simplified grammar for patterns is shown below:

\begin{lstlisting}
@Grammar
  Constant ::= 
    s = Str { Const(s) } 
  | i = Int { Const(i) }  
  | 'true' { Const(true) } 
  | 'false' { Const(false) }.
  EmptySeq ::= '}' { Const(Seq{}) }.
  HeadTail ::= h = Pattern '|' t = Pattern '}' { Pair(h,t) }.
  Pair ::= 'Seq{' (HeadTail | EmptySeq).
  Path ::= p = ('::' Name)* { p }.
  Pattern ::= 
    Constant 
  | n = Name NameTail^(n) 
  | Pair.
  NameTail(n) ::= 
    p = Path '[' s = Slots ']' { Object(Seq{n|p},s) } 
  | { Var(n) }.
  Slots ::= 
    s = Slot ss = (',' Slot)* { Seq{s | ss} } 
  | { Seq{} }.
  Slot ::= n = Name '=' p = Pattern { Slot(n,p) }.
end 
\end{lstlisting}Given an element and a pattern, matching is implemented as a mechanism
that constructs an environment of variable bindings. Each binding
associates a variable with a value such that substituting each variable
for its value in the pattern produces the original element. For example,
given the pattern:

\begin{lstlisting}
  C[x=10,y=v]
\end{lstlisting}and an instance o of C such that o.x = 10 and o.y = 20 then the environment
associating y with 20 allows the pattern to be equivalent to the element
o. The mechanism is implemented by defining a match operation for
each of the pattern classes. Each operation is called match and expects
two arguments: the value to be matched and the current variable binding
environment. The result of the match is either the variable binding
environment or null, if the match fails. For constants:

\begin{lstlisting}
context Const
  @Operation match(value,env)
    if value = self.value
    then env
    else null
    end
  end
\end{lstlisting}A variable pattern matches anything:

\begin{lstlisting}
context Var
  @Operation match(value,env)
    env->bind(name,value)
  end
\end{lstlisting}A pair pattern must match a sequence (and not an empty sequence since
that is treated as a constant) such that the head and tail of the
pattern matches the corresponding elements of the value:

\begin{lstlisting}
context Pair
  @Operation match(value,env)
    if value.isReallyKindOf(Seq(Element)) and 
       not value = Seq{}
    then 
      env := head.match(value->head,env);
      if env <> null
      then tail.match(value->tail,env)
      else env
      end
    else null
    end
  end
\end{lstlisting}An object pattern matches an instance of the appropriate class where
all the slot patterns match the appropriate slot values:

\begin{lstlisting}
context Object
  @Operation match(value,env)
    if value.isReallyKindOf(self.classifier())
    then 
      @For slot in slots do
        if env <> null
        then env := slot.match(value,env)
        end
      end;
      env
    else null
    end
  end
\end{lstlisting}Each slot matches the appropriate slot value:

\begin{lstlisting}
context Slot
  @Operation match(object,env)
    if object.hasSlot(name)
    then pattern.match(object.get(name),env)
    else null
    end
  end
\end{lstlisting}
\subsection{Documents}

\begin{figure}
\begin{center}

\includegraphics[width=12cm]{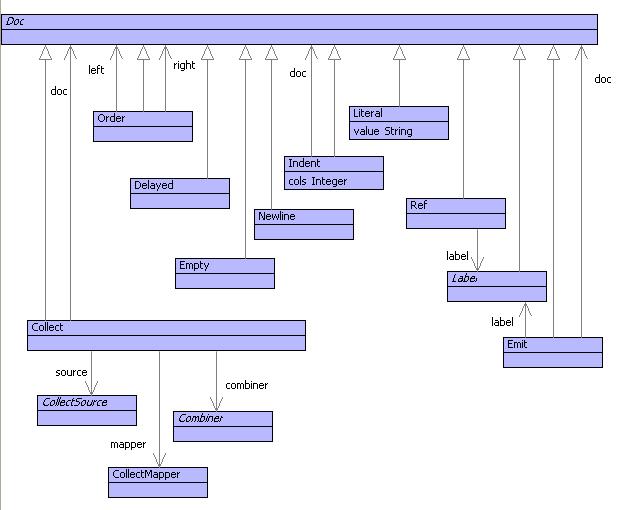}

\caption{Document Model\label{fig:Document-Model}}

\end{center}
\end{figure}

Once patterns have matched a collection of elements, document templates
are produced in the context of the matched environment. The document
model is shown in figure \ref{fig:Document-Model}. The model is used
in three ways: to represent document patterns in rules; to represent
document templates; to represent document displays. The difference
between these three uses is: patterns may contain delayed expressions;
templates may contain unresolved labels; document displays contain
no unresolved labels and no delayed expressions.

The classes in the model match the constructs in the language examples
from the previous section. The following classes are worth noting:
the source, mapper and combiner of a collect are classes that are
specialized in specific ways for different kinds of builtin collection
operations; Label is a class that is either specialized as a delayed
label or a literal label.

Delayed document components are used to allow arbitrary expressions
to be embedded. The expressions are to be evaluated when the document
is created in the context of the environment produced by successful
pattern matching. When the delayed document is created, the expression
is turned into an operation whose arguments correspond to the variable
names used in the expression. Delayed components include: source,
mapper and combiner of a collect; a delayed document; and a delayed
label.

The grammar for documents is shown below:

\begin{lstlisting}
@Grammar extends OCL::OCL.grammar
  Doc(FV) ::= d = AtomicDoc^(FV) DocTail^(FV,d).
  AtomicDoc(FV) ::= 
    Build^(FV) 
  | Delayed^(FV) 
  | Emit^(FV) 
  | Empty 
  | Newline 
  | Ref^(FV) 
  | Literal 
  | Indent^(FV) 
  | Label^(FV).
  DocTail(FV,d1) ::= 
    '+' d2 = Doc^(FV) { Order(d1,d2) } 
  | { d1 }.
  Build(FV) ::= 
    '{' s = Source^(FV) 
        m = Map^(FV) 
        c = Combiner d = Doc^(FV) 
    '}' {
       Collect(s,m,c,d) 
  }.
  Source(FV) ::= 
    LabelSource^(FV) 
  | DelayedSource^(FV).
  LabelSource(FV) ::= '[' e = Exp ']' op = DocOp^(FV,e) { 
    DelayedLabelSource(op) 
  }.
  Map(FV) ::= 
    '<' e = DelayedExp '>' op = DocOp^(FV,e) { 
       DelayedMapper(op) } 
  | 'id' { IdMap() }.
  DelayedSource(FV) ::= '<' e = Exp '>' op = DocOp^(FV,e) { 
    DelayedSource(op) 
  }.
  Combiner ::= 
   'nl' { CombineWithNewlines() } 
  | 'ignore' { Ignore() }.
  Label(FV) ::= '[' e = Exp ']' op = DocOp^(FV,e) { 
    DelayedLabel(op) 
  }.
  Delayed(FV) ::= '<' e = Exp '>' op = DocOp^(FV,e) { 
    Delayed(op)  
  }.
  Emit(FV) ::= 'emit' l = Label^(FV) d = Doc^(FV) { 
    Emit(l,d) 
  }.
  Empty ::= 'empty' { Empty() }.
  Newline ::= 'nl' { Newline() }.
  Ref(FV) ::= '!' l = Label^(FV) { Ref(l) }.
  Literal ::= s = Str { Literal(s) }.
  Indent(FV) ::= '->' '[' d = Doc^(FV) ']' { 
    Indent(2,d) 
  }.
  DocOp(FV,e) ::= { DocOp(FV,e) }.
end 
\end{lstlisting}The document grammar extends the OCL grammar in order to refer to
the Exp grammar rule for delayed document components. Many of the
grammar clauses for Doc are parameterized with respect to a set of
free variables. This feature allows delayed expressions to be transformed
into operations where the arguments of the operation are the intersection
of the free variables referenced in the expression and those bound
in the current pattern environment. For example, the rule:

\begin{lstlisting}
@Rule Class[name=n] -> "class " + <toUpper(n)> end
\end{lstlisting}contains a delayed expression: toUpper(n). The delayed expression
contains two free variables: toUpper and n. However, when the rule
fires, only one of the free variables (n) will be bound in the pattern
matching environment. The delayed expression is translated by the
parser into an operation:

\begin{lstlisting}
@Operation(n) toUpper(n) end
\end{lstlisting}where the arguments (n) are calculated as the intersection of the
free variables in the body \{toUpper,n\} and the variables bound by
the pattern \{n\}. Patterns implement an operation FV that calculates
the set of variables bound by the pattern; this set is supplied to
the Doc clauses. The set FV is ultimately used in the clause DocOp
that creates an instance of the class DocOp. A DocOp is just like
a normal operation except that its arguments are calculated by taking
the intersection of the FV and and free variables of its body.

The rest of the Doc grammar should be self explanatory. Note that
some classes are used that are not shown in figure \ref{fig:Document-Model}.
These are sub-classes of Label and collection classes; they capture
various special cases and associate them with special purpose syntax.

\subsection{Rules and Rule Bases}

\begin{figure}
\begin{center}

\includegraphics[width=12cm]{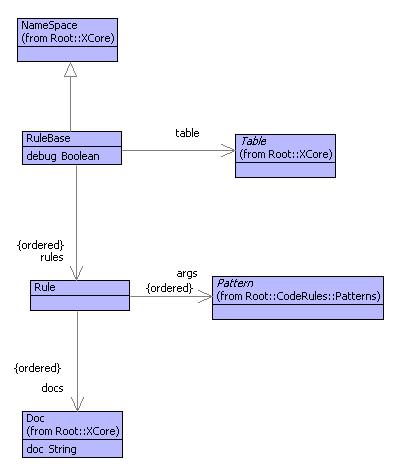}

\caption{Rule Model\label{fig:Rule-Model}}

\end{center}
\end{figure}

Figure \ref{fig:Rule-Model} shows the model of rules and rule bases.
A rule consists of a sequence of patterns and a sequence of documents.
A rule base is a sequence of rules that will be tried in turn when
the rule base is supplied with elements. A rule base has a table that
is used to hold associations between labels and documents when rules
are fired. On the resolution pass, labels are replaced with their
documents from the table. The table also allows particular documents
to be indexed by their name; for example only part of a document in
a large model can be re-generated.

The rest of this section describes how rules are fired and then the
resulting document is produced by resolving labels. Argument elements
are supplied to a rule base via the operation apply defined below. 

Each rule is applied to the args in turn until one is enabled by returning
a non-null environment. When this occurs the rule is fired, supplying
the environment to the rule. Notice that the environment is extended
with a binding for 'map' which allows the document to call the rule-base
again:

\begin{lstlisting}
context RuleBase
  @Operation apply(args)
    let done = false;
        result = null
    in @For rule in R when not done do
         let env = rule.match(args,Seq{})
         in if env <> null
            then
              let env = env->bind("map",
                   @Operation(.args)
                     self.apply(args)
                   end)
              in done := true; 
                 result := rule.fire(env,table)
              end
            end 
         end 
       end;
       if not done
       then self.error(formats("No rule for (~{,~;~S~}~%",Seq{args}))
       else result
       end
    end
  end
\end{lstlisting}Rules are matched using the match operation defined below which, in
turn, calls the match operation for each of the patterns:

\begin{lstlisting}
context Rule
  @Operation match(values,env)
    if args->size = values->size
    then 
      @For arg,value in args,values do
        if env <> null
        then env := arg.match(value,env)
        end
      end;
      env
    else null
    end
  end
\end{lstlisting}
\subsection{Forcing Delayed Documents}

A rule is fired by forcing its documents. Forcing a document causes
all of the delayed expressions to be evaluated and all 'emit' documents
to update the rule-base table:

\begin{lstlisting}
context Rule
  @Operation fire(env,table)
    let result = null
    in @For doc in docs do
         result := doc.force(env,table)
       end;
       result
    end
  end
\end{lstlisting}Forcing a document achieves two tasks:

\begin{itemize}
\item Components of a document may be delayed expressions that evaluate
to produce documents. These include expressions in < .. > and within
the {[} ... ] part of emit and label references. Forcing a document
evaluates the delayed expressions and returns the original document
with the delayed expressions replaced with the results.
\item Components of a document may be 'emit'-ed. Forcing a document updates
the document table with associations between the emit-label and the
emit-document (after delayed expressions have been evaluated).
\end{itemize}
Each document class defines a force operation that expects an environment
and a table. The environment is produces by the pattern matching and
contains values for free variables referenced in the delayed expressions.
The table is used to associate labels and documents produced by 'emit'.
Many of the force operations just force the sub-components, for example:

\begin{lstlisting}
context Collect
  @Operation force(env,table)
    // Just return the collect with the
    // components forced...
    Collect(
      source.force(env),
      mapper.force(env),
      combiner,
      doc.force(env,table))
  end
\end{lstlisting}Delayed expressions are implemented by associating documents with
operations. The arguments of the operation define the free variables
that must be supplied by the pattern matching environment. Components
with delayed expressions are implemented in the same way. Here is
the implementation of force for a delayed expression that occurs in
source code as <...>:

\begin{lstlisting}
context Delayed
  @Operation force(env,table)
    // Get the bindings from the environment that provide values
    // for the parameter names of the operation...
    let args = operation.paramNames()
                 ->collect(name | 
                    env->lookup(name.toString())) then
        // Force the delayed expression by supplying the argument
        // values...
        value = operation.invoke(self,args)
    in // Coerce the return value to be a document...
       @TypeCase(value)
         Doc do
           value
         end
         else Literal(value.toString())
       end
    end
  end
\end{lstlisting}Finally, an emit document updates the table:

\begin{lstlisting}
context Emit
  @Operation force(env,table)
        // Force the label...
    let label = label.force(env) then
        name = label.label();
        // Force the body...
        doc = doc.force(env,table)
    in // Extend the table with an entry for the label...
       if table.hasKey(name)
       then table.put(name,table.get(name) + Seq{doc})
       else table.put(name,Seq{doc})
       end;
       doc
    end
  end
\end{lstlisting}
\subsection{Displaying Documents}

Once rules have been fired and all documents have been forced, the
result is a table that associates labels with document templates.
A document template may contain label references; it is translated
to a document by replacing all the references. This is done using
the display operation of the rule base as shown below:

\begin{lstlisting}
@Operation display(label:String)
  // Display the document with the given label...
  if table.hasKey(label)
  then 
    // Use a string output buffer because there may be
    // lots of output...
    let buffer = Buffer(1000,true)
    in table.get(label)->at(0).display(0,table,buffer);
       buffer.toString()
    end
  else self.error("No label in table: " + label)
  end
end
\end{lstlisting}Each of the document classes defined an operation called display that
takes three arguments: the current level of indentation; the table
associating labels with document templates and an output buffer. Each
of the document classes implements the display operation differently.
The rest of this section describes how the documents are displayed.

The Collect class expects the collect source to produce a sequence
of documents, each of which is mapped and then combined to produce
a final single document that is displayed:

\begin{lstlisting}
@Operation display(indent,table,buffer)
  source.elements(table)->iterate(element doc = doc | 
    combiner.combine(doc,mapper.map(element)))
  .display(indent,table,buffer)
end
\end{lstlisting}A label reference requires that the label name is a key in the table.
The table should associate a collection of documents with the label
name; labels are used to reference the first element (collection documents
are used to combine multiple document entries in the table):

\begin{lstlisting}
@Operation display(indent,table,buffer)
  if table.hasKey(label)
  then table.get(label)->at(0).display(indent,table,buffer)
  else self.error("Label: cannot find label " + label)
  end
end
\end{lstlisting}A NewLine causes a new-line character to be added to the output buffer.
After each new-line, the current level of indentation is achieved
by adding the appropriate number of spaces:

\begin{lstlisting}
@Operation display(indent,table,buffer)
  buffer.add("\n"->at(0));
  buffer.append(formats("~V",Seq{indent}))
end
\end{lstlisting}Ordered documents just display the first and then the second:

\begin{lstlisting}
@Operation display(indent,table,buffer)
  left.display(indent,table,buffer);
  right.display(indent,table,buffer)
end
\end{lstlisting}

\chapter{The SECD Machine}

There are many different types of model used to analyse programs and
programming languages. These include models of source code, e.g. flow-graphs,
models of type systems, models of pre and post conditions, petri-nets
and various calculi.

Models of programs and their languages are important because designing
and implementing programs, as anyone knows who has tried it for real,
is a challenging task. The key features of what makes a program work,
are hidden away inside a black-box. Even if it was possible to look
inside the box, all that would be revealed is a bunch of electronics,
memory addresses, registers and binary data.

Programs are not designed in terms of the implementation platform
that is eventually used to run the code. A program is defined and
implemented in terms of models that describe various abstractions
over the implementation platform. The models make programming feasible
for humans.

A key model that is used when constructing programs, is the programming
language itself. When the code is eventually performed on the machine,
the program that was written by the human engineer is certainly not
executed directly by the hardware. Over time, languages have been
developed that provide abstractions over the implementation platform.

Having said that current programming languages (their libraries and
the various software platforms that are currently used) are abstractions;
they still must deal with many features of implementation detail that
allow the hardware to poke through. Take any serious programming language
manual and there will be a section describing how to make the thing
run as fast as possible. Use any commercial library (graphics libraries
being a case in point) and there will be many features that, whilst
they shield the user from the basic underlying hardware, still seem
fairly complex.

There is a grand history of models of programming languages. These
models are used to understand the fundamentals of what essential features
make a program work for an intended application. If you like, these
models are an abstraction of an abstraction, but don't let that put
you off; a familiarity with these models is a sure step in the direction
of being in-control of a wide range of programming notations.

One of the most important programming language models is the lambda-calculus.
It is essentialy a model of procedure-call (and method-call) based
languages and has be used to model languages including Pascal, Ada,
and Java. There are plenty of texts describing the lambda-calculus
and it is certainly not the intent of this book to provide a tutorial.
Many of the texts are quite theoretical in nature and should be ignored;
pick out some of the more practical looking texts if you are interested,
it is certainly worth it.

The rest of this section shows how the lambda-calculus can be modelled.
In particular, a machine is developed that runs the calculus. The
SECD machine is a fantastic tool to understand how programming languages
tick. It is simple and can be extended in a miriad of ways to model
useful abstractions of programming language execution.

The syntax of the lambda-calculus is defined as follows:

\begin{lstlisting}
E ::=               Expressions
  V                 Variables
| \V.E              Functions (of 1 arg)
| E E               Applications
\end{lstlisting}Where an expression is either a variable (just a name), a function
(1 arg and an expression body) or an application (of a function to
an arg). 

The execution of any calculus expression is also quite simple and
takes place with respect to an \textit{environment} that associates
variable names with values. To evaluate a variable, look up the variable
name in the environment. To evaluate a function, produce a closure
that associates the function with the current environment. To evaluate
an application, first evaluate the function and arg expressions to
produce a closure and a value. Evaluate the body of the closure-function
with respect to the closure-environment extended by associating the
closure-arg with the arg-value.

The execution is the essence of programming language execution where
procedure (method) calls evaluate by supplying the arguments and then
wherever the body of the procedure refers to the argument by name,
the supplied value is used instead. Of course there are many caveats
and special cases (not least order of evaluation and side-effects),
but these can all be dealt with by equipping the basic calculus with
the right machinery. 

\begin{figure}
\begin{center}

\includegraphics[width=12cm]{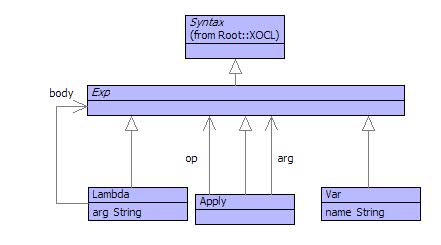}

\caption{Lambda Calculus\label{fig:Lambda-Calculus}}

\end{center}
\end{figure}

The lambda-calculus syntax is modelled in figure \ref{fig:Lambda-Calculus}.
The class Exp extends Syntax so that lambda expressions are self-evaluating.
The grammar is as follows:

\begin{lstlisting}
@Grammar
  Exp ::= Exp1 'end'.
  Exp1 ::= Lambda | a = Atom Composite^(a).
  Lambda ::= '\' a = Name '.' e = Exp1 { Lambda(a,e) }.
  Composite(a) ::= 
    b = Atom c = { Apply(a,b) } Composite^(c) 
  | b = Lambda { Apply(a,b) } | { a }.
  Atom ::= Var | Integer | '(' Exp1 ')'.
  Field ::= n = Name '=' e = Exp1 { Field(n,e) }.
  Var ::= n = Name { Var(n) }.
end 
\end{lstlisting}The definition of Composite provides an interesting feature: left
association. Applications are left associative in the lambda-calculus,
therefore the expression a b c is grouped as the application a b applied
to c, or equivalently (a b) c. The Composite rule is defined using
an argument a that accumulates the applications to the left.

\begin{figure}
\begin{center}

\includegraphics[width=12cm]{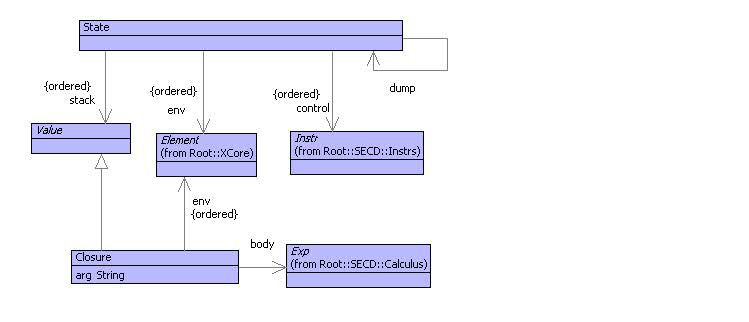}

\caption{Machine States\label{fig:Machine-States}}

\end{center}
\end{figure}

Execution of lambda-calculus expressions can be performed by the SECD
machine (whose states are shown in figure \ref{fig:Machine-States}),
so-called because it has 4 components:

\begin{enumerate}
\item The \textit{stack} S that is used to hold intermediate results from
expressions.
\item The \textit{environment} E that is used to associate variable names
with values.
\item The \textit{control} C that is used to hold sequences of machine instructions.
\item The \textit{dump} D that is used to hold resumptions.
\end{enumerate}
\begin{figure}
\begin{center}

\includegraphics[width=12cm]{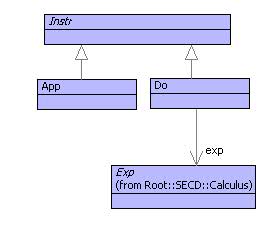}

\caption{Control Instructions\label{fig:Control-Instructions}}
\end{center}
\end{figure}

Execution of the machine is performed by case analysis on the instruction
at the head of the control. The instructions are shown in figure \ref{fig:Control-Instructions}.
Any expression is translated into an instruction using Do. the App
instruction is used to perform an application as shown below.

The best way to see what how the SECD machine works is to see it running.
The following trace shows the execution of an expression (\textbackslash{}v.
v v)(\textbackslash{}x.x). In order to make the output readable, the
following pretty-printers have been defined for the various classes:
States: (s,e,c,d); sequences: {[}x1,x2,x3]; pairs: h->t; App: @; Do:
ignored.

\begin{lstlisting}
(1) ([],[],[(\v.v v) (\x.x)],null)
(2) ([],[],[(\x.x),(\v.v v),@],null)
(3) ([<x,[],x>],[],[(\v.v v),@],null)
(4) ([<v,[],v v>,<x,[],x>],[],[@],null)
(5)   ([],[v-><x,[],x>],[v v],([],[],[],null))
(6)   ([],[v-><x,[],x>],[v,v,@],([],[],[],null))
(7)   ([<x,[],x>],[v-><x,[],x>],[v,@],([],[],[],null))
(8)   ([<x,[],x>,<x,[],x>],[v-><x,[],x>],[@],([],[],[],null))
(9)     ([],[x-><x,[],x>],[x],
          ([],[v-><x,[],x>],[],
            ([],[],[],null)))
(10)    ([<x,[],x>],[x-><x,[],x>],[],
          ([],[v-><x,[],x>],[],
            ([],[],[],null)))
(11)  ([<x,[],x>],[v-><x,[],x>],[],([],[],[],null))
(12)([<x,[],x>],[],[],null)
\end{lstlisting}The first thing to appreciate about the evaluation is roughly what
the expression is doing: the function \textbackslash{}v.v v is aplied
to the function \textbackslash{}x.x causing the function \textbackslash{}x.x
to be applied to itself producing the result \textbackslash{}x.x.
Now look at the trace and step through it to see how a machine evaluates
the expression. Each step in the trace is a new machine state produced
from the previous state by performing a transition. A transition looks
at the head of the control in the pre-state to determine the next
state.

Line (1) is the initial state. the expression is loaded onto the control.
Everything else is empty. Transition (1-2) performs the application
expression by unpacking it at the head of the control and adding an
@ instruction. Transition (2-3) evaluates the function \textbackslash{}x.x
to produce the closure at the head of the stack. Notice how the stack
is used to hold intermediate results; machine transitions consume
0 or more values at the head of the stack and produce 0 or 1 new values
at the head of the stack. Transition (3-4) evaluates the function
\textbackslash{}v. v v. Transition (4-5) performs the @ instruction
by setting up the evaluation of the body of the operator. The state
of the machine is saved in the dump so that the return value from
the application can be used. Notice how the envirnment in the new
state (5) contains a binding for the arg v. Transition (5-6) evaluates
the application v v. Transitions (6-9) evaluate the two references
to the variable v and then perform the application (again saving the
current state for return). Transition (9-10) evaluates the reference
to x and (10-11) returns a value since the control is exhausted. Transition
(11-12) perfoms the final return leaving the value of the original
expression at the head of the stack.

The rules of evaluation for the basic lambda-calculus are very simple.
They can be described using pattern matching and are defined in the
operation trans below that performs a single transition. A good understanding
of the rules of evaluation and their many variations is an excellent
basis for a deep understanding of how modelling can inform and enrich
programming.

\begin{lstlisting}
@Operation trans(state:State):State
  @Case state of
(1) State(s,e,Seq{Do(Var(n))|c},d) do
      State(Seq{e.lookup(n)|s},e,c,d)
    end
(2) State(s,e,Seq{Do(Apply(o,a))|c},d) do
      State(s,e,Seq{Do(a),Do(o)} + Seq{App()|c},d)
    end
(3) State(Seq{Closure(n,e1,b),a} + s,e2,Seq{App()|c},d) do
      State(Seq{},e1.bind(n,a),Seq{Do(b)},State(s,e2,c,d))
    end
(4) State(s,e,Seq{Do(Lambda(n,b))|c},d) do
      State(Seq{Closure(n,e,b)|s},e,c,d)
    end
(5) State(Seq{v},e1,Seq{},State(s,e2,c,d)) do
      State(Seq{v|s},e2,c,d)
    end
  end
end
\end{lstlisting}Transition (1) evaluates a variable by looking its name up in the
current environment and pushing the value on the stack. Transition
(2) unpacks an application at the head of the control. Transition
(3) performs an @ instruction at the head of the control. It expects
a closure above an argument at the head of the stack. The current
state is pushed onto the dump and a new state is created from the
closure body and environment. Transition (4) defines evaluation of
a function. Finally, transition (5) describes how a value is returned
from a completed function application (the body is exhausted).

\section{Extending the SECD Machine}

As it stands, the calculus and its execution by the SECD machine does
not look much like a standard programming language. Using just the
features you have seen, it is possible to encode virtually all standard
programming language concepts. Doing so is a bit like a cryptic crossword
- fiendishly difficult and spectacularly rewarding.

Rather than work with models of programs at the basic calculus level,
it is usual to add extensions. In fact, this is where the calculus
and its mechanical execution starts to be useful: if you have a requirement
to understand or design a new novel feature then see what you need
to do to add it to the calculus and SECD machine.

This section shows how the basic model can be extended by adding records,
integers and builtin operations such as +. A record is represented
as in {[}age=56, yearsService=10]. A tuple, or vector, is a record
where the fields have pre-determined names (usually represented as
numbers); for example a pair is {[}fst=1,snd=2] rather than (1,2).
This allows builtin operations to take tuples (represented as records)
as arguments: add{[}fst=1,snd=2] is the same as 1 + 2.

\begin{figure}
\begin{center}

\includegraphics[width=12cm]{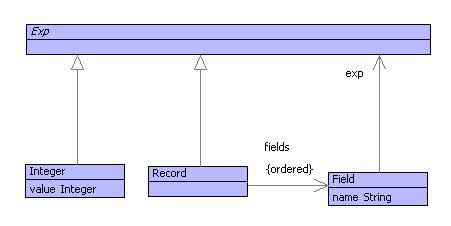}

\caption{Extended Syntax Model\label{fig:Extended-Syntax-Model}}

\end{center}
\end{figure}

The extensions to the syntax model are shown in figure \ref{fig:Extended-Syntax-Model}
and the grammar extension is shown below:

\begin{lstlisting}
Atom ::= Var | Record | Integer | '(' Exp1 ')'.
Record ::= '[' f = Field fs = (',' Field)* ']' { Record(Seq{f|fs}) }.
Field ::= n = Name '=' e = Exp1 { Field(n,e) }.
Integer ::= i = Int { Integer(i) }.
\end{lstlisting}The following shows an execution of the expression (\textbackslash{}x.\textbackslash{}y.add{[}fst=x,snd=y])
10 20. The builtin environment is shown as E. E contains a binding
for the name add to the builtin operation for +.

\begin{lstlisting}
([],E,[(\x.\y.add [fst=x,snd=y]) 10 20],null)
([],E,[20,(\x.\y.add [fst=x,snd=y]) 10,@],null)
([20],E,[(\x.\y.add [fst=x,snd=y]) 10,@],null)
([20],E,[10,(\x.\y.add [fst=x,snd=y]),@,@],null)
([10,20],E,[(\x.\y.add [fst=x,snd=y]),@,@],null)
([10,20],E,[\x.\y.add [fst=x,snd=y],@,@],null)
([<x,E,\y.add [fst=x,snd=y]>,10,20],E,[@,@],null)
  ([],E[x->10],[\y.add [fst=x,snd=y]],
    ([20],E,[@],null))
  ([<y,E[x->10],add [fst=x,snd=y]>],E[x->10],[],
    ([20],E,[@],null))
([<y,E[x->10],add [fst=x,snd=y]>,20],E,[@],null)
  ([],E[y->20,x->10],[add [fst=x,snd=y]],([],E,[],null))
  ([],E[y->20,x->10],[[fst=x,snd=y],add,@],([],E,[],null))
  ([],E[y->20,x->10],[x,y,{fst,snd},add,@],([],E,[],null))
  ([10],E[y->20,x->10],[y,{fst,snd},add,@],([],E,[],null))
  ([20,10],E[y->20,x->10],[{fst,snd},add,@],([],E,[],null))
  ([[fst=10,snd=20]],E[y->20,x->10],[add,@],([],E,[],null))
  ([!add,[fst=10,snd=20]],E[y->20,x->10],[@],([],E,[],null))
  ([30],E[y->20,x->10],[],([],E,[],null))
([30],E,[],null)
\end{lstlisting}

\chapter{Modelling with XML Data}

There is a general requirement for model data to be interchangeable
between tools. This is because organisations rely increasingly on
tool chains to support their business. No single tool can support
the whole process. Data generated by a modelling tool might hand the
information on to a code-generation tool or might hand the information
on to an engine that runs the models directly. In any case the different
tools in the tool chain need to agree on a format for the data.

In order to have different tools work with shared data there needs
to be a common serialization format. Data is serialized when it is
exported into a format that can be saved in a file. The format of
the data in the file needs to be known to all tools in the chain.
The format definition is often referred to as \textit{meta-data.}

The de-facto standard for data representation is XML. It is a textual
format that can be saved to files. Essentially, XML data is a tree
where each node in the tree is either raw text or is an element. An
element has a tag and some name/value attribute pairs. Each element
has a sequence of child nodes. Often you can think of the node tag
as the type of the node (for example: an employee record or a customer
transaction) and the attribute pairs as containing node properties
(for example: the employee name or the date of the transaction). The
children of the node usually encode relationships (for example: employee-to-department
or transaction-to-product).

XML meta-data (often also encoded as XML) defines the legal structure
of the XML documents. For example it defines the set of tags that
can be used in elements, the legal set of attributes for an element
with a given tag and the order in which child nodes can occur.

Both models and model instance data can be encoded using XML. For
example, the UML modelling language has an XML meta-data definition
called XMI that defines how UML models are serialized and therefore
shared between UML tools. 

When using a model-driven approach to system development it is important
to understand how to serialize and de-serialize model data. You may
be using a standard meta-data definition or be designing one of your
own. This chapter describes a number of language driven approaches
to using XML data.

\begin{figure}
\begin{center}

\includegraphics[width=12cm]{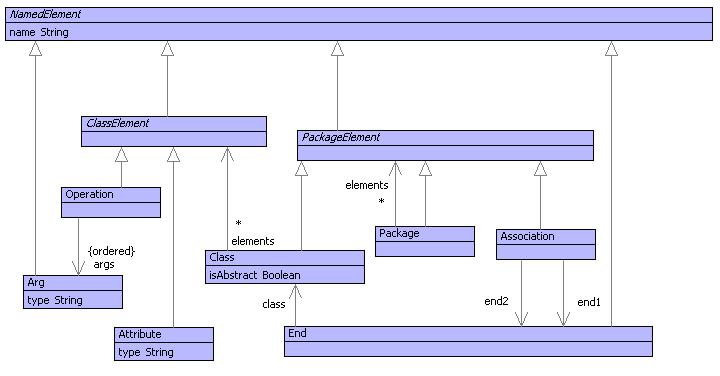}
\caption{Models\label{fig:Models}}

\end{center}
\end{figure}

Figure \ref{fig:Models} shows a simple UML-style modelling language
consisting of packages of classes and associations between them. This
chapter shows how instances of this language can be serialized as
XML data and how the resulting files can be read back into models.

\section{Generating XML Data}

Given an instance of the modelling language defined in figure \ref{fig:Models},
the translation to XML data is straightforward. Each class defines
an operation toXML that takes an output channel to which XML is written.
The format of XML data is standard, so it is useful to define a language
construct to support its generation. The following shows how the toXML
operation for Package is defined:

\begin{lstlisting}
@Operation toXML(out:OutputChannel)
  @XML(out)
    <Package name=self.name>
      @For e in elements do
        e.toXML(out)
      end
    </Package>
 end
end
\end{lstlisting}The XML construct is used to guide the generation of XML data to an
output channel. The output channel is supplied in ( and ) after @XML.
The body of the @XML construct defines the format of the generated
XML data. An XML element has the format:

\begin{lstlisting}
<TAG ATTS>
  CHILDREN
</TAG>
\end{lstlisting}where TAG is just a name, ATTS is a sequence of attribute value pairs,
CHILDREN is a sequence of child nodes and the final </TAG> should
match up with the corresponding opening tag. In the example, the tag
is Package, there is a single attribute called name. The value of
the name attribute (on the right of the =) is the name of the package. 

The @XML language construct allows XML elements to be written as they
will appear in the output, the characters will be send to the supplied
output channel. This makes it easy for the system to check that there
are no mistakes in the XML format and removes the need for the user
to have to use a format-statement or equivalent to describe how the
output should appear.

The children part of the @XML construct for packages is an XOCL @For
statement that iterates through the package elements and invokes their
toXML operations, supplying the same output channel.

\begin{figure}
\begin{center}

\includegraphics[width=12cm]{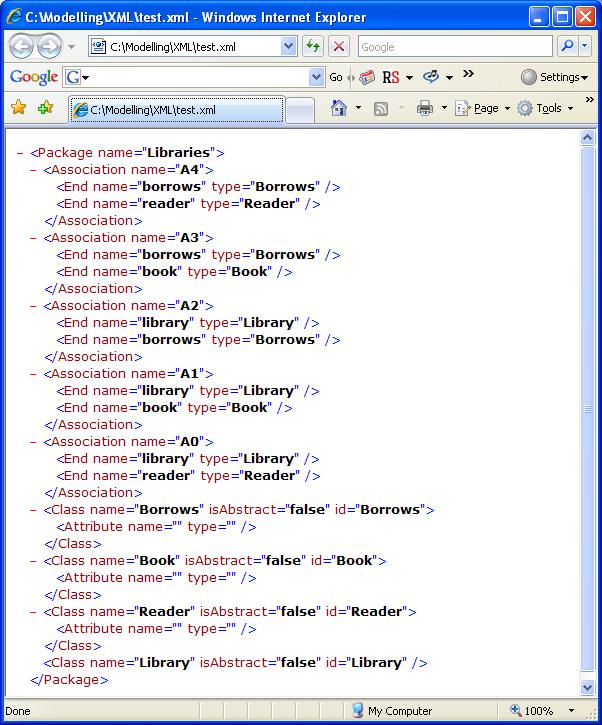}

\caption{A Simple Model Serialized as XML\label{fig:A-Simple-Model-as-XML}}

\end{center}
\end{figure}

Figure \ref{fig:A-Simple-Model-as-XML} shows a simple library model
serialized as XML using the toXML operations. Notice that the classes
attached to the ends of associations are encoded as the unique identifiers
attributes to the classes (in this case the class names are used as
the class identifiers, however in a real implementation some unique
string is usually used). The rest of this section defines the toXML
operations for all other classes in the modelling language.

Associations just wrap an association element around the two ends:

\begin{lstlisting}
@Operation toXML(out)
  @XML(out)
    <Association name=name>
      end1.toXML(out);
      end2.toXML(out)
    </Association>
  end
end
\end{lstlisting}An end generates a name and an identifier for the class that it attaches
to. The name of the class is used as the identifier:

\begin{lstlisting}
@Operation toXML(out)
  @XML(out)
    <End name=name type=class.name()/>
  end
end
\end{lstlisting}A class generates an element with attributes for the name, whether
it is abstract and its identifier. the children are generated by asking
each element of the class to generate some XML:

\begin{lstlisting}
@Operation toXML(out)
  @XML(out)
    <Class name=name isAbstract=isAbstract id=name>
      @For e in elements do
        e.toXML(out)
      end
    </Class>
  end
end
\end{lstlisting}The toXML operations for attributes, operations and arguments are
very similar to those describes above.

\section{An XML Generator}

The previous section uses a simple language construct to generate
XML output. The language construct allows the output from a model
to be described in XML format rather than using raw character output.
This makes it easier to focus on the structure of the output data
rather than the detail of the XML character formatting. This section
shows how the construct is implemented. 

\begin{figure}
\begin{center}

\includegraphics[width=12cm]{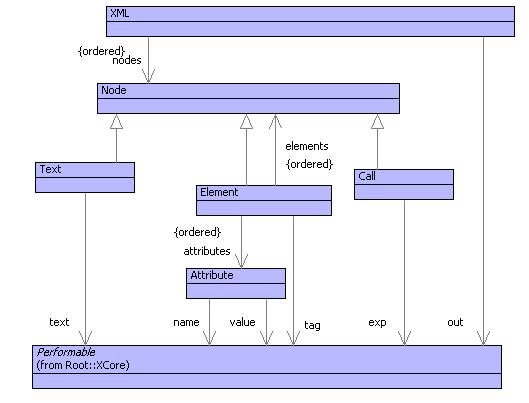}

\caption{The PrintXML Classes\label{fig:The-PrintXML-Classes}}

\end{center}
\end{figure}

Figure \ref{fig:The-PrintXML-Classes} shows the classes used to implement
the XML printing construct. An XML template is a performable expression
(out) and a sequence of nodes. The out expression will evaluate to
an output channel at run-time and a node is a syntax construct that
produces XML output to the channel. A text node is an expression that
produces raw text in the XML output. A call node is a general XOCL
expression; this allows any code to be embedded in the XML template
(such as an @For expression). An element is a syntax construct that
formats an XML element to the output channel. The tag and attribute
components are all XOCL expressions that produce strings. An element
node references sub-elements each of which are XML template nodes.

The XML template construct is parsed by the following grammar:

\begin{lstlisting}
@Grammar extends OCL::OCL.grammar
  AtomicElement ::= 
    '<' t = Tag A = Att* '/>' 
    { Element(t,A,Seq{}) }.
  Att ::= 
    n = AttName '=' v = Exp 
    { Attribute(n,v) }.
  AttName ::= 
     n = Name { OCL::StrExp(n) } 
   | '(' e = Exp ')' { e }.
  Call ::= e = Exp { Call(e) }.
  Element ::= 
      AtomicElement 
    | StructuredElement 
    | Text 
    | Call.
  Out ::= 
      '(' e = Exp ')' { e } 
    | { [| stdout |] }.
  StructuredElement ::= 
    '<' t1 = Tag A = Att* '>' 
      E = Element* 
    '</' t2 = Tag '>'  
    { Element(t1,A,E) }.
  Tag ::= 
      n = Name { OCL::StrExp(n) } 
    | '(' e = Exp ')' { e }.
  Text ::= 'Text' '(' e = Exp ')' { Text(e) }.
  XML ::= out = Out E = Element* { XML(out,E) }.
end
\end{lstlisting}The grammar extends the OCL grammar that provides the Exp rule; producing
an instance of XOCL::Performable. The grammar should be self explanatory;
it is worth noting the mechanism by which the template constructs
are populated by instances of Performable. The Out rule recognizes
an expression in parentheses or returns the expression (via the syntax
quotes {[}| and |]) referencing the standard input stdin. The Tag
and AttName rules transforms a recognized name to a string expression.

The grammar synthesizes an instance of the class XML, which is syntactic
sugar and transforms itself into basic XOCL code. The Package XML
template:

\begin{lstlisting}
@XML(out)
  <Package name=self.name>
    @For e in elements do
      e.toXML(out)
    end
  </Package>
end
\end{lstlisting}becomes the following XOCL code:

\begin{lstlisting}
format(out,"<~S",Seq{"Package"});
format(out," ~S='~X'",Seq{"name",self.name});
format(out,">");
@For x in elements do
  x.toXML(out)
end;
format(out,"</~S>",Seq{"Package"})
\end{lstlisting}
Each of the XML template classes defines an operation named desugar.
The XOCL execution engine knows that if it presented with an instance
of XOCL::Sugar it will not know directly how to perform the construct.
However, the construct will implement the desugar operation that can
be used to transform it into something the engine \textit{does} know
how to deal with. The rest of this section describes the desugar operations
defined by the various XML template classes.

The XML::desugar operation is as follows:

\begin{lstlisting}
@Operation desugar()
  nodes->iterate(n e = [| null |] | 
    [| <e>; <n.desugar(out)> |])
end
\end{lstlisting}Each of the XML nodes is desugared and sequenced using the ';' operator.
The Text::desugar operation is straightforward as it just sends the
text to the output:

\begin{lstlisting}
@Operation desugar(out)
  [| format(<out>,<text>) |]
end
\end{lstlisting}The Code::desugar operation is also straightforward:

\begin{lstlisting}
@Operation desugar(out)
  exp
end
\end{lstlisting}Finally, Element::desugar is where the main work is done. The action
is different depending on whether the sub-elements are empty or not.
In the case of no sub-elements:

\begin{lstlisting}
@Operation desugarNoElements(out)
  [| format(<out>,"<~S",Seq{<tag>});
     <attributes->iterate(a e = [| format(<out>,"/>",Seq{<tag>}) |] |
       [| <a.desugar(out)>; <e> |])>
  |]
end
\end{lstlisting}All the sub-elements are nodes and can be transformed to code via
their desugar operation:

\begin{lstlisting}
@Operation desugarWithElements(out)
  [| format(<out>,"<~S",Seq{<tag>});
     <attributes->iterate(a e = [| format(<out>,">") |] |
       [| <a.desugar(out)>; <e> |])>;
     <elements->iterate(e x = [| null |] |
       [| <x>; <e.desugar(out)> |])>;
     format(<out>,"</~S>",Seq{<tag>})
  |]
end
\end{lstlisting}
\section{Parsing XML}

When XML data is read in it must be recognized against a meta-data
description and then the corresponding model data can be synthesized
as a result. This is the same activity that occurs when a text language
is parsed and corresponding data structures are synthesized. The only
significant difference is that instead of parsing a sequence of characters
(or tokens) the structure to be parsed in the XML case is a tree of
elements. This section describes how the meta-data description for
an XML document can be expressed as a grammar including actions that
synthesize data. An XML document is recognized and synthesized using
a new language construct: an XML grammar; an XML grammar is defined
for the simple package modelling language defined in figure \ref{fig:Models}.
This section describes a language construct for expressing XML grammars
and shows how it is implemented.

A standard grammar is used by a parser to process an input sequence
of characters. A parser follows the rules defined by the grammar,
consuming text when the grammar specifies a terminal symbol. The parser
takes care to keep track of alternatives; when the parser encounters
a choice point, the current position is marked in the input so that
the parser can return to the position if the choice fails. The parser
encounters actions in the grammar rules, these synthesize elements.
The parse succeeds when there is no further grammar elements to be
processed; the parser returns the top-most syntheiszed element.

An XML grammar is very similar to a standard grammar with the following
differences:

\begin{itemize}
\item An XML parse processes an XML document which is a tree of elements.
The elements that are specified in the grammar and recognized in the
parse are elements, element attributes and element children. Unlike
text, an element consists of a starting tag (<TAG>) and an terminating
tag (</TAG>) with child elements in between. 
\item XML documents may include identifier references that are used to encode
a graph structure within the tree structure supported by XML. The
identifiers are associated with elements synthesized by the parse
and must be resolved by the parser.
\end{itemize}
The following is an XML grammar for the XML document produced in the
previous section:

\begin{lstlisting}
(1) @Grammar Models
(2)   Attribute ::=
(3)     <Attribute n=name t=type/>
(4)     { Attribute(n,t)
      }. 
      Class ::= 
        <Class n=name a=isAbstract id=id>
(5)       elements = ClassElement*
(6)     </Class> id := {
(7)        elements->iterate(e c = Class(n,a="true") | 
             c.addToElements(e))
        }.
(8) ClassElement ::= 
      Attribute 
    | Operation.
    Model ::= Package.
    Operation ::= 
      <Operation n=name>
        as = Arg*
      </Operation> {
        Operation(n,as)
    }.
    Package ::=
      <Package n=name>
        elements = PackageElement*
      </Package> { 
        elements->iterate(e p = Package(n) | 
          p.addToElements(e)) 
    }.
    PackageElement ::= 
      Package 
    | Class 
    | Association.
(9)  Association ::=
      <Association n=name>
        <End n1=name t1=type/>
        <End n2=name t2=type/>
      </Association> {
(10)    Association(n,End(n1,Ref(t1)),End(n2,Ref(t2)))
      }.
  end
\end{lstlisting}Line (1) introduces the grammar named Models. Line (2) is a typical
example of a grammar clause; it defines a rule for an attribute. Attributes
exist in the XML document as elements with two attributes: name and
type. Line (3) declares that an attribute takes the form of an element
with tag Attribute and with XML attributes name and type whose values
are references as n and t respectively. Line (4) is an action that
syntheisizes an attribute; the action may refer to any of the variables
that have been bound when the clause matches (in this case n and t).

Attribute is an example of an XML element where the children are of
no interest.Such elements are declared in grammar clauses by <TAG
... />. The clause for Class is an example where the children are
of interest. Such elements are declared as <TAG> ... </TAG>. Line
(5) declares that the children of a class element must match the elements
declared in ClassElement and there may be any number of such elements
(as defined by {*}).

Each class has a unique identifier supplied as the value of the attribute
id. Elsewhere in the XML document, classes may be referenced by association
ends. The references are made using the identifier of the class. An
XML grammar registers a synthesized element against a unique identifier
using the declarator :=. Line (6) shows how the class syntheiszed
by the Class clause is registered against the value of the variable
id. References to the identifier in association ends synthesized by
the parser will be automatically resolved as part of the parse. Line
(7) shows that a class is synthesized and the elements synthesized
by the ClassElement clause are each added to the resulting class.

Line (8) shows the definition ofthe classElement clause. A class may
contain attributes and operations. This clause shows how alternatives
are declared in clauses using | to sepaate the options. When the parser
tries to consume the next XML element against the ClassElement clause,
it will try Attribute, if that succeeds in recognizing an element
then the parse continues, otherwise the parse will backtrack and try
to recognize an Operation.

Line (9) shows the clause for associations. An association end in
the XML document references the type attached to the end via its class
identifier. An association end element references the class directly,
so the identifier must be resolved to the class (which is synthesized
elsewhere in the parse). Line (10) shows how the grammar can declare
that an identifier reference must be resolved. The class Ref is used
to construct an identifier reference. Providing that the identifier
in the reference is registered elsewhere in the parse (using := as
in line 6) then the parse will automatically replace the reference
Ref(id) with the model element registered against the id. 

The model for the grammar language is shown in figure XXX.

The grammar for XML grammars is shown below:

\begin{lstlisting}
@Grammar extends OCL::OCL.grammar 
    
   Action ::= '{' exp = Exp '}' { 
     Action(Seq{Exp(exp,exp.FV()->asSeq,null)}) 
   }.
   
   Any ::= 'ANY' { Any() }.
      
   AtomicElement ::= '<' tag = Tag as = Attribute* '/>' {
     Element(tag,as,Empty()) 
   }.
      
   Attribute ::= 
     var = Name 
     tag = AttributeTag 
     default = AttributeDefault {
       BindAtt(var,tag,default) 
   }.
      
   AttributeTag ::= 
     '=' Tag 
   | { "" }.
      
   AttributeDefault ::= 
     ':=' d = Exp { Exp(d) } 
   | { null }.

   Call ::= name = Name { Call(name) } .
      
   Clause ::= name = Name '::=' def = Disjunct '.' { 
     Clause(name,def) 
   }.
      
   ClauseAtom ::= 
     Element 
   | Empty 
   | Action 
   | Call 
   | Any 
   | Text 
   | Unordered 
   | '(' d = Disjunct ')' { Paren(d) }.
      
   Conjunct ::= p = ClauseBind qs = (Conjunct)* { 
     qs->iterate(q p = p | And(p,q)) 
   }.
      
   ClauseBind ::= 
     name = Name '=' p = Repeat { Bind(Seq{name},p) } 
   | ClauseUpdate.
      
   ClauseUpdate ::= 
     name = Name ':=' p = Repeat { Update(name,p) } 
   | Repeat.
      
   Children ::= 
     Disjunct 
   | { Empty() }.
      
   CompositeElement ::= 
     '<' tag = Tag attributes = Attributes '>' 
        children = Children 
     '</' Tag '>' { 
       Element(tag,attributes,children) 
   }.
      
   Disjunct ::= p = Conjunct qs = ('|' Disjunct)* { 
     qs->iterate(q p = p | Or(p,q)) 
   }.
      
   Element ::= 
     AtomicElement 
   | CompositeElement.
      
   Empty ::= 'EMPTY' { Empty() }.
      
   Grammar ::= name = Name clauses = Clause* 'end' { 
     XML::Parser::Grammar(name,Seq{},clauses).lift()
   }.
      
   Repeat ::= 
     p = Opt ('*' { Star(p) } 
   | '+' { Plus(p) } 
   | '#' { Star(p,true) } 
   | {p}).
      
   Opt ::= 
     '[' p = ClauseAtom ']' { Opt(p) } 
   | ClauseAtom.
      
   Tag ::= 
     Str 
   | Name.
      
   Text ::= 'TEXT' { Text() }.
      
   Unordered ::= 'Set' '{' UnorderedElements '}'.
      
   UnorderedElements ::= 
     p = ClauseBind 
     qs = (UnorderedElements)* { 
       qs->iterate(q p = p | Unordered(p,q))
   }.
end
\end{lstlisting}A parser for an XML grammar must process an XML document. If the document
conforms to one of the possible trees defined by the grammar then
the parse succeeds and the parser returns the value syntheisized by
the actions that have been performed. At any stage during the parse,
the parser maintains a parse-state containing all the information
necessary to continue from this point onwards. The elements of the
parse state make up the arguments of a parse operation defined for
each class in the XML grammar model. The parse-state contains the
following:

\begin{itemize}
\item The grammar. This is used to resolve calls to clauses.
\item A variable environment containing the current collection of variable/value
bindings. Each time the parse performs a binding of the form v = X,
the value of X is calculated (possibly as the result of calling a
clause) and the environment is extended with a binding for x. When
an action is performed, the variables bound in the environment are
made available for reference in the action expression.
\item An identifier binding containing the current collection of identifier/value
bindings. Each time the parse performs an identifier binding of the
form i := X, the value of X is calculated and the environment is extended
with a binding for identifier i. When the parse is complete, occurrences
of Ref(i) are replaced with the value of the identifier i from the
identifier environment
\item A stack of XML input elements. Each time a grammar element of the
form <TAG> is encountered, the stack is popped and the tag of the
top element must match TAG. If this fails then the parse backtracks
to the last choice point. Otherwise the children of the popped element
are pushed back on to the stack and the parse continues with the child
elements in the grammar.
\item A success continuation. This is an operation that represents what
to do next in the parse. Each time the parse succeeds, it continues
by calling the success continuation. The continuation operation is
supplied with parameters: a value; a variable environment; an identifier
environment; a stack of XML elements; and, a fail continuation. These
arguments represent the parse-state components that are required by
the rest of the parse and which may change.
\item A fail continuation. This is an operation that represents what to
do when the current parse fails. A parse fails when the expected tag
specified in the grammar does not match the supplied tag in the input
element. A parse continuation operation has no arguments; it is simply
called when the parse fails.
\end{itemize}
The parse-state is defined by arguments to the operation parse:

\begin{lstlisting}
context Pattern
  @AbstractOp parse(grammar,env,ids,elements,succ,fail)
  end
\end{lstlisting}The rest of this section describes the implementation of the parse
operation. The first two definitions are for Empty and Any. Both of
these clause patterns succeed; Empty does not consume any input whereas
Any consumes the next XML element. Both produce the value null:

\begin{lstlisting}
context Any
  @Operation parse(grammar,env,ids,elements,succ,fail)
    succ(null,env,ids,elements->tail,fail)
  end

context Empty
  @Operation parse(grammar,env,ids,elements,succ,fail)
    succ(null,env,ids,elements,fail)
  end
\end{lstlisting}An And pattern occurs in a clause when two patterns must occur in
sequence. The left pattern is performed first followed by the right
pattern. If it succeeds, the left pattern may have modified the parse-state;
this explains why the success continuation has arguments corresponding
to the state that may have changed:

\begin{lstlisting}
context And
  @Operation parse(grammar,env,ids,elements,succ,fail)
   left.parse(grammar,env,ids,elements,
     @Operation(ignore,env,ids,elements,fail)
       right.parse(grammar,env,ids,elements,succ,fail)
     end,
     fail)
  end
\end{lstlisting}An Or pattern occurs in a clause when there is a choice between two
patterns. The choice involves a left and a right pattern. The parse
continues with the left pattern. The right pattern is supplied as
the activity performed by a new fail continuation. Notice that the
current parse-state is \emph{closed-in} to the new fail continuation.
If the fail continuation is ever called, the parse will continue with
the parse-state that is current when the fail continuation is created
(not when the continuation is called):

\begin{lstlisting}
context Or
  @Operation parse(grammar,env,ids,elements,succ,fail)
    left.parse(grammar,env,ids,elements,succ,
      @Operation()
        right.parse(grammar,env,ids,elements,succ,fail)
      end)
  end
\end{lstlisting}An optional pattern can be implemented by Empty and Or:

\begin{lstlisting}
context Opt
  @Operation parse(grammar,env,ids,elements,succ,fail)
    Or(pattern,Empty()).parse(grammar,env,ids,elements,succ,fail)
  end
\end{lstlisting}When a clause name is called, the grammar is asked for the named clause,
a new parse-state is constructed and the clause body is parsed. Variables
are local to each clause, therefore when a clause is called the initial
variable environment is empty. A new success continuation is created
that passes the return value of the clause to the caller, but reinstates
the current variable environment:

\begin{lstlisting}
context Call
  @Operation parse(grammar,env,ids,elements,succ,fail)
    let clause = grammar.clauseNamed(name)
    in clause.body().parse(grammar,Seq{},ids,elements,
         @Operation(value,ignore,ids,elements,fail)
           succ(value,env,ids,elements,fail)
         end,
         fail)
    end
  end
\end{lstlisting}An action consists of an expression. An expression contains free variable
references. These come in two categories: those variables bound by
the current clause and global variables. An action expression encodes
the freely referenced variables as arguments of the operation that
implements the expression. Locally bound variables are found in the
variable environment. Globally bound variables are found by looking
the names up in the currently imported name spaces. The operation
Grammar::valueOfVar takes a name, an environment and returns the value
(whether it is locally or globally bound):

\begin{lstlisting}
context Action
  @Operation parse(grammar,env,ids,elements,succ,fail)
    let args = exp.args->collect(a | grammar.valueOfVar(a,env))
    in succ(exp.op.invoke(self,args),env,ids,elements,fail)
    end
  end
\end{lstlisting}Variables are bound as follows:

\begin{lstlisting}
context Bind
  @Operation parse(grammar,env,ids,elements,succ,fail)
    pattern.parse(grammar,env,ids,elements,
      @Operation(value,env,ids,elements,fail)
        succ(value,env->bind(names->head,value),ids,elements,fail)
      end,
      fail)
  end
\end{lstlisting}Update works like Bind except that the identifier is bound in a different
environment:

\begin{lstlisting}
context Update
  @Operation parse(grammar,env,ids,elements,succ,fail)
    pattern.parse(grammar,env,ids,elements,
      @Operation(value,env,ids,elements,fail)
        let id = env->lookup(name)
        in succ(value,env,ids->bind(id,value),elements,fail)
        end
      end,
      fail)
  end
\end{lstlisting}An XML element pattern requires that one or more input elements are
consumed. The parse succeeds when the tags of the elements match and
when the child element patterns match the child input elements. As
part of the match, the attribute values of the input element are bound
to the corresponding pattern attributes. The parser is defined below.
Notice how the attributes may specify a default value in the pattern;
if the corresponding attribute is not available in the input element
then the pattern variable is bound to the supplied default value (created
by evaluating an expression):

\begin{lstlisting}
context Element
  @Operation parse(grammar,env,ids,elements,succ,fail)
    if elements->isEmpty
    then fail()
    else
      let e = elements->head
      in if e.isKindOf(XML::Element) andthen e.tag() = tag
         then 
           @For p in attributes do
             @Find(a,e.attributes())
               when a.name = p.att()
               do env := env->bind(p.var(),a.value)
               else
                 if p.value() <> null
                 then env := env->bind(p.var(),p.value().op().invoke(self,Seq{}))
                 end
             end
           end;
           children.parse(grammar,env,ids,e.children(),
             @Operation(value,env,ids,ignore,fail)
               succ(value,env,ids,elements->tail,fail)
             end,
             fail)
         else fail()
         end
       end
     end
   end
\end{lstlisting}The {*} designator in clauses declares that the preceding underlying
pattern may occur any number of times. The return value from such
a pattern is the sequence of return values from each execution of
the underlying pattern. This is implemented as follows. Notice that
the failure continuation that is used for each execution of Star(pattern)
causes the parse to succeed, but supplies the empty sequence of results.
Therefore {*} cannot fail, 0 executions of the underlying pattern
is OK:

\begin{lstlisting}
context Star
  @Operation parse(grammar,env,ids,elements,succ,fail)
    pattern.parse(grammar,env,ids,elements,
      @Operation(value,env,ids,elements,ignore)
        Star(pattern).parse(grammar,env,ids,elements,
          @Operation(values,env,ids,elements,ignore)
            succ(Seq{value|values},env,ids,elements,fail)
          end,
          fail)
      end,
      @Operation()
        succ(Seq{},env,ids,elements,fail)
      end)
  end
\end{lstlisting}Finally, the child of an XML input element may be arbitrary text:

\begin{lstlisting}
context Text
  @Operation parse(grammar,env,ids,elements,succ,fail)
    if elements->isEmpty
    then fail()
    else
      let e = elements->head
      in if e.isKindOf(XML::Text)
         then succ(e.text,env,ids,elements,fail)
         else fail()
         end
      end
    end
  end
\end{lstlisting}It remains to define how to invoke the parser. A grammar has an operation
parse that is supplied with the name of the clause that will start
the parse and the name of the file containing the XML document. The
class IO::DomInputChannel takes an input channel as an argument and
returns an input channel that translates XML source text into an instance
of the XML model. The parse operation creates an initial parse-state
and then starts the parse by calling the starting clause:

\begin{lstlisting}
context Grammar
  @Operation parse(file,start)
    @WithOpenFile(fin <- file)
      let din = DOMInputChannel(fin) then
          doc = din.parse();
          succ = 
            @Operation(value,env,ids,elements,fail) 
              self.resolve(ids,value) 
            end;
          fail = @Operation() "FAIL" end 
      in Call(start).parse(self,Seq{},Seq{},Seq{doc.root},succ,fail)
      end
    end
  end
\end{lstlisting}
\section{Resolving References}

A successful parse returns a synthesized value. The value may contain
unresolved references to identifiers that were encountered during
the parse. An identifier is registered using the i := X construct
in a clause and may be used in a synthesized value by constructing
an instance Ref(i). Resolving the references involves replacing all
occurrences of ref(i) with the corresponding value of X.

Identifier resolution is performed by an operation Grammar::resolve.
It is supplied with an identifier environment and a synthesized value.
The ResolveRefs walker expects an identifier table (for efficiency),
so the environment is translated to a table and then the synthesized
value is walked:

\begin{lstlisting}
context Grammar
  @Operation resolve(ids,value)
    let table = Table(100)
    in @For id in ids->collect(pair | pair->head) do
         table.put(id,ids->lookup(id))
       end;
       ResolveRefs(table).walk(value,null)
    end
  end
\end{lstlisting}Resolving references is easy in a meta-circular environment because
engines can be defined that process the data as instances of very
general data types. ResolveRefs is a walker that traverses the data
and replaces all references (or reports an error if unregistered identifiers
are encountered). The rest of this section gives an overview of ResolveRefs:

\begin{lstlisting}
@Class ResolveRefs extends Walker

  // The refTable is supplied as a result from the parse. 
  // The table registers identifiers and elements...

  @Attribute refTable : Table end  

  // It may be desirable to limit the scope of the walk. 
  // The objPred is used to prevent the walker from 
  // descending into objects that may be large and are 
  // guaranteed not to contain identifier references...

  @Attribute objPred  : Operation = 
    @Operation(o) true end 
  end

  // A walker must always call initWalker as part 
  // of creation...
    
  @Constructor(refTable) ! 
    self.initWalker()
  end

  // The main work is done by walkObject. If the object 
  // is a registered identifier then return the result  
  // of walking associated element. Otherwise the reference 
  // is illegal. If the object is not a reference then 
  // objPred controls whether or not the walk descends 
  // into the object's slots...
    
  @Operation walkObject(o:Object,arg:Element):Element
    if o.isKindOf(Ref)
    then 
      if refTable.hasKey(o.id)
      then self.walk(refTable.get(o.id),arg)
      else self.error("Reference to undefined id: " + o.id)
      end
    elseif objPred(o)
    then super(o,arg)
    else o
    end
  end

  // The following operation is characteritic of the 
  // rest of the walker...
    
  @Operation walkSeq(s:SeqOfElement,arg:Element):Element 
    if not s->isEmpty
    then
      s->head := self.walk(s->head,arg);
      s->tail := self.walk(s->tail,arg)
    end;
    s
  end

  // More walking operations for basic data types...
     
end
\end{lstlisting}


\chapter{Interactive Applications}

An increasing number of interactive applications can be downloaded
onto devices such as mobile phones, PDAs, web-browsers and TV set-top
boxes. The applications involve presenting the user with information,
options, menus and buttons. The user typically enters information
by typing text and choosing amongst alternatives. An event is generated
by the user clicking a button or selecting from a menu. Once an event
is generated an engine that services the interactive application processes
the event, updates its internal state and then produces a new dialog
to present to the user.

The dialogs required by the interactive applications tend to be fairly
simple and are often used in conjunction with other applications such
as being broadcast together with TV video and audio content. The technology
used to construct the interactive application should be accessible
to as broad a spectrum of users as possible, including users whose
primary skill is not in developing interactive software applications.

Technologies used for interactive displays include Java-based technologies
such as MHP , HTML, JavaScript etc. These technologies are very platform
specific, they include a great deal of technical detail and are certainly
not approachable by a non-specialist. Furthermore, the general low-level
nature of the technologies does not enforce any standard look-and-feel
to interactive applications. The applications developed for a given
client (for example a single TV company) should have a common look
and feel that is difficult to enforce at such a low-level.

A common way to abstract from technical detail and to enforce a common
look-and-feel for a suite of applications it to develop a \emph{domain
specific language} (DSL) whose concepts match the expectations and
skill-levels of the intended users. A DSL for interactive applications
will include constructs for expressing textual content, buttons, choices
etc. The DSL leaves the rendering of the display features and the
event generation to a display engine that enforces a given look-and-feel.
The display engine can be replaced, changing the look-and-feel without
changing the DSL.

In addition to the DSL supporting high-level features for expressing
display content, it must provide some means for describing what the
application \emph{does}. Normally, application processing algorithms
are expressed at a low-level in program-code. If the DSL is designed
with an \emph{execution engine} then the same approach to abstraction
from rendering detail can be applied to abstraction from the operational
detail.

An executable DSL is a language for expressing complete applications
without detailed knowledge of implementation technologies. The xDSL
is a modelling language, instances of which are expressed as data.
An execution engine processes the data and runs the application. The
xDSL engine can be embedded within devices and other software applications
in order to run the models.

This chapter describes the design of an xDSL for interactive applications.
The xDSL has a textual syntax and is implemented in XMF. The rest
of this paper is structured as follows: section \ref{sec:Interactive-Application-Architecture}
describes an architecture for interactive applications based on an
xDSL; section \ref{sec:A-DSL-for-Interactive-Applications} describes
the xDSL from the point of view of the application developer, it outlines
the language features in terms of a simple application involving a
quiz; section \ref{sec:Implementation} describes how the xDSL is
implemented on XMF; section \ref{sec:Simulation} shows how the rendering
engine can be simulated and connected to the xDSL engine to simulate
the execution of an interactive application; section \ref{sec:XML-Representation-for}
shows how the application models can be serialized as XML; section
\ref{sec:Conclusion} concludes by reviewing the key features of the
approach and the technology used to develop the xDSL engine.

\section{Interactive Application Architecture\label{sec:Interactive-Application-Architecture}}

\begin{figure}
\begin{center}

\includegraphics[width=12cm]{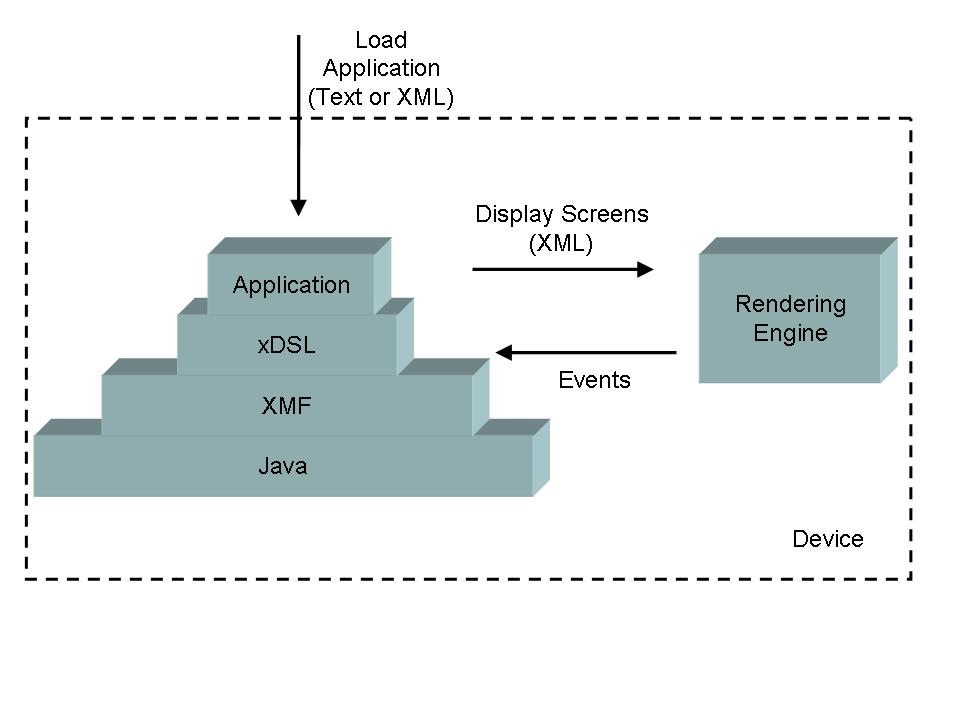}

\caption{Application Architecture\label{fig:Application-Architecture}}

\end{center}
\end{figure}

Figure \ref{fig:Application-Architecture} shows an overview of the
architecture for an interactive application xDSL. XMF is used as the
DSL implementation engine. XMF provides facilities for developing
text-based modelling languages and their associated execution engines.
The xDSL is written as a model in XMF, the applications are then loaded
onto the xDSL engine and executed.

The application generates display information in a general-purpose
format: in this case XML. The XML display information is sent to a
rendering engine. The engine understands the display features of the
DSL and interprets them in terms of the rendering technology. This
achieves a separation of concerns whereby the DSL can focus on the
information content and execution logic whereas the rendering engine
can focus on a standard way of displaying the information without
necessarily having to understand anything about the application that
is being rendered.

The rendering engine controls the hardware that interacts with the
user. The user generates events that are sent back to the xDSL engine.
In principle the events can be very detailed and can be encoded in
a suitable DSL. The example presented in this paper uses a very simple
encoding of events.

When the xDSL receives an event, it must process the data in an appropriate
way to produce more display information for the rendering engine.
The processing information is expressed in the application model running
on the xDSL engine. The display/event loop continues until the application
is terminated.

The architecture shown in figure \ref{fig:Application-Architecture}
has been used a number of times based on the XMF processing engine.
the development environment XMF-Mosaic is completely based upon a
number of rendering engines based on various features of Eclipse:
browsers, property editors and diagram editors. The architecture has
also been successfully used where XMF is deployed as a web-application
and the rendering engine is a standard web-browser (using various
combinations of HTML and GWT).

\section{A DSL for Interactive Applications\label{sec:A-DSL-for-Interactive-Applications}}

This section presents a simple interactive application expressed using
the xDSL and then shows it running. The following section shows how
the xDSL is implemented in XMF.

Textual languages are developed in XMF by extending the basic language
with new language features. XMF has a small basic language; the rest
is developed by extension. Each new language feature starts with a
'@' character; the feature may be used wherever any normal language
construct is expected. In this way, the XMF engine can be developed
into any special purpose DSL engine.

The following fragment shows the start of an interactive application
(@Model ...):

\begin{lstlisting}
@Model Quiz
  // The Quiz model describes an interactive application
  // for a TV quiz. Viewers are presented with a sequence
  // of questions and get a final score...
  score : Integer;
  // Screen definitions...
end
\end{lstlisting}Each model consists of a collection of screens. Each screen describes
how it is to be rendered and how it responds to events. For example,
the following screen starts the application. It places some text above
a button names Start. When the engine receives a Start event then
the application makes a transition to the screen named Question1:

\begin{lstlisting}
screen START()
  vertical
    text Welcome to the Quiz. Click the button to Start end
    button Start
      go Question1()
    end
  end
end
\end{lstlisting}Options are offered using a construct that lists the options (how
they are displayed is up to the rendering engine). The option group
is named; the name is used to refer to the particular option value
that the user selects when the event returns to the xDSL engine. This
is a typical way of encoding variable information during dialogs:
http does this and can be used to determine the values of fields and
choices on HTML screens. The first part of the Question1 screen uses
options as shown below:

\begin{lstlisting}
screen Question1()
  vertical
    text What is the capital of England? end
    options Choice
      option London; 
      option Paris;
      option Madrid;
    end
    // Question1 continues...
\end{lstlisting}Layout can be handled using many sophisticated schemes. A useful,
simple way to specify layout is to use horizontal and vertical flow,
where these can be nested. The options are displayed below the text
in the fragment given above. In the fragment below, the buttons for
Next and Quit are displayed next to each other (but below the options):

\begin{lstlisting}
    // ... Question1 continues...
    horizontal
      button Next
        // Next action continues...
      end
      button Quit
       go Quit()
      end
    end
  end
end
\end{lstlisting}The Next event is received after the user has chosen an option. If
the option is correct then the user is congratulated and the score
is incremented, otherwise the user is told the correct answer. In
both cases the dialog continues with the next question. 

Actions may be conditional, the conditional expression may refer to
choice variables, values of variables passed to screens and the current
state of the model instance. Actions may also produce displays (without
having to go to a new screen) this allows variables to be scoped locally
within an action %
\footnote{We really should have a let-construct and some local variables here
to show that the nested display has access to locally-scoped variables
over a user-transaction.%
}. In the following, the Next action has two local displays that are
used to respond to the choice:

\begin{lstlisting}
        // ...Next action continues...
        if Choice = "London" 
        then 
          display
            text Well Done end
            button Next
              score := score + 1;
              go Question2()
            end
          end
        else
          display
            text Wrong! Answer is London. end
            button Next
              go Question2()
          end
        end
      end
\end{lstlisting}That completes the overview of the DSL for interactions. It is simple
and closely matches the concepts required to define an interactive
application. It includes execution by encoding event handlers. It
deals with complexity by being simple and supporting nesting with
local scope and modularity. A non-expert in interactive software applications
should have no problems writing an application in this DSL.

The following shows run-fragment of this application. Since there
is no rendering engine attached, the XML is printed and the responses
encoded by hand:

\begin{lstlisting}
<Screen>
  <Vertical>
    <Text text=' Welcome to the Quiz. Click the button to Start '/>
    <Button name='Start'/>
  </Vertical>
</Screen>
Start                <-- Event from rendering engine
<Screen>
  <Vertical>
    <Text text=' What is the capital of England? '/>
    <Options name='Choice'>
      <Option name='London'/>
      <Option name='Paris'/>
      <Option name='Madrid'/>
    </Options>
    <Horizontal>
      <Button name='Next'/>
      <Button name='Quit'/>
    </Horizontal>
  </Vertical>
</Screen>
Next Choice=London   <-- Event from rendering engine
<Screen>
  <Text text=' Well Done '/>
  <Button name='Next'/>
</Screen>
Next                 <-- Event from rendering engine
\end{lstlisting}
\section{Implementation\label{sec:Implementation}}

The implementation of the xDSL has two main components: a syntax model
and a grammar that transforms text to instances of the syntax model;
and, a semantics model that defines an execution engine. The syntax
model defines a modeling language that defines an application \emph{type};
an instance of the type is defined by the semantics model. This is
a typical way to define a language: models represent things that can
be \emph{performed} in some sense. Performing the models produces
instances whose behaviour is expressed by the model. Another way to
think about this is that we aim to produce libraries of reusable interactive
applications (instances of the syntax model). A run-time occurrence
of an application is described by the semantic model.

XMF provides facilities for working with text including grammars,
XML parsers, XML formatters etc. The models have been developed using
the XMF development engine XMF-Mosaic and then run on the basic engine.

\subsection{Syntax}

The syntax for the DSL has two parts: the abstract syntax and the
concrete syntax. The abstract syntax consists of a collection of models
that are described below. The concrete syntax is defined by a collection
of grammars that are defined at the end of this section.

\begin{figure}
\begin{center}

\includegraphics[width=12cm]{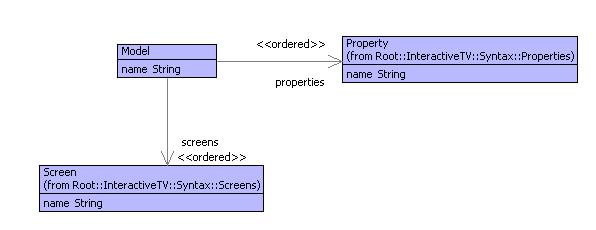}

\caption{Interactive Models\label{fig:Interactive-Models}}

\end{center}
\end{figure}

\begin{figure}
\begin{center}

\includegraphics[width=12cm]{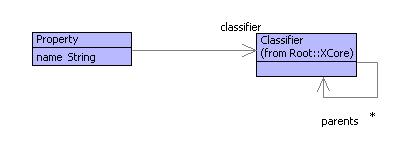}

\caption{Properties\label{fig:Properties}}

\end{center}
\end{figure}

Figure \ref{fig:Interactive-Models} shows the top-level model for
the interactive application language. A model consists of a collection
of properties and a collection of screens. Each property is defined
by the model in figure \ref{fig:Properties}; it has a name and a
classifier. XMF has a meta-model that provides features such as Class,
Object etc. A type is called a Classifier in XMF; Integer, String,
Set(Object) are all XMF classifiers.

\begin{figure}
\begin{center}

\includegraphics[width=12cm]{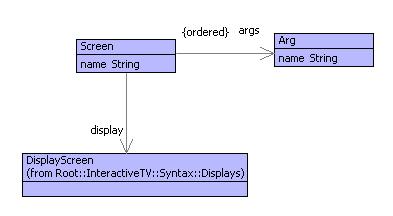}

\caption{Screens\label{fig:Screens}}

\end{center}
\end{figure}

\begin{figure}
\begin{center}

\includegraphics[width=12cm]{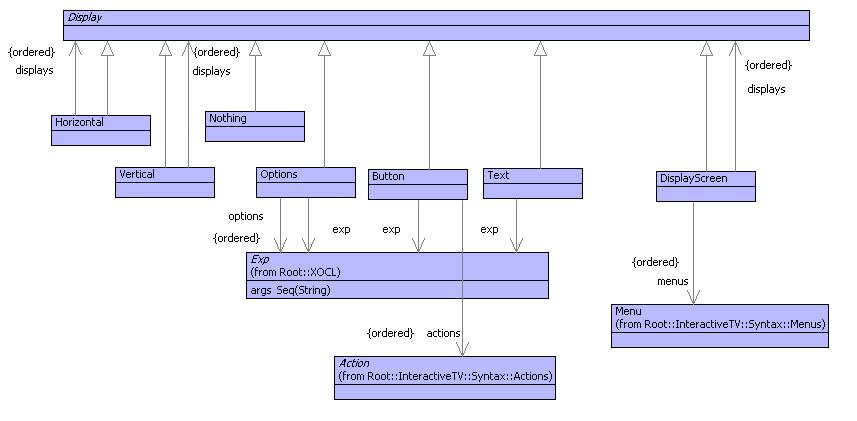}

\caption{Displays\label{fig:Displays}}

\end{center}
\end{figure}

Screens are shown in figure \ref{fig:Screens}. A screen has a collection
of arguments. An action may cause the application to make a transition
to a screen in which case the transition can supply argument values
to be used when calculating the display for the screen. Figure \ref{fig:Displays}
shows the model for displays. Each element of the displays model can
produce some XML output that is understood by the rendering engine.
In addition, some of the display elements are associated with actions
that will be performed when the appropriate event is received by the
xDSL engine.

The display elements of an application model refer to an XOCL class
called Exp. This is used wherever an executable fragment of code is
required. It allows features of the displays to be computed dynamically.
For example when a transition to a screen is made, the names of the
buttons may depend on the values of the arguments that are passed
to the screen. This is the reason why a button has an exp: it is used
to calculate the label on the button in terms of the variables that
are in scope at the time%
\footnote{Unfortunately no examples of this feature are given in the document.
However, imagine a list of voting options that will depend on the
current state of the system.%
}. The Exp class is a way of importing the models for XMF expressions
into the displays model.

\begin{figure}
\begin{center}
\includegraphics[width=12cm]{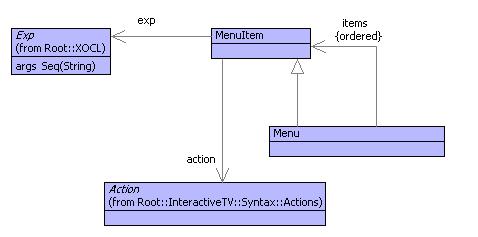}

\caption{Menus\label{fig:Menus}}
\end{center}
\end{figure}

\begin{figure}
\begin{center}

\includegraphics[width=12cm]{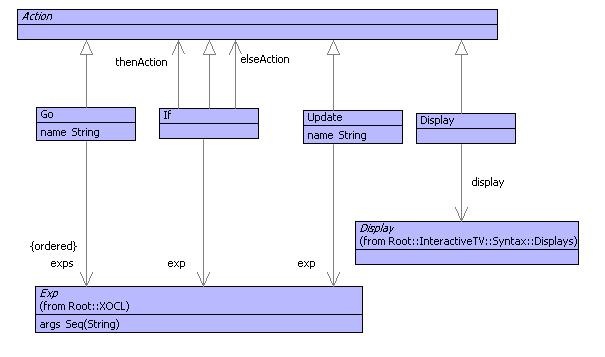}

\caption{Actions\label{fig:Actions}}

\end{center}
\end{figure}

Menus are shown in figure \ref{fig:Menus}, each menu has a label
that is computed (again depending on the context) and an action. Actions
are defined in figure \ref{fig:Actions}. An action is either a transition
to a new screen (Go), a conditional action (If), an update to a variable
currently in scope (Update) or a local display.

The developer of an interactive application does not directly create
instances of the abstract syntax model. The idea is that they write
in a text language that is parsed to synthesize instances of the syntax
model%
\footnote{Another way of doing this is to use some form of graphical notation.
XMF is designed to interface to EMF and GMF and therefore provide
execution engines for models defined in EMF. The interaction language
seems to lend itself more to a textual representation due to the actions
and nesting that needs to be expressed.%
}.

XMF allows any class to be extended with a grammar. The class then
defines a new syntax construct that is completely integrated into
the base language of XMF. Any number of classes can be added in this
way and new syntax classes can build on existing syntax classes. The
following is a simple example of a class definition that implements
a simple guarded expression:

\begin{lstlisting}
@NotNull [e1].m(e2,e3,e4)
  else e5
end
\end{lstlisting}where e1 evaluates to produce an object to which we want to send the
message m with args e2,e3 and e4. However, e1 might produce null in
which case we don't want to send the message, we want to do e5 instead.
This language construct is implemented as follows in XMF:

\begin{lstlisting}
@Class NotNull extends Sugar

  @Attribute exp       : String (?,!)      end
  @Attribute name      : String (?,!)      end
  @Attribute isMessage : Boolean (?,!)     end
  @Attribute args      : Seq(Performable)  end
  @Attribute error     : Performable (?,!) end
    
  @Constructor(exp,name,error) ! end
    
  @Constructor(exp,name,args,error) 
    self.isMessage := true
  end
    
  @Grammar extends OCL.grammar
    NotNull ::= 
      '[' e = Exp ']' '.' n = Name NotNullTail^(e,n) 'end'.
      
    NotNullTail(e,n) ::= 
      '(' as = NotNullArgs ')' x = NotNullElse { NotNull(e,n,as,x) }
    | x = NotNullElse { NotNull(e,n,x) }.
      
    NotNullArgs ::=
      e = Exp es = (',' Exp)* { Seq{e|es} }
    | { Seq{} }.
      
    NotNullElse ::=
      'else' Exp 
    | { [| null |] }.
      
  end
    
  @Operation desugar():Performable
    [| let notNullValue = <exp>
       in if notNullValue = null
          then <error>
          else <if isMessage
                then Send([| notNullValue |],name,args)
                else [| notNullValue.<name> |]
                end>
          end
       end
    |]
  end
    
end
\end{lstlisting}The key features of the NotNull class are as follows:

\begin{itemize}
\item The class extends Sugar which means that it is defining a new syntax
construct by providing an operation called 'desugar' whose responsibility
is to turn an instance of NotNull into program code.
\item The grammar definition extends the OCL grammar and thereby imports
all of the basic grammar-rule definitions. This provides the rule
for Exp which is the top-level grammar-rule for all language constructs.
\item Each grammar-rule consists of a name and a body. The rule may optionally
have arguments. The body consists of terminals (in ' and'), builtins
such as Name, rule-calls (possibly with arguments) and actions (inside
\{ and \}). The rule actions are any program expression, in most cases
they use class-constructors to create an instance of a named class.
\item The desugar operation uses lifting-quotes ({[}| and |]) to create
an instance of syntax-classes. The opposite of \emph{lifting} is \emph{dropping}
(< and >) used to calculate syntax by evaluating an program expression.
\end{itemize}
The rest of this section describes how the grammar feature of XMF
can be used to define the interaction language. A model consists of
a name followed by properties and screen definitions:

\begin{lstlisting}
context Model
  @Grammar extends Property.grammar, Screen.grammar
     Model ::= n = Name ps = Property* ss = Screen* 'end' {
        Model(n,ps,ss)
      }.
  end
\end{lstlisting}A property is a name and a simple expression (that cannot include
the ';' operator). The property-rule action uses an interesting feature
of syntax classes that allows the expression to be dropped into the
syntax element and is thereby evaluated to produce a classifier for
the property type:

\begin{lstlisting}
context Property extends OCL::OCL.grammar
  @Grammar extends OCL.grammar
    Property ::= n = Name ':' e = SimpleExp ';' {
      Property(n,Drop(e))
    }.
  end
\end{lstlisting}A screen has a name, arguments, menus and display elements. The rule
for screen arguments shows how optional elements are processed: it
returns either a non-empty sequence of names Seq\{a|as\} (head followed
by tail) or the empty sequence Seq\{\}:

\begin{lstlisting}
context Screen
  @Grammar extends Menu.grammar, Display.grammar
    Screen ::= 
      'screen' n = Name '(' as = ScreenArgs ')' 
         ms = Menu* 
         ds = Display* 
      'end' { Screen(n,as,DisplayScreen(ms,ds)) }.
    ScreenArgs ::=
      a = Name as = (',' Name)* { Seq{a|as} }
    | { Seq{} }.
  end
\end{lstlisting}A menu is shown below. This shows how expressions are captured in
data. The rule for a menu item name returns an instance of the class
Exp that is used in data to wrap an evaluable expression. There are
two forms of construction for Exp: Exp(e) and Exp(e,V,null). In both
cases e is an instance of a syntax class. In the second case V is
a collection of variable names that occur freely in e. The values
of variables in V can be supplied when the expression is evaluated
(via keyApply as shown below).

Another interesting feature of the menu item name rule is the use
of 'lift' to transform a data element (in this case a string n) into
an expression whose evaluation produces the original data element:

\begin{lstlisting}
context Menu
  @Grammar extends OCL.grammar
    Menu ::= 'menu' n = MenuItemName is = MenuItem* 'end' {
      Menu(n,is)
    }.
    MenuItemName ::= 
      n = Name { Exp(n.lift()) }
    | e = SimpleExp { Exp(e,e.FV(),null) }.
    MenuItem ::=
      Menu
    | 'item' n = MenuItemName a = Action 'end' { MenuItem(n,a) }.
  end
\end{lstlisting}Since Display is an abstract class, the grammar-rule for Display is
a list of concrete alternatives:

\begin{lstlisting}
context Display
  @Grammar extends Action.grammar, OCL.grammar
    Display ::=
        Text
      | Button
      | Options
      | Horizontal
      | Vertical.
    Text ::= 'text' t = Char* 'end' { 
      Text(Exp(t.asString().lift())) 
    }.
    Button ::= 
      'button' n = ComputedName 
        as = Action* 
      'end' { Button(n,as) }.
    ComputedName ::=
      n = Name { Exp(n.lift()) }
    | e = SimpleExp { Exp(e,e.FV(),null) }.
    Options ::= 
      'options' n = ComputedName 
        os = Option* 
      'end' { Options(n,os) }.
    Option ::=
      'option' n = Name ';' { n }.
    Horizontal ::=
      'horizontal'
        ds = Display*
      'end' { Horizontal(ds) }.
    Vertical ::=
      'vertical'
        ds = Display*
      'end' { Vertical(ds) }.
  end
\end{lstlisting}Action is another example of an abstract class:

\begin{lstlisting}
contxt Action
  @Grammar extends OCL.grammar
    Action ::=
        UpdateProperty
      | IfAction
      | Go
      | DisplayScreen.
      DisplayScreen ::= 
        'display' 
          ms = Menu* 
          ds = Display* 
        'end' { DisplayScreen(ms,ds) }.
      UpdateProperty ::=
        n = Name ':=' e = SimpleExp ';' {
          Update(n,Exp(e,e.FV(),null))  
      }.
      Go ::= 'go' n = Name '(' as = GoArgs ')' { Go(n,as) }.
      GoArgs ::= 
        e = GoArg es = (',' GoArg)* { Seq{e|es} }
      | { Seq{} }.
      GoArg ::= e = SimpleExp { Exp(e) }.
      IfAction ::=
        'if' e = SimpleExp
        'then' d = Action
        IfActionTail^(e,d).
      IfActionTail(e,d1) ::=
        'else' d2 = Action 'end' { If(Exp(e,e.FV(),null),d1,d2) }
      | 'end' { If(Exp(e,e.FV(),null),d1,null) }. 
  end
\end{lstlisting}
\subsection{Semantics}

\begin{figure}
\begin{center}

\includegraphics[width=12cm]{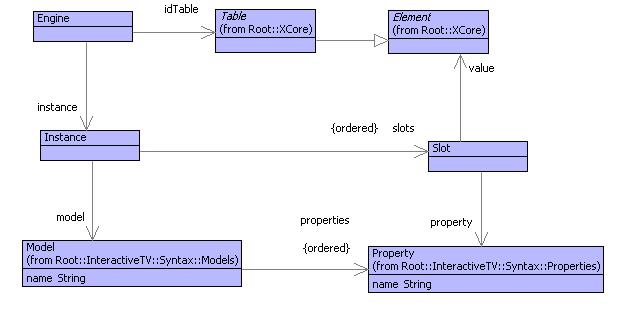}

\caption{Semantics Model\label{fig:Semantics-Model}}

\end{center}
\end{figure}

The semantics of the interactive application modelling language defines
an instance model for the syntax and also defines an execution engine
for the instances. The instance model is shown in figure \ref{fig:Semantics-Model}.
An instance of a Model is defined by the class Instance, it should
have a slot for each property of the model. The value of each slot
should have a type that is defined by the classifier of the corresponding
property.

The class Engine defines the execution engine. An engine controls
an instance and maintains an id-table that maps event ids to handlers
as shown below. The idea is that each time an event occurs, a handler
from the id-table is used to produce a new XML display that is sent
to the rendering engine. The XML data is calculated from the information
in the model and the current state of the instance slots.

The rest of this section defines the engine execution semantics; the
following section shows how the engine is connected to a rendering
engine. The engine processes a screen-transition using the 'go' operation
defined below. XMF supports many types of input and output channel;
this operation shows an example of a string output channel used to
capture and then return the XML output data as a string:

\begin{lstlisting}
context Engine
  @Operation go(screen:String,args:Seq(Element))
    let sout = StringOutputChannel()
    in instance.go(screen,args,self,sout);
       sout.getString()
    end
  end
\end{lstlisting}An instance handles a screen-transition by looking up the screen in
its model. If the screen exists then it is requested to display itself
with the supplied argument values:

\begin{lstlisting}
context Instance
  @Operation go(screen:String,args:Seq(Element),engine:Engine,out:OutputChannel)
    @NotNull [model.indexScreensByName(screen,null)].display(self,args,engine,out)
      else self.error("No screen called " + screen)
    end
  end
\end{lstlisting}A screen delegates the 'display' message to its underlying display
component. The screen argument names are bound to the argument values
to produce an \emph{environment of bindings} using the 'env' operation:

\begin{lstlisting}
context Screen
  @Operation display(instance,args,engine,out)
    display.display(instance,self.env(instance,args),engine,out)
  end
context Screen    
  @Operation env(instance,values)
    let env = args.zip(values)
    in instance.slots()->iterate(slot env = env | 
         env.bind(slot.property().name(),slot.value()))
    end
  end
\end{lstlisting}Each display element type : Button; Text; Horizontal; Vertical; and,
Options implements a 'display' operation that writes XML data to the
supplied output channel. As a side effect, if any of the elements
have event handling actions then the engine is updated with an appropriate
handler for the event when it is received from the rendering engine.

Each 'display' operation shows the use of an XMF language feature
@XML ... end that is used to write XML output to a channel. The construct
has the form:

\begin{lstlisting}
@XML(out)
  <TAG ATTS>
    .. program code ...
  </TAG>
end
\end{lstlisting}where the literal XML data is written to the supplied output channel.
In-between writing the starting tag and ending tag, an arbitrary program
is processed.

\begin{lstlisting}
context DisplayScreen
  @Operation display(instance,env,engine,out)
    @XML(out)
      <Screen>
        @For menu in menus do
          menu.display(instance,env,engine,out)
        end
        @For display in displays do
          display.display(instance,env,engine,out)
        end
      </Screen>
    end
  end
\end{lstlisting}The 'display' operation for text shows an example of the shortened
form of @XML ... end and also the use of the 'keyApply' operation
of the Exp class. The 'env' argument supplied to 'display' contains
bindings for all variables in scope. The 'keyApply' operation performs
the expression in the context of these variables:

\begin{lstlisting}
context Text
  @Operation display(instance,env,engine,out)
    @XML(out)
      <"Text" text=exp.keyApply(env)/>
    end
  end
\end{lstlisting}A button contains an action that is used to handle the event arising
from the user pressing the button in the rendering engine. The 'display'
operation for Button shows how an event handler is registered in the
engine. The arguments passed to 'registerActions' are the context
required to perform the actions when the event associated with 'id'
is received:

\begin{lstlisting}
context Button
  @Operation display(instance,env,engine,out)
    let id = exp.keyApply(env)
    in engine.registerActions(id,instance,env,actions);
       @XML(out)
         <Button name=id/>
       end
    end
  end
\end{lstlisting}Horizontal and Vertical are similar:

\begin{lstlisting}
context Horizontal
  @Operation display(instance,env,engine,out)
    @XML(out)
      <Horizontal>
        @For display in displays do
          display.display(instance,env,engine,out)
        end
      </Horizontal>
    end
  end
\end{lstlisting}the 'display' operation for Options shows an example of interleaving
of XML and program code:

\begin{lstlisting}
context Options
  @Operation display(instance,env,engine,out)
    @XML(out)
      <Options name=exp.keyApply(env)>
        @For option in options do
          @XML(out)
            <Option name=option/>
          end
        end
      </Options>
    end
  end
\end{lstlisting}The 'registerActions' operation of Engine must define a handler for
an event. The definition associates the event identifier 'id' with
an operation in the id-table of the engine. Actions are performed
using their 'perform' operation which expects to receive arguments
that include the current environment of variable bindings. The variables
available to an action include all those bound by selecting options
on the display. These display-bound variables are supplied to the
handler (in the same way that http works) as an environment 'env2':

\begin{lstlisting}
contxt Engine
  @Operation registerActions(id,instance,env1,actions)
    idTable.put(id,
      @Operation(env2)
        let value = null
        in @For action in actions do
             value := action.perform(instance,env2 + env1,self)
           end;
           value
        end
      end)
  end
\end{lstlisting}There are four types of action: If; Update; Go; and, Display. Each
action produces a result and the last action performed should return
an XML string to be sent to the rendering engine. If performs one
of two actions (or nothing) depending on the outcome of a test:

\begin{lstlisting}
context If
  @Operation perform(instance,env,engine)
    if exp.keyApply(env + instance.env())
    then thenAction.perform(instance,env,engine)
    else @NotNull [elseAction].perform(instance,env,engine) end
    end
  end
\end{lstlisting}An update changes the value of a variable currently in scope. The
variables in scope are: the slots of the instance; the current screen
arguments. The following operation checks whether the named variable
is a slot and updates the instance appropriately, or updates the current
environment:;

\begin{lstlisting}
context Update
  @Operation perform(instance,env,engine)
    @NotNull [instance.getSlot(name)].setValue(exp.keyApply(env + instance.env()))
      else env.set(name,exp.keyApply(env + instance.env()))
    end
  end
\end{lstlisting}Go makes a transition to a new screen. The screen will produce the
XML output. Notice that the current 'env' is not supplied to the 'go'
operation; therefore any variables currently in scope are not available
to the target screen unless their values are passed as arguments:

\begin{lstlisting}
context Go
  @Operation perform(instance,env,engine)
    engine.go(name,exps->collect(exp | exp.keyApply(env)))
  end
\end{lstlisting}Display is a way of locally displaying a screen without losing the
variables that are currently in scope:

\begin{lstlisting}
context Display
  @Operation perform(instance,env,engine)
    let sout = StringOutputChannel()
    in display.perform(instance,env,engine,sout);
       sout.getString()
    end
  end
\end{lstlisting}
\subsection{Handling Events}

Events occur when the user interacts with the rendering engine, for
example by pressing a button. When the event occurs, the current screen
may contain any number option groups. Each option group is named and
offers a number of alternative values. The selected option may affect
the behaviour of the engine in terms of variable updates and screen
transitions. Therefore, the event sent from the rendering engine to
the xDSL engine must encode the value of any option variables currently
displayed.

In addition there may be any number of ways an event can be raised:
menu selection or button press. Each must be uniquely identified and
the event must supply the identifier of the event that occurred. 

An event is defined to have a format that starts with the event id
and is followed by any number of option variable/value pairs:

\begin{lstlisting}
<ID> <VAR>=<VALUE> ... <VAR>=<VALUE>
\end{lstlisting}The event is encoded as a string and must be decoded by the engine.
This is easily done by defining an event as a grammar-rule:

\begin{lstlisting}
context Engine
  @Grammar
    Event ::= n = Name e = Binding* { Seq{n|e} }.
    Binding ::= n = Name '=' v = Name { Seq{n|v} }.
  end
\end{lstlisting}When an event is received by the engine it is supplied to 'getDisplay'
which calculates a new XML display string for the rendering engine.
The operation uses the grammar defined above to synthesize a pair
Seq\{id|env\} containing the event id and an environment of option-group
variable bindings. If the id is bound in the id-table then the handler
is supplied with the environment:

\begin{lstlisting}
context Engine
  @Operation getDisplay(event:String)
    let pair = Engine.grammar.parseString(event,"Event",Seq{}) then
        id = pair->head;
        env = pair->tail
    in @TableGet handler = idTable[id] do
         idTable.clear();
         handler(env)
       else self.error("No handler for " + name)
       end
    end
  end
\end{lstlisting}
\section{Simulation\label{sec:Simulation}}

Figure \ref{fig:Application-Architecture} shows the architecture
of an interactive application. The rendering engine is external to
the design of an xDSL; the relationship between the two is defined
by the XML schema for the display language and the format of event
strings. However, it is useful to be able to simulate the rendering
engine in order to test the xDSL engine. This can be done by setting
up a simple test harness for a pair of data consumers and linking
the xDSL engine with a rendering engine simulation that reads events
strings in from the terminal.

The following class implements a data producer-consumer pair:

\begin{lstlisting}
@Class Consumer

  @Attribute filter : Operation end
  @Attribute other  : Consumer (!) end
    
  @Constructor(filter) ! end
    
  @Operation consume(data)
    other.consume(filter(data))
  end
end
\end{lstlisting}The filter operation is used to generate data that is supplied to
the other consumer. If a pair of Consumer instances are linked together
then the data will bounce back and forth as required. The following
operation creates a filter for the xDSL engine:

\begin{lstlisting}
@Operation mk_xDSL_filter(model:Model)
  let engine = Engine(model.new())
  in @Operation(event)
       engine.getDisplay(event)
     end
  end
end
\end{lstlisting}The following filter operation simulates the rendering engine. It
does so by pretty-printing the XML to the standard-output. An XML
string can be transformed into an XML tree using the 'asXML' operation
defined for String. The standard-input is flushed and a line containing
the event is read and returned:

\begin{lstlisting}
@Operation renderFilter(xml:String)
  xml.asXML().pprint(stdout);
  "".println();
  stdin.skipWhiteSpace();
  stdin.readLine().stripTrailingWhiteSpace()
end 
\end{lstlisting}Given a model 'model', the following code produces, and starts, a
simulation:

\begin{lstlisting}
@Operation simulate(model:Model)
  let eConsumer = Consumer(mk_xDSL_filter(model));
      dConsumer = Consumer(renderFilter)
  in eConsumer.setOther(dConsumer);
     dConsumer.setOther(eConsumer);
     eConsumer.consume("START")
  end
end
\end{lstlisting}
\section{XML Representation for Applications\label{sec:XML-Representation-for}}

A requirement for interactive applications is to be able to dynamically
update the content and to be able to transfer the content from remote
locations in a standard format. The application describes in this
paper is completely represented in data. This means that, although
the application is executable, it can easily be serialized, send over
a communications link, and then uploaded onto the device that is running
the xDSL engine.

XMF provides support for encoding any data elements as XML. There
is a basic markup provided for all XMF data; the mechanisms for which
can easily be extended to provide bespoke XML encoding. Using the
basic mechanisms, a model can be encoded as follows:

\begin{lstlisting}
@WithOpenFile(fout -> "c:/model.xml")
  let xout = XMLOutputChannel(fout,NameSpaceXMLFormatter())
  in xout.writeValue(model)
  end
end
\end{lstlisting}%
\begin{figure}
\begin{center}

\includegraphics[width=12cm]{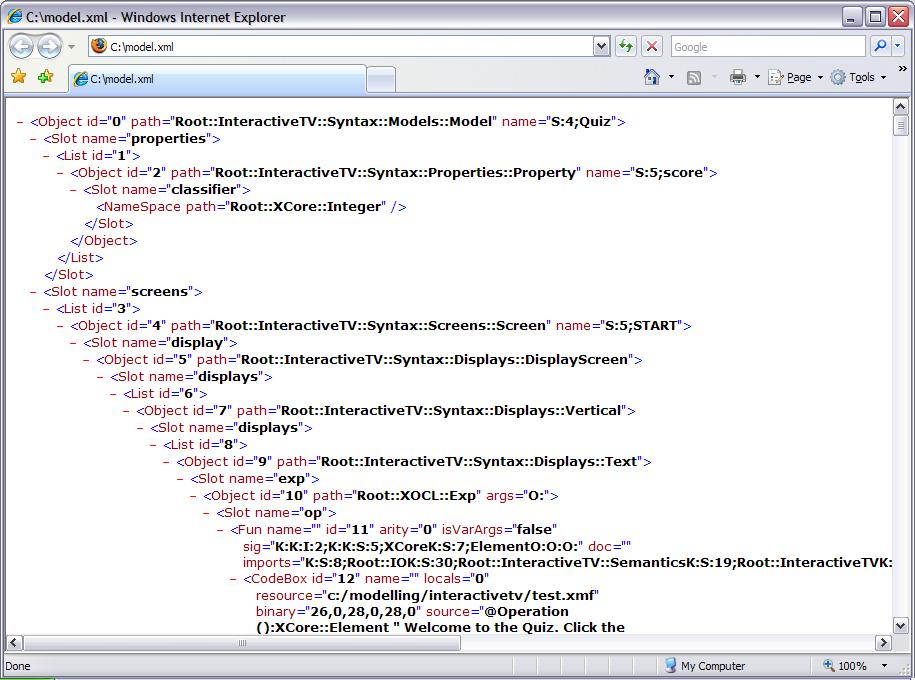}

\caption{A Serialized Model\label{fig:A-Serialized-Model}}

\end{center}
\end{figure}

The resulting output is produced in an XML file that is shown in figure
\ref{fig:A-Serialized-Model}. The XML markup shown in the figure
is the default vanilla-flavour markup provided by XMF. It is possible
to add adapters to an XML output channel that filter the data as it
is processed and thereby use domain specific markup. So instead of
seeing Object and Slot as element tags, we might use Screen and Button.

The XML can be read back in using the following code:

\begin{lstlisting}
@WithOpenFile(fin <- ModelFile)
  let xin = XMLInputChannel(fin,NameSpaceXMLInflater())
  in xin.parse()
  end
end
\end{lstlisting}
\section{Conclusion\label{sec:Conclusion}}

This chapter has described an approach to modelling systems whereby
an engine for an executable domain specific language (xDSL) is used
to develop and run the applications. The xDSL is designed to use concepts
that are suited to the application domain which allows the language
to abstract away from implementation details; the language is then
executable, can be used with different implementation platform technologies,
and is suitable for use by people whose primary skill lies in the
application domain rather than the implementation technology.

A method for developing xDSLs has been shown that involves a separation
of concerns between syntax elements that describe type-level features
of a model, and semantics elements that define run-time features of
an application. Experience has shown that this separation is natural
and allows the xDSL developer to effectively manage the process.

We have shown how a textual syntax can be added to an xDSL. In practice,
most xDSLs will require a concrete syntax. The precise form of this
will depend on the nature of the application and who the intended
users are. Sometimes, a purely graphical syntax is appropriate (for
example UML class-diagrams). Other times a purely textual syntax works
best, especially where executable features are involved and when complexity
can be controlled effectively using textual nesting. Often there is
scope for a mixture of the two where different aspects of the models
are constructed using graphical or textual syntax.

Modelling all features of a language has a number of benefits that
arise because everything is \emph{reified}. Reification involves representing
everything as data as opposed to program code or transient non-tangible
artifacts such as system events. Once everything is reified, many
types of analysis are possible including well-formedness checking,
type checking, application of style rules. It becomes easy to capture
and apply patterns and to perform refactoring. All features of an
application can be transformed to a target technology. 

Modelling actions is particularly important here; often actions are
left as unprocessed strings of program code which makes it very difficult
to analyze and run as part of an xDSL engine. The application given
in this paper has shown that it is straightforward to model actions
and to integrate them into the execution engine for an xDSL. By following
a few basic guidelines in terms of variable scope and control flow,
actions are easy to implement and are completely integrated into the
xDSL, its analysis and transformation.

\chapter{Choosing Between Alternatives}

Processing information is a linear activity. Listen to anyone describing
how something is performed and you will hear phrases like: 'first
get an X, turn it into a Y and feed it into the Z'. This is great
because mechanical processing of models is a linear activity too;
there is a close correspondence between real-world processes and their
modelled equivalent.

However, things are often not that simple. On closer inspection a
process can often be more like: 'select an X, find a way to turn it
into a Y and feed it into one of the Zs'. In this case processing
is not so obviously linear. What happens if the X that is selected
turns out to be the wrong choice when trying to make a Y? Can we go
back and try another X? If no Z is available who is at fault: the
X or the choice of Y?

The world is full of choices. Modelling the world often needs to take
account of choice. Once choice is admitted into a process, the \textit{scope}
of the choice becomes an important issue. How far can processing proceed
before a particular choice cannot be undone? In a linear activity,
undoing a choice within a particular scope is called \textit{backtracking,}
i.e. returning to the last choice point, selecting an alternative
and re-processing from that point.

Choice (or equivalently \textit{non-determinism}) is a property of
a model. Backtracking is a property of the execution engine for the
model. Backtracking is a way of implementing choice as a linear activity;
it is an important and often occurring pattern. This section is about
how to implement the pattern.

\section{A Choice Pattern}

Choice tends to occur as a pattern involving:

\begin{itemize}
\item A collection of elements to choose from (or alternative actions).
\item Something to do next with the selected element.
\item Something to backtrack with (a failure).
\end{itemize}
Each use of this pattern tends to be slightly different in the detail.
But the principle is generally the same. The pattern can be encoded
as follows:

\begin{lstlisting}
(1) @Operation select(S,next,fail)
(2)  if S->isEmpty
(3)  then fail()
     else
(4)  let e = S->sel then
(5)      S = s->excluding(e)
(6)  in next(x,@Operation() select(S,next,fail) end)
     end
   end
\end{lstlisting}The operation select (1) encodes the pattern. It takes 3 arguments:
S a set of elements to choose from, the activity to do next and a
fail for backtracking. The next activity is encoded as an operation
that is expecting an element and a fail. The fail argument is encoded
as an operation that takes no arguments and jumps to the last choice
point to choose an alternative.

The idea is that the next operation contains all future processing
for a selected element. Cruicially, the next operation may be invoked
more than once with different choices from S. Lines (4-6) show this
occurring, where an element e is selected from S and supplied to next.
The failure argument supplied to next is a new choice point; if it
is ever invoked then select is called again with fewer options to
choose from, \textit{but the same next and fail}. Therefore, select
can cause the same next argument to be invoked multiple times with
different elements chosen from S.

If select is ever invoked with no options (2) then the initial failure
argument is chosen. This allows chaining of different uses of select.

A slight variation on this pattern is the inclusion of a predicate
that controls whether the chosen element is valid or not:

\begin{lstlisting}
(1) @Operation select(S,next,pred,fail)
(2)  if S->isEmpty
(3)  then fail()
     else
(4)  let e = S->sel then
(5)      S = s->excluding(e)
(6)  in if pred(e)
(7)     then next(x,@Operation() select(S,next,pred,fail) end)
(8)     else select(S,next,pred,fail)
(9)     end
     end
   end
\end{lstlisting}The predicate is used in (6) to test the element. If pred is satisfied
then next is invoked as before (7), otherwise alternatives are tried
(8).

This pattern can be encoded as a language construct. The basic mechanism
involves repeated selection from a supplied set of alternatives. The
grammar below uses a locally defined recursive operation to continually
backtrack though the elements of a supplied set:

\begin{lstlisting}
@Grammar extends OCL::OCL.grammar
  Select ::= 
    '(' e = Name ',' f = Name ')' 'from' s = Exp 
        test = ('when' Exp | [| true |]) 'do' 
        body = Exp 
    'else' fail = Exp 'end' 
  { [|
      @Letrec
        select = 
          @Operation(S,<f>)
            if S->isEmpty
            then <Var(f)>()
            else 
              let <e> = S->sel then
                  S = S->excluding(<Var(e)>) then
                  <f> = @Operation() select(S,<Var(f)>) end
              in if <test>
                 then <body>
                 else <f>()
                 end
              end
            end
          end
       in select(<s>,@Operation() <fail> end)
       end
     |] }.
end
\end{lstlisting}Instead of defining a selection operation each time choice is used,
the selection construct can be used as follows:

\begin{lstlisting}
@Select(e,f) from S do
  next(e,f)
else fail()
end
\end{lstlisting}The variable e is initially bound to an element selected from the
set S and the body of the construct (after the do) is performed. The
variable f is bound to an operation; is f is ever invoked (with 0
args) then another element of S is selected and the body is performed
again. Subsequent uses of f have the same effect until S is exhausted
at which point the alternative part of the construct (after the else)
is performed.

\begin{figure}
\begin{center}

\includegraphics[width=12cm]{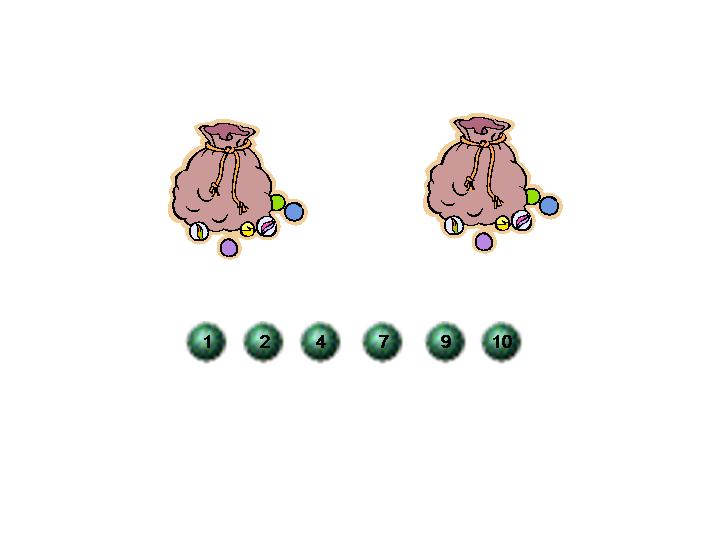}

\caption{\label{fig:Two-sacks-with}Two sacks with numbered balls}

\end{center}
\end{figure}

Consider a situation, shown in figure \ref{fig:Two-sacks-with} where
two sacks contain an arbitrary collection of numbered balls. Balls
are selected from each sack in turn and placed in a line. The idea
is to wind up with a sequence of balls whose numbers form an ascending
sequence.

Since it is not permitted to look inside the sacks when selecting
a ball, things can go wrong at some point during the process; although
it is always possible to select the balls in a way that achieves the
desired outcome. If things go wrong then balls can be put back (balls
are appropriately marked as to which sack they came from), and the
selection restarted from any point.

The sacks can be modelled as follows:

\begin{lstlisting}
@Class Sack
  @Attribute balls : Set(Integer) end
  @Attribute other : Sack (!,?) end
  @Constructor(balls) ! end
end
\end{lstlisting}where two sacks are set up as follows:

\begin{lstlisting}
let s1 = Sack(Set{1,7,3,5,9});
    s2 = Sack(Set{8,2,6,4,10})
in s1.setOther(s2);
   s2.setOther(s1)
end
\end{lstlisting}The process of selection is to be modelled as an operation on Sack
that builds a sequence of integers (balls) by ping-ponging between
the receiver and 'other', each optionally adding an integer. What
happens when there are balls left to choose, but which cannot be added
to the sequence because their numbers are too low? Is it always possible
to select so that this situation does not occur? Inspection of the
rules shows that this is not the case (the contents of the other sack
cannot be inspected). Therefore, a stalemate situation can occur,
therefore backtracking is a candidate modelling technique.

The rest of this section builds up the implementation of an operation
defined by Sack for ordering the elements selected from two sacks.
The operation is called with no arguments:

\begin{lstlisting}
@Operation order()
(1)  self.order(Seq{},
(2)    @Operation(orderedBalls,next,fail) 
(3)      other.order(orderedBalls, 
(4)        @Operation(orderedBalls,next2,fail) 
(5)          next(orderedBalls,next2,fail) 
(6)        end,fail) 
       end,
(7)    @Operation() 
         "Error!" 
       end)
end
\end{lstlisting}The ordering operation needs to initialise the ordered balls and create
an initial next and fail. Line (1) calls a second Sack operation called
order that has these as arguments. The initial collection of ordered
balls is Seq\{\} (1), the initial next operation and fail are supplied
in lines (2) and (7).

In this example, each next takes 3 arguments: the current sequence
of ordered balls, a next for the other sack and a fail. At line (2)
the initial next calls the ordering operation for the other sack (3)
passing it a next at (4) that ping-pongs back.

The initial fail (7) returns an error value because it should always
be possible to order the balls, i.e. this should never occur.

The second order operation is defined as follows:

\begin{lstlisting}
@Operation order(orderedBalls,next,fail)
(1)  if not other.allChosen(orderedBalls)
     then 
(2)    self.order(orderedBalls,balls,next,
(3)      @Operation() 
           next(orderedBalls,
(4)          @Operation(orderedBalls,next,fail) 
(5)            self.order(orderedBalls,next,fail) 
             end,fail) 
         end)
(6)  else self.order(orderedBalls,balls,next,fail)
     end
end
\end{lstlisting}If the other sack is empty (1) then the selection continues with the
current sack (7). If the other sack is not empty then there is a choice:
add an element from the current sack (2) and continue or just continue.
This alternative is encoded by extending the fail (3) so that if things
go wrong, the curren sack is skipped. If it is skipped then play resumes
with the current sack due to the next continuation (4-5).

The final operation uses Select to continually select from the available
balls:

\begin{lstlisting}
@Operation order(orderedBalls,balls,next,fail)
  if balls->isEmpty
  then orderedBalls
  else
    @Select(ball,f) from balls 
      when not orderedBalls->exists(b | b > ball) do
        next(orderedBalls + Seq{ball},
          @Operation(orderedBalls,next,fail) 
            let balls = balls->excluding(ball)
            in self.order(orderedBalls,balls,next,fail)
            end
          end,f)
    else fail()
    end
  end
end
\end{lstlisting}
\section{Finding a Route}

\begin{figure}
\begin{center}

\includegraphics[width=12cm]{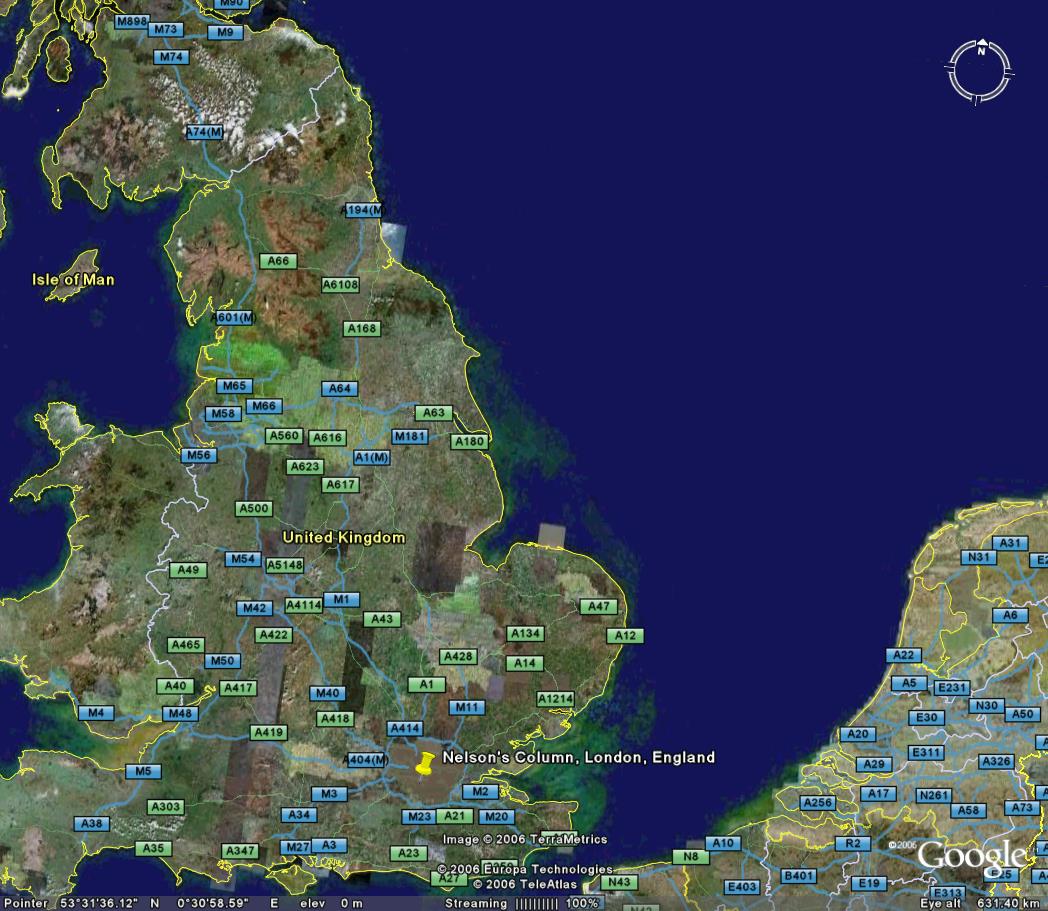}

\caption{\label{fig:Routes-in-the}Routes in the UK}

\end{center}
\end{figure}

Figure \ref{fig:Routes-in-the} shows a map of the UK with locations
and roads. One way of modelling this information is using the graph
construct defined in section \ref{sub:Graphs}:

\begin{lstlisting}
@Graph(Routes,City,Road)
    Birmingham()
    Brighton()
    Bristol()
      A48 -> Swansea
    Carlisle()
      A74 -> Glasgow
      A69 -> Newcastle
    Glasgow()
      M8 -> Edinburgh
      M80 -> Perth
    London()
      M23 -> Brighton
      A12 -> Ipswich
      A41 -> Oxford
      A3 -> Portsmouth
      M4 -> Bristol
      M1()->Leeds
      A1()->Peterborough
      M11()->Cambridge
    Hull()
    Ipswich()
    Leeds()
      M62()->Bradford
      M62 -> Hull
      A1()->Newcastle
    Oxford()
      M40 -> Birmingham
    Perth()
      M90 -> Edinburgh
    Portsmouth()
    Swansea()
    Bradford()
      M62()->Manchester
    Newcastle()
      A1()->Edinburgh
    Manchester()
      M61 -> Preston
    Preston()
      M6 -> Carlisle
    Peterborough()
      a1 -> Newcastle
    Cambridge()
    Edinburgh() 
      A91()-> Dundee
    Dundee()
      A956()->Aberdeen
    Aberdeen()
end
\end{lstlisting}Each node of the graph is a city. Each edge of the graph is a road
labelled with its name. For example from London it is possible to
go to Brighton using the M23 (and vice versa), and possible to go
to Bristol using the M4.

Consider how to find a route from one city to another. Suppose you
want to get from London to Carlisle. It is possible via the M1, A1
and then A69. Alternatively, you could go A1, A1, A69. In principle
there are many routes between two cities. Ignoring distance and other
criteria that might lead to selecting one route over another, the
only rule that should be applied is that a route does not pass through
the same city twice.

How might the construction of a route be modelled? Given a start and
end city, if there is a road linking the two then a route is found.
Otherwise, any road from the start (or end) can be selected leading
to a new start (or end) point. Given the new start (or end) then the
process can be repeated until it can be shown that no route exists
(remember you cannot go through the same city twice) or a route is
found. Given that choices are made, if no route can be found then
backtracking can b employed to return and make an alternative choice.

The class Routes extends Graph with an operation for finding a route:

\begin{lstlisting}
(1) @Operation route(source:String,target:String)
(2)   self.route(Set{},source,target,
(3)     @Operation(usedRoads,path,fail) 
(4)       path 
        end,
(5)     @Operation() 
         "NONE" 
       end)
   end
\end{lstlisting}Given the names of the source and target cities (1), a route is consructed
by calling another operation with a set of used roads, the source,
target, a next (3) and a fail (5). The initial fail is invoked when
no route exists, it returns the special no-route-available value NONE.

A next expects a set of used roads, a path and a fail. The path links
the source and target cities and is a sequence of road names.

The second route operation is defined below:

\begin{lstlisting}
@Operation path(usedRoads,source,target,next,fail)
 if source = target
 then succ(usedRoads,Seq{},fail)
 else
  @Select(road,otherRoad) 
   from self.edgesIncidentOn(source) - usedRoads do 
    self.pathAfter(
     usedRoads,source,target,road,otherRoad,next,fail)
   else 
    @Select(road,otherRoad) 
     from self.edgesIncidentOn(target) - usedRoads do
      self.pathBefore(
       usedRoads,source,target,road,otherRoad,next,fail)
     else fail()
    end
  end
 end
end
\end{lstlisting}\begin{lstlisting}
@Operation pathAfter(usedRoads,source,target,road,otherRoad,succ,fail)
       self.path(usedRoads->including(road),road.otherEnd(source),target,
         @Operation(usedRoads,path,fail)
           succ(usedRoads,Seq{road.label()} + path,fail)
         end,
         otherRoad)
     end
@Operation pathBefore(usedRoads,source,target,road,otherRoad,succ,fail)
       self.path(usedRoads->including(road),source,road.otherEnd(target),
         @Operation(usedRoads,path,fail)
           succ(usedRoads,path + Seq{road.label()},fail)
         end,
         otherRoad)
     end
\end{lstlisting}
\section{Explanation and NoGood Sets}

Patterns of search (as described above) can be encoded using language
constructs. Typically these involve success and failure continuations
that maintain the state of the search and provide a convenient mechanism
for branching. Continuations make it easy to backtrack when it is
found that an incorrect branch has been selected. These techniques
work well for finding a configuration amongst a number of alternatives
such that the configuration satisfies a given condition. However,
extra work is required to explain the lack of a satisfactory configuration. 

One technique is to invert the condition that selects a correct configuration
and to use it to explain why no satisfactory configuration could be
found. Instead of selecting a single satisfactory configuration, such
an inversion produces a collection of configurations, each of which
has a problem. Such configurations are sometimes referred to as \emph{noGood}
sets. This section describes an application that uses search to select
a configuration of business components; noGood sets are used to explain
why a solution cannot be found.

Consider developing a business plan. The business must set out its
\emph{goals} in terms of what the business wants to achieve. Each
goal is supported by a collection of \emph{tactics} each of which
describes an activity or approach that claims to achieve the goal.
In addition, a goal may be decomposed into a collection of child goals;
achieving all the child goals individually is the same as achieving
the overall goal. In order to implement a given tactic it is necessary
to allocate some resource. Resource is provided by organizational
\emph{units}, which may be company departments or individuals.

A goal may be \emph{achieved} if it has a supporting tactic that is
adequately resourced; or may be achieved if all of its children are
achieved by the same rule. It is important that a company does not
over-commit its resources, therefore a goal can only be achieved if
the unit resource limits are not exceeded.

Here is a high-level goal for a telephone banking business:

\begin{center}
\includegraphics[width=5cm]{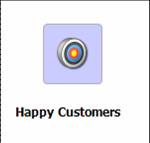}
\end{center}

The business will
be successful if it has happy customers. This goal can be decomposed
into any number of sub-goals:

\begin{center}
\includegraphics[width=5cm]{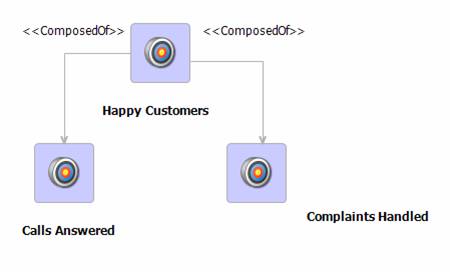}
\end{center}

where
a customer is happy if their calls are answered effectively and their
complaints are handled. Each of the sub-goals has any number of tactics
that will achieve the goal:

\begin{center}
\includegraphics[width=5cm]{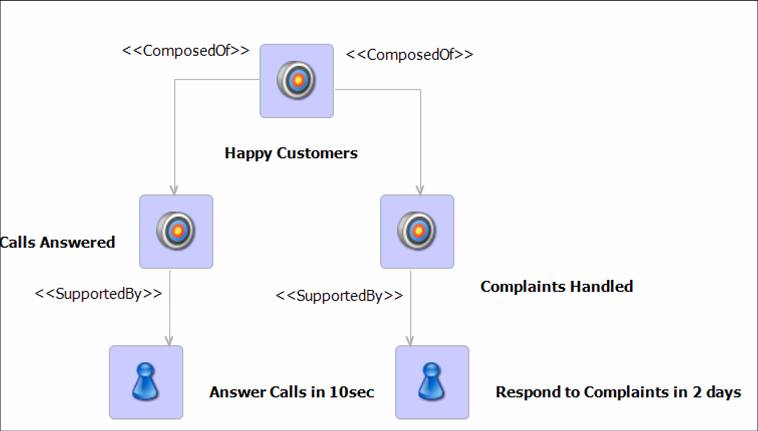}
\end{center}

Finally, the tactics need to be resourced by an organization unit.
A unit may resource more than one tactic and a tactic may be resourced
by more than one unit. In this case the unit is a telephone call centre:

\begin{center}
\includegraphics[width=5cm]{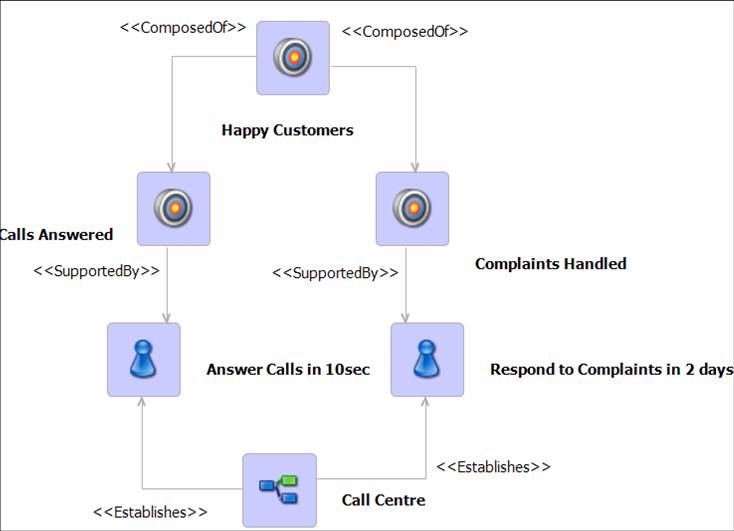}
\end{center}

The models shown above can be used as part of a business as a decision
support solution. Each model describes the goals the business intends
to achieve and how the goals are to be implemented and resourced.
The solution provides feedback in terms of the options for resourcing
the tactics that achieve the goals. If the goals cannot be achieved
then the solution will provide feedback on why the different resourcing
options fail.

The rest of this section is in two parts: the first part describes
how an engine is defined that finds a resourcing solution; the second
part describes how noGood sets are used to provide feedback when there
is no solution.

\begin{figure}
\begin{center}

\includegraphics[width=12cm]{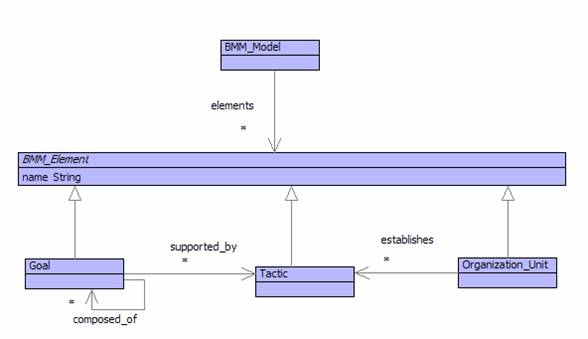}

\caption{The Business Motivation Model \label{fig:The-Business-Motivation}}

\end{center}
\end{figure}

\begin{figure}
\begin{center}

\includegraphics[width=12cm]{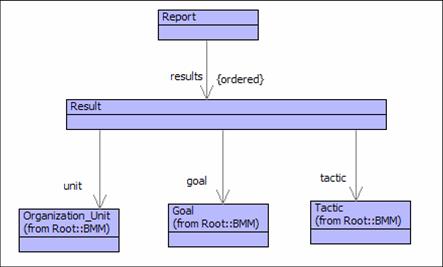}

\caption{Configuration Results\label{fig:Configuration-Results}}

\end{center}
\end{figure}

Figure \ref{fig:The-Business-Motivation} shows a model of the business
concepts used by the solution. Figure \ref{fig:Configuration-Results}
shows a report model that is produced when there is a solution consisting
of a collection of results each of which associates a goal, a tactic
and an organization unit.

The following rules are applied to find a solution. A goal g is achievable
when:

\begin{enumerate}
\item g is supported by at least one tactic that is established by at least
one organization unit; or
\item g has at least one sub-goal and each sub-goal is achieved; and
\item the number of tactics supported by each unit does not exceed the resource
for the unit.
\end{enumerate}
The rules are implemented as an operation called achievable that is
defined below. The arguments of the operation form the context for
the achievable-engine. It is supplied with a goal, a table of resources,
the current results and continuations for success and failure. The
table of resources is an a-list mapping unit names to the current
amount of resource available for that unit. The results are just a
collection of result instances. 

\begin{lstlisting}
context BMM_Model
  @Operation achievable(goal,resources,results,succ,fail)
    @Select tactic:Tactic from goal.getSupported_by() do
      @Select unit:Organization_Unit from self.getElements() 
        when unit.establishes(tactic)
        do @Result(goal,tactic,unit) 
             @Use(unit) end
           end
      end
    else
      @Unless goal.getComposed_of().isEmpty() do 
        @ForAll child in goal.getComposed_of() do
          self.achievable(child,resources,results,succ,fail)
        end
      end
    end
  end
\end{lstlisting}The success and fail continuations are used to implement a backtracking
mechanism. The rules are implemented by the achievable-engine in terms
of search. A choice-point arises when there is more than one tactic
that could be chosen to support a goal and when there is more than
one unit that could be chosen to establish a tactic. In both cases,
one of the options is chosen and a choice point is created that will
be used if ever the search fails. 

Failure occurs when a goal cannot be achieved because it has no tactics
or a tactic has no establishing unit. The resource associated with
a unit is reduced by 1 each time the unit is used to establish a tactic.
Once the count reaches 0 the unit cannot be used again; a subsequent
attempt to select the unit causes backtracking.

The achievable-engine uses language constructs that are specifically
designed to support the business solution. These constructs are: Select
which is used to select from a collection; Result which is used to
produce a result; Use which is used to reduce the amount of available
resource for a unit; Unless which is used to place a guard on an action;
and, ForAll which is used to require a condition holds for all element
of a collection. These language constructs use the arguments of the
engine as 'globals' and are implemented as follows.

The language construct for Select uses the selection operation to
try each element of a collection:

\begin{lstlisting}
context Root
  @Class Select extends Sugar
    @Attribute var        : String end
    @Attribute type       : Performable end
    @Attribute collection : Performable end
    @Attribute body       : Performable end
    @Attribute guard      : Performable end
    @Attribute alt        : Performable end
    @Constructor(var,type,collection,guard,body,alt) ! end
    @Grammar extends OCL::OCL.grammar
      Select ::= 
        v = Name ':' t = Exp 'from' c = Exp g = Guard 'do' 
          b = Exp a = Alt 'end' {
            Select(v,t,c,g,b,a)
      }.
      Guard ::=
        'when' Exp
      | { [| true |] }.
      Alt ::= 
        'else' Exp
      | { [| fail() |] }.
    end
  end
\end{lstlisting}The desugar operation for Select builds up a choice-point through
the fail continuations:

\begin{lstlisting}
context Select
  @Operation desugar() 
    [| let // Filter on the type of the elements...
           C = <collection>.asSeq()->select(x | 
             x.isKindOf(<type>)) then
           // Filter using the guard...
           C = C->select(<var> | <guard>) then
           // Set up the action to perform...
           action = 
             @Operation(<var>,fail) 
               <body> 
             end then
           // The final fail performs the alternative action...
           finalFail = 
             @Operation() 
               <alt> 
             end then
           // Build up the choice-tree using fails...
           fail = C->iterate(x fail = finalFail | 
                    @Operation() 
                      action(x,fail) 
                    end)
       in // Perform the first choice...
          fail()
       end |]
  end
\end{lstlisting}The Result construct adds a new result to the current collection of
results. This establishes a new collection of results for the scope
of the Result body:

\begin{lstlisting}
@Class Result extends Sugar
    @Attribute goal   : Performable end
    @Attribute tactic : Performable end
    @Attribute unit   : Performable end
    @Attribute body   : Performable end
    @Constructor(goal,tactic,unit,body) ! end
    @Grammar extends OCL::OCL.grammar
      Result ::= '(' g = Exp ',' t = Exp ',' u = Exp ')' 
        b = Exp 'end' {
          Root::Result(g,t,u,b)
      }.
    end
    @Operation desugar()
      [| let result = Result(<goal>,<tactic>,<unit>) then
             results = results->including(result)
         in <body>
         end
      |]
    end
  end
\end{lstlisting}When a unit is used to establish a tactic, the number of resources
associated with the unit must be reduced by one. However a unit cannot
use more resource than it owns, therefore if the resources are exhausted
for the required unit then engine uses fail to select an alternative:

\begin{lstlisting}
@Class Use extends Sugar
    @Attribute unit : Performable end
    @Attribute body : Performable end
    @Constructor(unit,body) ! end
    @Grammar extends OCL::OCL.grammar
      Use ::= '(' u = Exp ')' b = Body 'end' {
        Use(u,b)
      }.
      Body ::=
        Exp
      | { [| succ(resources,results,fail) |] }.
    end
    @Operation desugar()
      [| let name = <unit>.getName() then
             resource = resources.lookup(name)
         in if resource = 0
            then fail()
            else 
              let resources = resources.bind(name,resource - 1)
              in <body>
              end
            end
         end |]
    end
  end
\end{lstlisting}The Unless construct simply places a guard on an action; if the guard
fails then an alternative is found using the fail continuation:

\begin{lstlisting}
@Class Unless extends Sugar
    @Attribute guard : Performable end
    @Attribute body : Performable end
    @Constructor(guard,body) ! end
    @Grammar extends OCL::OCL.grammar
      Unless ::= g = Exp 'do' b = Exp 'end' {
        Unless(g,b)
      }.
    end
    @Operation desugar()
      [| if <guard>
         then fail()
         else <body>
         end
      |]
    end
  end
\end{lstlisting}The ForAll construct requires a condition to hold for all elements
of a collection. This is achieved using the success continuation;
each element of the collection is selected and the success continuation
returns to the collection to try the next element:

\begin{lstlisting}
@Class ForAll extends Sugar
    @Attribute var        : String end
    @Attribute collection : Performable end
    @Attribute body       : Performable end
    @Constructor(var,collection,body) ! end
    @Grammar extends OCL::OCL.grammar
      ForAll ::= v = Name 'in' c = Exp 'do' b = Exp 'end' {
        ForAll(v,c,b)
      }.
    end
    @Operation desugar()
      [| let next = 
               @Operation(<var>,resources,results,succ,fail) 
                 <body> 
               end
         in <collection> ->iterate(x succ = succ |
              @Operation(resources,results,fail)
                next(x,resources,results,succ,fail)
              end )
         end |]
    end
  end
\end{lstlisting}That concludes the language constructs for the achievable-engine.
The search is controlled using continuations so that all possible
combinations of goal, tactic and unit are tried to see if they are
compatible. When an illegal combination is encountered, the fail continuation
is used to backtrack. Therefore, the engine will find a suitable collection
of results if they exist. 

When no suitable combination of business elements exists, the engine
returns no value. It simply reports that the goal cannot be achieved.
The rest of this section describes how a noGood sets can be constructed
to explain the reasons why a solution cannot be found.

A noGood set is a combination of goals, tactics and units that invalidates
the achievable rules defined earlier. The rules are invalidated when
a goal has no tactics, when all tactics for a goal lack suitable organization
units, or when all the sub-goals of a goal invalidate the rules. When
no valid solution can be found for a goal, it is safe to assume that
at least one noGood set exists for the goal; there may bemore than
one noGood set and it is useful to report all of them.

The following operation constructs the noGood sets for a goal:

\begin{lstlisting}
context BMM_Model
  @Operation noGoods(goal)
    self.noTacticsOrChildren(goal) +
    self.noEstablishingUnits(goal) +
    self.overusedUnits(goal) +
    self.failingChildren(goal)
  end
\end{lstlisting}There are four reasons for a noGood set. Each reason is implemented
as a separate operation. The first operation constructs a singleton
set Unachievable(g) when the goal g has no tactics and no children:

\begin{lstlisting}
context BMM_Model
  @Operation noTacticsOrChildren(goal)
    @Cmp [Set] Set{Unachievable(goal)} where
      ? goal.noChildren() and 
        goal.noTactics() 
    end
  end
\end{lstlisting}If a goal has a tactic with no establishing units then a noGood set
contains Unestablished(g,t):

\begin{lstlisting}
context BMM_Model
  @Operation noEstablishingUnits(goal)
    @Cmp [Set] Set{Unestablished(goal,tactic)} where
      tactic <- goal.tactics(), 
      ? self.noEstablishingUnit(tactic)
    end 
\end{lstlisting}If a goal has a tactic with a unit then, since we know that there
are no good sets, the unit must be overused (or the noGood set is
incomplete). The following operation records allocations that will
become illegal when combined with other elements:

\begin{lstlisting}
contxt BMM_Model
  @Operation overusedUnits(goal)
    @Cmp [Set] Set{Allocation(goal,tactic,unit)} where
      tactic <- goal.tactics(), 
      unit <- self.establishingUnits(tactic)
    end
\end{lstlisting}Finally, each of the children of a goal must fail in a noGood set.
The following operation calculates the noGood sets for each child
and then combines them:

\begin{lstlisting}
context BMM_Model
  @Operation failingChildren(goal)    
    @Iterate1 child <- goal.children() 
      initially NG = Set{Set{}} do
        @Cmp [Set] options1 + options2 where
          options1 <- self.noGoods(child),
          options2 <- NG
        end
      else Set{}
    end
  end
\end{lstlisting}Finaly, each noGood set must be interpreted to produce an explanation
of why a good set could not be produced. The following operation translates
the noGood sets into strings that explain why each fails:

\begin{lstlisting}
context BMM_Model
  @Operation noGoodDisplays(NGs)
    @Cmp [Set] u.goal.getName() + " has no tactics or children." where 
       NG <- NGs,
       u:Unachievable <- NG
     end +
     @Cmp [Set] u.goal.getName() + 
            " cannot be established by " + 
            u.tactic.getName() + 
            " because it has no supporting units." where 
       NG <- NGs,
       u:Unestablished <- NG 
     end +
     @Cmp [Set] o.getName() + " is overused." where 
       NG <- NGs,
       o:Organization_Unit <- self.getElements().asSeq(), 
       ? @Cmp a where 
           a:Allocation <- NG, 
           ? a.unit = o
         end->size > self.resources().lookup(o.getName()) 
     end
   end
\end{lstlisting}

\chapter{A Command Language}

When modelling applications there is often a requirement for some
kind of command language. UML, for example, has state machines where
the actions on the transitions are written in an action language and
describe what happens to the state of the system when the transitions
are fired. One way of providing support for actions in a model is
to allow the user to write commands in their favourite programming
language; the commands are represented in the model as strings. This
is a very weak form of command language, since the strings cannot
be processed easily by the modelling engines. Typically they cannot
be performed at all until the models are exported as program code,
where the command strings are simply embedded in the output. 

More flexibility is provided by modelling the command language. The
model instances can then be interpreted by an engine as part of performing
the overall model. Furthermore, since the commands are modelled, they
can be analysed in various ways and transformed to different target
implementation languages rather than being limited to one.

This chapter is concerned with writing interpreters (translators and
compilers) for a command language. When defining an interpretive semantics
for a language there is often a similar language at hand (at least
the language in which the interpreter is written).There is a choice
choice as to whether the syntax structures are directly interpreted
or whether they are translated to syntax phrases in the existing language.
The advantage of a translational approach is that the execution engine
for the target language already exists and is probably more efficient
than any interpreter we could write for the source language.

Compilers are translators for languages where the target language
is usually highly optimized and very different from the source language.
In principle, however, there is no essential difference between a
translator and a compiler. A language that uses the translational approach
to target a sub-language of itself is often referred to as \emph{desugaring}
the source language.

\section{XCom - A Simple Command Language}

XCom is a simple command language with values that are either records
or are atomic. An atomic data value is a string, integer or boolean.
A record is a collection of named values. XCom is block-structured
where blocks contain type definitions and value definitions. XCom
has simple control structures: conditional statements and loops. The
following is a simple example XCom program that builds a list of even
numbers from 2 to 100:

\begin{lstlisting}
begin
  type Pair is head tail end
  type Nil is end
  value length is 100 end
  value list is new Nil end
  while length > 0 do
    begin
      if length % 2 = 0
      then
        begin
          value pair is new Pair end
          pair.head := length;
          pair.tail := list;
          list := pair;
        end
      end;
      length := length - 1;
    end
end
\end{lstlisting}XCom expressions evaluate to produce XCom values. Values are defined
in the Values package and which is the 
semantic domain for XCom. Values are either atomic: integers and
booleans, or are records. We use a simple representation for records:
a sequence of values indexed by names. XCom records are created by
instantiating XCom record types. A record type is a sequence of names.
Types raise an interesting design issue: should the types be included
as part of the semantic domain since evaluation of certain XCom program
phrases give rise to types that are used later in the execution to
produce records. The answer to the question involves the phase distinction
that occurs between static analysis (or execution)
and dynamic execution. Types are often viewed
as occurring only during static analysis; although this is not always
the case.

We will show how the semantics of XCom can be defined with and without
dynamic types. All XCom values are instances of sub-classes of the class
Value.

\begin{lstlisting}
context Bool
  @Operation binAnd(Bool(b))
    Bool(value and b)
  end
context Bool
  @Operation binOr(Bool(b))
    Bool(value or b)
  end
context Int
  @Operation binAdd(Int(n))
    Int(value + n)
  end
\end{lstlisting}Record types are sequences of names. Records are sequences of values
indexed by names; the names are found by navigating to the

type of the record:

\begin{lstlisting}
context Record
  @Operation lookup(name:String)
    fields->at(type.names->indexOf(name))
  end
context Record
  @Operation update(name:String,value:Element)
    fields->setAt(type.names->indexOf(name),value)
  end
\end{lstlisting}A new record is produced by performing a new
expression. The type to instantiate is given as a string. An alternative
representation for types in new expressions
would be to permit an arbitrary expression that
evaluates to produce a type. This design choice would rule out static
typing and force the language to have dynamic types. We wish to use
XCom to illustrate the difference between dynamic and static types
in semantic definitions so we use strings to name types in 
new expressions:

The concrete syntax of expressions is defined by the XBNF grammar
for the class Exp:

\begin{lstlisting}
@Grammar 
  // Start at Exp. Logical operators bind weakest.
  Exp ::= e = ArithExp [ op = LogicalOp l = Exp { BinExp(op,e,l) } ].
  LogicalOp ::= 'and' { "and" } | 'or' { "or" }.
  // The '.' for field ref binds tighter than '+' etc.
  ArithExp ::= e = FieldRef [ op = ArithOp a = FieldRef { BinExp(op,e,a) } ].
  ArithOp ::= '+' { "+" }.
  // A field reference '.' optionally follows an atomic expression.
  FieldRef ::= e = Atom ('.' n = Name { FieldRef(e,n) } | { e }).
  // Atomic expressions can be arbitrary exps if in ( and ).
  Atom ::= Const | Var | New | '(' Exp ')'.
  Const ::= IntConst | BoolConst.
  IntConst ::= i = Int { Const(i) }.
  BoolConst ::= 'true' { Const(true) } | 'false' { Const(false) }.
  Var ::= n = Name { Var(n) }.
  New ::= 'new' n = Name { New(n) }.
end
\end{lstlisting}
XCom statements are used to:

\begin{itemize}
\item Introduce new names associated with either types or values.
\item Control the flow of execution.
\item Perform side effects on records.
\end{itemize}
A block (as in Pascal or C) contains local definitions. Names introduced
in a block are available for the rest of the statements in the block
(including sub-blocks) but are not available when control exits from
the block. A declaration introduces either a type or a value binding.
A type declaration associates a type name with a sequence of field
names. To keep things simple we don't associate fields with types.
A value declaration associates a name with a new value. The value
is produced by performing an expression at run-time. A while statement
involves a test and a body. An if statement involves a test, a then-part
and an else-part:

\begin{lstlisting}
@Grammar extends Exp.grammar
  Statement ::= 
    Block 
  | Declaration 
  | While 
  | If 
  | Update 
  | FieldUpdate.
  Block ::= 'begin' s = Statement* 'end' { Block(s) }.
  Declaration ::= 
    TypeDeclaration 
  | ValueDeclaration.
  TypeDeclaration ::= 
    'type' n = Name 'is' ns = Name* 'end' 
    { TypeDeclaration(n,ns) }.
  ValueDeclaration ::= 
    'value' n = Name 'is' e = Exp 'end' 
    { ValueDeclaration(n,e) }.
  FieldUpdate ::= 
    e = Exp '.' n = Name ':=' v = Exp ';' 
    { FieldUpdate(e,n,v) }.
  While ::= 
    'while' e = Exp 'do' s = Statement 'end' 
    { While(e,s) }.
  If ::= 
    'if' e = Exp 
    'then' s1 = Statement 
    'else' s2 = Statement 
    'end' 
    { If(e,s1,s2) }.
  Update ::= 
    n = Name ':=' e = Exp ';' 
    { Update(n,e) }.
end
\end{lstlisting}
\section{An Evaluator for XCom}

As described in the introduction we are interested in defining XCom
operational semantics. We will do this in a number of different ways
in the rest of this note. The first, and possibly most straightforward,
approach is to define an interpreter for XCom
in the XOCL language. This involves writing an 
eval operation for each of the XCom syntax classes. The 
eval operation must be parameterized with respect to any context
information that is required to perform the evaluation. An XCom program
p is then evaluated in a context 
e by: p.eval(e).

\subsection{Evaluating Expressions}

Expression evaluation is defined by adding eval operations to each
class. This section defines the eval operation:

\begin{lstlisting}
context Const
  @Operation eval(env)
    @TypeCase(value)
      Boolean do Bool(value) end
      Integer do Int(value) end
    end
  end
\end{lstlisting}Evaluation of a variable involves looking up the current value. The
value is found in the current context of evaluation: this must contain
associations between variable names and their values:

\begin{lstlisting}
context Var
  @Operation eval(env)
    env.lookup(name)
  end
\end{lstlisting}Evaluation of a binary expression involves evaluation of the sub-expressions
and then selecting an operation based on the operation name. The following
shows how XCom semantics is completely based on XOCL semantics since
+ in XCom is performed by + in XOCL:

\begin{lstlisting}
context BinExp
  @Operation eval(env)
    @Case op of
      "and" do left.eval(env).binAnd(right.eval(env)) end
      "or" do left.eval(env).binOr(right.eval(env)) end
      "+" do left.eval(env).binAdd(right.eval(env)) end
    end
  end
\end{lstlisting}Creation of new records is performed by evaluaing a new expression.
The interpreter has dynamic types so the type to instantiate is found
by looking up the type name in the current environment:

\begin{lstlisting}
context New
  @Operation eval(env)
    env.lookup(type).new()
  end
\end{lstlisting}Field reference is defined as follows:

\begin{lstlisting}
context FieldRef
  @Operation eval(env)
    value.eval(env).lookup(name)
  end
\end{lstlisting}
\subsection{Evaluating Statements}

XCom statements are performed in order to introduce new names, control
flow or to update a record field. Statements are defined to evaluate
in a context and must observe the rules of scope that require variables
are local to the block that introduces them. The context of execution
is an environment; evaluation of a statement may update the supplied
environment, so statement evaluation returns an environment:

\begin{lstlisting}
  context ValueDeclaration
    @Operation eval(env)
      env.bind(name,value.eval(env))
    end
\end{lstlisting}A type declaration extends the supplied environment with a new type:

\begin{lstlisting}
context TypeDeclaration
  @Operation eval(env)
    env.bind(name,Type(names))
  end
\end{lstlisting}A block must preserve the supplied environment when its evaluation
is complete. Each statement in the block is performed in turn and
may update the current environment:

\begin{lstlisting}
context Block
  @Operation eval(originalEnv)
    let env = originalEnv
    in @For statement in statements do
         env := statement.eval(env)
       end
    end;
    originalEnv
  end
\end{lstlisting}A while statement continually performs the body while the test expression
returns true. A while body is equivalent to a block; so any updates
to the supplied environment that are performed by the while body are
discarded on exit:

\begin{lstlisting}
context While
  @Operation eval(originalEnv)
    let env = orginalEnv
    in @While test.eval(env).value do
         env := body.eval(env)
       end;
       originalEnv
    end
  end
\end{lstlisting}An if statement conditionally performs one of its sub-statements:

\begin{lstlisting}
context If
  @Operation eval(env)
    if test.eval(env).value
    then thenPart.eval(env)
    else elsePart.eval(env)
    end
  end
\end{lstlisting}\begin{lstlisting}
context FieldUpdate
  @Operation eval(env)
    record.eval(env).update(name,value.eval(env))
  end
context Update
  @Operation eval(env)
    env.update(name,value.eval(env))
  end
\end{lstlisting}
\section{A Translator for XCom with Run-Time Types}

The previous section defines an interpreter for XCom. This is an appealing
way to define the operational semantics of a language because the
rules of evaluation work directly on the abstract syntax structures.
However the resulting interpreter can often be very inefficient. Furthermore,
an interpreter can lead to an \emph{evaluation phase distinction}.

Suppose that XCom is to be embedded in XOCL. XOCL has its own interpretive
mechanism (the XMF VM); at the boundary between XOCL and XCom the
XOCL interpretive mechanism must hand over to the XCom interpreter
-- the XCom code that is performed is a data structure, a completely
alien format to the VM. This phase distinction can lead to problems
when using standard tools, such as save and load mechanisms, with
respect to the new language. For example a mechanism that can save
XOCL code to disk cannot be used to save XCom code to disk (it can,
however, be used to save the XCom interpreter to disk).

An alternative strategy is to translate the source code of XCom to
a language for which we have an efficient implementation. No new interpretive
mechanism is required and no phase distinction arises. Translation
provides the opportunity for static analysis (since translation is
performed prior to executing the program). As we mentioned earlier,
static analysis can translate out any type information from XCom programs;
the resulting program does not require run-time types.

Since static analysis requires a little more work, this section describes
a simple translation from XCom to XOCL that results in run-time types;
the subsequent section showshow this can be extended to analyse types
statically and remove them from the semantic domain.

\subsection{Translating Expressions}

Translation is defined by adding a new operation desugar\} to each
sbatract syntax class. There is no static analysis, so the operation
does not require any arguments. The result of the operation is a value
of type Performable which is the type of elements that can be executed
by the XMF execution engine. An XCom constant is translated to an
XOCL constant:

\begin{lstlisting}
context Const
  @Operation desugar1():Performable
    value.lift()
  end
\end{lstlisting}An XCom binary expression is translated to an XOCL binary expression.
Note that the sub-expressions are also translated:

\begin{lstlisting}
context BinExp
  @Operation desugar1():Performable
    @Case op of
      "and" do [| <left.desugar1()> and <right.desugar1()> |] end
      "or" do [| <left.desugar1()> and <right.desugar1()> |] end
      "+" do [| <left.desugar1()> + <right.desugar1()> |] end
    end
  end
\end{lstlisting}An XCom new expression involves a type name. Types will be bound to
the appropriate variable name in the resulting XOCL program; so the
result of translation is just a message new sent to the value of the
variable whose name is the type name:

\begin{lstlisting}
context New
  @Operation desugar1():Performable
    [| <OCL::Var(type)>.new() |]
  end
\end{lstlisting}XCom variables are translated to XOCL variables:

\begin{lstlisting}
context Var
  @Operation desugar1():Performable
    OCL::Var(name)
  end
\end{lstlisting}XCom field references are translated to the appropriate call on a
record:

\begin{lstlisting}
context FieldRef
  @Operation desugar1():Performable
    [| <value.desugar1()>.ref(<StrExp(name)>) |]
  end
\end{lstlisting}
\subsection{Translating Statements}

An XCom statement can involve local blocks. The equivalent XOCL expression
that provides local definitions is let. A let expression consists
of a name, a value expression and a body expression. Thus, in order
to translate an XCom declaration to an XOCL let we need to be passed
the body of the let. This leads to a translational style for XCom
commands called \emph{continuation passing} where each desugar1 operation
is supplied with the XOCL command that will be performed next.

A type declaration is translated to a local definition for the type
name. Note that the expression names.lift() translates the sequence
of names to an expression that, when performed, produces the same
sequence of names: list is a means of performing evaluation in reverse:

\begin{lstlisting}
context TypeDeclaration
  @Operation desugar1(next:Performable):Performable
    [| let <name> = Type(<names.lift()>) 
       in <next> 
       end |]
  end
\end{lstlisting}A value declaration is translated to a local decinition:

\begin{lstlisting}
context ValueDeclaration
  @Operation desugar1(next:Performable):Performable
    [| let <name> = <value.desugar1()> 
       in <next> 
       end |]
  end
\end{lstlisting}A block requires each sub-statement to be translated in turn. Continuation
passing allows us to chain together the sequence of statements and
nest the local definitions appropriately. The following auxiliary
operation is used to implement block-translation:

\begin{lstlisting}
context Statements
  @Operation desugar1(statements,next:Performable):Performable
    @Case statements of
      Seq{} do 
        next 
      end
      Seq{statement | statements} do 
        statement.desugar1(desugar1(statements,next)) 
      end
    end
  end
\end{lstlisting}Translation of a block requires that the XOCL local definitions are
kept local. Therefore, the sub-statements are translated by chaining
them together and with a final continuation of null. Placing the result
in sequence with next ensures that any definitions
are local to the block.

\begin{lstlisting}
context Block
  @Operation desugar1(next:Performable):Performable
    [| <desugar1(statements,[| null |])> ; 
    <next> |]
end
\end{lstlisting}A while statement is translated to the equivalent expression in XOCL:

\begin{lstlisting}
context While
  @Operation desugar1(next:Performable):Performable
    [| @While <test.desugar1()>.value do
         <body.desugar1([|null|])>
       end;
       <next> |]
end
\end{lstlisting}An if statement is translated to an equivalent expression in XOCL:

\begin{lstlisting}
context If
  @Operation desugar1(next:Performable):Performable
    [| if <test.desugar1()>.value
       then <thenPart.desugar1(next)>
       else <elsePart.desugar1(next)>
       end |]
end
\end{lstlisting}\begin{lstlisting}
context FieldUpdate
@Operation desugar1(next:Performable):Performable
  [| <record.desugar1()>.update(<StrExp(name)>,<value.desugar1()>);
     <next> |]
end
context Update
  @Operation desugar1(next:Performable):Performable
    [| <name> := <value.desugar1()>;
       <next> |]
end
\end{lstlisting}
\section{A Translator for XCom without Run-Time Types}

It is usual for languages to have a static (or \emph{compile time})
phase and a dynamic (or \emph{run time}) phase. Many operational features
of the language can be performed statically. This includes type analysis:
checking that types are defined before they are used and allocating
appropriate structures when instances of types are created. This section
shows how the translator for XCom to XOCL from the previous section
can be modified so that type analysis is performed and so that types
do not occur at run-time.

\subsection{Translating Expressions}

Since types will no longer occur at run-time we will simplify the
semantic domain slightly and represent records as \emph{a-lists}.
An a-list is a sequence of pairs, the first element of each pair is
a key and the second element is a value. In this case a record is
an a-list where the keys are field name strings. XOCL provides operations
defined on sequences that are to be used as a-lists: {\tt l->lookup(key)}
and {\tt l->set(key,value)}.

The context for static analysis is a type environment. Types now occur
at translation time instead of run-time therefore that portion of
the run-time context that would contain associations between type
names and types occurs during translation. Translation of a constant
is as for desugar1. Translation of binary expressions is as for desugar1
except that all translation is performed by desugar2:

\begin{lstlisting}
context BinExp
  @Operation desugar2(typeEnv:Env):Performable
    @Case op of
      "and" do 
        [| <left.desugar2(typeEnv)> and
           <right.desugar2(typeEnv)> |] 
      end
      "or" do 
        [| <left.desugar2(typeEnv)> or 
           <right.desugar2(typeEnv)> |] 
      end 
      "+" do 
        [| <left.desugar2(typeEnv)> + 
           <right.desugar2(typeEnv)> |] end
      end
  end
\end{lstlisting}Translation of a variable is as before. A new expression involves
a reference to a type name. The types occur at translation time and
therefore part of the evaluation of new can occur during translation.
The type should occur in the supplied type environment; the type contains
the sequence of field names. The result of translation is an XOCL
expression that constructs an a-list based on the names of the fields
in the type. The initial value for each field is null:

\begin{lstlisting}
context New
  @Operation desugar2(typeEnv:Env):Performable
    if typeEnv.binds(type)
    then
      let type = typeEnv.lookup(type)
      in type.names->iterate(name exp = [| Seq{} |] | 
          [| <exp>->bind(<StrExp(name)>,null) |])
      end
    else self.error("Unknown type " + type)
    end
  end
\end{lstlisting}A field reference expression is translated to an a-list lookup expression:

\begin{lstlisting}
context FieldRef
  @Operation desugar2(typeEnv:Env):Performable
    [| <value.desugar2(typeEnv)>->lookup(<StrExp(name)>) |] 
  end
\end{lstlisting}
\subsection{Translating Statements}

A statement may contain a local type definition. We have already discussed
continuation passing with respect to desugar1 where the context for
translation includes the next XOCL expression to perform. The desugar2
operation cannot be supplied with the next XOCL expression because
this will depend on whether or not the current statement extends the
type environment. Therefore, in desugar2 the continuation is an operation
that is awaiting a type environment and produces the next XOCL expression.
A type declaration binds the type at translation time and supplies
the extended type environment to the continuation:

\begin{lstlisting}
context TypeDeclaration
  @Operation desugar2(typeEnv:Env,next:Operation):Performable
    next(typeEnv.bind(name,Type(names)))
  end
\end{lstlisting}A value declaration introduces a new local definition; the body is
created by supplying the unchanged type environment to the continuation:

\begin{lstlisting}
context ValueDeclaration
  @Operation desugar2(typeEnv:Env,next:Operation):Performable
    [| let <name> = <value.desugar2(typeEnv)> 
       in <next(typeEnv)> 
       end |]
  end
\end{lstlisting}Translation of a block involves translation of a sequence of sub-statements.
The following auxiliary operation ensures that the continuations are
chained together correctly:

\begin{lstlisting}
context Statements
  @Operation desugar2(statements,typeEnv,next):Performable
    @Case statements of
      Seq{} do 
        next(typeEnv)
      end
      Seq{statement | statements} do 
        statement.desugar2(
          typeEnv,
          @Operation(typeEnv)
            desugar2(statements,typeEnv,next)
          end) 
      end
    end
  end
\end{lstlisting}A block is translated to a sequence of statements where local definitions
are implemented using nested let expressions in XOCL. The locality
of the definitions is maintained by sequencing the block statements
and the continuation expression:

\begin{lstlisting}
context Block
  @Operation desugar2(typeEnv:Env,next:Operation):Performable
    [| <desugar2(
         statements,
         typeEnv,
         @Operation(ignore) 
           [| null |] 
         end)>;
       <next(typeEnv)> |]
  end
\end{lstlisting}A while statement is translated so that the XOCL expression is in
sequence with the expression produced by the contintuation:

\begin{lstlisting}
context While
  @Operation desugar2(typeEnv:Env,next:Operation):Performable
    [| @While <test.desugar2(typeEnv)>.value do
         <body.desugar2(typeEnv,@Operation(typeEnv) [| null |] end)>
       end;
       <next(typeEnv)> |]
  end
\end{lstlisting}The if statement is translated to an equivalent XOCL expression:

\begin{lstlisting}
context If
  @Operation desugar2(typeEnv:Env,next:Operation):Performable
    [| if <test.desugar2(typeEnv)>.value
       then <thenPart.desugar2(typeEnv,next)>
       else <elsePart.desugar2(typeEnv,next)>
       end |]
  end
\end{lstlisting}\begin{lstlisting}
context FieldUpdate
  @Operation desugar2(typeEnv:Env,next:Operation):Performable
    [| <record.desugar2(typeEnv)>.update(
         <StrExp(name)>,
         <value.desugar2(typeEnv)>);
       <next(typeEnv)> |]
  end

context Update
  @Operation desugar2(typeEnv:Env,next:Operation):Performable
    [| <name> := <value.desugar2(typeEnv)>;
       <next(typeEnv)> |]
  end
\end{lstlisting}
\section{Compiling XCom}

The previous section shows how to perform static type anslysis while
translating XCom to XOCL. XOCL is then translated to XMF VM instructions
by the XOCL compiler (another translation process). The result is
that XCom cannot to anything that XOCL cannot do. Whilst this is not
a serious restriction, there may be times where a new language wishes
to translate directly to the XMF VM that XOCL does not support. This
section shows how XCom can be translated directly to XMF VM instructions.

\subsection{Compiling Expressions}

\begin{lstlisting}
context Exp
  @AbstractOp compile(typeEnv:Env,valueEnv:Seq(String)):Seq(Instr)
  end
\end{lstlisting}\begin{lstlisting}
context Const
  @Operation compile(typeEnv,valueEnv)
    @TypeCase(value)
      Boolean do 
        if value 
        then Seq{PushTrue()} 
        else Seq{PushFalse()} 
        end 
      end
      Integer do 
        Seq{PushInteger(value)}
      end
    end
  end
\end{lstlisting}\begin{lstlisting}
context Var
  @Operation compile(typeEnv,valueEnv)
    let index = valueEnv->indexOf(name)
    in if index < 0
       then self.error("Unbound variable " + name)
       else Seq{LocalRef(index)}
       end
    end
  end
\end{lstlisting}\begin{lstlisting}
context BinExp
  @Operation compile(typeEnv,valueEnv):Seq(Instr)
    left.compile(typeEnv,valueEnv) + 
    right.compile(typeEnv,valueEnv) +
    @Case op of
      "and" do Seq{And()} end
      "or" do Seq{Or()} end
      "+" do Seq{Add()} end
    end
  end
\end{lstlisting}\begin{lstlisting}
context New
  @Operation compile(typeEnv,valueEnv):Seq(Instr)
    self.desugar2(typeEnv).compile()
  end
\end{lstlisting}\begin{lstlisting}
context FieldRef
  @Operation compile(typeEnv,valueEnv):Seq(Instr)
    Seq{StartCall(),
        PushStr(name)} +
    value.compile(typeExp,valueExp) +
    Seq{Send("lookup",1)}
  end
\end{lstlisting}
\subsection{Compiling Statements}

\begin{lstlisting}
context Statement
  @AbstractOp compile(typeEnv:Env,varEnv:Seq(String),next:Operation):Seq(Instr)
  end
\end{lstlisting}\begin{lstlisting}
context TypeDeclaration
  @Operation compile(typeEnv,varEnv,next)
    next(typeEnv.bind(name,Type(names)),varEnv)
  end
\end{lstlisting}\begin{lstlisting}
context ValueDeclaration
  @Operation compile(typeEnv,varEnv,next)
    value.compile(typeEnv,varEnv) +
    Seq{SetLocal(name,varEnv->size),
        Pop()} +
    next(typeEnv,varEnv + Seq{name})
  end
\end{lstlisting}\begin{lstlisting}
context Statements
  @Operation compile(statements,typeEnv,varEnv,next)
    @Case statements of
      Seq{} do 
        next(typeEnv,varEnv) 
      end
      Seq{statement | statements} do 
        statement.compile(
          typeEnv,
          varEnv,
          @Operation(typeEnv,varEnv)
            compile(statements,typeEnv,varEnv,next)
          end) 
     end
   end
  end
\end{lstlisting}\begin{lstlisting}
context Block
  @Operation compile(typeEnv,varEnv,next)
    compile(
     statements,
     typeEnv,
     varEnv,
       @Operation(localTypeEnv,localVarEnv) 
         next(typeEnv,varEnv) 
       end)
   end
\end{lstlisting}\begin{lstlisting}
context While
  @Operation compile(typeEnv,varEnv,next)
    Seq{Noop("START")} +
    test.compile(typeEnv,varEnv) +
    Seq{SkipFalse("END")} +
    body.compile(typeEnv,varEnv,
      @Operation(typeEnv,varEnv) 
        Seq{} 
      end) +
    Seq{Skip("START")} +
    Seq{Noop("END")} +
    next(typeEnv,varEnv)
  end
\end{lstlisting}

\section{Flow Graphs}

The previous section has shown how to represent the abstract syntax
of XCom as textual concrete syntax. We would like to compare this
to a graphical concrete syntax for XCom and show how the XCom programs
can be developed graphically rather than textually.

\emph{Flow Graphs} are a standard way of representing imperative programs.
Flow graph nodes are labelled with either program statements or boolean
expressions. Flow graph edges are labelled with next if the edge represents
the control flow between statements, true if the edge represents the
control flow arising when a test succeeds, and false if the edge represents
the flow from a test failure.

\begin{figure}
\begin{center}

\includegraphics[width=12cm]{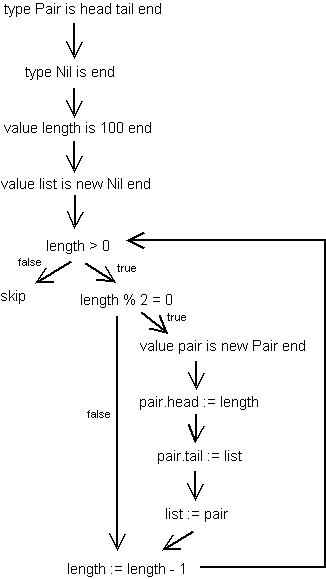}

\caption{\label{fig:An-Example-Flow}An Example Flow Graph}

\end{center}
\end{figure}

Figure \ref{fig:An-Example-Flow}shows an example flow graph corresponding
to the original example XCom program. Note that a while loop is represented
as a test node whose true edge eventually leads back to the test node.
XMF provides a packages called Graph that provides a basic collection
of graph structures and operations. A graph node is created: Node(l)
where l is the label on the graph node. A graph edge is created: 
Edge(l,s,t) where l is the label on the edge,
s is the source node for the edge and 
t is the target node for the edge. A graph is created: 
Graph(N,E) where N is a set of nodes and
E is a set of edges. The graph shown in figure
\ref{fig:An-Example-Flow} is constructed as follows where classes
Node and Edge have been
specialized appropriately:

\begin{lstlisting}
let n1 = Statement("type Pair is head tail end");
n2 = Statement("type Nil is end");
n3 = Statement("value length is 100 end");
n4 = Statement("value list is new Nil end");
n5 = Guard("length > 0");
n6 = Guard("length % 2 = 0");
n7 = Statement("value pair is new Pair end");
n8 = Statement("pair.head := length;");
n9 = Statement("pair.tail := list;");
n10 = Statement("list := pair;");
n11 = Statement("length := length - 1;");
n12 = Statement("skip") then
e1 = Next(n1,n2);
e2 = Next(n2,n3);
e3 = Next(n3,n4);
e4 = Next(n4,n5);
e5 = True(n5,n6);
e6 = True(n6,n7);
e7 = Next(n7,n8);
e8 = Next(n8,n9);
e9 = Next(n9,n10);
e11 = Next(n10,n11);
e12 = Next(n11,n5);
e13 = False(n6,n11);
e14 = False(n5,n12)
in Graph(Set{n1,n2,n3,n4,n5,n6,n7,n8,n9,n10,n11,n12},
Set{e1,e2,e3,e4,e5,e6,e7,e8,e9,10,e11,e12,e13,e14})
end
\end{lstlisting}Graphs provide an operation reduce that takes
a node n and a sub-graph 
H such that G.reduce(n,H) produces a new
graph that is the result of removing H from
G, adding  n to the new
graph and re-linking to n any edges left dangling
when H is removed.

\subsection{Flow Graph Patterns}

Flow graphs can represent the control flow of a wide variety of different
languages and are used to analyse the structure of programs. The basic
elements of flow graphs are very low level - just nodes (representing
statements and expressions) and edges (representing flow between computational
states and the outcome of tests).

The low-level nature of flow-graphs gives rise to a weak correspondence
language features. This is done by identifying patterns in flow-graphs;
wherever the pattern occurs, it can be replaced by a single node.
The choice of patterns depends on the collection of program features
that we intend to work with. For the purposes of XCom we will capture
the following patterns: P is a pair of statements
performed in sequence; C is an if statement
and  W is a while-loop. 

\begin{figure}
\begin{center}

\includegraphics[width=12cm]{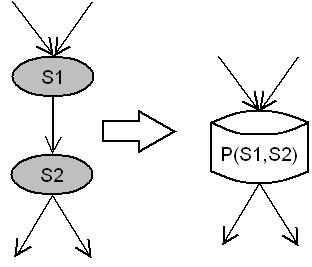}

\caption{A P-Reduction Pattern\label{fig:A-P-Reduction-Pattern}}

\end{center}
\end{figure}

The rest of this section shows how the pattern matching facilities
of XMF can be used to detect these patterns. We define a flow-graph
aspect to XMF graphs by adding reduction operations
to the class Graph. The simplest case is detecting
and reducing a P node. This occurs when two
statements are linked via a next edge. The
reduction is shown in figure \ref{fig:A-P-Reduction-Pattern}.

\begin{figure}
\begin{center}

\includegraphics[width=12cm]{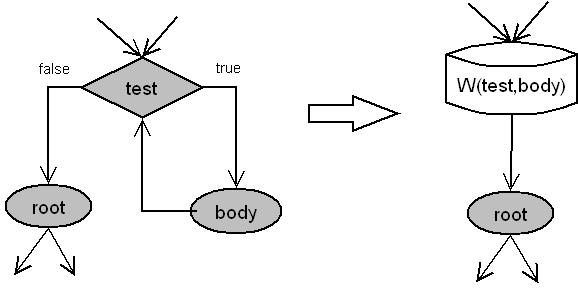}

\caption{A W-Reduction Pattern\label{fig:A-W-Reduction-Pattern}}

\end{center}
\end{figure}

\begin{lstlisting}
context Graph
  @Operation reduceP():Graph
    @Case self of
      // Find two nodes with a then-edge
      // between them, replace with a P
      // node...
      Graph(N->including(n1 = Statement())
             ->including(n2 = Statement()),
            E->including(e = Next())
        when e.source() = n1 and
             e.target() = n2) do
        self.reduce(P(e),Graph(Set{n1,n2},Set{e}))
      end
      // Otherwise, leave the receiver unchanged...
      else self
    end
  end
\end{lstlisting}A W pattern corresponds to a
while loop and occurs when a test node has a true outcome that is
a body statement whose next leads back to the test. The false outcome
leads to some root node for the rest of execution. The reduction involves
replacing the test and its body with a W node
and adding a new next edge from the W node
to the root. The reduction is shown in figure \ref{fig:A-W-Reduction-Pattern}.

\begin{figure}
\begin{center}

\includegraphics[width=12cm]{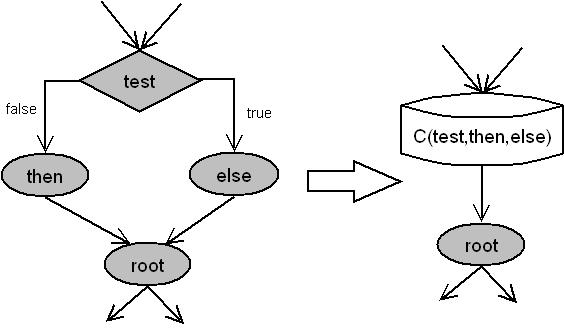}

\caption{A C-Reduction Pattern\label{fig:A-C-Reduction-Pattern}}

\end{center}
\end{figure}

\begin{lstlisting}
context Graph
  @Operation reduceW():Graph
    @Case self of
      // Find a test node and a statement that are linked
      // via a true outcome and a loop back to the test.
      // Replace them with a W node ...
      Graph(N->including(test = Guard())
             ->including(body = Statement())
             ->including(root),
            E->including(enter = True())
             ->including(loop = Next())
             ->including(exit = False())
        when enter.source() = test and
             enter.target() = body and
             exit.source() = test and
             exit.target() = root and
             loop.source() = body and
             loop.target() = test) do 
        let Wnode = W(test,body);
            H = Graph(Set{test,body),Set{enter,loop,exit}) then
            newEdge = Next(Wnode,root)
        in self.reduce(Wnode,H).addEdge(newEdge)
        end
      end
      // Otherwise, leave the receiver unchanged...
      else self
    end
  end
\end{lstlisting}A C pattern corresponds to an if statement and occurs when there is
a test node leading to two nodes both of which lead to a single root
node. This forms a diamond pattern with the test node at one end.
The test, then and else parts of the graph are removed and replaced
with a C node. The reduction is shown in figure \ref{fig:A-C-Reduction-Pattern}.

\begin{lstlisting}
context Graph
  @Operation reduceC():Graph
    @Case self of
      Graph(N->including(test = Guard())
             ->including(thenNode)
             ->including(elseNode)
             ->including(root),
            E->including(trueEdge = True())
             ->including(falseEdge = False())
             ->including(endThen = Next())
             ->including(endElse = Next())
        when trueEdge.source() = test and
             trueEdge.target() = thenNode and
             falseEdge.source() = test and
             falseEdge.target() = elseNode and
             endThen.source() = thenNode and
             endThen.target() = root and
             endElse.source() = elseNode and
             endElse.target() = root) do
       let Cnode = C(trueEdge,falseEdge);
           H = Graph(Set{test,thenNode,elseNode},
                     Set(trueEdge,falseEdge,endThen,endElse}) then
           e = Next(Cnode,root)
       in self.reduce(Cnode,H).addEdge(e)
       end
     end
     else self
    end
  end
\end{lstlisting}Equality on graphs is defined in terms of the labels on the nodes
and edges. Two graphs G and H are equal when they have the same number
of nodes, when the respective sets of labels are equal, when the edges
of G correspond to the edges of 
H in terms of the source and target nodes and when the respective
sets of edge labels are equal. 

Graph reduction can involve more than one step. For example, certain
configurations of graph require a P reduction
to occur before a C reduction can occur. In
order to reduce a graph fully we continually perform a reduction until
no further reductions can take place:

\begin{lstlisting}
context Graph
  @Operation reduce():Graph
    let G = self.reduceP().reduceW().reduceC()
    in if G.equals(self)
       then G
       else G.reduce()
       end
    end
  end
\end{lstlisting}

\appendices
\addcontentsline{toc}{part}{Appendix}

\chapter{Implementation Issues}

\section{Working with Files}

XMF is a dynamic engine that runs program code loaded from files.
Files can be loaded at any time and the definitions come into immediate
effect. It is usual for the files to be loaded form the top-level
command loop (although loading files from an executing program is
also possible). Files have a '.xmf.' extension and have a specific
format:

\begin{lstlisting}
<PARSER_IMPORT>*

<IMPORT>*

(<DEFINITIONS> | <COMMAND>)*
\end{lstlisting}The parse imports are used to place syntax classes into scope for
the rest of the file. If a syntax class C is in scope then the file
may contain constructs of the form @C ... end. When the XMF parser
encounters such a construct it looks up the syntax classes in scope,
finds a class named C and uses the grammar of C to parse the construct
and to synthesize a performable element. Each parse import has the
form:

\begin{lstlisting}
parserImport <PATH>;
\end{lstlisting}where a path references a name-space containing syntax classes. Typically
a file will contain a single parser-import for the name-space XOCL;
this name-space contains all the syntax classes for XOCL definitions
such as For, Class, Package, attribute etc.

The imports are used to place named elements in scope for the rest
of the file. Each import references a name-space whose named elements
can be referenced within the file without further qualification. Each
import has the form:

\begin{lstlisting}
import <PATH>;
\end{lstlisting}A definition is used to add an element to a container; typically adding
a named element to a name-space. A definition has the form:

\begin{lstlisting}
context <PATH>
  <EXP>
\end{lstlisting}where the path references a name-space and the expression evaluates
to produce a named element. On loading the file the expression is
evaluated, added to the name-space and then initialised (causing any
internal references to be resolved). Typically, definitions are used
to create class and package definitions. 

Definitions are also used to add operations to classes. It is usual
to spread the definition of a package and a class over several files,
each file contains the definition of various related features. For
example a package may be defined in a file, its classes in several
files and then the operations for each class in several further files.
When doing this, care should be taken to load the files in an appropriate
order so that the name-spaces referenced in a definition are available
when the file is loaded.

Commands in files are just XOCL expressions terminated by a ';'. Typically
commands are used for their side-effect.

Here is a typical file:

\begin{lstlisting}
parserImport XOCL;

import MyModels;
import MyPackage;

// Add a new class. Notice that MyClass
// is imported and needs no further
// qualification...

context MyPackage
  @Class NewClass extends MyClass
    @Attribute x : MyClass end
  end

// Extend MyModels::MyPackage::MyClass...

context MyClass
  @Operation myOp()
    // Some Body
  end
\end{lstlisting}A source file is compiled using the operation:

\begin{lstlisting}
Compiler::compileFile(String,Boolean,Boolean)
\end{lstlisting}This operation takes three arguments: the path to the source file
and two boolean values both of which should be true. If the file compiles
correctly then the resulting binary file has an extension '.o' and
can be loaded using String::loadBin(). The top-level command loop
has some useful commands that make it easy to compile and load files.
The paths are supplied without file extensions:

\begin{lstlisting}
?c  <PATH>     // Compile the source file.
?cl <PATH>     // Compile then load the source file.
?l  <PATH>     // Load gthe binary file.
?c             // Compile the last supplied path.
?l             // Load the last supplied path.
\end{lstlisting}
\section{Manifests}

XOCL applications tend to run to a number of files. It is possible
to write a single file containing directives to compile and load the
application files in the correct order, however this becomes difficult
to manage since paths have to be absolute etc. The manifest language
construct works like a make-file: it is used to list all the files
in a module and all the sub-modules. The manifest uses paths relative
to a root so that the module is easily relocatable. In addition manifests
include some useful directives that can be used to control the compilation
and loading of the module.

A manifest file is called Manifest.xmf and has the following structure:

\begin{lstlisting}
parserImport Manifests;

@Manifest <NAME>
  <DIRECTIVE>*
end;
\end{lstlisting}The name of the manifest is just used for documentation. A manifest
file occurs in a directory, all directives in the manifest are relative
to this directory. A directive refers either to a file or a directory.
A file directive has the following form:

\begin{lstlisting}
@File [^<LOADTIME>] <PATH> end
\end{lstlisting}The optional load-time directive specifies when the file should be
loaded. A manifest is used to both compile and load files; load-time
directives are one of:

CompileTime; RunTime; Both; None. The path to the file is relative
to the directory containing the manifest and should be specified with
no file extension. Whe the manifest is used to compile entries, a
file is compiled if the source (.xmf) is out of date with respect
to the binary (.o). Once compiled if the load-time directive is specified
as CompileTime or Both then the file is loaded. When the manifest
is used to load entries, the binary is loaded providing that the load-time
directive is not specified or is RunTime or Both.

A directory directive has the following form:

\begin{lstlisting}
@Ref [^<LOADTIME>] <PATH> end
\end{lstlisting}where the path references a sub-directory that contains a manifest
file. On compilation, the referenced manifest is loaded and its entries
compiled. On loading, the referenced manifest is loaded and its entries
are loaded.

The following is a typical manifest:

\begin{lstlisting}
parserImport Manifests;

@Manifest MyModule
  @File MyPackage end
  @File MyTopLevelClass end
  @File SomeOperations end
  @Ref MyFirstSubPackage end
  @Ref MySecondSubPackage end
end;
\end{lstlisting}Manifests can be used under program control:

\begin{lstlisting}
// Load the manifest file...
let m = "someDir/Manifest.o".loadBin()
in // Use the manifest to compile the entries.
   // Supply the directory containing the 
   // manifest...
   m.build("someDir");
   // Use the manifest to load the entries...
   m.load("someDir")
end
\end{lstlisting}The top-level command loop provides commands that make it easy to
use manifests. In the following, <PATH> refers to a directory containing
a manifest file:

\begin{lstlisting}
?m b  <PATH>   // Compile entries.
?m l  <PATH>   // Load entries.
?m d  <PATH>   // Delete binaries.
?m bl <PATH>   // Build then load entries.
\end{lstlisting}
\section{XMF Startup Arguments}

XMF can be started from a batch file or from Java. In both cases initialisation
arguments are supplied to the VM as it starts up. These arguments
are detailed below. Subsequent sections describe how these arguments
are supplied to the different ways of starting XMF.

\begin{itemize}
\item -arg {\tt <NAME>:<VALUE>} The global object named xmf contains a collection
of name/value pairs that are supplied via this initialzation argument
designator. The name and the value are separated using :. The collection
of supplied arguments can be referenced as xmf.startupArgs(). Given
a name n for an argument, the value is found as {\tt xmf.startupArgs()->lookup(n)}.
\item -filename {\tt <FILE>}. Specifies a binary file containing a replacement
for the XMF top-level command loop when running the evaluator image
(see below). When starting the evaluator image, if this argument designator
is supplied then the contents are loaded and executed instead of starting
the command loop. When the file completes then the XMF VM terminates.
\item -heapSize {\tt <SIZE>}. The size of the XMF heap is set using this initialisation
argument designator. The size should be a value in XMF VM words and
must be sufficiently large to accommodate any specified image bearing
in mind that the Java VM will manage two heaps (for garbage collection)
of the specified size. A standard size is 10000.
\item -initfile {\tt <FILE>}. When the eval.img image starts (see below) it will,
by default, start a top-level command loop. Before the loop starts
the supplied binary file is loaded.
\item -image {\tt <FILE>}. When the XMF VM starts it loads a saved XMF image into
the heap and continues to run the image from the point at which the
image was saved. This argument designator specified the file containing
the image. For the standard command-line XMF engine, use Images/eval.img.
\item -port {\tt <INT>}. Specifies a port that is used by the XMV VM for socket
connections.
\item -stackSize {\tt <SIZE>}. Specifies the size of the XMF VM stack in machine
words. XMF implements tail calling, so it is not usually necessary
to specify this argument designator.
\end{itemize}
A minimal startup is:

\begin{lstlisting}
-image %XMFHOME%\Images\eval.img -heapSize 10000
\end{lstlisting}
\section{Starting XMF using a Batch File}

There is an example startup file bin/xos.bat for starting the XMF
VM with the top-level command loop defined in eval.img. Run this supplying
the value of XMFHOME to start a top-level command loop:

\begin{lstlisting}
%XMFHOME%\bin\xos %XMFHOME%
\end{lstlisting}or by supplying a filename to replace the top-level loop:

\begin{lstlisting}
%XMFHOME%\bin\xos %XMFHOME% -filename <FILE>
\end{lstlisting}
\section{Starting XMF from Java}

The XMF VM can be started from Java as follows:

\begin{lstlisting}
String[] args = {"-heapSize","10000","-image",XMFHOME + "/Images/eval.img"};
XOS.OperatingSystem os = new XOS.OperatingSystem();
os.init(args)
\end{lstlisting}Different combinations of arguments can be supplied as required.

\section{Rebuilding XMF}

Rebuilding XMF is in two stages (for mainly historical reasons). Assuming
you have all the binaries recompiled (see the next section):

\begin{lstlisting}
%XMFHOME%\bin\makexmf %XMFHOME%
\end{lstlisting}The above creates a new image in XMFHOME\textbackslash{}Images\textbackslash{}xmf.img.
This is a basic image that supports XOCL etc. The next step extends
this image by loading libraries most of which are essential to run
the top-level command loop:

\begin{lstlisting}
%XMFHOME%\bin\makexmf %XMFHOME%
\end{lstlisting}A fresh XMFHOME\textbackslash{}Images\textbackslash{}eval.img is created.

\section{Recompiling XMF}

XMF can be recompiled only from a valid XMF top-level command loop
which requires a valid eval.img. This is done as follows:

\begin{lstlisting}
C:\xmf>bin\xos .
[ bin/xos .         ]
[ Starting XOS ]
[ Load ./Images/eval.img ]

XMF Copyright (C) Xactium Ltd, 2003-2007.

Version 1.9

Type ?h for top level help.

[1] XMF> ?m b .
[ Loading ./Manifest.o...................... 0:0:0:109 ms ]
[ Loading ./Kernel/Manifest.o............... 0:0:0:125 ms ]
[ ./Kernel/Boot.o......................... is up to date. ]
[ ./Kernel/Kernel.o....................... is up to date. ]
[ ./Kernel/Doc.o.......................... is up to date. ]
[ ./Kernel/Attribute.o.................... is up to date. ]
[ ./Kernel/Bind.o......................... is up to date. ]
[ ./Kernel/Boolean.o...................... is up to date. ]
[ ./Kernel/Boot.o......................... is up to date. ]
[ ./Kernel/Buffer.o....................... is up to date. ]
[ ./Kernel/Class.o........................ is up to date. ]
[ ./Kernel/Classifier.o................... is up to date. ]
// Lots more compilation printout....
\end{lstlisting}If you modify the XMF sources in any way then you will need to recompile
and the rebuild to create a fresh image. The rebuilsing process is
describes in the previous section.

\section{Debugging}

Operations can be traced so that information is printed out when the
operation is entered and when it is exited. The output is intended
to reflect the call nesting. Use Operation::trace() to set tracing
and Operation::untrace() to remove tracing. Use Container::traceAll()
to trace all operations in a name-space and Container::untraceAll()
to remove all tracing.

\section{Boolean Operations}

The following list shows the main operations defined by the datatype
Boolean:

\begin{lstlisting}
Boolean::booland(other:Boolean):Boolean
Boolean::boolor(other:Boolean):Boolean
Boolean::boolnot():Boolean
\end{lstlisting}
\section{Float Operations}

The following list shows the main operations defined by the datatype
Float:

\begin{lstlisting}
Float::abs():Integer
Float::add(other::Float):Float
Float::ceiling():Integer
Float::cos():Float
Float::floor():Integer
Float::random():Float
Float::sun():Float
Float::slash(other:Float):Float
\end{lstlisting}
\section{Integer Operations}

The following list shows the main operations defined by the datatype
Integer:

\begin{lstlisting}
Integer::add(other:Integer):Integer
Integer::bit(offset:Integer):Integer
Integer::byte(offset:Integer):Integer
Integer::div(other:Integer):Integer
Integer::greater(other:Integer):Integer
Integer::isAlphaChar():Boolean
Integer::isLowerCaseChar():Boolean
Integer::isnewLineChar():Boolean
Integer::isNumericChar():Boolean
Integer::isUpperCaseChar():Boolean
Integer::isWhiteSpaceChar():Boolean
Integer::less(other:Integer):Boolean
Integer::lsh(offset):Integer
Integer::max(other:Integer):Integer
Integer::min(other:Integer):Integer
Integer::mod(other:Integer):Integer
Integer::mul(other:Integer):Integer
Integer::pow(power:Integer):Integer
Integer::random(upper:Integer):Integer
Integer::rsh(offset:Integer):Integer
Integer::slash(Integer):Float
Integer::sqrt():Float
Integer::sub(other:Integer):Integer
Integer::to(upper:Integer):Seq(Integer)
\end{lstlisting}
\section{Sequence Operations}

The following list shows the main operations defined by the datatype
Seq(Element):

\begin{lstlisting}
Seq(Element)::append(other:Seq(Element)):Seq(Element)
Seq(Element)::asSeq():Seq(Element)
Seq(Element)::asSet()::Set(Element)
Seq(Element)::asString():String
Seq(Element)::asVector():Vector
Seq(Element)::assoc(key):Seq(Element)
Seq(Element)::at(i:Integer)
Seq(Element)::bind(key,value):Seq(Element)
Seq(Element)::butLast():Seq(Element)
Seq(Element)::collect(pred:Operation):Seq(Element)
Seq(Element)::contains(element):Boolean
Seq(Element)::delete(element):Seq(Element)
Seq(Element)::drop(n:Integer):Seq(Element)
Seq(Element)::excluding(element):Seq(Element)
Seq(Element)::exists(pred:Operation):Boolean
Seq(Element)::flatten():Seq(Element)
Seq(Element)::forAll(pred:Operation):Boolean
Seq(Element)::hasPrefix(p:Seq(Element)):Boolean
Seq(Element)::head()
Seq(Element)::includes(Element):Boolean
Seq(Element)::includesAll(other:Seq(Element):Boolean
Seq(Element)::including(element):Seq(Element)
Seq(Element)::indexOf(element):Integer
Seq(Element)::insertAt(element,index:Integer):Seq(Element)
Seq(Element)::intersection(other:Seq(Element)):Seq(Element)
Seq(Element)::isEmpty():Boolean
Seq(Element)::isProperSequence():Boolean
Seq(Element)::iter(iter:Operation,start)
Seq(Element)::last()
Seq(Element)::lookup(key)
Seq(Element)::map(name:String):Seq(Element)
Seq(Element)::max():Integer
Seq(Element)::mul(other:Seq(Element)):Seq(Element)
Seq(Element)::prepend(element):Seq(Element)
Seq(Element)::qsort():Seq(Element)
Seq(Element)::sortNamedElements:Seq(Element)
Seq(Element)::subSequence()
Seq(Element)::subst(new,old,all):Seq(Element)
Seq(Element)::tail()
Seq(Element)::take(n:Integer):Seq(Element)
Seq(Element)::unbind(key):Seq(Element)
Seq(Element)::zip(other:Seq(Element)):Seq(Element)
\end{lstlisting}
\section{Set Operations}

The following list shows themain operations defined by the datatype
Set(Element):

\begin{lstlisting}
Set(Element)::asSeq():Seq(Element)
Set(Element)::asSet()::Set(Element)
Set(Element)::collect(pred:Operation):Set(Element)
Set(Element)::contains(element):Boolean
Set(Element)::excluding(element):Set(Element)
Set(Element)::exists(pred:Operation):Boolean
Set(Element)::flatten():Set(Element)
Set(Element)::forAll(pred:Operation):Boolean
Set(Element)::includes(Element):Boolean
Set(Element)::includesAll(other:Set(Element):Boolean
Set(Element)::including(element):Set(Element)
Set(Element)::intersection(other:Set(Element)):Set(Element)
Set(Element)::isEmpty():Boolean
Set(Element)::iter(iter:Operation,start)
Set(Element)::map(name:String):Set(Element)
Set(Element)::max():Integer
Set(Element)::mul(other:Set(Element)):Set(Element)
Set(Element)::power():Set(Element)
Set(Element)::sel()
Set(Element)::size():Integer
Set(Element)::union(other::Set(Element)):Set(Element)
\end{lstlisting}
\section{String Operations}

The following list shows the main operations defined by the datatype
String:

\begin{lstlisting}
String::asBool():Boolean
String::asFloat():Float
String::asInt():Integer
String::asSeq():Seq(Integer)
String::asXML():XML::Element
String::at(i:Integer):Integer
String::fileExists():Boolean
String::greater(other:String):Boolean
String::hasPrefix(p:String):Boolean
String::hasSuffix(s:String):Boolean
String::isDir():Boolean
String::loadBin()
String::lowerCase():String
String::lowerCaseInitialLetter():String
String::mkDir():Boolean
String::size():Integer
String::splitBy(chars:String,start:Integer,last:Integer):Seq(String)
String::subst(new:String,old:String,all:Boolean):String
String::toLower():String
String::toUpper():String
\end{lstlisting}

\section{OCL Syntax Classes\label{sec:OCL-Syntax-Classes}}

This section defines the classes that make up the Object Command Language
(the core part of XOCL). They are defined in the package OCL. The
syntax classes are grouped roughly in terms of their basic properties
and usage.

\begin{figure}
\includegraphics[scale=0.6]{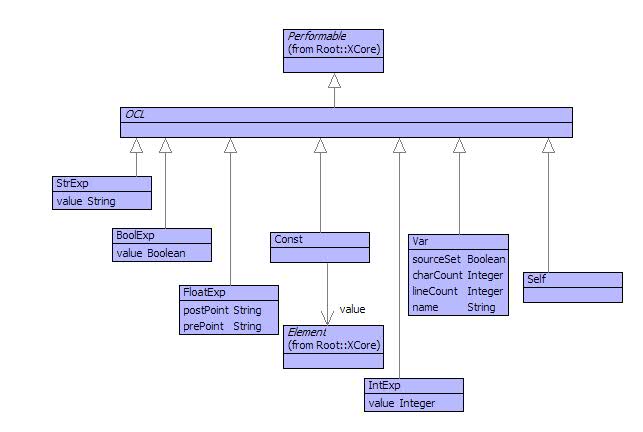}
\caption{Atomic Expressions\label{fig:Atomic-Expressions}}
\end{figure}

\begin{figure}
\includegraphics[scale=0.75,angle=90]{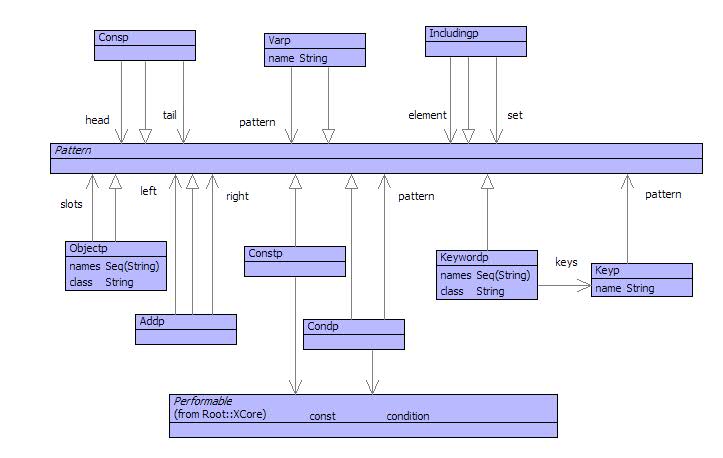}
\caption{Pattern Expressions\label{fig:Pattern-Expressions}}
\end{figure}

Figure \ref{fig:Atomic-Expressions} shows classes for atomic expressions.
Figure \ref{fig:Pattern-Expressions} shows the classes for patterns
that occur in argument positions of operation expressions (and in
@Case expressions). Briefly, Consp is a pair pattern, Varp is a simple
variable, Includingp is a set pattern, Objectp is a pattern corresponding
to a constructor, Addp is a sequence concatenation or set union pattern,
Constp is a constant (where the constant is the result of evaluating
an expression), Condp is a guard and Keywordp is a pattern matching
objects where the slots are given as keys.

\begin{figure}
\includegraphics[scale=0.75]{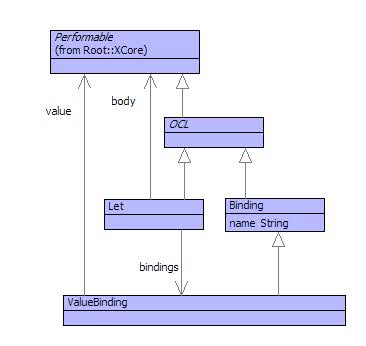}
\caption{Let Expressions\label{fig:Let-Expressions}}
\end{figure}

\begin{figure}
\includegraphics[scale=0.75,angle=90]{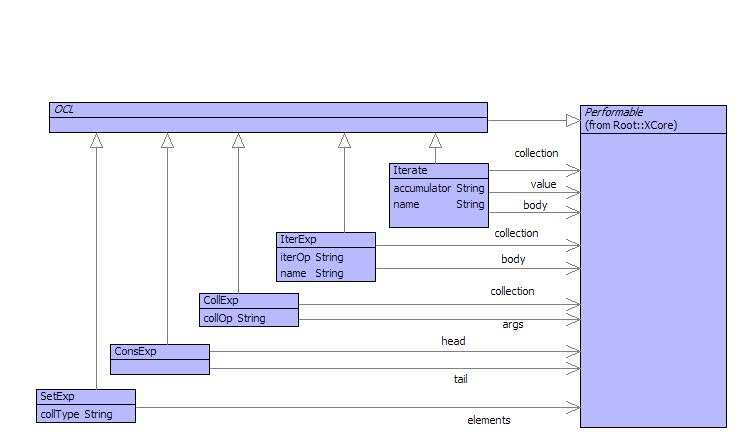}
\caption{Iteration Expressions\label{fig:Iteration-Expressions}}
\end{figure}

Figure \ref{fig:Let-Expressions} shows the syntax of let-expressions.
Figure \ref{fig:Iteration-Expressions} shows the classes for iteration
expressions. Both set and sequence expressions are represented as
instances of SetExp with the collType set to Set or Seq respectively.
A CollExp supports all concrete expressions of the form {\tt S->x(a,b,c)}
and {\tt S->x} where the arguments to the latter default to the empty sequence.
IterExp represents all concrete expressions of the form {\tt S->n(x | e)}
where n is exists, forAll, select, collect or reject.

\begin{figure}
\includegraphics[scale=0.75,angle=90]{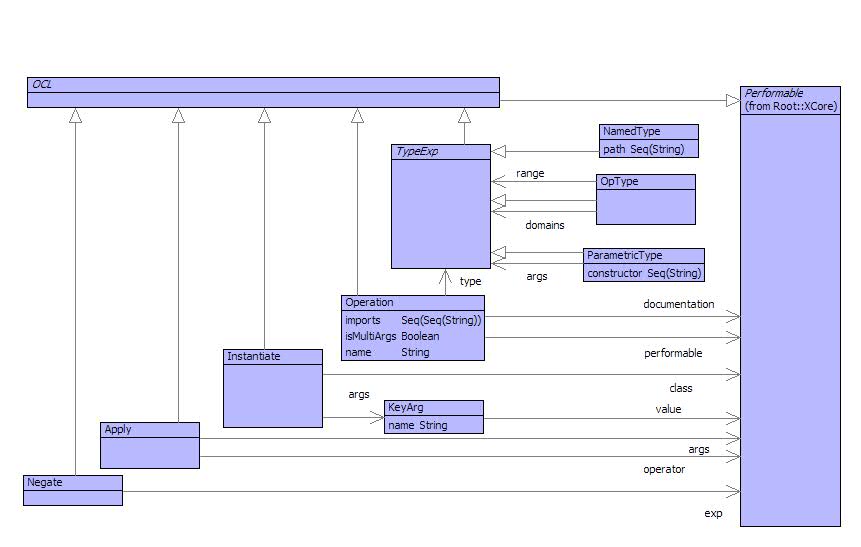}
\caption{Operation Expressions\label{fig:Operation-Expressions}}
\end{figure}

\begin{figure}
\includegraphics[scale=0.75,angle=90]{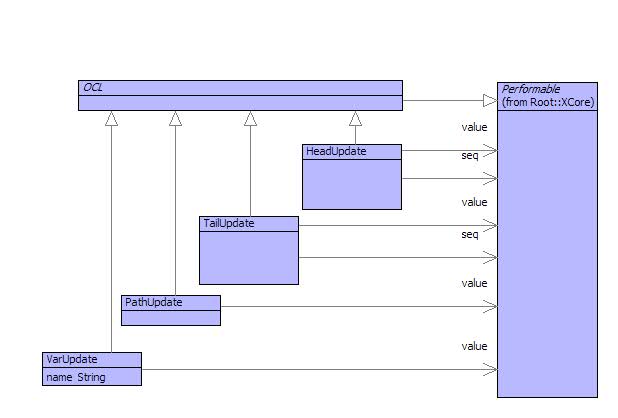}
\caption{Update Expressions\label{fig:Update-Expressions}}
\end{figure}

Figure \ref{fig:Operation-Expressions} shows the classes for operations.
Note that operation arguments are not shown but are an attribute of
type Seq(Pattern). Figure\ref{fig:Update-Expressions} shows the classes
for update expressions. Updates can be variable, x := e, path P::Q::x
:= e, or head and tail: {\tt s->head := e} and {\tt s->tail := e}. 

\begin{figure}
\includegraphics[scale=0.75,angle=90]{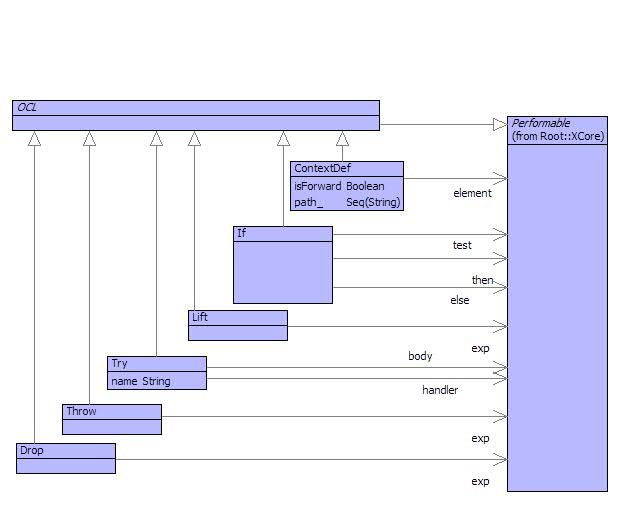}
\caption{Misc Expressions\label{fig:Misc-Expressions}}
\end{figure}

Figure \ref{fig:Misc-Expressions} shows the rest of the expression
types.

\backmatter
\end{document}